%% file: lawther_swift_z2_quasars.tex
\documentclass[a4paper,fleqn,usenatbib,useAMS]{mnras}

\usepackage{graphicx}
\usepackage{color}
\usepackage{epsfig}
\usepackage{wasysym}
\usepackage[english]{babel}
\usepackage[latin1,applemac]{inputenc}
\usepackage[T1]{fontenc}
\usepackage{aecompl}
\usepackage{deluxetable}
\usepackage{multirow}
\usepackage{amsmath,amssymb}
\usepackage{hyperref}

\usepackage{xspace}

\newcommand{\aox}{$\alpha_{\mathrm{ox}}$\xspace}

\newcommand{\Fuv}{$F_\nu(2500\text{\,\AA})$\xspace}
\newcommand{\Fx}{$F_\nu(2\text{\,keV})$\xspace}
\newcommand{\Luv}{$L_\nu(2500\text{\,\AA})$\xspace}
\newcommand{\Lx}{$L_\nu(2\text{\,keV})$\xspace}

\title[\emph{Swift} SEDs of $z\approx2$ quasars.]{A catalog of optical to X-ray spectral energy distributions of $z\approx2$ quasars observed with \emph{Swift}. \textrm{I}: First results}

\author[Lawther, D., Vestergaard, M., Raimundo, S., Grupe, D.]{Lawther, D.,$^1$ Vestergaard, M.,$^1$ $^2$ Raimundo, S.,$^1$  Grupe, D.,$^3$\\
	$^1$ Dark Cosmology Centre, Niels Bohr Institute, University of Copenhagen.\\
	$^2$ Steward Observatory, University of Arizona, 933 N. Cherry Avenue, 85721 Tucson, AZ, USA.\\
	$^3$ Department of Earth and Space Sciences, Morehead State University, 235 Martindale Dr., Morehead, KY 40351, USA}

\begin{document}

\maketitle

\begin{abstract}
We present the \emph{Swift} optical to X-ray spectral energy distributions (SEDs) of 44 quasars at redshifts $z\approx2$ observed by \emph{Swift}, part of a larger program to establish and characterize the optical through X-ray SEDs of moderate-redshift quasars. Here we outline our analysis approach and present preliminary analysis and results for the first third of the full quasar sample. Not all quasars in the sample are detected in X-rays; all of the X-ray detected objects so far are radio loud. As expected for radio loud objects, they are X-ray bright relative to radio-quiet quasars of comparable optical luminosities, with an average \aox$=1.39\pm0.03$ (where \aox\,is the power-law slope connecting the monochromatic flux at 2500 \AA\,and at 2 keV), and display hard X-ray spectra. We find integrated 3000 \AA\,- 25 keV accretion luminosities of between $0.7\times10^{46}$ erg s$^{-1}$ and $5.2\times10^{47}$  erg s$^{-1}$. Based on single-epoch spectroscopic virial black hole mass estimates, we find that these quasars are accreting at substantial Eddington fractions, $0.1\apprle L/L_{\mathrm{Edd}}\apprle1$.
\end{abstract}

\begin{keywords}
	quasars: general -- astronomical databases: catalogs
\end{keywords}

\input{sec1introduction}

\input{sec2sample}

\input{sec3swiftdata}

\input{sec4sdss}

\input{sec5sed}

\input{sec6analysis}

\input{sec7discussion}

\paragraph*{Acknowledgements: } 

We thank the anonymous referee for helpful feedback with regards to the X-ray modeling, and for a thorough report that improved the clarity of the manuscript as a whole.

This work took advantage of the Danish \emph{Swift} guaranteed time program. We thank local \emph{Swift} affiliate Daniele Malesani, and the \emph{Swift} helpdesk at the University of Leicester, for their assistance in obtaining and processing the data. DL, MV and SR gratefully acknowledge support from the Danish Council for Independent Research via grant no. DFF 4002-00275. 

Funding for SDSS-III has been provided by the Alfred P. Sloan Foundation, the Participating Institutions, the National Science Foundation, and the U.S. Department of Energy Office of Science. The SDSS-III web site is http://www.sdss3.org/.SDSS-III is managed by the Astrophysical Research Consortium for the Participating Institutions of the SDSS-III Collaboration including the University of Arizona, the Brazilian Participation Group, Brookhaven National Laboratory, Carnegie Mellon University, University of Florida, the French Participation Group, the German Participation Group, Harvard University, the Instituto de Astrofisica de Canarias, the Michigan State/Notre Dame/JINA Participation Group, Johns Hopkins University, Lawrence Berkeley National Laboratory, Max Planck Institute for Astrophysics, Max Planck Institute for Extraterrestrial Physics, New Mexico State University, New York University, Ohio State University, Pennsylvania State University, University of Portsmouth, Princeton University, the Spanish Participation Group, University of Tokyo, University of Utah, Vanderbilt University, University of Virginia, University of Washington, and Yale University.

\bibliographystyle{mnras}
\bibliography{swift_paper}

\clearpage

\input{tables/sample_table}
\clearpage
\input{tables/xrt_table}
\clearpage
\input{tables/uvot_table}
\clearpage
\input{tables/uvot_table_mags}
\clearpage
\input{tables/detection_table}
\clearpage
\input{tables/analysis_table}
\clearpage
\input{tables/gamma_table}
\clearpage
\input{tables/uvcorr_analysis_table}

\clearpage
\appendix

\input{appendixA_sed}

\end{document}

%% file: sec1introduction.tex
\section{Introduction}\label{sec:introduction}

The extreme energy output of quasars is almost certainly due to accretion of gas onto a central black hole \citep[e.g.,][]{LyndenBell1971,Lynden-Bell1978}. According to the unification paradigm \citep[e.g.,][]{Urry1995}, unobscured quasars are an intrinsically luminous subset of Active Galactic Nuclei (AGN) for which we have a direct view of the continuum source and of the broad emission-line region. However, we still lack a detailed explanation of how the accretion physics of the central engine, together with reprocessing of the continuum emission on small scales and in the host galaxy, produces the observed spectral energy distributions (SEDs) of quasars.

Standard models for the UV-optical continuum source posit a geometrically thin, optically thick accretion disk around a black hole \citep[so-called $\alpha$-disk models,][]{Shakura1973}. The `big blue bump' feature observed in the UV-optical energy range is consistent with partially reprocessed thermal emission from such a disk \citep[e.g.,][]{Siemiginowska1995,Kishimoto2008}. For $\alpha$-disk models, the main intrinsic parameters that determine the disk emission are the mass ($M_\mathrm{BH}$), spin, and mass accretion rate ($\dot{M}_\mathrm{BH}$) of the black hole. For $M_\mathrm{BH}\approx10^9M_{\astrosun}$ quasars, the peak emission temperature for an $\alpha$-disk model corresponds to UV or extreme-UV (hereafter, EUV) energies \citep[e.g.,][]{Krolik1988,Davis2011}. This peak is difficult to determine observationally for individual quasars, due to strong EUV absorption by Galactic gas. Studies of composite spectra find evidence of a spectral turnover at roughly 1000 \AA\,in the rest frame \citep[e.g.,][]{Shang2005,Barger2010,Shull2012} (but see also \citet{Capellupo2016} for a recent study identifying some quasars with a lower-energy turnover).

The $\alpha$-disk model predicts very little hard X-ray emission for accretion onto supermassive black holes. Additional components are therefore necessary to explain the observed X-ray spectra, as follows. To first order, the hard X-ray SEDs follow power-law functions \citep[e.g.,][]{Nandra1994}. Recent studies using the \emph{NuSTAR} X-ray telescope have greatly improved our knowledge of the high-energy cutoff of this continuum emission, measuring cutoff energies of $\sim$ 50--250 keV \citep[e.g.,][]{Brenneman2014,Fabian2015}\footnote{Interestingly, an unusually high cutoff energy $\sim$720 keV is found for the X-ray continuum of NGC 5506 \citep{Matt2015}.. This X-ray continuum may be due to inverse-Compton upscattering of accretion disk photons by hot ($kT_e\sim100$ keV) electrons in an optically thin corona or inner disk region \citep{Zdiarski1996,Ghosh2016}. Analyses of X-ray variation timescales, and of X-ray reverberation lags due to reflection in the accretion flow, suggest that the coronas have sizes of order 3--10 times the gravitational radius of the black hole \citep[e.g.,][]{Emmanoulopoulos2014,Uttley2014,Fabian2015}}. Radio-loud quasars (hereafter, RLQs) differ from radio-quiet quasars (RQQs) in that they are more X-ray luminous \citep{Zamorani1981}, have harder X-ray spectra \citep[e.g.,][]{Elvis1994}, and display a correlation between X-ray and radio-core flux \citep{Tananbaum1983}. These results suggest the presence of an additional X-ray emitting component for RLQs. Indeed, extended X-ray emission from the radio jet is observed for some RLQs \citep[e.g.,][]{Worrall2009}. In addition to the hard X-ray continuum, many AGN display a `soft excess' component\footnote{The soft excess component can only be unambiguously identified at $z\apprle0.4$ using current X-ray telescopes.} below rest-frame 1 keV \citep{Porquet2004,Piconcelli2005}, possibly due to upscattering of UV photons in the warm ($kT_e\approx1$ keV) atmosphere of the accretion disk \citep[e.g.,][]{Czerny1987,Haardt1991}.

In summary, UV-optical and X-ray observations support a scenario where a fraction of the available accretion energy is reprocessed by Compton upscattering in one or more regions of hot, diffuse electron gas. The remainder of the accretion energy not consumed by the black hole is either emitted as thermal radiation, or (for RLQs) channeled into the radio jet. The relative emission strengths of these components can be estimated using observations of optical to X-ray SED. For 47 quasars spanning $0.01<z<3.3$, the pioneering work of \citet{Elvis1994} reveals flux deviations of up to 1 dex from the mean SED for individual objects, when normalized in the near-infrared. \citet{Richards2006} compile a larger SED catalog while expanding the spectral coverage to the mid-IR, while \citet{Elvis2012} present an X-ray selected SED catalog (in contrast to \citet{Elvis1994}, who select quasars primarily based on UV excess). While these authors attribute some of the dispersion in quasar SED properties to reddening and host galaxy contamination, other shape variations may be due to changes in the accretion physics of the central engine. In particular, variations in the ratio of UV to X-ray luminosity \citep[as parameterized by \aox,][]{Tananbaum1979} may be due to the physical relationship between the accretion disk and the X-ray corona \citep{Lusso2016}. At $z<0.4$, SED studies that include UV-optical spectroscopy find several correlations between broad emission line widths and spectral shape parameters \citep{Jin2012}. These correlations suggest that the black hole mass, accretion rate, and Eddington ratio largely determine the SED shape, with the X-ray spectrum steepening and the SED becoming more disk-dominated at higher Eddington ratios; the latter is also found by \citep{Grupe2010}.

Quasars are variable in luminosity over rest-frame timescales of days \citep[in the X-ray, e.g.,][]{Gibson2012} to years \citep[e.g.,][]{VandenBerk2004,Kaspi1996,Kaspi2007}. This can introduce uncertainty to SED measurements if there are time delays between the observation of the different spectral regions. In fact, \citet{Jin2012} find that the bolometric correction for a given AGN correlates with the Eddington ratio of that object, while Kilerci Eser and Vestergaard (2017, in prep.) find that the bolometric corrections display temporal variation of order 10\% (up to 100\% in extreme cases) for individual AGN. These findings emphasize the importance of simultaneous multi-band observations for accurate measurement of the bolometric luminosity and study of the Eddington ratio. X-ray observatories such as \emph{XMM-Newton} \citep{Jansen2001} and \emph{Swift} \citep{Roming2005} include UV-optical detectors, allowing contemporaneous observation of the optical to X-ray SED. The \emph{Swift} Gamma-ray Burst Explorer telescope has been used to study low-redshift AGN \citep{Grupe2010}, and to study a large, serendipitously observed sample of AGN at $z<5.5$ \citep{Wu2012}. We are currently observing a sample of 133 quasars using \emph{Swift}, carefully selected to probe a diverse set of quasar properties at $z\approx2$ (\S \ref{sec:sample}). Here we present and characterize the SEDs of 44 quasars with Swift detections or bona-fide upper limits. We describe the data processing procedure for our \emph{Swift} XRT and UVOT data in \S \ref{sec:swiftdata}. We include data from the \emph{Sloan} Digital Sky Survey to support our analysis (\S \ref{sec:sdss}). We present the SEDs (\S \ref{sec:sdss}) and provide a preliminary statistical analysis of this sample subset (\S \ref{sec:analysis}), including a characterization of the UV to X-ray SED shapes of these quasars as parameterized by the UV-optical to X-ray spectral index \aox \citep{Tananbaum1979}. In the same section we present a preliminary study of their mass accretion rates and Eddington ratios. We use a cosmology with $H_0=67.48\pm0.98$ and $\Omega_{\mathrm{m}}=0.313\pm0.013$ throughout \citep{Planck2016}.

%% file: sec2sample.tex
\section{Sample Selection and Observations}\label{sec:sample}

\subsection{Sample Selection}

\begin{figure}
	\centering
	\includegraphics[width=\linewidth]{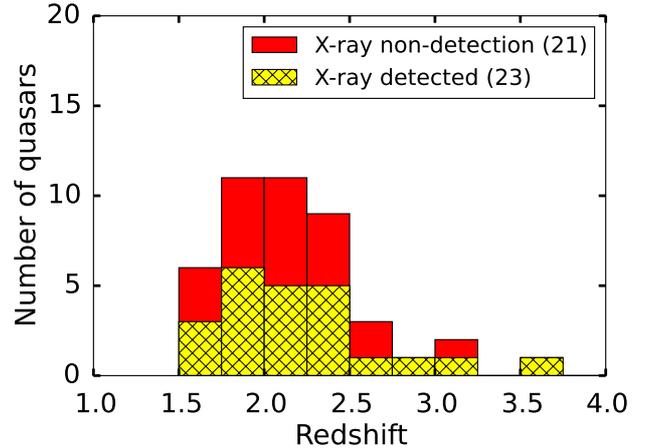}
	\caption{The redshift distribution of quasars in the current sample. Stacked histogram: the bar height shows the total number of quasars at a given redshift, while the areas of the hash-marked yellow and plain red regions show the distribution of X-ray detections and non-detections, respectively.}
	\label{fig:z_histogram}
\end{figure}

Our $z\approx2$ quasar sample is selected to span a wide range in luminosity, radio strength, and (when possible) radio spectral index, thereby allowing us to test for dependencies of the SED shape on these properties. We use the quasar sample presented by \citet{Vestergaard2000}, containing roughly equal numbers of RLQs and RLQs, the majority of which are pair-matched in redshift and in absolute \emph{V}-band magnitude. The sample quasars fulfill the following criteria:

\emph{a)} Redshift $z\ge1.5$, to ensure that the Lyman-$\alpha$ emission line is observable in ground-based spectroscopy.

\emph{b)} V-band absolute magnitudes that satisfy $-29.5\,$mag$ < M_V < -26.5$ mag. 

\emph{c)} V-band apparent magnitudes $m_V<20$, so as to ensure that high signal-to-noise spectroscopy can be obtained within reasonable exposure times.

\emph{d)} Strongly variable objects (e.g., blazars) are excluded.

\emph{e)} Objects identified as broad absorption line (BAL) quasars as per 1998 are also excluded by \citet{Vestergaard2000}. Note that two sample objects (Q1227+120 and Q2350-007) were subsequently found to be BAL or mini-BAL quasars.

The \citet{Vestergaard2000} sample does not include the most luminous quasars at a given redshift, as the original intention was to study objects with a range of `population-typical' luminosities. For the RLQs, measurements of the radio fluxes, the radio spectral index, and the radio core fraction at 5 GHz are compiled by \citet{Vestergaard2000}. The sample spans the full range of radio spectral indices, $\alpha_{5000}^{1400}$, typically observed for bright quasars. Assuming that $\alpha_{5000}^{1400}$ can be interpreted as a measure of our viewing angle relative to the radio jet \citep[e.g.,][]{Jarvis2006}, the RLQ sample thus spans the full range of viewing angles, $i$, for unobscured (`type 1') quasars ($5^\circ\apprle i\apprle45^{\circ}$,  \citet{Barthel1989}). To increase the statistical power of our study, we also include 16 quasars from the Sloan Digital Sky Survey Data Release 7 (SDSS DR7) that fulfill $z>1.6$ and $m_V<18$ mag. These 16 quasars have been observed in X-rays prior to our observing campaign, and are selected to be sufficiently X-ray bright as to be detectable with \emph{Swift} for modest exposure times. We refer to objects from the \citet{Vestergaard2000} sample using the name convention Q$xxxx\pm xxx$, while J$xxxx\pm xxx$ denote the additional SDSS DR7 quasars.

At present, our full sample comprises 133 quasars in the redshift interval $1.5<z<3.6$, with a roughly equal number of RLQs and RQQs. This paper presents an analysis of 44 quasars from our sample, for which the initially requested \emph{Swift} observations are complete as per 2015 September: we either have a secure X-ray detection, or find that the X-ray count rate is so low that additional X-ray observations would not yield a detection (as defined in \S \ref{sec:method_xrt_detection}). Basic information for the 44 quasars in the current subsample is listed in Table \ref{tab:sample_selection}; their redshift distribution is shown in Figure \ref{fig:z_histogram}. An expanded analysis of the full sample will be presented in future work.

\subsection{\emph{Swift} Observations}

The \emph{Swift} Gamma-ray Burst Explorer satellite \citep{Gehrels2004} is equipped with two Narrow Field Instruments: the X-ray Telescope \citep[XRT,][]{Burrows2005}, and the UV/Optical Telescope \citep[UVOT,][]{Roming2005}. In this study we utilize simultaneously observed XRT (0.3-10 keV) and UVOT data. \emph{Swift} observed the current sample between 2010 June -- 2015 August; most of the observations were performed during 2013-2015.  As our targets have not been observed in the X-ray regime prior to this study, the necessary XRT exposure times are estimated based on the \emph{V}-band fluxes, assuming an optical to X-ray flux ratio \aox=1.35, typical for unobscured quasars \citep[e.g.,][]{Laor1997,Jin2012}. For objects where these initial observations did not yield a secure X-ray detection, we are currently obtaining additional observations, unless the object appeared to be strongly absorbed (i.e., have \aox$\apprge2$), in which case additional observations would not provide a secure detection. The total UVOT exposure time per object is approximately equal to the total XRT exposure time. However, as we require multi-band photometry in order to determine the UV/optical spectral energy distribution, the total UVOT exposure time is distributed across several UVOT imaging filters. All XRT observations were performed in photon-counting (PC) mode \citep{Hill2004}.

% As per standard procedure for \emph{Swift} fill-in targets, the quasars were observed for at least 80\% of the requested exposure time.

The quasars were observed by \emph{Swift} as `fill-in' observations (i.e., observed while the telescope was not responding to a Gamma-ray Burst trigger or performing other time-critical observations) as part of the Danish \emph{Swift} program. Most of the quasars were observed multiple times so as to fulfill the requested exposure time. For bright quasars, the typical intrinsic source variability is expected to be of order 10\% and to take place on timescales of months to years in the rest-frame \citep[e.g.,][]{Kaspi2007}. The observations for individual objects were performed over rest-frame time intervals not exceeding two months. As a result, we do not expect strong source variability to occur between individual \emph{Swift} observations. Indeed, we see no significant variability in the data (\S \ref{sec:method_uvot}). 

%% file: sec3swiftdata.tex
\section{Processing of \emph{Swift} XRT and UVOT data}\label{sec:swiftdata}

\subsection{XRT Data}\label{sec:method_xrt}

The basic processing steps are as follows. We first select X-ray events from the XRT event files in a source and a background region, respectively (\S \ref{sec:method_xrt_eventselection}). We then establish whether an X-ray source was detected at the $3\sigma$ level (\S \ref{sec:method_xrt_detection}), and finally fit a model to the X-ray data, and measure X-ray fluxes and photon indices of the best-fitting model (\S \ref{sec:analysis_xrayfitting}). For non-detections, in the last step,  we only measure an upper-limit flux (\S \ref{sec:method_xrt_upperlims}) and a hardness ratio (\S \ref{sec:analysis_HR}). These basic measurements are presented in Table \ref{tab:xrtdata}.

\subsubsection{Event Selection}\label{sec:method_xrt_eventselection}

We use the \emph{HEASARC} tool \emph{`xrtpipeline'} to perform basic pipeline processing on the raw XRT data for each individual observation. We select events using the default grading threshold (0-12) for the XRT PC mode. For each observation, we generate separate event files for the source and background regions using the \emph{`xselect'} tool. We define a default source region of radius 20 pixels centered on the quasar coordinates, and an annulus of width 60 pixels for background extraction. We resize these default regions (or change the shape of the background region) if visual inspection of the XRT image reveals a contaminating source. We discard events with energies outside the 0.3 keV to 10.0 keV window within which the XRT is sensitive, along with events that took place outside the `good time intervals' for each observation, e.g., events that took place while the spacecraft was slewing. As none of our objects have an X-ray flux larger than 0.1 counts s$^{-1}$, photon pile-up is not an issue\footnote{For the XRT detector in photon-counting mode, a correction for pile-up is advised at count rates of 0.5 counts s$^{-1}$ or higher (\url{http://www.swift.ac.uk/analysis/xrt/pileup.php}). The three highest-flux objects in our sample, with count rates of between 0.02 and 0.07 counts s$^{-1}$, show no evidence of pile-up.}.

We use the task `\emph{xrtmkarf}' to generate Auxiliary Response Files (ARFs) for each observation. The information in these files is used by XSPEC to account for the effective area, quantum efficiency, and PSF shape of the instrument. For the extraction of X-ray spectra we use the Redistribution Matrix File, located in the HEASARC calibration database\footnote{As all observations took place later than 2007, we use the latest XRT RMF file, \emph{(swxpc0to12s6\_20070901v011.rmf)}}, to account for the finite energy resolution of the detector. As a final preparation for the X-ray analysis, we use the task \emph{`grppha'} to group the events in bins with at least one event per bin; this is necessary for model fitting using the Cash statistic \citep{Cash1979} to be robust (\S \ref{sec:analysis_xrayfitting}).

\subsubsection{Establishing Source Detections}\label{sec:method_xrt_detection}

To establish whether a source is detected in our \emph{XRT} observations, we generate 99.73\% ($3\sigma$) confidence intervals for the source count rate given the observed number of counts using the Bayesian formalism of \citet{Kraft1991}, which assumes that the observed source count-rate is Poisson-distributed. This method produces confidence intervals that are valid for photon-counting experiments even if the background-subtracted source counts are negative; this can happen if the background signal in the source aperture fluctuates downwards (relative to the background extraction region). We use a flat non-negative prior probability on the mean count rate. Following \citet{Kraft1991}, we define a source as non-detected if the $3\sigma$ lower bound on the source count rate (as given by the lower confidence interval) equals zero. For non-detections, we propagate the upper $3\sigma$ confidence interval on the source count rate to an upper limit on the source flux, as described in \S \ref{sec:method_xrt_upperlims}. We list the number of background-subtracted XRT counts $N_{\mathrm{sub}}$ and the XRT detection status for each quasar in Table \ref{tab:xrtdata}. We also provide the detection significance, $\sigma$, calculated based on the total number of counts detected in the background and source apertures.

\subsubsection{Spectral Modeling for XRT Detected Quasars}\label{sec:analysis_xrayfitting}

We use the XSPEC software \citep{Arnaud1996} to model the X-ray spectra of each quasar. Although long-exposure observations of AGN often reveal multiple emission components, to first order the X-ray spectra of quasars resembles a power-law \citep[e.g.,][]{Nandra1994}. The observations in this study have too few background-subtracted XRT counts for us to perform a detailed study of the X-ray spectral shape. We therefore model the X-ray emission of each quasar as an absorbed power-law function\footnote{This model is expressed in \emph{XSPEC} modeling syntax as wabs(\emph{nH})$\,\cdot\,$zpowerlw(\emph{$z,K,\Gamma$}).}, 

\begin{equation}\label{eq:powerlaw_photonindex}
A(E)=K\left[E(1+z)\right]^{-\Gamma}e^{-N_H\sigma(E)}
\end{equation}

where $A(E)$ is the count-rate in units of photons keV$^{-1}$ cm$^{-2}$ s$^{-1}$ at energy $E$, and $\Gamma$ is the photon index.\footnote{In the X-ray literature, the spectral index $\alpha$ is occasionally used, where $f_\nu\propto(\nu/\nu_0)^{-\alpha}$. This implies that $\Gamma=\alpha+1$.} The exponential term represents Galactic absorption. We adopt the Galactic column density toward the quasar from the work of \citet{Kalberla2005}, as retrieved from the HEASOFT $N_H$ task, and use the photoelectric absorption cross-sections $\sigma(E)$ calculated by \citet{Morrison1983}. Our observations contain too few detected X-ray photons for Gaussian statistics to be applicable. Instead, we find the best-fitting model by minimizing the Cash statistic \citep{Cash1979}, which is the appropriate maximum-likelihood statistic for Poisson-distributed data. The Cash statistic performs optimally when the data are binned such that each bin has at least one count - we ensured this using the task \emph{`grppha'} (\S \ref{sec:method_xrt_eventselection}).

\begin{figure*}
	\centering
	\includegraphics[width=0.35\linewidth,angle=270]{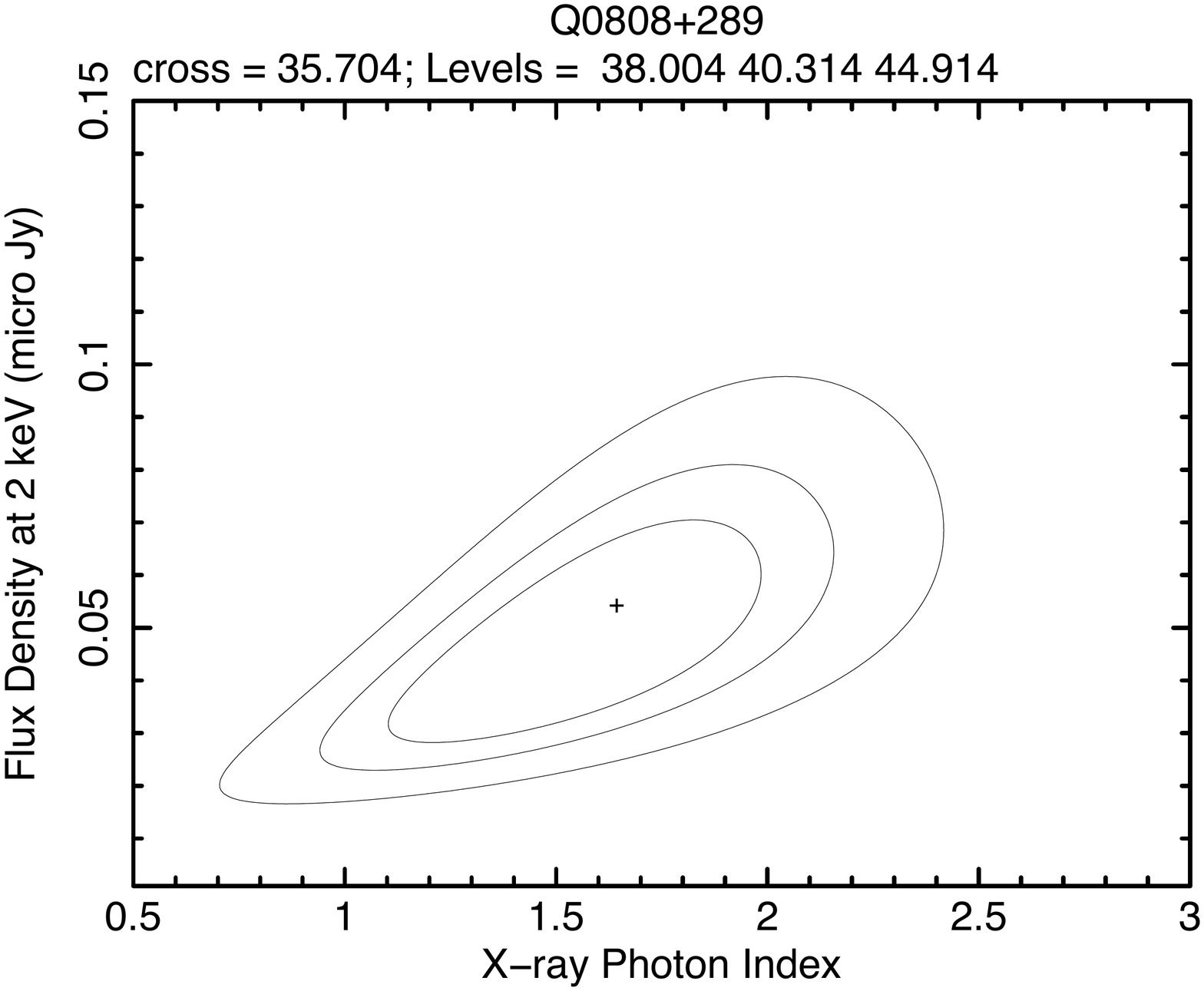}
	\includegraphics[width=0.35\linewidth,angle=270]{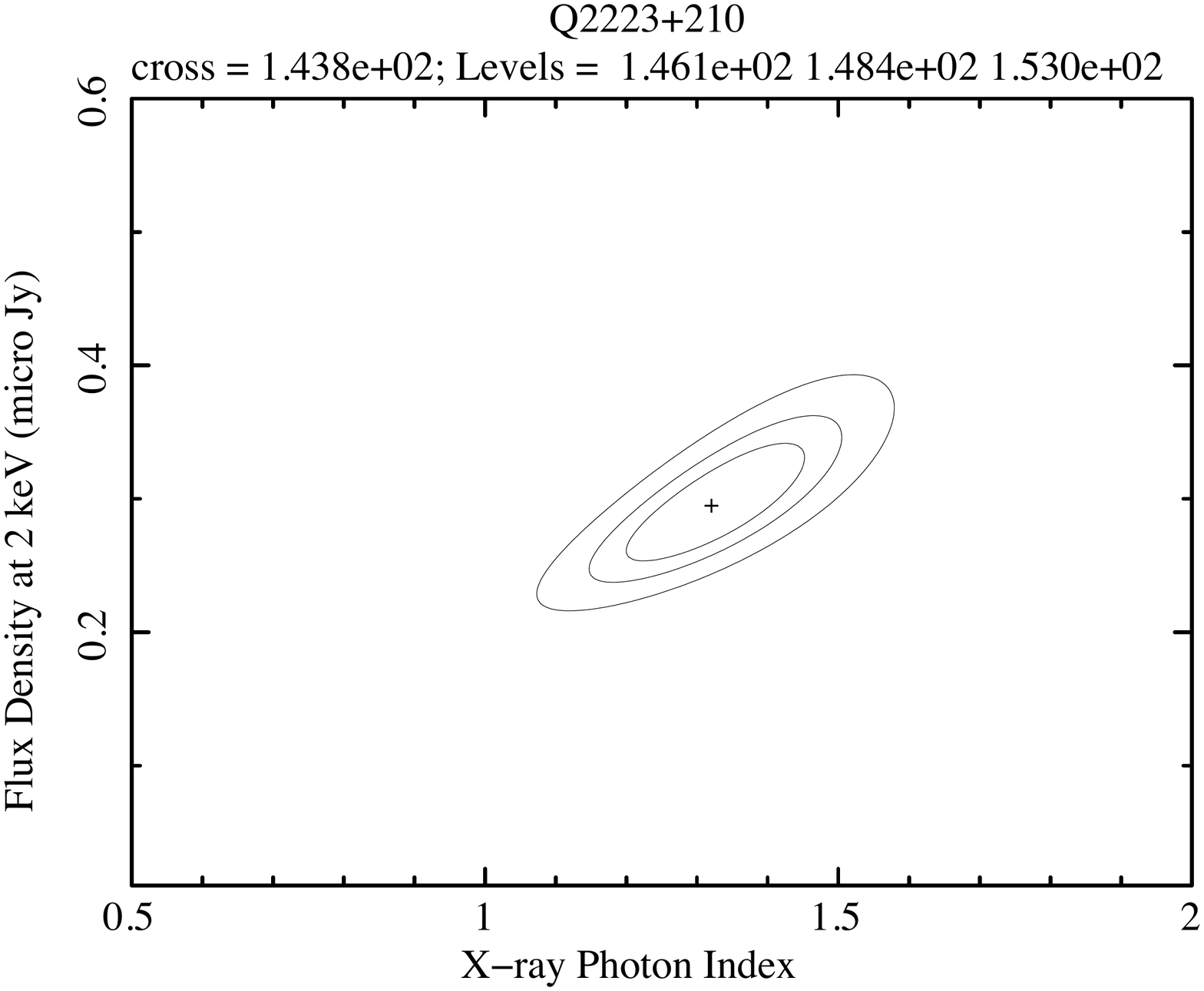}
	\includegraphics[width=0.35\linewidth,angle=270]{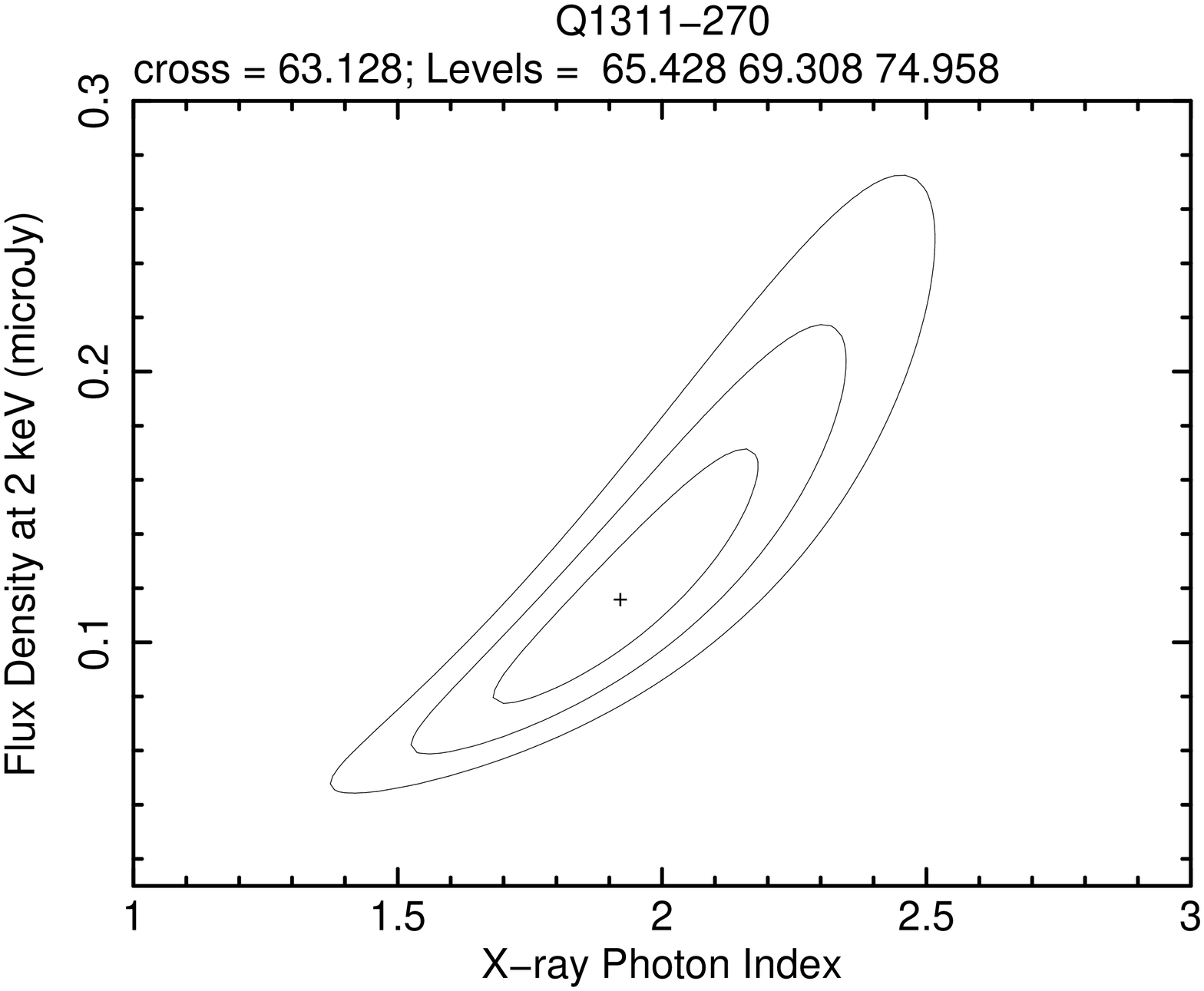}
	\includegraphics[width=0.35\linewidth,angle=270]{figures/Q2251+244_contour}
	\caption{Confidence region contours for the photon index and normalization, for XRT modeling of four quasars. The best-fit solution for $\Gamma$ and $F_\nu$(2 keV) is marked with a cross symbol, while the curves show confidence regions corresponding to the $1\sigma$, $2\sigma$ and $3\sigma$ levels. 'Cross' and 'Levels' denote the Cash statistic for the best-fit solution and for the three contour levels. Q0808+289 has the smallest number of background-subtracted counts for all X-ray detected quasars in the present sample, $N_\mathrm{sub}\approx27$, while Q2223+210 has $N_\mathrm{sub}\approx283$. Q1311-270 and Q2251+244 have the largest and smallest best-fit values of $\Gamma$ (softest and hardest spectra), respectively.}
	\label{fig:Gamma_contours}
\end{figure*}

Our modeling yields the best-fit value of the normalization $K$ and of the photon index $\Gamma$. We use the XSPEC \emph{`error'} task to estimate the $1\sigma$ confidence intervals on these parameters. We use the best-fitting model directly to calculate the integrated X-ray flux between 0.3 keV and 10 keV in the observed frame, $F$(0.3-10 keV), and the flux density at 2 keV in the quasar rest-frame, $F_\nu$(2 keV). These flux measurements are presented in Table \ref{tab:xrtdata}. All listed fluxes are unabsorbed values (i.e., we fit the model including the Galactic absorption component, but calculate the fluxes we would receive for the unabsorbed power law component alone). To investigate the parameter space, we use the XSPEC task \emph{`steppar'} to generate two-dimensional confidence regions ($\Gamma$ versus $F_\nu$(2 keV)) for each X-ray detected quasar. In general, the spectral index and normalization are not fully independent: a smaller value of $\Gamma$ (i.e., a harder spectrum) allows a higher $F_\nu$ (Figure \ref{fig:Gamma_contours}). We therefore advise caution in using our measurements to investigate any putative relationship between $\Gamma$ and the X-ray luminosity. % However, as the X-ray bands observed by \emph{Swift} only account for a small fraction of the bolometric luminosity (Section \ref{sec:analysis_integrated_luminosity_Lbol}), an investigation of the relationship between the Eddington ratio and $\Gamma$ \citep[e.g.,][]{Brightman2013} should not be strongly effected.}

We are in the process of obtaining follow-up \emph{Swift} observations for objects in the full quasar sample which are marginally detected (with source counts $N_\mathrm{sub}\approx15$) during the initial set of observations (\S \ref{sec:sample}). The quasars scheduled for follow-up observations are not included in the current analysis. We do, however, include very faint objects for which a follow-up observation of reasonable duration would not yield a secure XRT detection (\S \ref{sec:sample}). This results in a bimodal distribution of $N_\mathrm{sub}$: all detected objects have $N_\mathrm{sub}>25$, while all non-detections have $N_\mathrm{sub}<6$.

\subsubsection{Limiting Fluxes for Non-Detected Quasars}\label{sec:method_xrt_upperlims}

For non-detections, as defined in \S \ref{sec:method_xrt_detection}, we follow the approach of \citet{Wu2012} to establish upper limits on $F$(0.3-10 keV) and $F_\nu$(2 keV). Again, we use XSPEC to model the X-ray spectrum as an absorbed power-law (Equation \ref{eq:powerlaw_photonindex}). This allows us to include the per-pixel exposure map and the calibration information, as contained in the ARF file generated for the observation. For non-detections we hold the value of the photon index constant at $\Gamma=1.91$, a typical value for AGN X-ray spectra (e.g., \citet{Young2009}, \citet{Jin2012}). This yields values of $F$(0.3-10 keV) and $F_{\nu}$(2 keV) corresponding to the background-subtracted count-rate, which in these cases are below our detection limit. To obtain $3\sigma$ upper limits on $F_{\nu}$(2 keV), we scale the best-fitting Galactic absorption-corrected \Fx value as:

\begin{equation}\label{eq:xrt_limiting_flux}
F_{\mathrm{lim}}=F_\lambda(2\text{keV})\frac{N_\mathrm{sub}}{N_{\mathrm{lim}}}
\end{equation}
\vspace{0.1cm}

where $N_{\mathrm{sub}}$ is the observed number of background-subtracted source counts, and $N_{\mathrm{lim}}$ is the number of counts corresponding to the $3\sigma$ upper limit on the mean source count-rate (as determined in \S \ref{sec:method_xrt_detection}).

For objects with very small (or negative) background-subtracted total counts, \emph{XSPEC} is unable to model the data. For these objects we calculate the ratio of photon flux to detected XRT counts, and use this to correct the upper-limit source counts, so as to account for the same calibration information used in our XSPEC processing. Following \citet{Wu2012}, in practice we obtain this correction by processing the event file with the same tool (\emph{xrtmkarf}) that we use to generate ARF files for the XSPEC analysis, as it produces the flux-to-counts ratio as part of its output. We convert the corrected limiting count-rate to a limiting \Fx using the HEASOFT task PIMMS\footnote{\url{http://heasarc.gsfc.nasa.gov/docs/software/tools/pimms.html}}, assuming a power-law spectrum with $\Gamma=1.91$.

\subsubsection{Sensitivity to Intrinsic Absorption}

Our observations are not deep enough, and do not cover enough of the soft X-ray regime for $z\sim2$, to accurately measure the intrinsic absorption column density; we establish this using the following test. For observations with $N_\mathrm{sub}>50$, we fit an alternative model\footnote{This model is expressed in \emph{XSPEC} modeling syntax as wabs(\emph{nH})$\,\cdot\,$zwabs(\emph{nH$_{\mathrm{int}}$})$\,\cdot\,$zpowerlw(\emph{$z,K,\Gamma$}).}: 

\begin{equation}\label{eq:powerlaw_photonindex_intabs}
	A(E)=K\left[E(1+z)\right]^{-\Gamma_\mathrm{abs}}e^{-N_H\sigma(E)}e^{-N_{H,\mathrm{int}}\sigma(E(1+z))}
\end{equation}
\vspace{0.1cm}

\begin{figure}
	\centering
	\includegraphics[width=\linewidth]{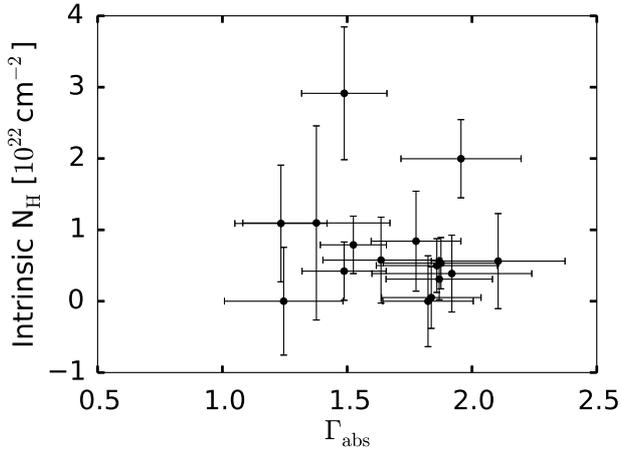}
	\caption{The best-fit value of the H\textsc{I} column density $N_{H\mathrm{,int}}$ is shown as a function of the X-ray photon index, $\Gamma_\mathrm{abs}$, as measured for models including an intrinsic absorber component, for quasars with $N_{\mathrm{sub}}>50$. Note the large uncertainties on $N_{H\mathrm{,int}}$: for many objects, the intrinsic column density is consistent with zero in our modeling.}
	\label{fig:gamma_vs_gamma}
\end{figure}

The intrinsic column density $N_{H,\mathrm{int}}$ is an additional free parameter in this model compared to Equation \ref{eq:powerlaw_photonindex}. The best-fit values of $N_{H\mathrm{,int}}$ have large uncertainties and are consistent with zero for most objects, i.e., the intrinsic column density is not well-constrained by our data (Figure \ref{fig:gamma_vs_gamma}). For this reason we adopt the simpler model of Equation \ref{eq:powerlaw_photonindex} (i.e., no intrinsic absorption) for all quasars in the remainder of this work.

\subsection{UVOT data}\label{sec:data_reduction_uvot}\label{sec:method_uvot}

\subsubsection{Filter Selection}

We are ultimately interested in estimating the integrated continuum energy output of the accretion disk, along with the flux density at rest-frame $2500$ \AA\,- the latter is needed for the calculation of \aox as defined in \citet{Tananbaum1979} (\S \ref{sec:analysis_alphaox}). Quasar spectra bluewards of rest-frame $1000$ \AA\, generally suffer Ly-$\alpha$ forest absorption, and therefore do not represent the intrinsic continuum. We therefore select all UVOT filters that cover rest-frame wavelengths redwards of Ly-$\alpha$, where we expect the emission to be dominated by the unabsorbed continuum. For the highest-redshift objects in our sample, only the UVOT \emph{V} filter samples rest-frame wavelengths longer than 1000 \AA. In these cases, we split the observing time between the \emph{U}, \emph{B} and \emph{V} filters, so as to at least partially constrain the UV spectral shape. Quasars selected from SDSS DR7 were originally observed as part of another Danish \emph{Swift} program, with different filter selection criteria. However, all these observations include the \emph{Swift} \emph{V}, \emph{B} and \emph{U} filters, which cover the desired spectral regions.

\subsubsection{Data Processing}

We first combine the data from multiple observations to obtain one UVOT image per bandpass filter for each quasar. We then perform aperture photometry to extract the source and background fluxes, respectively. Lastly, we correct the measured fluxes for Galactic reddening.

Each UVOT observation consists of multiple snapshot observations (detector readouts) with exposure times of a few seconds. The imaging data for each filter are delivered as multi-extension FITS files containing all snapshots for that filter. We use the task `\emph{uvotimsum}' once on each `Level 2' image to produce a combined image for that filter. We visually inspect this image, checking for pointing offsets and other issues (such as satellite trails) with the images. Most of the quasars in our sample were observed by \emph{Swift} multiple times. Aperture photometry was performed on each of the single-observation images in order to check for significant flux variations between observations. For quasars detected in the individual exposures, we do not see flux variations at a $3\sigma$ level; however, we are only sensitive to variations of around 0.2 mag or greater, and only for those quasars in our sample with the brightest apparent magnitudes ($m_V\approx17.5$). Finally, we use the task `\emph{uvotimsum}' to produce a stacked image of all observations of a given quasar in a given filter.

\subsubsection{Photometric Measurements}\label{sec:method_uvot_extraction}

We perform aperture photometry using the task `\emph{uvotsource}'. We use a circular source aperture of radius 3", which maximizes the signal-to-noise in the extraction region \citep{Poole2008,Breeveld2010}, and a circular background aperture with a radius of approximately 30". We adjust the background aperture size and positioning on a per-image basis so as to sample the sky background near the quasar's image coordinates while avoiding contaminating sources. Since the UVOT photometric system is calibrated using a 5" aperture, we apply an aperture correction, again using the `\emph{uvotsource}' task, in order to obtain fully calibrated source apparent magnitudes. The `\emph{uvotsource}' task also provides the background-subtracted flux density at the bandpass pivot wavelength, assuming a GRB-like power-law source spectrum \citep{Poole2008}, also suitable for quasars in the UV-optical. If the quasar is not detected at a significance of $5\sigma$ or greater in the combined image, we instead record the $5\sigma$ upper limit on the flux density. We correct the observed flux densities for Galactic dust extinction using values of $E(B-V)$ given by \citet{Schlafly2011}. We present the UVOT extinction-corrected flux densities for our quasar sample in Table \ref{tab:uvotfilters}, and the apparent magnitudes on the UVOT photometric system \citep[similar to the Johnson system;][]{Poole2008} in Table \ref{tab:uvotfilters_magnitudes}.

\subsection{Summary of \emph{Swift} Data}

We present the number of detections versus non-detections in our XRT and UVOT data in Table \ref{tab:detection_summary}. Not all objects have XRT detections (21 are undetected), or are detected in all UVOT filters (2 objects are undetected in all filters, while 7 objects are only detected in two filters). To better establish the shape of the UV continuum, we include additional photometric data in our UV analysis as available, as detailed in \S \ref{sec:sdss}.

%% file: sec4sdss.tex
\section{Supplementary SDSS UV-optical Data}\label{sec:sdss}

As the UVOT bandpasses do not cover rest-frame wavelengths longer than $\approx2000$ \AA\,at $z\approx2$, we include archival \emph{Sloan Digital Sky Survey} \citep[SDSS-III,][]{Eisenstein2011} photometry in our analysis so as to extend our spectral coverage. Of the present sample of 44 quasars, 35 objects have SDSS photometry; 23 objects also have SDSS spectroscopy. The SDSS data are processed using the latest SDSS photometric pipeline (Data Release 12, \citet{Alam2015}). 

The SDSS contains observations of the targets at time separations of months to years with respect to the \emph{Swift} observations. As we are interested in the instantaneous SEDs of individual objects, flux variability must be accounted for. We follow the approach of \citet{Wu2012} to create `pseudo-simultaneous' photometric data points by rescaling the flux levels of the SDSS photometry to match the UVOT flux density, assuming that the spectral shape is constant. We quantify the uncertainty due to possible spectral shape variation in \S \ref{sec:analysis_uv_optical_modeling}. For the lowest-redshift quasars in our sample ($z\apprle1.9$), we prefer to match the observed \emph{Swift U} band flux (central wavelength 3465 \AA) and the SDSS \emph{u} band (central wavelength 3551 \AA), as they sample very similar wavelength ranges. For objects with $z\ge1.9$ the \emph{U} and \emph{u} bands are strongly absorbed by the Lyman-$\alpha$ forest, and may not display the same flux variations as the unabsorbed continuum. We therefore match the fluxes to either the \emph{Swift} \emph{B} or \emph{V} bands for $z\apprge1.9$ objects, choosing whichever bandpass we infer to be more continuum-dominated at a given redshift.

To determine the necessary rescaling, we fit the SDSS broad-band photometry with a power-law model, omitting any SDSS bands that sample strong quasar emission lines (\S \ref{sec:analysis_uv_optical_modeling}). We use this model to estimate the flux density, as observed by the SDSS, at the pivot wavelength of the \emph{Swift} UVOT bandpass selected for matching. Based thereon, we rescale the SDSS photometric data to the flux level at the time of the UVOT observations. 

For the $z=3.53$ quasar Q1442+101, all of the UVOT bands cover spectral regions absorbed by the Lyman-$\alpha$ forest. In these cases our best option is to scale the SDSS photometry based on the \emph{Swift} \emph{V} band and the SDSS \emph{g} band. We consider this particular scaling to be highly uncertain, and therefore regard the subsequent modeling as only a rough estimate of the UV continuum level. 

%% file: sec5sed.tex
\section{Determining the UV to X-ray Spectral Energy Distribution}\label{sec:sed}

For black hole masses of order $10^{9}$ $M_{\astrosun}$, typical of bright quasars, standard $\alpha$-disk models predict that the inner accretion disk emission peaks in the far-UV or extreme-UV (EUV) \citep[e.g.,][]{Shakura1973}. The X-ray emission is commonly thought to be reprocessed UV continuum emission, along with a contribution from the jet for RLQ (\S \ref{sec:introduction}). Thus, by studying the continuum emission on each side of the unobservable EUV region, and assuming a shape for the EUV SED, we can constrain the energy released by the central accretion process (albeit with some uncertainty involved, \S \ref{sec:analysis_integrated_luminosity}). In this calculation, we do not include the IR emission in our estimate of the accretion energy, as the IR emission is due to dust outside the central regions, heated by the central continuum emission \citep[e.g.,][]{Barvainis1987,Hughes1993}. In the following we describe our modeling of the quasar UV continuum using UVOT and SDSS data (\S \ref{sec:analysis_uv_optical_modeling}), and present the observed SEDs covering the rest-frame UV and X-ray regimes based on our data and modeling (\S \ref{sec:sed_presentation}).

\subsection{Modeling the UV Quasar Continuum}\label{sec:analysis_uv_optical_modeling} 

Here, we model the UV continuum emission over the rest-frame wavelength interval 1000 \AA\, - 3000 \AA. The dominant spectral component in the rest-frame UV for these quasars is the continuum emission due to the accretion disk. We expect the host galaxy contamination of the UV continuum to be low: luminous quasars are known to reside in massive elliptical galaxies at $z\apprle0.25$ \citep{Dunlop2003}, which are faint emitters below rest-frame 4000 \AA\,\citep[e.g.,][]{Kinney1996}. At higher redshift there is evidence for quasar host galaxies being actively star-forming, however, the AGN contribution is still brighter than the host galaxy by $\sim2$ mag in the U band \citep{Floyd2012}. The broad emission lines, on the other hand, can contribute significantly to the broad-band flux if they are covered by the UVOT bandpasses. Absorption features such as the Lyman-$\alpha$ forest bluewards of 1200 \AA\,also complicate the measurement of the underlying continuum flux.

\begin{figure*}
	\centering
	\includegraphics[width=0.49\linewidth]{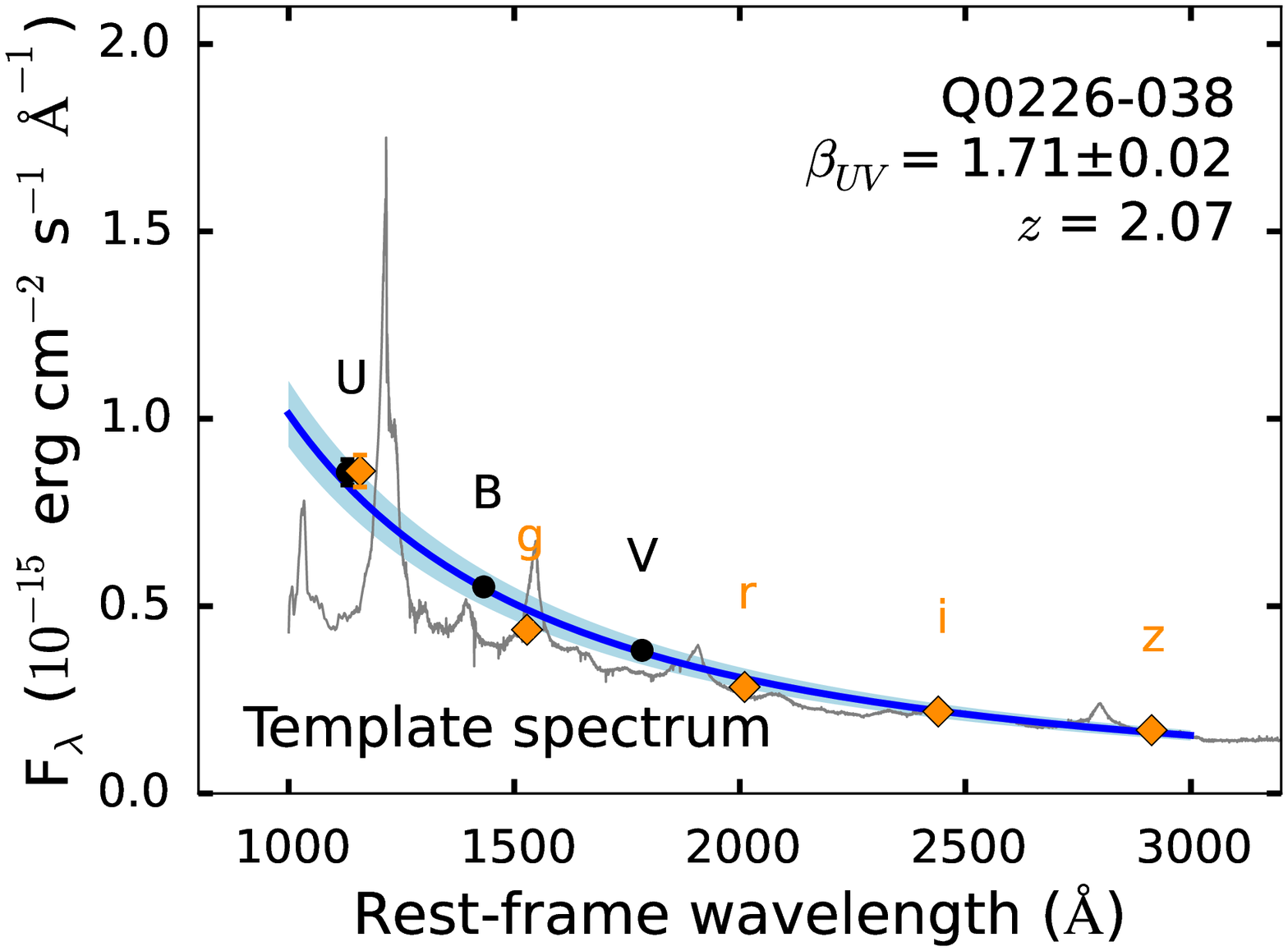}
	\includegraphics[width=0.49\linewidth]{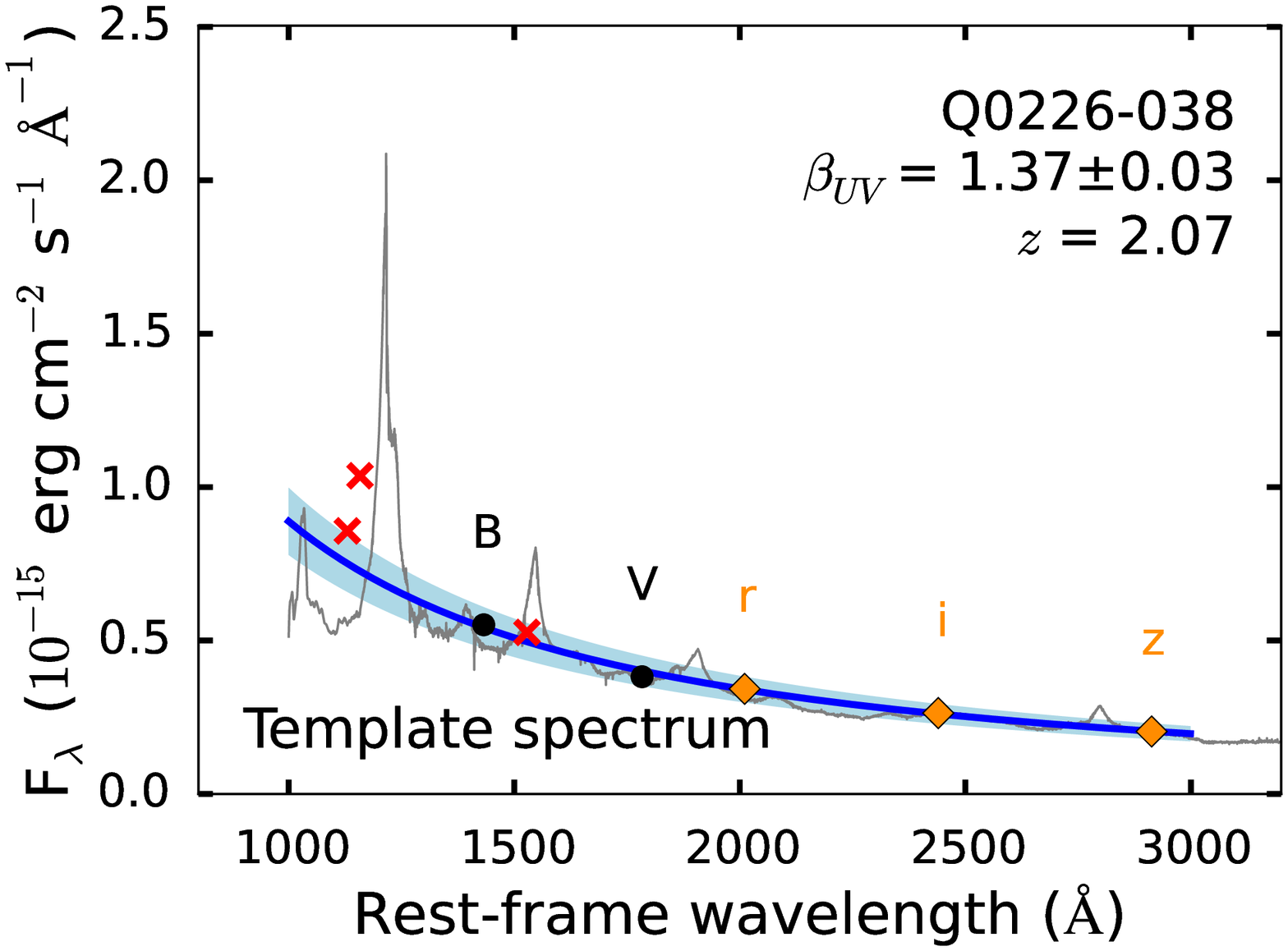}
	\caption{Steps in our modeling of the UV/optical spectrum for quasar Q0226-038. The dark blue curve and light blue shaded area shows our power-law continuum model and its uncertainty, fitted to the UVOT broad band photometric data (black points) and to the SDSS photometry (gold squares). The composite spectrum of bright $1.1<z<2.1$ quasars compiled by \citet{Selsing2016}, scaled to match the model flux at 2500 \AA, is included for illustrative purposes (grey curves). \emph{Left panel:} a fit to all available photometric data, including the UVOT \emph{U} and SDSS \emph{u} bands, which cover the Lyman-$\alpha$ emission line, and the SDSS \emph{g} filter, which is superimposed on the \ion{C}{iv} broad emission line. \emph{Right panel:} final model after removing the aforementioned emission-line contaminated bandpasses (red crosses). Note that the rescaling of the SDSS data points changes slightly between the initial and the final model. This is due to the \emph{g} filter no longer being used in the rescaling of SDSS photometry for the final fit.}
	\label{fig:uvot_powerlaw_fit}
\end{figure*}

For all 42 UVOT-detected quasars, we have broad-band photometry in at least two bandpasses (\S \ref{sec:data_reduction_uvot}). For 23 quasars with supplementary SDSS spectroscopy (\S \ref{sec:sdss}), we use these spectra to ascertain which of the available photometric bandpasses suffer absorption or emission line contamination. For objects without SDSS spectra, we use the composite quasar spectrum of \citet{Selsing2016} as a guide. This template spectrum is comprised of bright quasars with little host galaxy contribution, as expected for our $z\approx2$ sample. We perform an initial power law model fit using all available photometric bands (Figure \ref{fig:uvot_powerlaw_fit}, left panel). If the flux density in a given band lies outside the $1\sigma$ uncertainty range of this initial model, and we see emission or absorption features in the spectrum at that bandpass, we exclude it from the final model (Figure \ref{fig:uvot_powerlaw_fit}, right panel). For quasars with little usable photometric data, we exclude data points only if we believe that they introduce a gross error into our estimate of the integrated UV flux. E.g., we exclude the UVOT \emph{U} band for quasar Q0249-184 to avoid fitting an unrealistically red model continuum (Figure \ref{fig:sed8}).

We fit a power-law model\footnote{Equivalently, some authors fit a power-law model in wavelength space, $F(\lambda)=A_\lambda\lambda^{-\beta}$. Given that $F(\nu)=F\lambda(dnu/d\lambda)$, the relation between the power-law indices is $\beta=\alpha-2$.}, $F_{\lambda} = A\lambda^{-\beta_{\mathrm{UV}}}$, to the measured UVOT and the rescaled SDSS fluxes. Here, $F_\lambda$ denotes the flux density at wavelength $\lambda$, while $\beta_\mathrm{UV}$ is the UV spectral index. We perform a non-linear least-squares fit of this model to the available photometric data based on the Levenberg-Marquardt algorithm, using the \emph{scipy.optimize} package\footnote{SciPy: Open Source Scientific Tools for Python, 2001-, \url{http://www.scipy.org/} [Online; accessed 2015-10-27]}. Figure \ref{fig:uvot_powerlaw_fit} shows a typical case for a quasar with SDSS photometry. The left panel shows a model fitted to all available photometric data, while the right panel shows our final model, for which we reject the UVOT \emph{U} and SDSS \emph{u} bands in accordance with the criteria outlined above. We tabulate the model values of $\beta_\mathrm{UV}$ and \Fuv in Table \ref{tab:analysis}. The photometric data and best-fit models for our quasars are shown in Appendix \ref{appendix:sed} (Figures \ref{fig:sed1} to \ref{fig:sed15}, right panels). 

For three objects we only have a single usable photometric data point according to the above criteria (namely, Q0249-184, Q0445+097 and Q2212-299). To obtain a guideline estimate of the continuum emission in this case, we adopt a power law model with a spectral index equal to the canonical value $\beta_{UV}=1.5$, scaled to this single photometric data point. For three quasars, only two UV photometric measurements are usable (Q0238+100, Q0458-020 and Q0504+030). In these cases, we make a guideline estimate of the UV SED by connecting the two photometry points with a power law function. We exclude all six of the aforementioned quasars from the analyses of \S \ref{sec:analysis}, as we regard their UV continuum models as very crude guideline estimates, unfit for further analysis. 

Our flux rescaling of the SDSS data (\S \ref{sec:sdss}) assumes that the UV spectral index does not vary significantly on timescales of years. In fact, quasar UV-optical spectral index variations have been observed, of order $\Delta\beta_{\mathrm{UV}}\approx0.2$ \citep{Pu2006,Schmidt2012,Bian2012,Zhang2013}. To quantify the `worst-case' uncertainty due to spectral index variation, we artificially steepen the spectral index of the SDSS photometry by $\Delta\beta_{\mathrm{UV}}=0.2$ for quasar Q0038-019, leaving the UVOT data unchanged, and repeat our UV modeling (including the flux rescaling). This quasar is a fairly typical case for our SDSS sample, with two UVOT bandpasses and four SDSS bandpasses used to model the continuum. We find that the spectral index of the joint continuum model steepens significantly ($\Delta\beta_{\mathrm{UV}}=0.19$), while the integrated UV luminosity changes by less than 1\%. Although modeling based on UVOT photometry alone would negate the issue of spectral variability, the resulting SEDs would lack near-UV data for most sample quasars. We therefore prefer to include the SDSS photometry where available. A major advantage of observing with \emph{Swift} is the simultaneity between X-ray and UV-optical observations. We therefore do not use SDSS data to determine the UV-optical SED in cases where we lack a UVOT detection; in such cases we are ignorant of the appropriate flux rescaling for the SDSS photometry.

\subsection{UV to X-ray Spectral Energy Distributions}\label{sec:sed_presentation}

The rest-frame UV to X-ray SEDs of two quasars are shown in Figure \ref{fig:sed_mainpaper}. The remainder of our sample SEDs are presented in Appendix \ref{appendix:sed}. In the left panels we show $\nu L_\nu$ as a function of $\nu$. To represent the observed XRT data in units of physical flux, we show the `unfolded' XRT spectrum, i.e., the measured number of counts in a given bin, scaled by the ratio of the incident model to the model convolved with the instrumental response function. The X-ray data points are rebinned for clarity. Note that unfolded spectra of this type are model-dependent visualizations of X-ray data; we find them useful for presentation purposes, but they are not suitable for further spectral analyses. We also show the best-fit X-ray model including galactic absorption, along with the assumed underlying power law continuum corrected for Galactic absorption. In the right-hand panels we show a detailed view of our UV photometry, along with the best-fit continuum model, and its $1\sigma$ uncertainty (indicated by a shaded region). Quasar J082328.62+061146.07 (Figure \ref{fig:sed_mainpaper}, top) is securely detected by both the UVOT and the XRT. Quasar J014725.50-101439.11 (Figure \ref{fig:sed_mainpaper}, bottom) is an example of an XRT non-detection, for which we present the 3$\sigma$ limiting X-ray luminosity.

We include the SDSS spectra (rescaled by the same factor as for the SDSS photometric data) in the SED figures (Appendix \ref{appendix:sed}) for comparison purposes. For many objects, our model continua overestimate the true continuum level. This is due to the contributions of emission lines (e.g., the blended \ion{Fe}{ii} and Balmer emission features redwards of 2000 \AA) even in continuum-dominated bandpasses; see \S \ref{sec:analysis_integrated_luminosity_uverror}. Apart from this, we find significant flux offsets between the SDSS spectra and the SDSS photometry for several objects observed by the SDSS-III BOSS campaign \citep{Dawson2013}. In most cases, we are able to mitigate these offsets by applying the flux recalibration given by \citet{Margala2015}; in Appendix \ref{appendix:sed} we discuss the targets for which this is successful.

\def\sedplotsize{0.495\linewidth}
\begin{figure*}
	\centering
	\includegraphics[width=\sedplotsize]{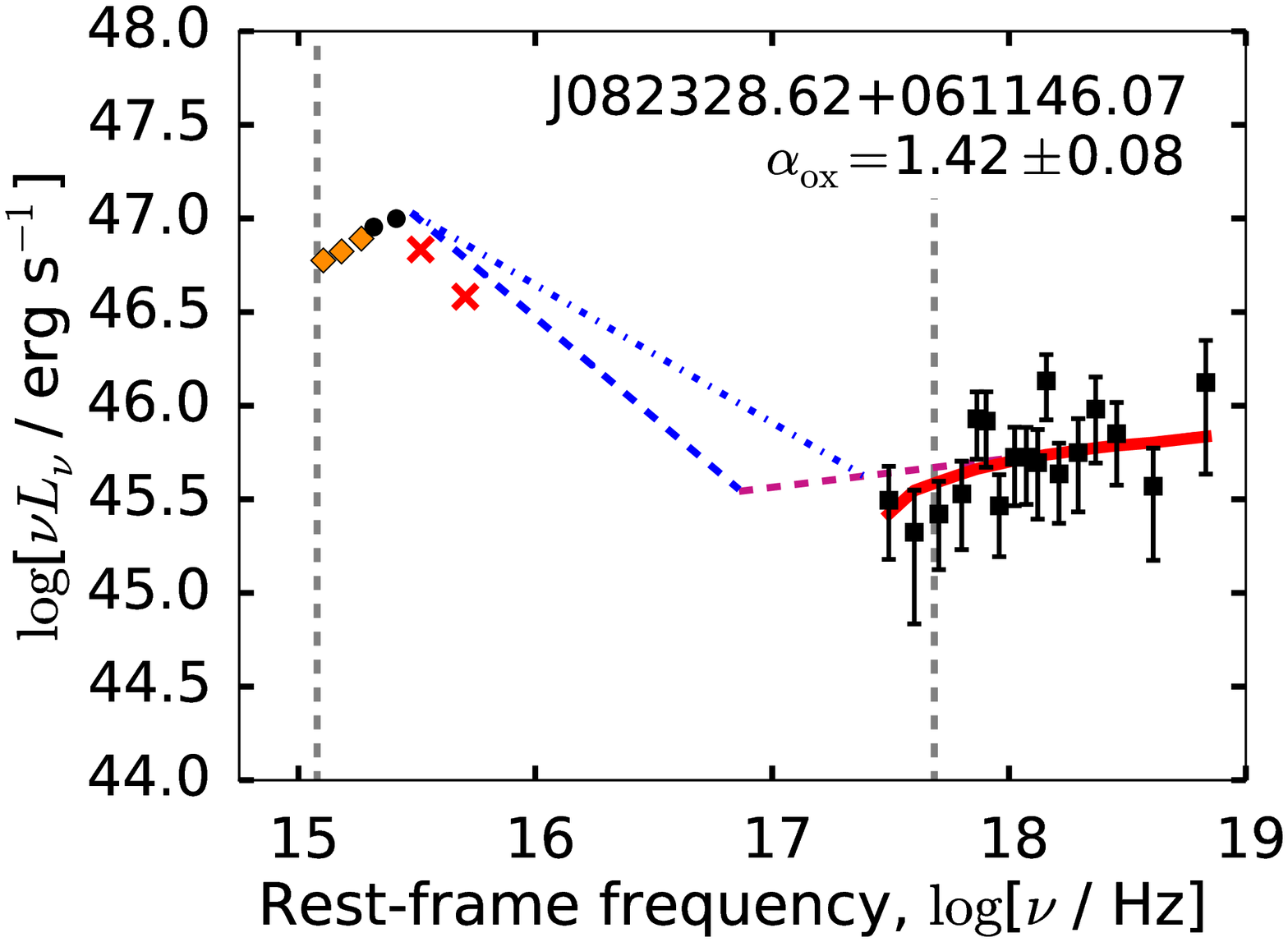}
	\includegraphics[width=\sedplotsize]{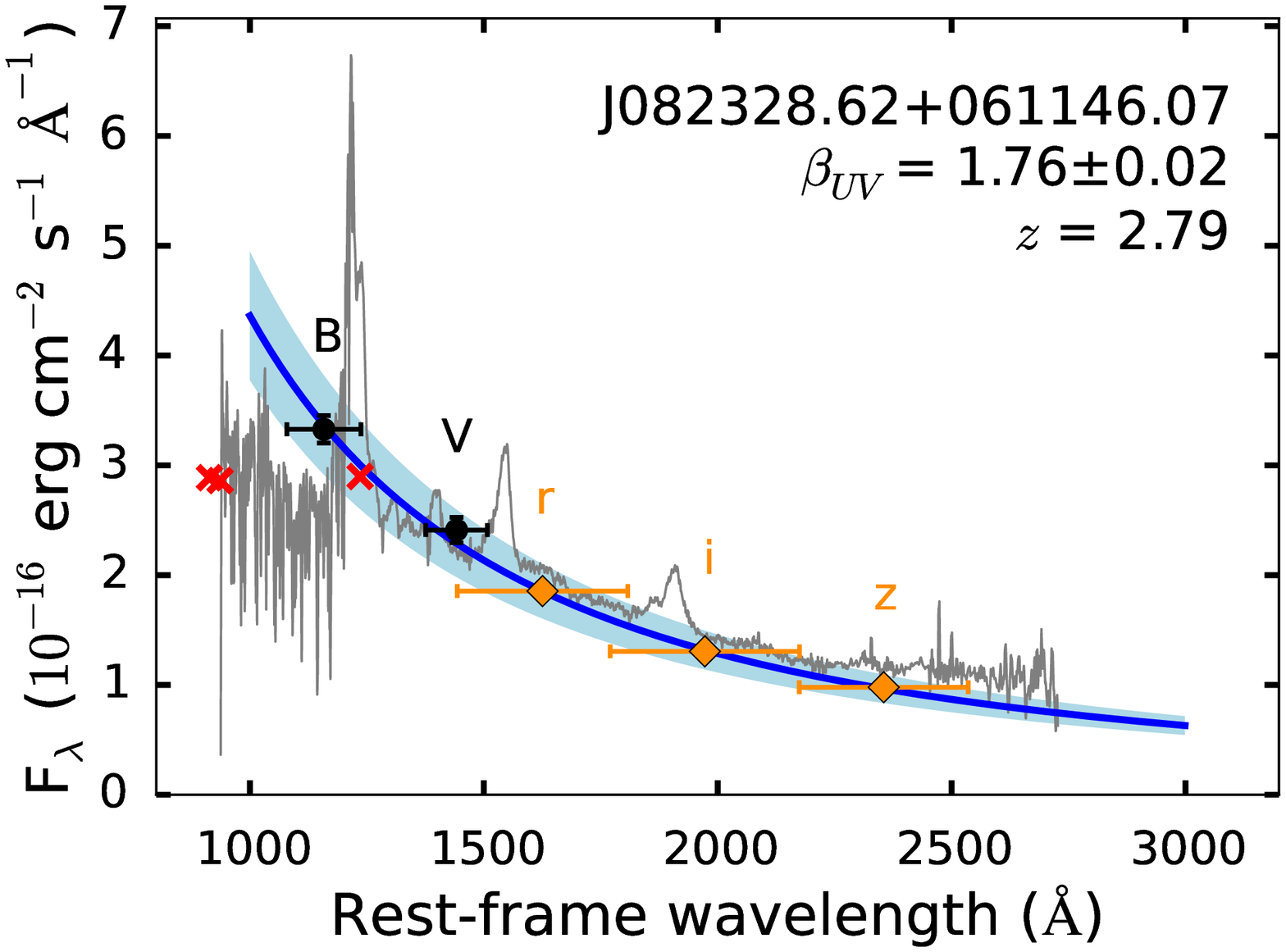}
	\includegraphics[width=\sedplotsize]{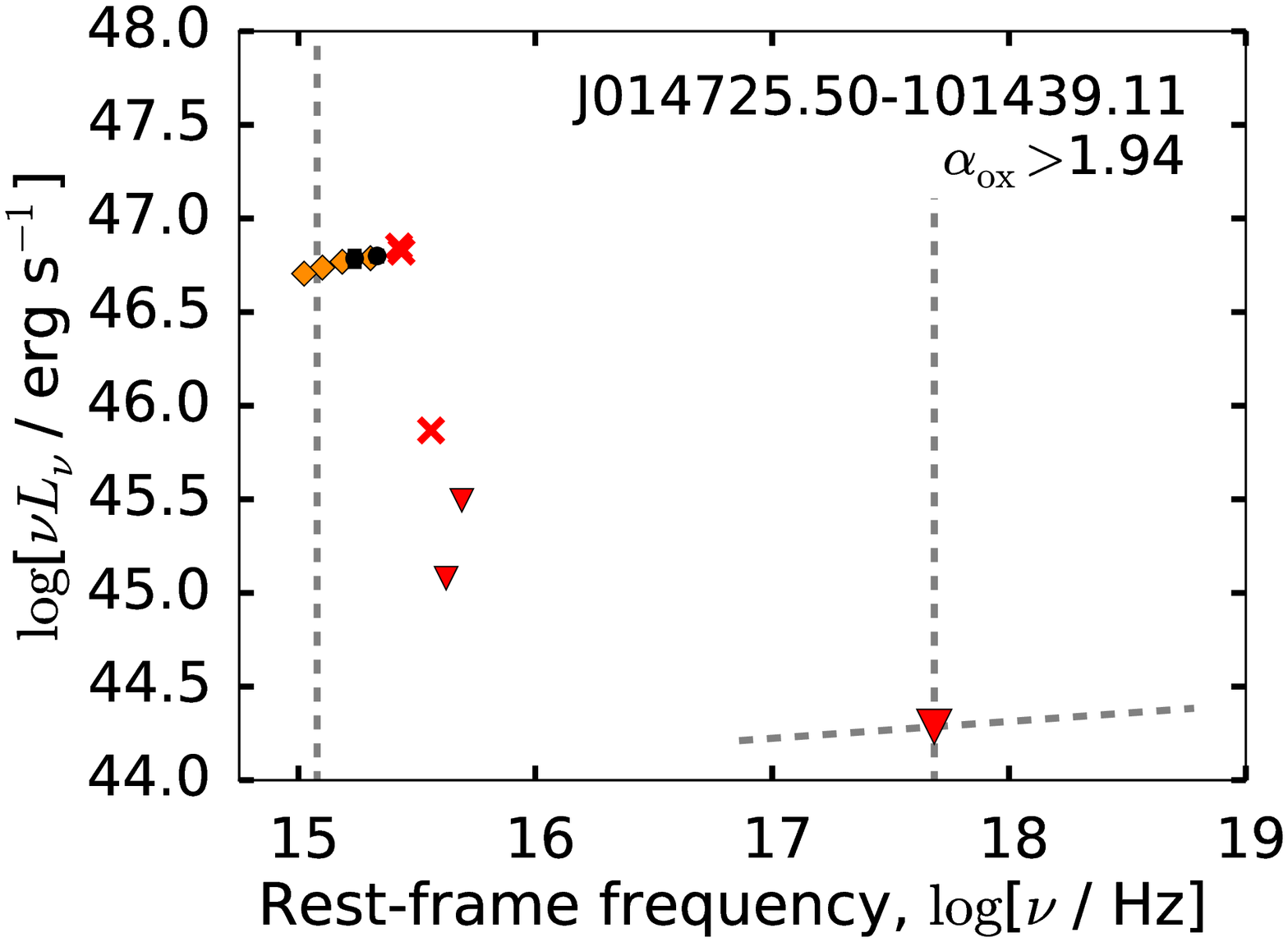}
	\includegraphics[width=\sedplotsize]{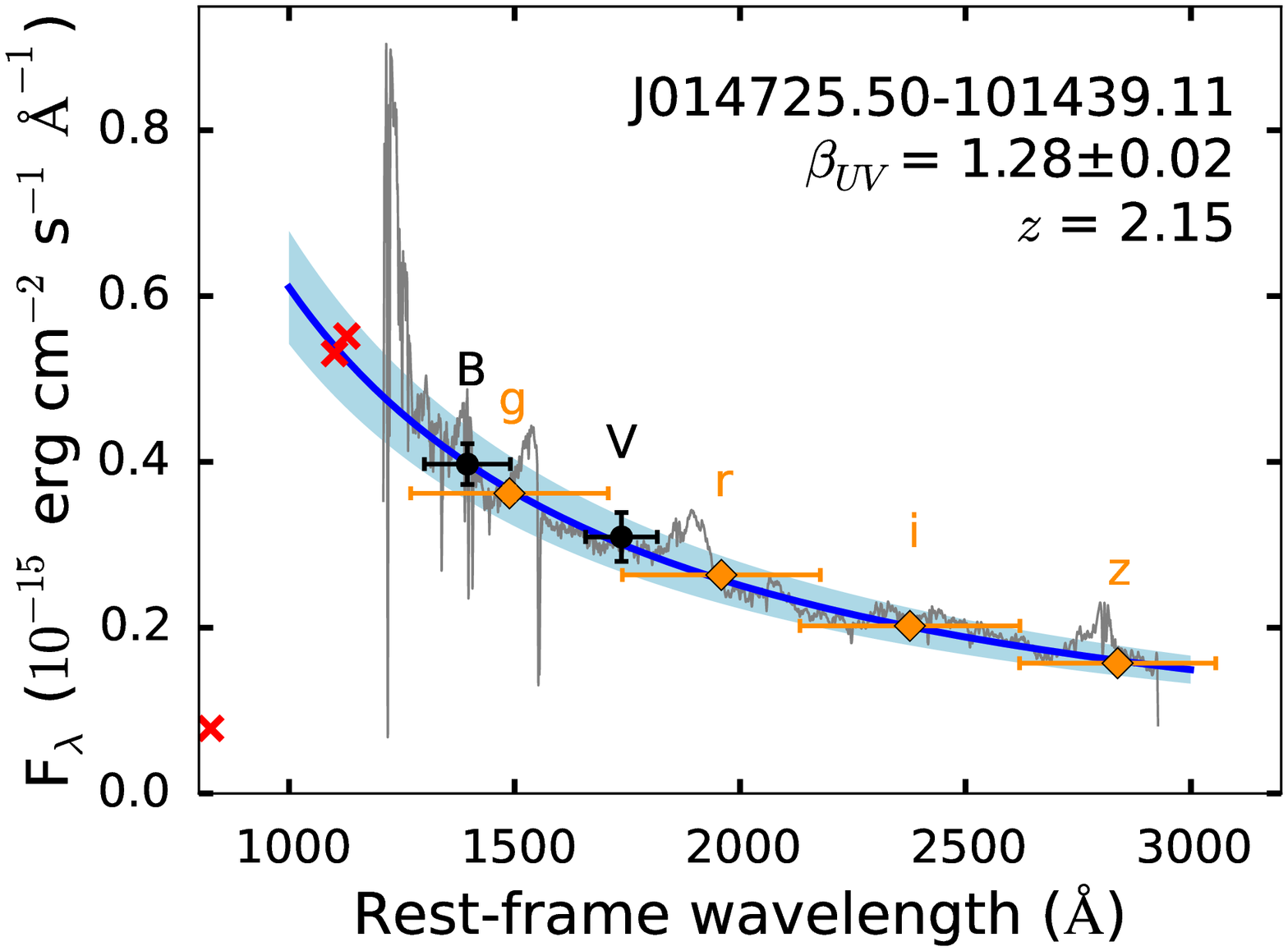}
	\caption{SEDs for two quasars in our sample: one X-ray detected object (upper panels), and one X-ray non-detection (lower panels). \textbf{Left panels:} Rest-frame UV to X-ray SEDs. The leftmost and rightmost vertical dashed lines indicate the frequencies corresponding to 2500 \AA\,and 2 keV, respectively.  \textbf{Right panels:} Detailed view of the UV data and continuum modeling. \textbf{Symbols:} Black squares: `unfolded' XRT spectrum (\S \ref{sec:sed_presentation}) with 1$\sigma$ uncertainties. Solid red curve: X-ray model including Galactic absorption. Dashed magenta curve: intrinsic, absorption-corrected X-ray model. Red triangles: $3\sigma$ upper limit fluxes (for X-ray upper limits, gray dashed line illustrates a $\Gamma=1.91$ power-law). Black dots: Galactic absorption-corrected UVOT photometry. Orange squares: rescaled SDSS broadband photometry (\S \ref{sec:sdss}). Red crosses: photometric data points excluded from modeling (\S \ref{sec:analysis_uv_optical_modeling}). Blue dashed and dash-dotted lines: EUV interpolations (\S \ref{sec:analysis_integrated_luminosity}), connecting with the X-ray model at 0.3 keV and 1 keV, respectively. Grey curve: SDSS spectrum, if available - otherwise, composite quasar spectrum of \citet{Selsing2016}, scaled to the model flux at 2500 \AA. In the right panels, the horizontal bars represent the Full Width at Half Maximum of the filter bandpasses.}
	\label{fig:sed_mainpaper}
\end{figure*}

%% file: sec6analysis.tex
\section{Measurements and Analysis}\label{sec:analysis}

Here we present some fundamental measurements of the quasar SEDs that help describe the current subsample, including estimates of the integrated UV to X-ray luminosity (\S \ref{sec:analysis_integrated_luminosity}). All X-ray detected objects are radio loud, while all but one of the X-ray non-detections are radio quiet. This is unsurprising, as RLQs tend to be brighter in the X-ray than RQQs at a given optical luminosity \citep{Zamorani1981}. We will address the remainder of the sample in future work, pending completion of the \emph{Swift} observations.

\subsection{UV and X-ray Luminosities}

\begin{figure*}
\centering
\includegraphics[width=0.49\linewidth]{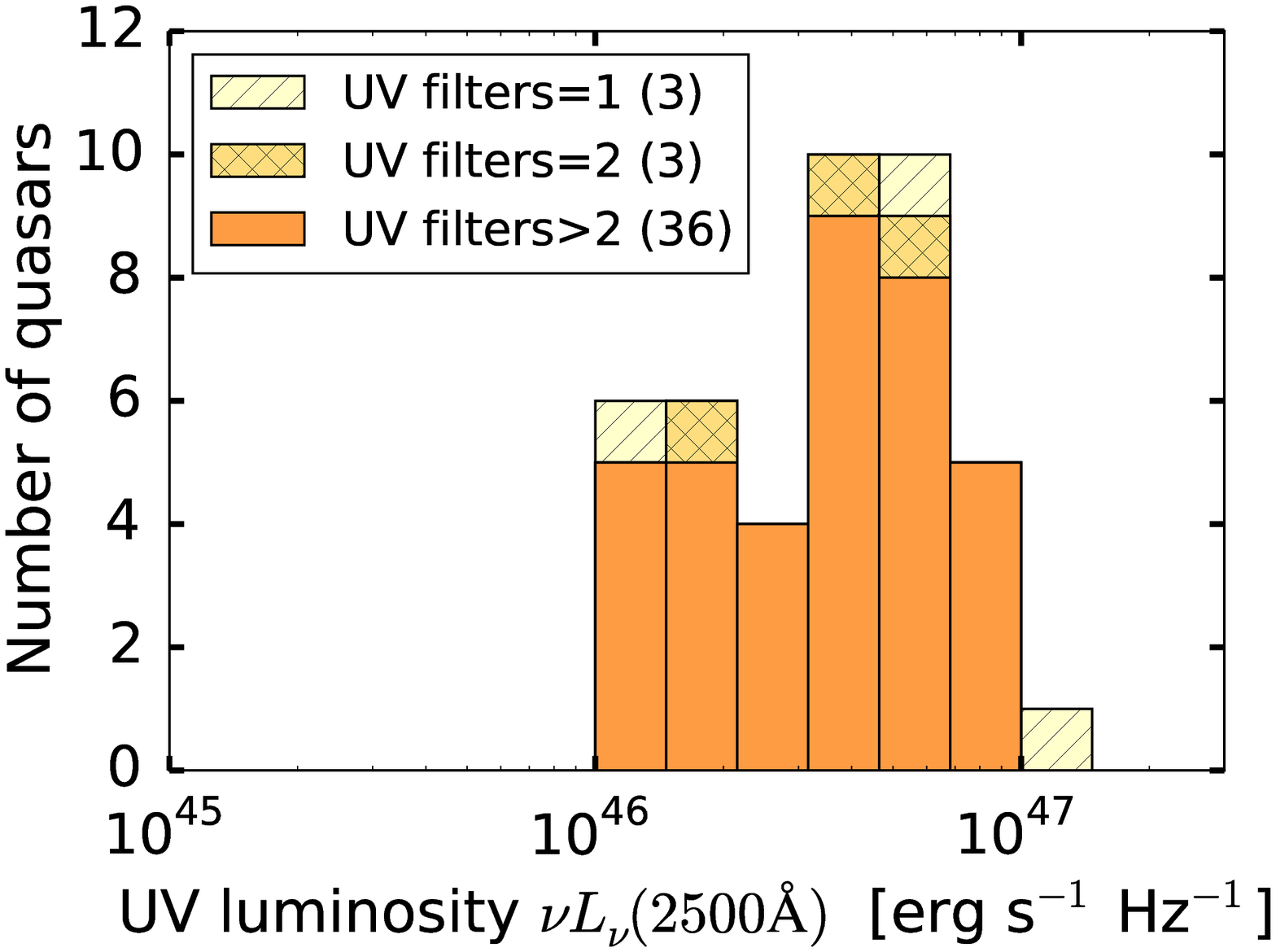}
\includegraphics[width=0.49\linewidth]{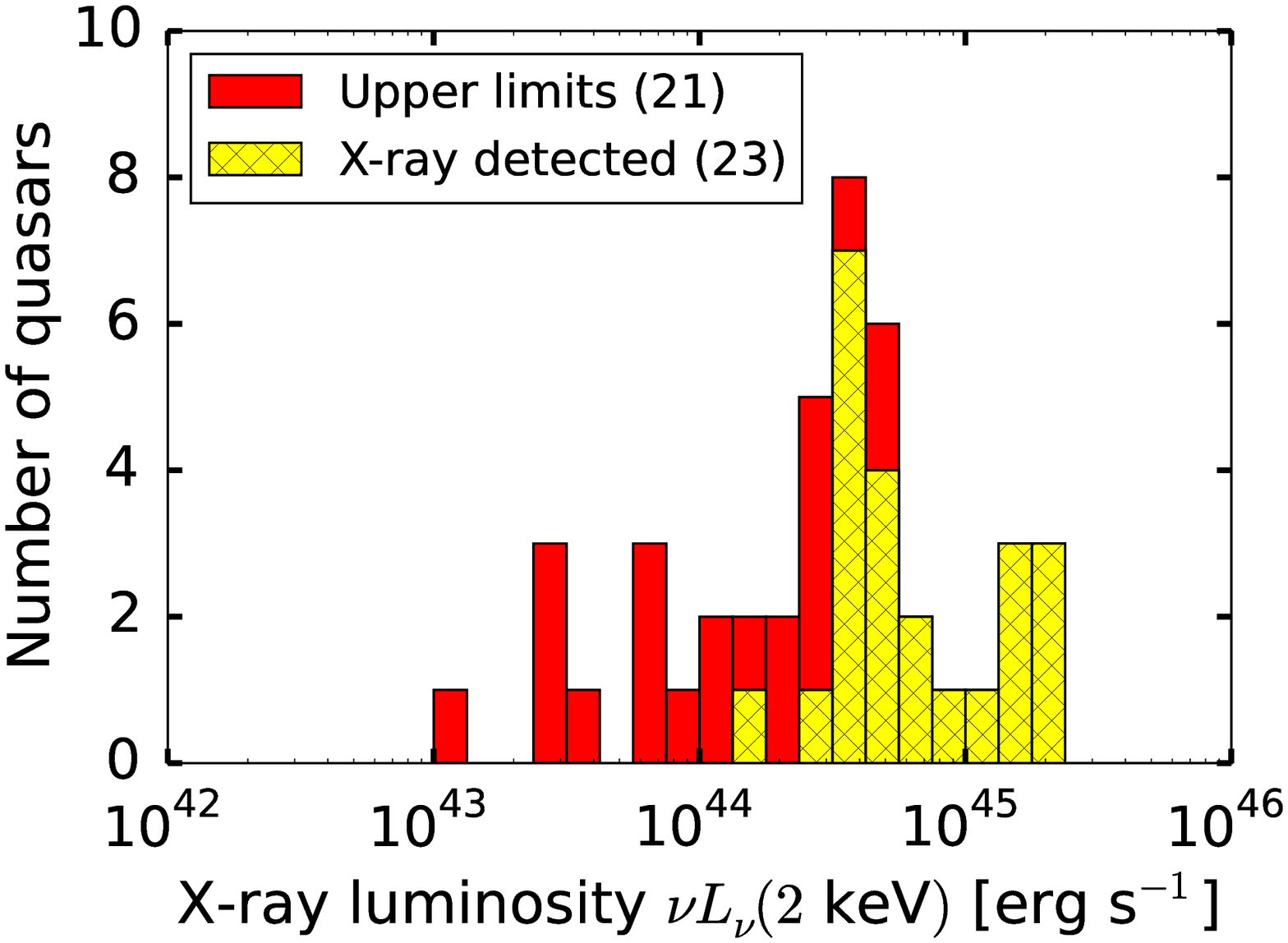}
\caption{\emph{Left:} Distribution of $\nu$\Luv for our sample, based on the continuum modeling (\S \ref{sec:analysis_uv_optical_modeling}). Only quasars detected in three or more UV bandpasses (solid orange histogram) are included in our further analysis (\S \ref{sec:analysis_alphaox}). \emph{Right:} Distribution of $\nu$\Lx. Upper limits for non-detections are at the $3\sigma$ level (solid yellow histogram). This is a stacked histograms (see caption of Figure \ref{fig:z_histogram})}
\label{fig:nuLdnu_histograms}
\end{figure*}

Our sample spans a full range $1.0\le$\Luv$/(10^{46}$ erg s$^{-1})\le9.8$ (Figure \ref{fig:nuLdnu_histograms}, left panel), with an average monochromatic UV luminosity of $\langle\nu$\Luv$\rangle=4.2\times10^{46}$ erg s$^{-1}$. Two quasars are not detected in any UVOT bands; we make no attempt to determine limiting values of \Luv for these objects, as we lack simultaneous UV data with which to rescale the SDSS photometry. Six quasars are detected in only one or two UV bandpasses, making the determination of \Luv uncertain (\S \ref{sec:analysis_uv_optical_modeling}). In the X-ray data we detect 23 quasars, and present $3\sigma$ upper limits for 21 objects (Figure \ref{fig:nuLdnu_histograms}, right panel). The average $\nu$\Lx for detected objects is $8.3\times10^{44}$ erg s$^{-1}$. All X-ray detected objects have $\nu L_\nu$(2 keV)$>10^{44}$ erg s$^{-1}$. The average integrated rest-frame 0.3 keV - 10 keV luminosity for X-ray detected quasars is $2.0\times10^{46}$ erg s$^{-1}$. Thus, our sample is comparable in terms of X-ray luminosity to the brightest quasars in the \emph{Swift} sample presented by \citet{Wu2012}.

\subsection{The X-ray Hardness Ratio}\label{sec:analysis_HR}

\begin{figure*}
	\centering
	\includegraphics[width=0.49\linewidth]{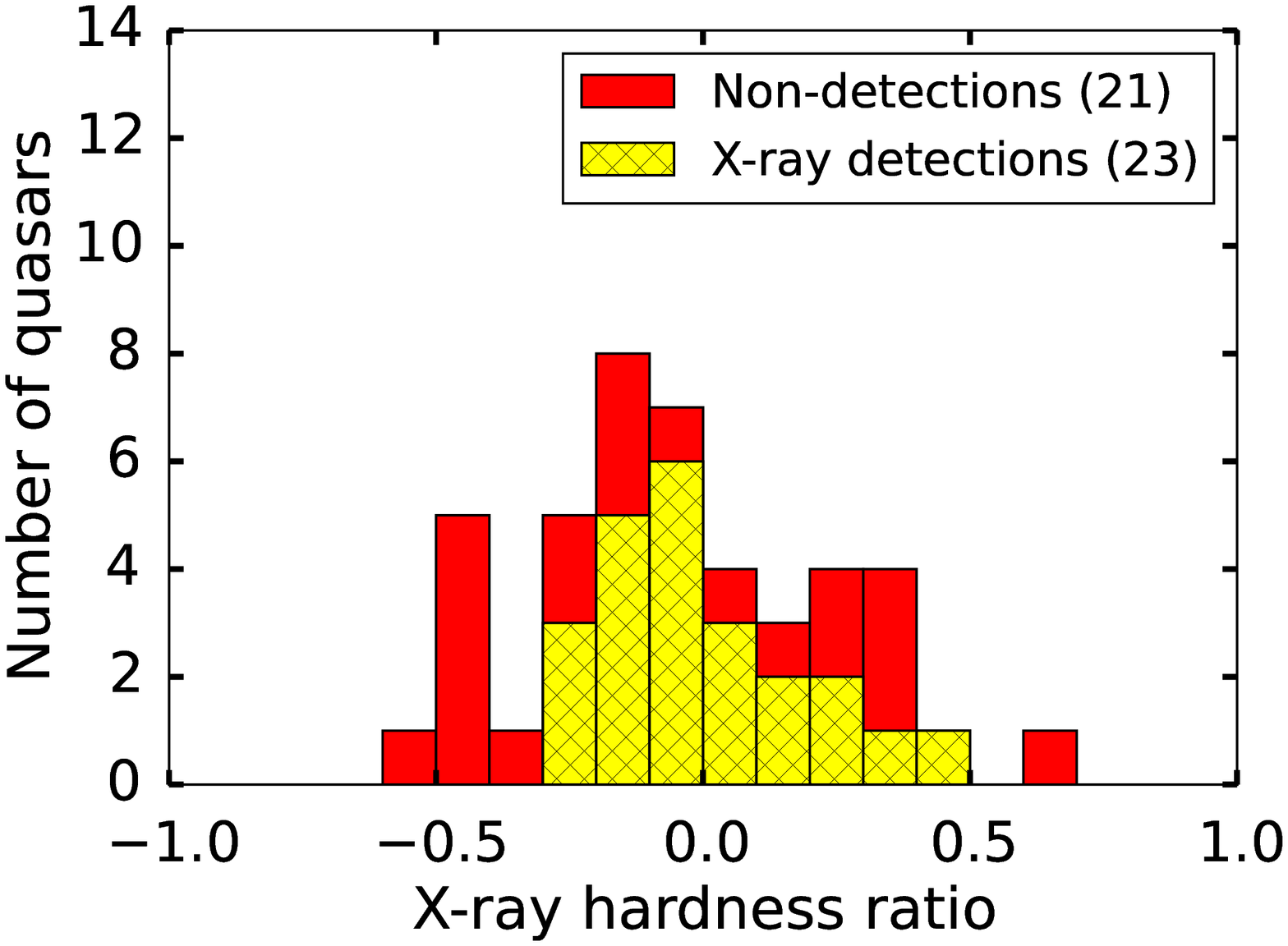}
	\includegraphics[width=0.49\linewidth]{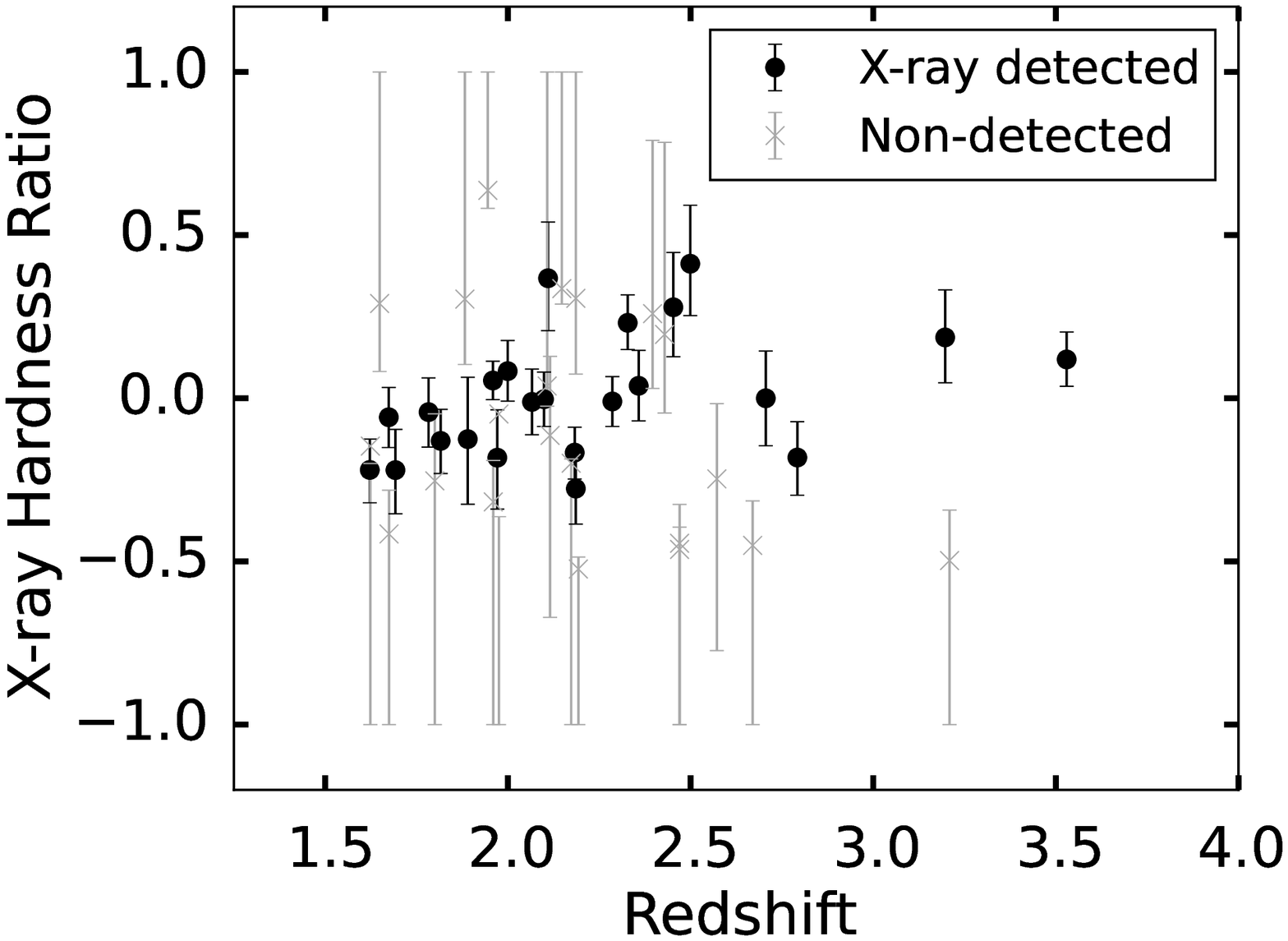}
	\caption{\emph{Left:} Distribution of the X-ray hardness ratio for our sample. Here, the soft band is 0.3 keV - 1.5 keV, while the hard band is the 1.5 keV - 10 keV range. Larger values of $HR$ imply harder X-ray spectra. This is a stacked histogram (see caption of Figure \ref{fig:z_histogram}). \emph{Right:} Redshift distribution of the hardness ratios. The mean hardness ratio is $\langle HR\rangle=0.01$ with a standard deviation of 0.19 for X-ray detections (black points), $\langle HR\rangle=-0.08$ with a standard deviation of 0.34 for non-detections (gray crosses).}
	\label{fig:HR}
\end{figure*}

The X-ray hardness ratio ($HR$) is a crude representation of the X-ray spectral slope, useful for observations with few detected X-ray counts, for which the photon index $\Gamma$ is not well-determined. We calculate the hardness ratio for each quasar as $HR=(H-S)/(H+S)$. Here, $H$ and $S$ are the background-subtracted hard-band (observed-frame 1.5 keV - 10 keV) and soft-band (0.3 keV - 1.5 keV) source aperture counts, respectively. We use the C- and Fortran-based software \emph{Bayesian Estimation of Hardness Ratios}\footnote{\url{http://hea-www.harvard.edu/AstroStat/BEHR/}} (BEHR) \citep{Park2006} to determine the uncertainties on $HR$, choosing a flat prior distribution. For X-ray detections we find a mean hardness ratio $\langle HR\rangle=0.01\pm0.01$, with a sample standard deviation $\sigma_{HR}=0.19$ (Figure \ref{fig:HR}, left panel). For non-detections we find $\langle HR\rangle=-0.08\pm0.04$ and $\sigma_{HR}=0.34$. For the majority of the X-ray non-detections, the hardness ratio is poorly determined, with uncertainties consistent with the extreme values ($HR=\pm1$, Figure \ref{fig:HR}, right panel). A possible weak tendency for a higher hardness ratio at higher redshift (Figure \ref{fig:HR}, right panel) is likely due to the soft excess becoming increasingly redshifted outside the \emph{Swift} observable window at higher $z$. It is difficult to determine how common the soft excess feature is at $z\sim2$, due to the redshifting of the relevant energies into the unobservable EUV spectral region. However, if there is a soft excess component, its high-energy tail would contribute to the XRT soft band for quasars at $z\sim1.5$, but less so for higher-redshift objects. \citet{Piconcelli2005} find evidence for a soft excess in 39 (of 40) $M_B\apprle23$ AGN, including a few objects at $z\sim1.5$.

\subsection{The X-ray Photon Index}\label{sec:analysis_Gamma}

\begin{figure*}
	\centering
	\includegraphics[width=0.49\linewidth]{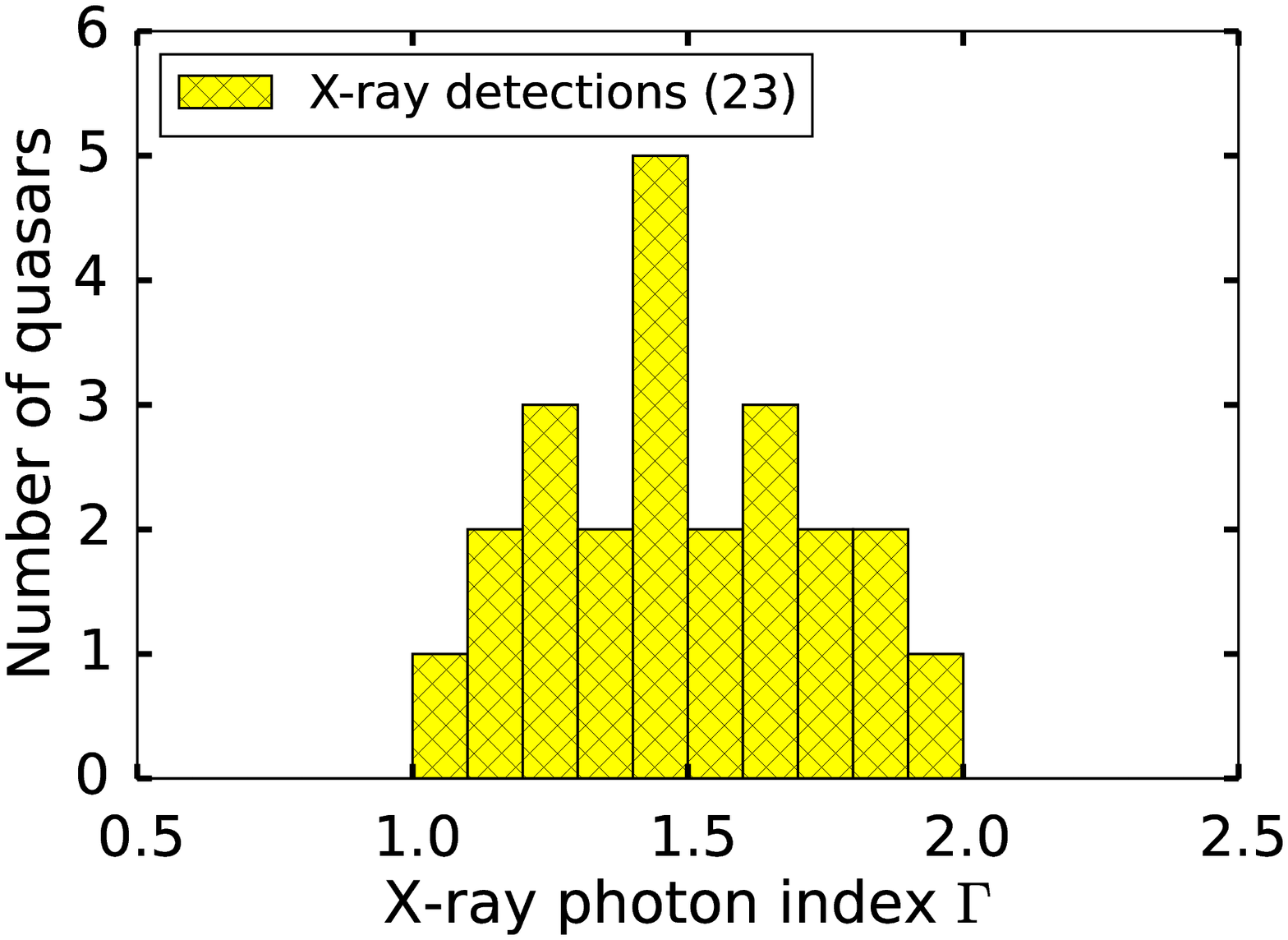}
	\includegraphics[width=0.49\linewidth]{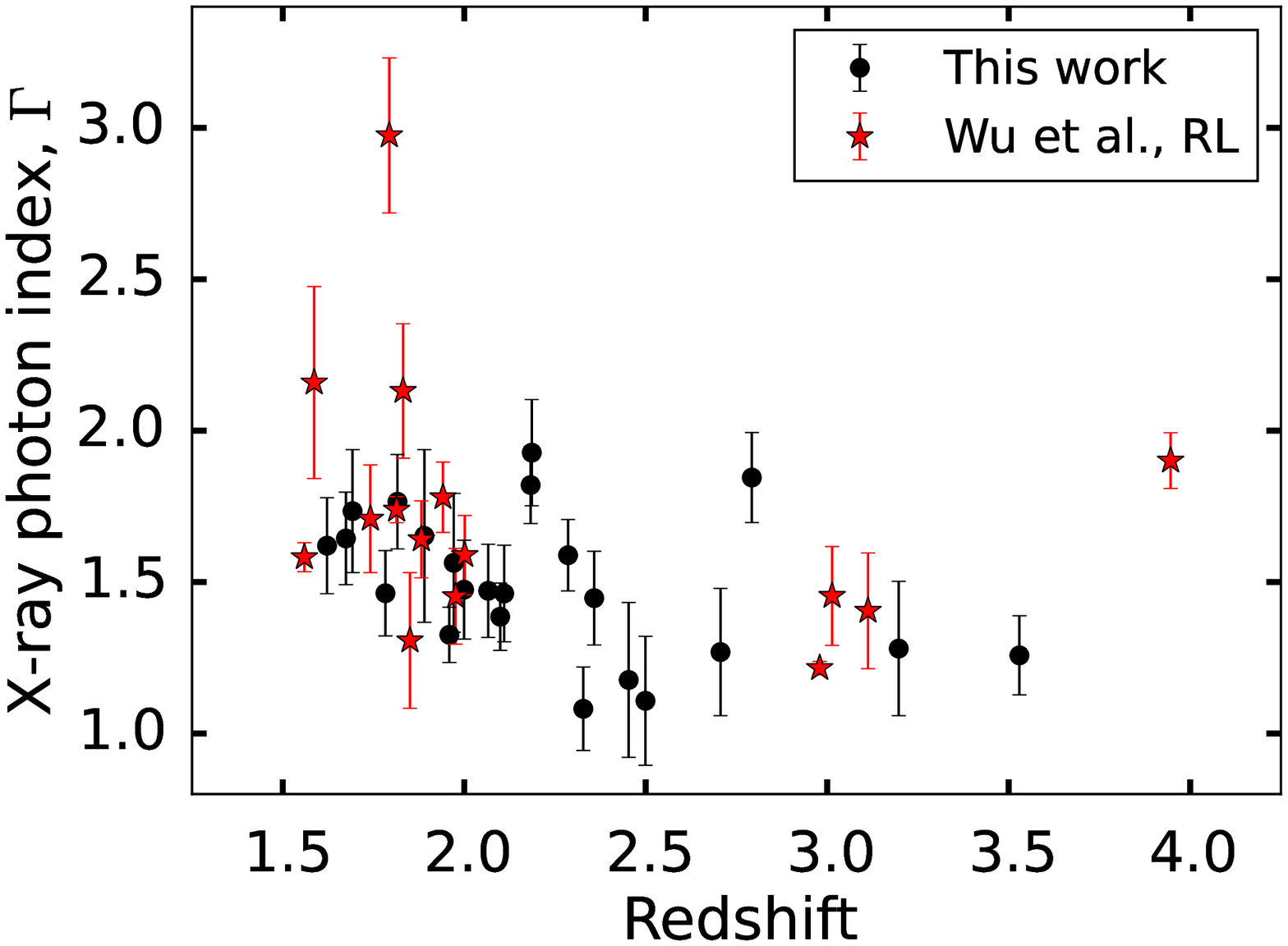}
	\caption{\emph{Left panel:} Distribution of the X-ray photon index $\Gamma$ for X-ray detected quasars, measured across the observed frame energy range 0.3 keV - 10 keV. The average uncertainty on individual measurements of $\Gamma$ is 0.15 for observations with $N_\mathrm{sub}>50$ (orange dotted area), and 0.23 for observations with $20<N_\mathrm{sub}<50$. This is a stacked histogram (see caption of Figure \ref{fig:z_histogram}). \emph{Right panel:} Redshift distribution of $\Gamma$ for X-ray detected objects in our sample (black points), and for radio loud $z>1.5$ quasars in the sample of \citet{Wu2012} (red stars).}
	\label{fig:Gamma}
\end{figure*}

For X-ray detected quasars, we find an average value $\langle\Gamma\rangle=1.46\pm0.05$, with a sample standard deviation of $0.23$, and a full range $1.08\le\Gamma\le1.93$. The distribution of $\Gamma$ is fairly symmetric around the mean (Figure \ref{fig:Gamma}, left panel), in agreement with previous studies of both RLQs and RQQs \citep[e.g.,][]{Scott2011,Wu2012}; we find a median photon index $\Gamma=1.47$, in agreement with our mean value. We see a weak tendency towards decreasing $\Gamma$ with increasing redshift (Figure \ref{fig:Gamma}, right panel). While this tendency is not highly significant for the current sample (Spearman's rank correlation $p=3\%$), it is in agreement with the findings of \citet{Scott2011} for a larger sample of $z<6$ AGN. Similarly to the weak trend discussed in \ref{sec:analysis_HR}, the cause of this trend in our sample may be the redshifting of the soft excess out of the XRT observing window.

We find a smaller $\langle\Gamma\rangle$ than that of any other study of RLQ of which we are aware (Table \ref{tab:Gamma_literature}). While the RLQ presented by \citet{Scott2011} are fainter than our sample in terms of the integrated luminosity between 2 keV - 10 keV, those presented by \citet{Wu2012} have comparable luminosities, and those of \citet{Page2005} are more luminous. Thus, this discrepancy is not purely a luminosity effect. The use of different energy ranges for which $\Gamma$ is measured for the different studies complicates this issue. However, the discrepancy persists if we limit our X-ray modeling to the rest-frame 2-10 keV energy range to allow a direct comparison with the findings of \citet{Page2005} and \citet{Reeves2000}: for this energy range, we find $\langle\Gamma\rangle=1.23\pm0.12$. Given our moderate sample size (23 detected RLQ), the low $\langle\Gamma\rangle$ may simply be a statistical anomaly. Indeed, the discrepancy between our sample and that of \citet{Wu2012} is driven by just three objects with $\Gamma>2$ in their sample (Figure \ref{fig:Gamma}). We discuss a possible explanation for this discrepancy, should it be real, in \S \ref{sec:discussion}.

\subsection{The UV-optical Spectral Index}\label{sec:analysis_UV_beta}

For quasars with at least three photometric data points used in the modeling, we find an average spectral index $\langle\beta_{UV}\rangle=1.41\pm0.05$, with a sample standard deviation of $\sigma_{\beta}=0.33$. This is somewhat redder than the $\beta_{UV}=1.56$ measured for the SDSS composite quasar spectrum \citep{VandenBerk2001}. It is, however, consistent with the average value $\langle\beta_{UV}\rangle=1.34\pm0.15$ found by \citet{Carballo1999} for RLQ. We note that our UV-optical measurements, and those of Carballo et al., are based on broadband photometry. \citet{Vestergaard2003} presents measurements of the UV continuum slope, based on spectroscopic data, for 22 quasars in the current sample. These measurements are on average steeper (bluer) than our $\beta_\mathrm{UV}$ by $0.32\pm0.11$. This indicates that our modeling is affected by broad emission line contamination of the photometric bandpasses, despite our strategy of excluding bandpasses with strong broad-line contributions (\S \ref{sec:analysis_uv_optical_modeling}). In particular, the presence of Balmer continuum emission and of the multitude of \ion{Fe}{ii} emission lines in the rest-frame near-UV causes a systematic overestimation of the continuum level and a flattening of the measured spectral index. In \S \ref{sec:analysis_integrated_luminosity} we quantify the resulting systematic uncertainty introduced into our accretion luminosity estimates.

\subsection{The X-ray to UV Spectral Index \aox}\label{sec:analysis_alphaox}

\begin{figure}
	\centering
	\includegraphics[width=\linewidth]{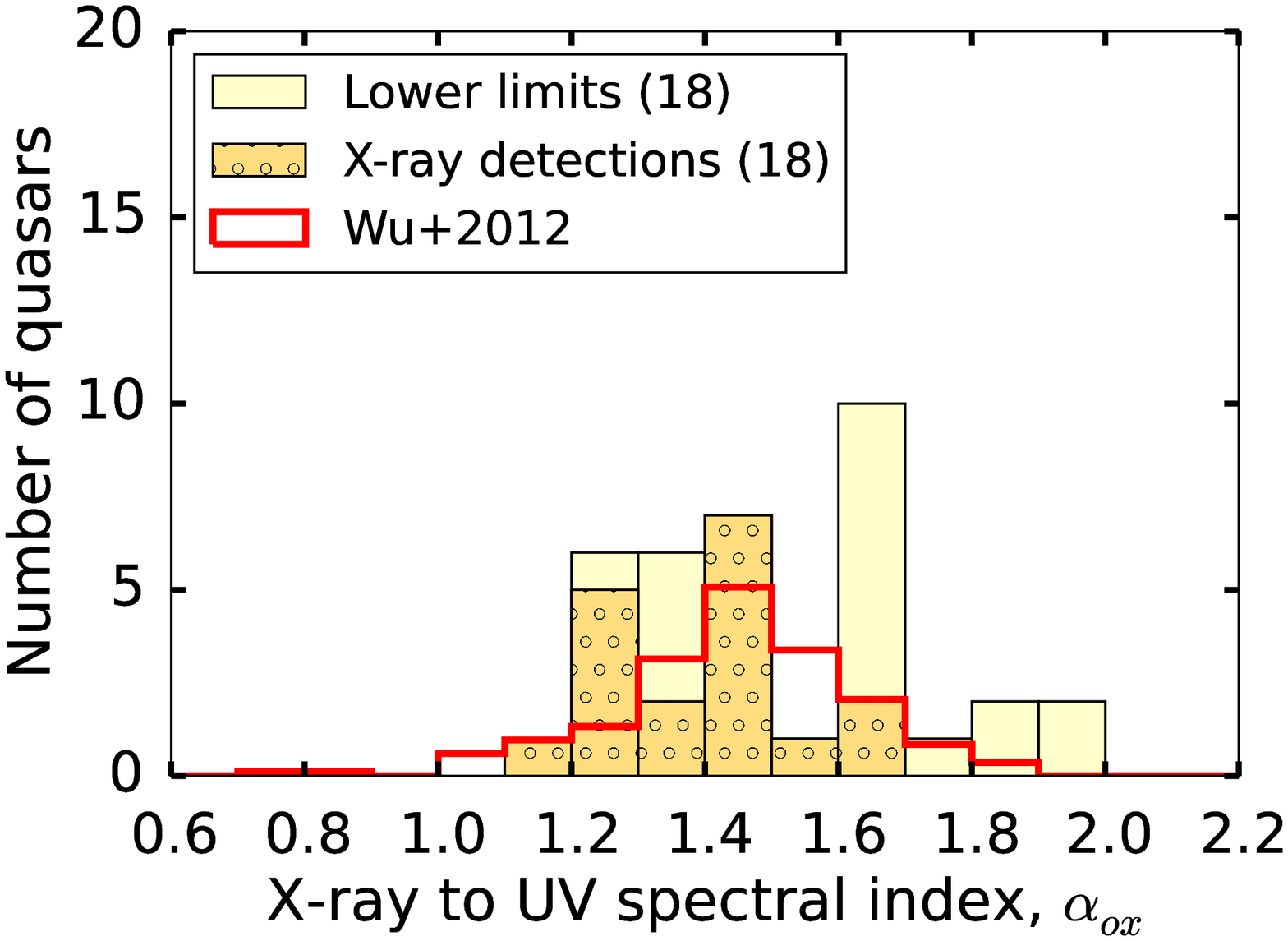}
	\caption{Distribution of \aox (and $3\sigma$ upper limits for X-ray non-detections) for all quasars with reliable UV continuum fits (\S \ref{sec:analysis_alphaox}). This is a stacked histogram (see caption of Figure \ref{fig:z_histogram}). The red histogram shows the distribution of \aox for objects detected in both UV and X-ray data at $z>1.5$ in the sample of \citet{Wu2012}, comprising both radio quiet and radio loud quasars. This histogram is normalized to have the same total area as that of our X-ray detected sample.} 
	\label{fig:aox_histogram}
\end{figure}

We measure the X-ray to UV spectral index \aox as defined by \citet{Tananbaum1979}, or its $3\sigma$ lower limit, for each quasar for which we have a UVOT detection (Table \ref{tab:analysis}). In the following discussion, we only include quasars for which we have three or more photometry data points suitable for our UV continuum modeling (36 objects, \S \ref{sec:analysis_uv_optical_modeling}). We find an average value of $\langle\alpha_\mathrm{ox}\rangle=1.39\pm0.03$ for the X-ray detected objects, with a sample standard deviation of 0.12 (Figure \ref{fig:aox_histogram}).  Assuming that those upper limits with \aox$\approx2$ are close to the true values, our sample covers a similar range of \aox as the $z>1.5$ subsample of \citet{Wu2012}, comprising both RLQs and RQQs, as shown in Figure \ref{fig:aox_histogram}. The mean value of \aox for the RLQ subsample is in rough agreement with previous studies of RLQs (\citealt{Wu2013}, $\langle\alpha_\mathrm{ox}\rangle=1.35\pm0.05$, \citealt{Miller2011}, $\langle\alpha_\mathrm{ox}\rangle=1.37\pm0.03$). Previous studies tend to find a steeper \aox for RQQs than for RLQs. For example, \citet{Steffen2006} find $\langle$\aox$\rangle=1.71\pm0.02$ for RQQs with comparable UV-optical luminosities to our sample. The upper limits on \aox obtained for RQQs in our sample are consistent with this: we find an average limiting value of \aox$>1.62$.

At low redshift, \aox correlates with $\Gamma$ (as measured in the observed XRT band, 0.3--10 keV) in the sense that AGN that are X-ray faint relative to the optical-UV tend to have softer X-ray SEDs \citep{Atlee2009,Grupe2010}. We find no such correlation in our $z\approx2$ sample, for which we model the SED at rest-frame energies $\apprge0.75$ keV. As suggested by \citet{Wu2012}, the observed trend at low redshift is likely driven by the soft X-ray excess component. The soft excess may be due to Compton-upscattered UV emission from the accretion disk, and may therefore be stronger when the accretion disk is brighter relative to the coronal emission (i.e., for higher values of \aox). The lack of this trend in our data is then explained by the redshifting of the soft excess outside our spectral window.

\begin{figure}
	\centering
	\includegraphics[width=\linewidth]{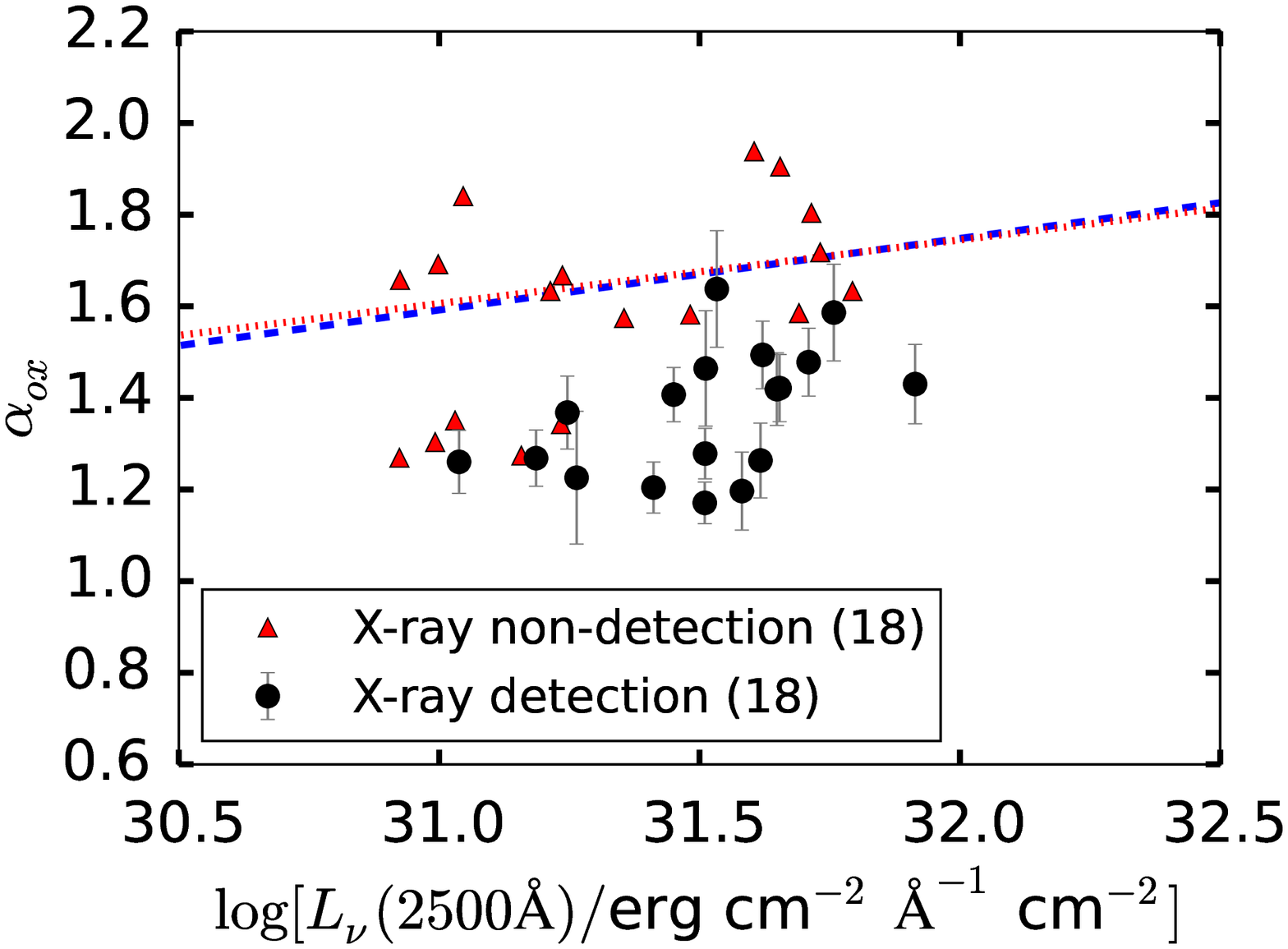}
	\caption{The SED shape, as represented by \aox, as a function of \Luv. The dashed blue line shows the \Luv-\aox sample of \citet{Wu2012}, which contains both RLQs and RQQs, but is predominately radio-quiet. The red dotted line shows the \Luv-\aox relation found by \citet{Strateva2005} for RQQs. We note that the distribution of our X-ray detections (which are radio-loud) is offset from the relations previously established for RQQs.}
	\label{fig:aox_luminosity}
\end{figure}

The relationship between \aox and \Luv for RQQs has been studied by several authors \citep[e.g.,][]{Tananbaum1979,Wilkes1994,Strateva2005,Wu2012}. These authors find a non-linear relationship between the UV and X-ray luminosities, with more UV-luminous AGN being relatively weaker X-ray emitters. Given the small dynamic range in \Luv for X-ray detections in our current sample, it is unsurprising that this trend does not exist in our data (a generalized Kendall's $\tau$ test gives the probability $p=0.17$ of observing this distribution given no intrinsic relation between \Luv and \aox). Here, we simply note that our X-ray detections are offset from the previously established relationships between \Luv and \aox for RQQs (Figure \ref{fig:aox_luminosity}), as expected for RLQ as they tend to be X-ray bright \citep[e.g.,][]{Zamorani1981,Miller2011}.

\subsection{Integrated Luminosities}\label{sec:analysis_integrated_luminosity}

\subsubsection{X-ray and UV Luminosities:}

As a measure of the X-ray luminosity of the quasars, we integrate the power-law model corrected for Galactic absorption over the rest-frame energy range 1 keV to 25 keV, $L_X$(1-25 keV). We select this energy range as it requires minimal extrapolation from the observed spectrum, given the redshift range of our sample. For ease of comparison with other work, we also provide two alternative estimates of the X-ray luminosity: namely, $L_X$ (0.3-10 keV), integrated from 0.3 keV to 10 keV in the rest-frame (which implies an extrapolation of our model fit into the soft X-rays, and is therefore systematically underestimated if these quasars have a soft excess component), and $L_X$ (XRT Band), integrated from 0.3 keV to 10 keV in the observed frame. The latter involves no extrapolation, but is not directly comparable between objects at different redshifts. Similarly, we estimate the quasar's UV luminosity $L_{\mathrm{UV}}$ by integrating the UV continuum model over the rest-frame interval 1000 \AA\,-- 3000 \AA. These measurements of $L_\mathrm{UV}$ and $L_X$, along with their uncertainties as propagated from the $1\sigma$ errors on the respective model parameters, are tabulated in Table \ref{tab:analysis}. 

\subsubsection{Extreme-UV and Total Luminosities:}

The extreme-UV (EUV) is not directly observable due to strong absorption by the neutral hydrogen and helium in the Milky Way, in the intergalactic medium, and by gas in the quasar host galaxy. This spectral region may harbor the peak energy output of the accretion disk emission feature (\S \ref{sec:introduction}), and must therefore be included in an estimate of the accretion luminosity. Physically motivated modeling of the EUV energy output generally depends on parameters that are not strongly constrained by broadband photometric observations. For example, the accretion disk model used by \citet{Done2012} requires a determination of the black hole mass (or spin). It also requires that the thermal emission peak is constrained by the observations, which is not true of $z\approx2$ quasars observed with \emph{Swift}. Similarly, the truncated power-law model employed by \citet{Korista1997} depends on an inner-disk cutoff temperature that is not constrained by broadband data. For the low-redshift Seyfert galaxy NGC 5548, Kilerci Eser and coworkers find that a simple linear interpolation between the observed UV and X-ray luminosities yields an integrated luminosity roughly halfway between that predicted by the \citet{Done2012} and the \citet{Korista1997} models. The difference between the interpolated luminosity and either of the model predictions is $23\%$ \citep[][ in prep.]{KilerciEser2014,KilerciEser2017}. Due to our ignorance of the EUV SED shape, and following \citet{Grupe2010} and \citet{Wu2012}, we make guideline estimates of the EUV luminosities, $L_\mathrm{EUV}$, by interpolating over the EUV region using a power law function. Due to Lyman-$\alpha$ forest absorption and/or Lyman-$\alpha$ emission in the shortest-wavelength UVOT bandpass, we do not use this data point directly in the interpolation. Instead we use our UV continuum model (\S \ref{sec:analysis_uv_optical_modeling}). However, due to emission line contamination, our UV model systematically overestimates the continuum luminosity. We quantify the resulting systematic uncertainty on the EUV luminosity and on the integrated X-ray to UV luminosity in \S \ref{sec:analysis_integrated_luminosity_uverror}.

Several previous studies find a spectral turnover at roughly 1000 \AA\,\citep{Shang2005,Barger2010,Shull2012,Stevans2014}. We therefore interpolate between the 1000 \AA\,model luminosity and the 1 keV unabsorbed X-ray model luminosity (blue dash-dotted curves, Figures \ref{fig:sed_mainpaper} and \ref{fig:sed1}-\ref{fig:sed15}). We also interpolate between 1000 \AA\, and 0.3 keV, providing an alternative estimate of $L_\mathrm{EUV}$ (blue dashed lines). The latter interpolation may be more appropriate for quasars that lack the soft X-ray excess component; for these $z\approx2$ quasars, we cannot determine whether the soft excess component is present based on the XRT data. In any case, due to our ignorance of the EUV SED, the resulting integrated  luminosities are order-of-magnitude estimates. The work of \citet[][in prep.]{KilerciEser2017} suggests that the uncertainty on $L_\mathrm{EUV}$ due to model assumptions is of order 25\%.

\begin{figure}
	\centering
	\includegraphics[width=\linewidth]{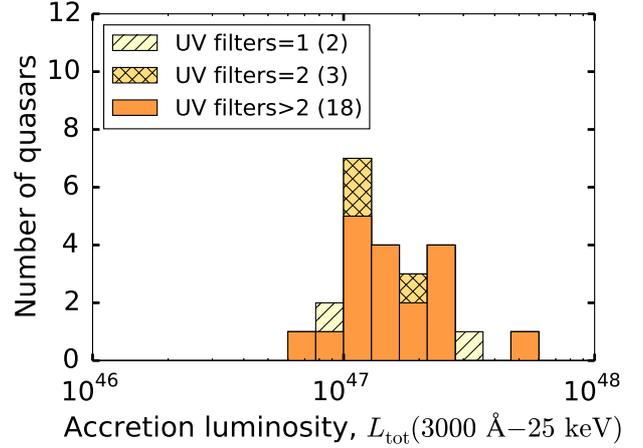}
	\caption{Stacked histogram of $L_{\mathrm{tot}}$ (3000 \AA\,-- 25 keV) for all sample quasars with both XRT and UVOT detections. Quasars with only one or two UV photometric data-points suitable for continuum fitting (\S \ref{sec:analysis_uv_optical_modeling}) are shown as striped and hashed regions, respectively.}
	\label{fig:L_tot}
\end{figure}

We tabulate $L_\mathrm{EUV}$, along with the estimated total UV to X-ray luminosity $L_\mathrm{tot}=L_\mathrm{UV}+L_\mathrm{EUV}+L_{X(1-25)}$, in Table \ref{tab:analysis}. The $1\sigma$ uncertainties on $L_\mathrm{EUV}$ for X-ray detected quasars are calculated by extrapolating between the $1\sigma$ limiting values of the 1000 \AA\,and 1 keV flux densities. The average value of $L_\mathrm{tot}$ for X-ray detected objects in our sample is $1.8\times10^{47}$ erg s$^{-1}$,  with a full span between $L_\mathrm{tot}=6.3\times10^{46}$ erg s$^{-1}$ and $L_\mathrm{tot}=4.5\times10^{47}$ erg s$^{-1}$ (Figure \ref{fig:L_tot}). Thus, the X-ray detected objects in our sample are comparable in terms of accretion luminosity with the more luminous objects presented by \citet{Wu2012}. 

\subsubsection{Systematic uncertainty on $L_{\mathrm{tot}}$ due to UV emission line contribution}\label{sec:analysis_integrated_luminosity_uverror}

\begin{figure}
	\centering
	\includegraphics[width=\linewidth]{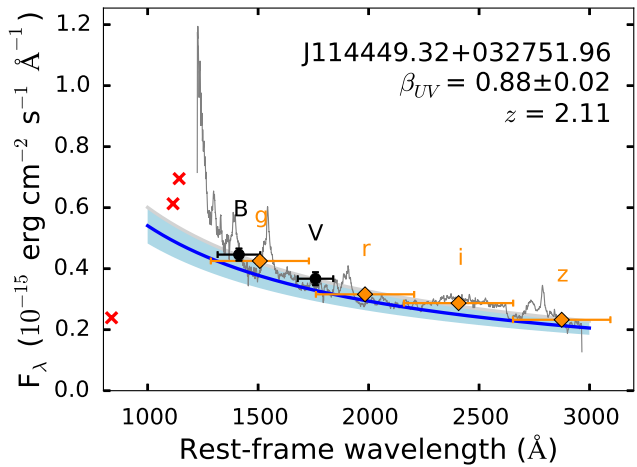}
	\caption{Rescaling of continuum model to match the flux level of the continuum dominated `window' near 1450 \AA. For this quasar, both the UVOT (black points) and SDSS (orange diamonds) photometry bandpasses suffer some emission line contamination. The horizontal error bars indicate the width of the photometric bandpasses. We decrease the model flux (blue curve) by 10\%, at which point it roughly matches the SDSS spectroscopy (dark gray curve) at 1450 \AA. The blue shaded region shows the uncertainty on the rescaled continuum level. The original continuum model fit, before rescaling, is shown as a light gray curve near the upper edge of the blue shaded region.}
	\label{fig:uvot_model_rescaling}
\end{figure}

While we exclude data points that we believe to be strongly contaminated by emission lines from our UV continuum modeling (\S \ref{sec:analysis_uv_optical_modeling}), the remaining bands also have an emission line contribution. In particular, the broad, blended \ion{Fe}{ii} emission lines and Balmer continuum produce a `pseudo-continuum' feature at rest-frame 2000 \AA\,-- 4000 \AA. We estimate the resulting systematic uncertainty as follows. The narrow spectral region around 1450 \AA\, is thought to be almost free of emission line flux \citep[e.g.,][]{VandenBerk2001,Selsing2016}. We therefore adjust the scaling of the UV power-law continuum model until its flux density around 1450 \AA\,roughly matches the continuum flux in this spectral region (Figure \ref{fig:uvot_model_rescaling}). For objects lacking SDSS spectroscopy, we perform this test using the \citet{Selsing2016} quasar template, scaled to the initial continuum model flux at 2500 \AA. We find that the systematic overestimation of the UV continuum flux is approximately 25\% in the worst cases, and 11\% on average. The resulting overestimation of $L_\mathrm{tot}$ is 18\% in the worst cases, and 8\% on average. While the main purpose of this exercise is to obtain an estimate of the \emph{average} systematic uncertainty, we nevertheless provide preliminary corrections to the UV, EUV and total luminosities (Table \ref{tab:analysis_uvcorr}). These corrections are somewhat subjective in nature, and should be used with caution. Note that the UV-optical models shown in Appendix \ref{appendix:sed} are the original model fits, before applying this correction.

\subsubsection{Bolometric Luminosity Estimates Based on UV Data}\label{sec:analysis_integrated_luminosity_Lbol}

\begin{figure}
\centering
\includegraphics[width=\linewidth]{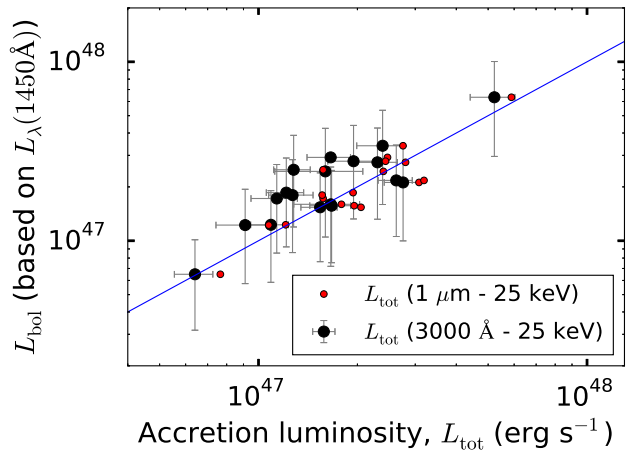}
\caption{Bolometric luminosity estimates based on $L_\lambda(1450\,\mathrm{\AA})$, using the average bolometric correction determined by \citet{Runnoe2012}, as a function of the accretion luminosity $L_{\mathrm{tot}}$ (3000 \AA\,-- 25 keV) (black points), as determined using our \emph{Swift} observations, for X-ray detected quasars in our sample. The blue line illustrates $L_{\mathrm{tot}}=L{\mathrm{bol}}$. Extrapolating the UV power-law continuum to 1 $\mu$m and calculating $L_{\mathrm{tot}}$ (1 $\mu$m -- 25 keV) provides a better agreement with $L_{\mathrm{bol}}$ (red points).}
\label{fig:Ltot_vs_L_BC}
\end{figure}

We make guideline estimates of the bolometric (rest-frame 1 $\mu$m - $8$ keV) luminosities, $L_{\mathrm{bol}}$, of our sample, using the average bolometric correction presented by \citet{Runnoe2012} for a sample of RLQs and RQQs at $z<1.4$ with $\log[L_{\mathrm{bol}}/\mathrm{erg\,s}^{-1}]<47.3$ and assuming isotropic emission. We note that, while the sample presented by \citet{Richards2006} is better matched in terms of redshift to our sample, their bolometric corrections include the infrared spectral region (1 $\mu$m -- 100 $\mu$m), which we do not wish to include in our estimate of the accretion luminosity (\S \ref{sec:sed}). We estimate $L_{\mathrm{bol}}$ using the specific luminosity at rest-frame 1450 \AA, using Equation 9 of \citet{Runnoe2012}. For X-ray detected quasars, the integrated accretion luminosity $L_{\mathrm{tot}}$ agrees with the estimated $L_{\mathrm{bol}}$ to within the $1\sigma$ level for all quasars (Figure \ref{fig:Ltot_vs_L_BC}, black points). However, the $L_{\mathrm{bol}}$ (1 $\mu$m - $8$ keV) estimates are on average 26\% larger than $L_{\mathrm{tot}}$ (3000 \AA\,-- 25 keV) as inferred from our \emph{Swift} data. This is likely due to the smaller spectral window covered by our $L_{\mathrm{tot}}$ measurement. Indeed, if we extrapolate our UV power-law continuum model to $1\mu $m, we find that $L_{\mathrm{bol}}$ is on average only 1\% larger than $L$ (1 $\mu$m -- 25 keV) (Figure \ref{fig:Ltot_vs_L_BC}, red points). We use $L_{\mathrm{bol}}$ to make guideline estimates of the Eddington luminosity for the X-ray non-detections in \S \ref{sec:analysis_eddington_ratios}.

\subsection{Mass Accretion Rates}\label{sec:analysis_accretion_rates}

The mass accretion rate, $\dot{M}$, represents the instantaneous growth rate of the black hole. We estimate $\dot{M}$ for a subset of our sample. Namely, those quasars for which we have spectroscopy covering the \ion{C}{iv} broad emission line, allowing estimation of the black hole mass, $M_{BH}$, and for which we have SDSS photometry, allowing a determination of the optical luminosity $L_{\mathrm{opt}}$. Given these requirements, we estimate $\dot{M}$ for a total of 34 quasars, as follows.

Quasar accretion disks are traditionally modeled as geometrically thin, optically thick $\alpha$-disks (\S \ref{sec:introduction}). For such models, \citet{Raimundo2012} (following \citealt{Davis2011}) show that $\dot{M}$ depends on $M_{BH}$ and $L_{\mathrm{opt}}$ as

\begin{equation}\label{eq:accretion_rate}
\begin{split}
\dot{M}=1.53M_{\astrosun}\mathrm{yr}^{-1}\left(\frac{\nu L_\nu(\mathrm{opt})}{10^{45}\cos(i)\mathrm{\,erg\, s}^{-1}}\right)^{3/2}\\
\left(\frac{M_{BH}}{10^8M_{\astrosun}}\right)^{-1}\frac{\lambda(\mathrm{opt})^2}{(4392\mathrm{\AA})^2}.
\end{split}
\end{equation}
\vspace{0.1cm}

The optical specific luminosity, $L_\nu(\mathrm{opt})$, can be measured at any optical wavelength $\lambda(\mathrm{opt})$. However, $\dot{M}$ has an additional dependence on the black hole spin, which becomes stronger at wavelengths shorter than 4000 \AA\,\citep[as illustrated by Figure 1 of][]{Davis2011}. To avoid extrapolation of our continuum model beyond the wavelength coverage of the data, we use $\nu L_\nu$ as measured at the pivot wavelength of the SDSS $z$ bandpass (pivot wavelength $\sim$2000--3500 \AA\,in the rest frame). This bandpass is our reddest photometric data point, and thus minimizes the dependence of $\dot{M}$ on black hole spin, although this dependence is not negligible at these wavelengths; we will quantify this uncertainty in future work, upon completion of our \emph{Swift} observing program. We also require estimates of $M_{BH}$ and the accretion disk inclination $i$, as detailed below.

\paragraph*{Black Hole Mass Estimates:} We use the scaling relationship presented by \citet{Vestergaard2006} (their equation 7) to calculate single-epoch spectroscopic black hole mass estimates, based on the \ion{C}{iv} FWHM \citep{Vestergaard2000b} and the monochromatic continuum luminosity $L_\lambda$ (1350~\AA). This luminosity is determined by extrapolating the  power-law continuum with slope and normalization at 1550~\AA, as presented by \citet{Vestergaard2003} for our sample quasars. For the 12 quasars in our SDSS sample (\S \ref{sec:introduction}), we measure $L_\lambda$ (1350~\AA) and the \ion{C}{iv} FWHM in the SDSS spectroscopy; if the \ion{Mg}{ii} broad emission line is also covered by the SDSS spectra, we use a variance-weighted average of $M_{BH}$ calculated using \ion{C}{iv} and \ion{Mg}{ii} (\citealt{Vestergaard2009}, Equation 1), with the variance based on the uncertainties of the spectral measurements. Our sample has an average black hole mass of $\langle M_{BH}\rangle=(5.0\pm0.7)\times10^9M_{\astrosun}$, with a full range in black hole mass of $0.6\le M_\mathrm{BH}/10^9M_{\astrosun}\le18.0$, similar to that of $z\approx2$ quasars in the Large Bright Quasar Survey \citep{Vestergaard2009}. Typical uncertainties on $M_{BH}$ are of order 0.6 dex, and are dominated by the scatter of the \citet{Vestergaard2006} scaling relationship with respect to the $M_{BH}$ determined by reverberation mapping, and by the uncertainty on the normalization of the reverberation-mapping mass scale itself, as determined by \citet{Onken2004}.

\paragraph*{Inclination Angle Estimates:} Equation \ref{eq:accretion_rate} contains a dependency on the disc inclination, $i$, as this determines the solid angle subtended by the disc on the sky, and therefore affects the observed luminosity. For the RLQs in our sample, we are able to make a crude estimate of $i$, as follows. If the radio jet is launched perpendicularly to the disk plane, and the jet does not twist, the radio jet inclination is a proxy for the accretion disk inclination. Here we use the radio spectral index, $\alpha_R$, as an indicator of the radio jet inclination \citep[e.g.,][]{Jarvis2006}. \citet{Padovani1992} apply a radio emission beaming model to the observed luminosity functions of RLQ and radio galaxies. In this scenario, steep-spectrum radio sources (SSS, defined as sources with radio spectral index $\alpha_R<-0.6$)  are the unbeamed parent population of flat-spectrum sources (FSS, $\alpha_R>-0.6$), but become obscured in the UV-optical (and are therefore seen as radio galaxies) for the largest inclinations. They find that flat-spectrum radio sources  are aligned with the line-of-sight to within $14^{\circ}$, while steep spectrum sources  are aligned at $14^{\circ}\le i\le40^{\circ}$, with an expectation value of $\langle i\rangle=28^{\circ}$. Based thereon, we assume accretion disk inclinations $i=10^\circ$ for FSS and $i=28^{\circ}$ for SSS. We calculate $\alpha_R$ between 408 MHz and 5 GHz if these measurements are available, or between 1400 MHz and 5 GHz otherwise (Table \ref{tab:sample_selection}). For our SDSS RLQ sample (4 objects), we only have single-band radio data, at 1.4 GHz as observed by FIRST \citep{Becker1995}, and therefore lack a measurement of $\alpha_R$. For the RQQs, and for the four SDSS RLQs (for which we only have single-band radio data and therefore lack measurements of $\alpha_R$), we assume $\langle i\rangle=28^{\circ}$, the average value of $i$ for a distribution of random orientations satisfying $0^{\circ}\le i\le40^{\circ}$. Likewise, for sources for which the radio spectra are known to peak at GHz frequencies, we assume $i=28^\circ$, as their radio SED shape is thought to be linked to their environment or youth, and not primarily to inclination \citep[e.g.,][]{Fanti1995,Orienti2016}. In practice the uncertainty on $\dot{M}$ is dominated by the $M_{BH}$ uncertainty, and therefore these guideline estimates of $i$ suffice for the current purpose.

\paragraph*{Mass Accretion Rate Estimates:} Using Equation \ref{eq:accretion_rate}, the average mass accretion rate for the 34 quasars examined here is $\langle\dot{M}\rangle=(6.7\pm1.3)M_{\astrosun}$ yr$^{-1}$, with a full range of $1.0<\dot{M}/(M_{\astrosun}\mathrm{yr}^{-1})<33.1$. We find no significant difference between the distributions of $\dot{M}$ for RLQs ($\langle\dot{M}_{\mathrm{RLQ}}\rangle=(7.3\pm1.7)M_{\astrosun}$ yr$^{-1}$) and RQQs ($\langle\dot{M}_{\mathrm{RQQ}}\rangle=(5.9\pm1.9)M_{\astrosun}$ yr$^{-1}$). The distribution is asymmetric, with more than half the sample having $\dot{M}\le6$ $M_{\astrosun}$ yr$^{-1}$ (Figure \ref{fig:Mdot_histogram}); the median $\dot{M}$ is just 3.3 $M_{\astrosun}$ yr$^{-1}$. We note that Equation \ref{eq:accretion_rate} is derived assuming a purely thermal accretion disk spectrum without emission lines. Existing analyses of high-redshift quasar spectroscopy that include decomposition into continuum and emission-line components \citep{Dietrich2003,Selsing2016} indicate that the typical emission line contribution to $L_\nu(\mathrm{opt})$ is of order 10\%-20\%, primarily due to the Balmer continuum and \ion{Fe}{ii} emission; our preliminary analysis of the overestimation of the power-law continuum level indicates a similar emission line contribution (\S \ref{sec:analysis_integrated_luminosity_uverror}). We calculate alternative estimates of $\dot{M}$ assuming a 20\% emission line contribution to the measured $L_{\mathrm{opt}}$, which decreases the average mass accretion rate to $\langle\dot{M}\rangle=(4.8\pm0.9) M_{\astrosun}$ yr$^{-1}$. For comparison, \citet{Davis2011} find $\langle\dot{M}\rangle=2.4M_{\astrosun}$ yr$^{-1}$ for $z\apprle2$ Palomar Green quasars with $M_{\mathrm{BH}}>10^9M_{\astrosun}$, similar to the quasars studied here, while \citet{Capellupo2015} find $\langle\dot{M}\rangle=5.1M_{\astrosun}$ yr$^{-1}$ for quasars with $M_{\mathrm{BH}}>10^9M_{\astrosun}$ in their SDSS-selected, $1.45\le z\le1.65$ sample. In summary, our sample quasars have similar mass accretion rates to previous studies of quasars with comparable $M_{BH}$ values; we note that all of these studies base their estimates on variants of Equation \ref{eq:accretion_rate}, and therefore share any systematic uncertainties inherent to the thin-disk approximation.

\begin{figure}
	\centering
	\includegraphics[width=\linewidth]{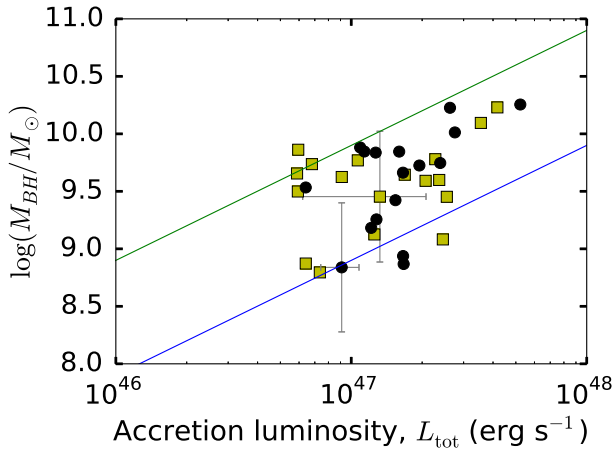}
	\caption{Accretion luminosities, $L_{\mathrm{tot}}$ (3000 \AA\,-- 25 keV) versus black hole masses for our sample. Black points are X-ray detected quasars, while yellow squares show X-ray non-detections, for which we use bolometric corrections to estimate $L_{\mathrm{tot}}$ (\S \ref{sec:analysis_integrated_luminosity_Lbol}). The green and blue lines trace Eddington luminosity ratios of $\lambda=0.1$ and $\lambda=1$, respectively. For clarity, we only show representative error bars on one X-ray detected object and one non-detection.
	}
	\label{fig:eddington}
\end{figure}

\begin{figure}
\centering
\includegraphics[width=\linewidth]{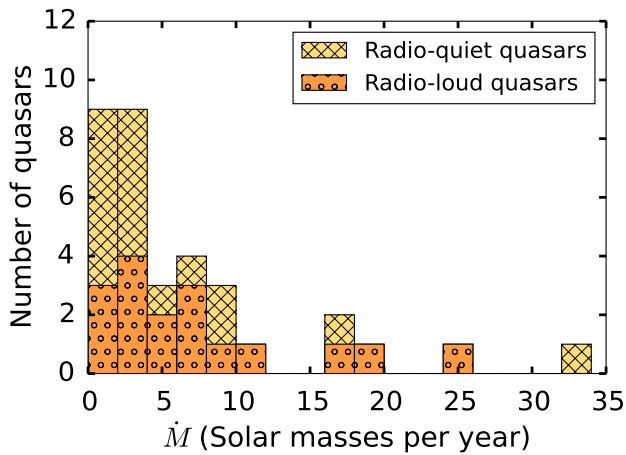}
\caption{The distribution of $\dot{M}$ for our sample, as estimated using Equation \ref{eq:accretion_rate}. The correction for emission line contamination of $L_{\mathrm{opt}}$ (\S \ref{sec:analysis_accretion_rates}) is not included here. Stacked histogram (as defined in Figure \ref{fig:z_histogram}).}
\label{fig:Mdot_histogram}
\end{figure}

\subsection{Eddington Luminosity Ratios}\label{sec:analysis_eddington_ratios} 

For spherically symmetric accretion of ionized hydrogen, the Eddington luminosity, given by $L_{\mathrm{Edd}}=4\pi Gm_pM_{\mathrm{BH}}c\sigma_T^{-1}$, is the limiting luminosity for which radiation pressure balances gravitational attraction\footnote{The equivalent limit for a thin-disk geometry may in some cases be larger than $L_{\mathrm{Edd}}$ by up to a factor $\sim5$ \citep{Abolmasov2015}. As we do not perform detailed modeling of the accretion disks here, we assume spherical symmetry, as is standard practice when calculating $L_{\mathrm{Edd}}$ for large samples of AGN \citep[e.g.,][]{Trakhtenbrot2012}.}; here, $m_p$ is the proton mass, $\sigma_T$ is the cross-section for Thomson scattering, $G$ is the gravitational constant, and $c$ is the speed of light. The Eddington luminosity ratio, $\lambda=L_{\mathrm{tot}}/L_{\mathrm{Edd}}$, is thus a measure of how close to the theoretical limiting luminosity a given supermassive black hole is accreting. We use our estimates of $L_{\mathrm{tot}}$ (1 $\mu$m -- 25 keV) (\S \ref{sec:analysis_integrated_luminosity}) and of  $M_{BH}$ (\S \ref{sec:analysis_accretion_rates}) to calculate $\lambda$ for X-ray detected quasars with SDSS photometry (21 objects in total), and find an average $\langle\lambda\rangle=0.52\pm0.10$, with a full range $0.11\le\lambda\le1.75$. All objects are consistent with sub-Eddington accretion to within the $1\sigma$ uncertainties on $\lambda$ (as indicated by the blue line in Figure \ref{fig:eddington}). For the X-ray non-detected quasars, we instead calculate the Eddington ratio using the estimated bolometric luminosity, $L_{\mathrm{bol}}$ (\S \ref{sec:analysis_integrated_luminosity_Lbol}). For these quasars we find $\langle\lambda\rangle=0.45\pm0.09$, consistent with the average value for X-ray detected quasars. As our X-ray non-detections are mostly RQQs, we thus find that RQQs and RLQs in our sample are accreting at similar Eddington fractions.

%% file: sec7discussion.tex
\section{Discussion and Summary}\label{sec:discussion}

We present the first results of an ongoing observing campaign of redshift $\sim2$ quasars with \emph{Swift}. The full sample of quasars will span a large range of redshifts, radio loudnesses, and, for the RLQs, accretion disk orientation angles. All X-ray detected objects presented here are radio-loud. We estimate the accretion luminosity for each X-ray detected quasar (\S \ref{sec:analysis_integrated_luminosity}), finding an average value of $L_\mathrm{tot}$(3000 \AA\,-- 25 keV) $=1.8\times10^{47}$ erg s$^{-1}$, similar to the brightest quasars studied by \citet{Wu2012}; our RLQs appear to be accreting close to the Eddington limit (\S \ref{sec:analysis_accretion_rates}). Our UV-optical spectral indices, based on broadband photometry, are somewhat flatter (redder) than those typically seen for spectroscopically studied SDSS quasars. This is due to emission line contamination in the photometric bandpasses (\S \ref{sec:analysis_UV_beta}), which we will address in detail in future work. We find an average UV-optical to X-ray spectral index $\langle\alpha_\mathrm{ox}\rangle=1.39\pm0.03$ for the X-ray detected sources, consistent with that found for RLQs in the literature (\S \ref{sec:analysis_alphaox}). While we do not detect any RQQs in the X-ray in the current sample, the lower limits derived for \aox are consistent with previous studies of RQQs of comparable luminosities. In summary, we find that the quasars in this preliminary sample display broadband SED shapes typical of quasars at $z\approx2$. 

We do, however, find the RLQs in our sample to have unusually hard X-ray spectra on average, compared to similar quasar samples in the literature (\S \ref{sec:analysis_Gamma}). This is likely due to an over-representation of flat-spectrum radio sources (FSS) in the current sample. According to the \citet{Padovani1992} study discussed in \S \ref{sec:analysis_accretion_rates}, RLQ are seen as blazars for $\phi<5^{\circ}$, as FSS for $5^{\circ}<\phi<15^{\circ}$, and as steep-spectrum sources for $15^{\circ}<\phi<40^{\circ}$. Thus, in a sample of bright, non-blazar-like RLQ with otherwise randomly drawn inclination angles, we would expect only around 13\% of the quasars to be FSS. In fact, based on the radio observations listed in Table \ref{tab:sample_selection}, at least 52\% of our X-ray detected objects are FSS; we lack multi-band radio data for two objects. We find $\langle\Gamma\rangle=1.37\pm0.06$ for the FSS, compared to $\langle\Gamma\rangle=1.60\pm0.10$ for the remainder of the sample, although the sample size is currently too small to conclusively show a bimodality in $\Gamma$. Similarly, \citet{Worrall1989} find that FSS sources display smaller $\Gamma$ than do other $z>0.8$ RLQs. It is well-known that RLQs as a class show harder X-ray SEDs than do RQQs at comparable luminosities \citep[e.g.,][]{Williams1992,Page2005,Scott2011}. This suggests that the X-ray emission due to the radio jet has a harder SED than that of the corona. This is consistent with the jet component possibly being more dominant for FSS relative to other RLQs, due to geometric effects, e.g., relativistic beaming of the jet X-ray emission for these small-angle sources.

Our sample is well-suited to the study of the radiative efficiency of quasar accretion due to the simultaneous nature of the \emph{Swift} X-ray and UV observations, the relatively low level of dust obscuration, and to the availability of radio data for RLQs, which allows estimation of the accretion disk inclination angle. We will present a more detailed analysis of these quasars, including an analysis of their radiative efficiencies, in the context of a larger sample of quasars including X-ray detected RQQs, upon completion of our \emph{Swift} observing program.

%% file: tables/sample_table.tex
\begin{deluxetable}{lrrrllllllr}
	%\rotate
	\tablecolumns{11}
	\tablewidth{0pc}
	\tabletypesize{\footnotesize}
	\tablecaption{Quasars in our current sample.\label{tab:sample_selection}}
	\tablehead{
		\colhead{Object} & \colhead{RA} & \colhead{Dec.} & \colhead{Redshift} & \colhead{$m_V$}& \colhead{Radio} & \colhead{$F_{408\mathrm{MHz}}$} & \colhead{$F_{1.4\mathrm{GHz}}$} &\colhead{$F_\mathrm{5GHz}$} & \colhead{$\alpha_{5000}^{1400}$} & \colhead{SDSS} \\
		\colhead{name} & \colhead{[deg]} & \colhead{[deg]} & \colhead{} & \colhead{[mag]} & \colhead{class} & \colhead{[mJy]} & \colhead{[mJy]} & \colhead{[mJy]} & \colhead{ } & \colhead{phot.?}\\
		\colhead{(1)} & \colhead{(2)} & \colhead{(3)} & \colhead{(4)} & \colhead{(5)} & \colhead{(6)} & \colhead{(7)} & \colhead{(8)} & \colhead{(9)} & \colhead{(10)} & \colhead{(11)}
	}
	\startdata
      J014725.50-101439.11 &  26.8562   & -10.2442   &    2.14777 & 17.41& RQQ & --- & --- & --- & ---  & \checkmark    \\
      J082328.62+061146.07 & 125.869    &   6.19613  &    2.81524 & 18.0 & RLQ & --- & (67.1) & --- & --- & \checkmark  \\
      J094853.60+085514.40 & 147.223    &   8.92067  &    1.98179 & 17.3 & RLQ & --- & (33.1) & --- & --- & \checkmark  \\
      J104915.44-011038.18 & 162.312    &  -1.1772   &    2.12943 & 17.9 & RQQ & --- & --- & --- & --- & \checkmark  \\
      J110607.48-173113.60 & 166.531    & -17.5204   &    2.54925 & 17.8 & RQQ & --- & --- & --- & --- & \checkmark  \\
      J111159.70+023719.76 & 167.999    &   2.62215  &    1.88247 & 17.8 & RQQ & --- & --- & --- & --- & \checkmark  \\
      J112542.30+000101.33 & 171.426    &   0.017037 &    1.69230 & 17.6 & RLQ & --- & (29.9) & --- & --- & \checkmark  \\
      J114449.32+032751.96 & 176.206    &   3.46443  &    2.10035 & 17.7 & RQQ & --- & --- & --- & --- & \checkmark  \\
\enddata
\end{deluxetable}

%% file: tables/xrt_table.tex
\begin{deluxetable}{lcccccccccrcc}
	\rotate
	\tablecolumns{13}
	\tablewidth{0pc}
	\tabletypesize{\footnotesize}
	\tablecaption{\emph{Swift} XRT observation log and X-ray modeling. \label{tab:xrtdata}}
	\tablehead{
		\colhead{Object} & \colhead{$N_{obs}$\tablenotemark{a}} & XRT exp. & \colhead{XRT} & $\sigma_{\mathrm{det}}$ & \colhead{$F_{0.3-10}$} & 
		\colhead{$F_\nu$(2 keV)} & \colhead{$\Gamma$} & \colhead{$N_H$} &\colhead{$N_\mathrm{sub}$} & 
		\colhead{$HR$} & \colhead{$N_{\mathrm{soft}}$} & \colhead{$N_{\mathrm{soft}}$} \\
		\colhead{name} & \colhead{} & \colhead{time [s]} & \colhead{detect?} & \colhead{} & \colhead{} & \colhead{} &\colhead{[$10^{22}$ cm$^{-2}$]} & \colhead{} & \colhead{} & \colhead{} & \colhead{} & \colhead{(background)}\\
		\colhead{(1)} & \colhead{(2)} & \colhead{(3)} & \colhead{(4)} & \colhead{(5)} & \colhead{(6)} & \colhead{(7)} & \colhead{(8)} & \colhead{(9)} & \colhead{(10)}
		& \colhead{(11)} & \colhead{(12)} & \colhead{(13)}}
	\startdata
	J014725.50-101439.11 & 1  &  6650 &  (P)       &-0.2 &$\le0.63 $                 & $  \le0.03 $      & ---                      & 0.0268 & -0.4 & ---                       & 0 & 1.2 \\%30(1.2) \\
	J082328.62+061146.07 & 3  & 11490 & \checkmark & 8.7 &$ 2.94 ^{+0.30 }_{-0.36}$  & $ 5.40 \pm 0.98 $ & $1.85 ^{+0.18} _{-0.17}$ & 0.0229 & 84.4 & $-0.18 ^{+0.11} _{-0.12}$ & 55 & 5.1\\ %128(5.1) \\
	J094853.60+085514.40 & 5  &  5139 & \checkmark & 6.3 &$ 4.05 ^{+0.54 }_{-1.16}$  & $ 4.27 \pm 1.03 $ & $1.56 ^{+0.22} _{-0.22}$ & 0.0252 & 42.4 & $-0.18 ^{+0.15} _{-0.16}$ & 27 & 1.3\\ %32(1.3) \\
	J104915.44-011038.18 & 2  & 11001 &            & 1.5 &$ \le0.54 $                & $ \le0.91 $       & ---                      & 0.0363 & 5.1  & $-0.11 ^{+0.24} _{-0.56}$ & 6 & 3.1\\ %77(3.1) \\
	J110607.48-173113.60 & 1  &  5769 &            & 1.8 &$ \le1.16 $                & $ \le2.21 $       & ---                      & 0.0439 & 5.4  & $-0.25 ^{+0.23} _{-0.53}$ & 5 & 1.6\\ %39(1.6) \\
	J111159.70+023719.76 & 2  &  5711 &            & 1.0 &$ \le1.26 $                & $ \le1.97 $       & ---                      & 0.0367 & 2.3  & $0.30 ^{+0.70} _{-0.20}$  & 2 & 1.7\\ %42(1.7) \\
	J112542.30+000101.33 & 1  &  5364 & \checkmark & 7.5 &$ 5.25 ^{+0.69}_{-1.06 }$  & $ 6.42 \pm 1.18 $ & $1.73 ^{+0.19} _{-0.19}$ & 0.0297 & 59.4 & $-0.22 ^{+0.12} _{-0.13}$ & 38 & 1.7\\ %42(1.7) \\
	J114449.32+032751.96 & 2  &  5507 &            & 0.9 &$ \le0.97 $                & $ \le1.62 $       & ---                      & 0.0199 & 2.2  & $0.04 ^{+0.96} _{-0.06}$  & 3 & 2.0\\
\enddata
\end{deluxetable}

%% file: tables/uvot_table.tex
\begin{deluxetable}{lcccccccc}
	\tablecolumns{9}
	\tablewidth{0pc}
	\tabletypesize{\footnotesize}
	\tablecaption{\emph{Swift} UVOT flux measurements.\label{tab:uvotfilters}}
	\tablehead{
		\colhead{Object} & \colhead{E(B-V)} & \colhead{$F_\lambda$} & \colhead{$F_\lambda$} & \colhead{$F_\lambda$} & \colhead{$F_\lambda$} & \colhead{$F_\lambda$} & \colhead{$F_\lambda$} & \colhead{SDSS matching}\\
		\colhead{name} & \colhead{} & \colhead{UW1} & \colhead{UW2} & \colhead{UM2} & \colhead{U} & \colhead{B} & \colhead{V} & \colhead{bandpass}\\
		\colhead{(1)} & \colhead{(2)} & \colhead{(3)} & \colhead{(4)} & \colhead{(5)} & \colhead{(6)} & \colhead{(7)} & \colhead{(8)} & \colhead{(9)}\\
		}
	\startdata
	J0147-1014\tablenotemark{a} & $0.0303\pm0.0009$ &  $7.85\pm1.21$ &  $<4.48$ &  $<1.48$ & $53.07\pm3.06$ & $39.73\pm2.45$ & $30.94\pm2.95$ & \emph{B} \\      
	J0823+0611\tablenotemark{a} & $0.0224\pm0.0003$ & --- & --- & $24.92\pm1.30$ & $28.87\pm1.25$ & $33.28\pm1.25$ & $24.11\pm1.16$ & \emph{B} \\
	J0948+0855\tablenotemark{a} & $0.0258\pm0.0005$ & $49.25\pm2.76$ & $10.37\pm1.87$ & $9.37\pm1.02$ & $78.18\pm3.33$ & $60.05\pm2.57$ & $46.51\pm2.29$ & \emph{B} \\
	J1049-0110\tablenotemark{a} & $0.0349\pm0.0013$ & $<6.50$ & $<2.73$ & $<1.37$ & $<8.30$ & $<11.01$ & $<15.81$ & --- \\
	J1106-1731\tablenotemark{a} & $0.0395\pm0.0011$ & --- & --- & --- & $28.51\pm1.16$ & $44.95\pm1.51$ & $33.52\pm1.30$ & \emph{V} \\
	J1111+0237\tablenotemark{a} & $0.033 \pm0.0012$ & $52.32\pm2.75$ & --- & $52.38\pm2.66$ & $65.61\pm2.83$ & $43.24\pm1.86$ & $30.32\pm1.84$ & \emph{B} \\
	J1125+0001\tablenotemark{a} & $0.0258\pm0.0003$ & $45.14\pm2.42$ & --- & --- & $79.27\pm3.25$ & $46.91\pm1.77$ & $27.31\pm1.40$ & \emph{V} \\
	J1144+0327\tablenotemark{a} & $0.0186\pm0.0005$ & $23.91\pm1.46$ & --- & $3.61\pm0.56$ & $61.29\pm2.85$ & $44.61\pm1.99$ & $36.70\pm2.12$ & \emph{B} \\
\enddata
	\tablecomments{(1) Object name. (2) The differential extinction E(B-V) is adopted from the work of \citet{Schlafly2011}, and is the mean value of all measurements within a 2 arcminute aperture centered on the quasar coordinates. (3,4,5,6,7,8) The effective flux density at the pivot wavelength in each UVOT bandpass, in unit of $10^{17}$ erg s$^{-1}$ \AA$^{-1}$, assuming a power law source SED over the bandpass (\S \ref{sec:method_uvot_extraction}). The 1$\sigma$ uncertainties on $F_\lambda$ are listed for detected quasars. The 5$\sigma$ upper limits are listed for non-detections. Filters that were not used for a given observation are marked with "---". (9) The UVOT filter bandpass used for rescaling of the SDSS photometric fluxes (\S \ref{sec:sdss}). Quasars which lack SDSS photometry, or which are not detected in UVOT imaging, are marked with "---".}
	\tablenotetext{a}{For typographical reasons, we shorten the names of the SDSS quasars in this table. See Table \ref{tab:sample_selection} for the full names.}
	%\tablenotetext{a}{The differential extinction E(B-V) is adopted from the work of \citet{Schlafly2011}, and is the mean value of all measurements within a 2 arcminute aperture centered on the quasar coordinates.}
	%\tablenotetext{b}{The effective flux density at the bandpass pivot wavelength, in unit of $10^{17}$ erg s$^{-1}$ \AA$^{-1}$, assuming a power law source SED over the bandpass (\S \ref{sec:method_uvot_extraction}). The 1$\sigma$ uncertainties on $F_\lambda$ are listed for detected quasars. The 5$\sigma$ upper limits are listed for non-detections. Filters that were not used for a given observation are marked with "---".}
	%\tablenotetext{c}{The UVOT filter bandpass used for rescaling of the SDSS photometric fluxes (\S \ref{sec:sdss}). Quasars which lack SDSS photometry, or which are not detected in UVOT imaging, are marked with "---".}
\end{deluxetable}

%% file: tables/uvot_table_mags.tex
\begin{deluxetable}{lccccccc}
	\tablecolumns{7}
	\tablewidth{0pc}
	\tabletypesize{\footnotesize}
	\tablecaption{\emph{Swift} UVOT apparent magnitudes.\label{tab:uvotfilters_magnitudes}}
	\tablehead{
		\colhead{Object} & \colhead{$m_{UW1}$} & \colhead{$m_{UW1}$} & \colhead{$m_{UM2}$} & \colhead{$m_U$} & \colhead{$m_B$} & \colhead{$m_V$} \\
		\colhead{(1)} & \colhead{(2)} & \colhead{(3)} & \colhead{(4)} & \colhead{(5)} & \colhead{(6)} & \colhead{(7)}
		}
	\startdata
	% name                 UW1              UW2              UM2        U                B                 V
	J014725.50-101439.11 & $19.26\pm0.17$ & $<20.19$       &  $<21.24$ & $17.05\pm0.06$ & $18.03\pm0.07$ & $17.71\pm0.10$  \\      
	J082328.62+061146.07 & ---        & ---        & $18.18\pm0.06$ & $17.72\pm0.05$ & $18.22\pm0.04$ & $17.98\pm0.05$  \\
	J094853.60+085514.40 & $17.27\pm0.06$ & $19.28\pm0.20$ & $19.24\pm0.12$ & $16.64\pm0.05$ & $17.58\pm0.05$ & $17.26\pm0.05$ \\
	J104915.44-011038.18 & $<19.47$        & $<20.73$        & $<21.33$ & $<19.07$ & $<19.42$ & $<18.44$ \\
	J110607.48-173113.60 &  ---           & ---            & ---     & $17.73\pm0.04$ & $17.90\pm0.04$ & $17.62\pm0.04$ \\
	J111159.70+023719.76 & $17.20\pm0.06$ & ---            & $17.37\pm0.06$ & $16.83\pm0.05$ & $17.94\pm0.05$ & $17.73\pm0.07$ \\
	J112542.30+000101.33 & $17.36\pm0.06$ & ---            & ---     & $16.62\pm0.04$ & $17.85\pm0.04$ & $17.84\pm0.06$ \\
	J114449.32+032751.96 & $18.05\pm0.07$ & ---            & $20.27\pm0.17$ & $16.90\pm0.05$ & $17.91\pm0.05$ & $17.52\pm0.06$ \\
\enddata
	\tablecomments{(1) Object name. (2,3,4,5,6,7) Apparent magnitudes on the UVOT photometric system. The 1$\sigma$ uncertainties are listed for detected quasars. The 5$\sigma$ upper limits are listed for non-detections. Filters that were not used for a given observation are marked with "---". }
	%\tablenotetext{a}{Apparent magnitudes on the UVOT photometric system. The 1$\sigma$ uncertainties are listed for detected quasars. The 5$\sigma$ upper limits are listed for non-detections. Filters that were not used for a given observation are marked with "---".}
\end{deluxetable}

%% file: tables/detection_table.tex
\begin{deluxetable}{lcccc}
	\tablecolumns{9}
	\tablewidth{0pt}
	\tabletypesize{\footnotesize}
	\tablecaption{Summary of \emph{Swift} detections for our sample.\label{tab:detection_summary}}
	\tablehead{
		\colhead{} & \colhead{UVOT} & \colhead{Detected in 2} & \colhead{Detected in $\ge3$} & \colhead{Sum\tablenotemark{a}}\\
		\colhead{} & \colhead{non-detection} & \colhead{UVOT filters} & \colhead{UVOT filters}}
	\startdata
	XRT non-detection    & 2 & 3 & 16 & 21 \\
	XRT detection        & 0 & 4 & 19 & 23 \\
	Sum\tablenotemark{b} & 2 & 7 & 35 & (sample of 44) \\
	\enddata
	\tablenotetext{a}{Number of XRT detections and non-detections, irrespective of UVOT detection status.}
	\tablenotetext{b}{Number of objects with a given UVOT detection status, irrespective of XRT detection status. All detected objects were detected in two or more bandpasses.}
\end{deluxetable}

%% file: tables/analysis_table.tex
% beta_table,F2500_dnu,aox_table,L_int_XRT,L_int_UVOT,L_estimated_UV

\begin{deluxetable}{lcccccccccc}
	\rotate
	\tablecolumns{11}
	\tablewidth{0pc}
	\tabletypesize{\footnotesize}
	\tablecaption{Integrated luminosities and SED diagnostics.\label{tab:analysis}}
	\tablehead{
		\colhead{Object} & \colhead{$\beta_\mathrm{UV}$} & \colhead{$F_\nu$(2500 \AA)} & \colhead{\aox} & \colhead{$L_{\mathrm{X}}$}& \colhead{$L_{\mathrm{X}}$\tablenotemark{e}}& \colhead{$L_{\mathrm{X}}$} & \colhead{$L_\mathrm{UV}$} & \colhead{$L_\mathrm{EUV}$} & \colhead{$L_\mathrm{EUV}$} & \colhead{$L_\mathrm{tot}$} \\
		\colhead{name} & \colhead{} & \colhead{} & \colhead{} & \colhead{2--10 keV} & \colhead{(0.3--10 keV)} & \colhead{(1--25 keV)} & \colhead{(1000--3000\AA)} & \colhead{(1000\AA--0.3keV)} & \colhead{(1000\AA--1keV)} & \colhead{(3000\AA--25keV)}\\
		\colhead{(1)} & \colhead{(2)} & \colhead{(3)} & \colhead{(4)} & \colhead{(5)} & \colhead{(6)} & \colhead{(7)} & \colhead{(8)} & \colhead{(9)} & \colhead{(10)} & \colhead{(11)}
		}
\startdata  % name                  beta               F_nu2500         alpha_ox  LXobs LX0310 L1-25  L_NUV L_FUV03 L_FUV1 L_tot
J0147-1014\tablenotemark{a} & $ 1.28 \pm 0.02$ & $ 3.90 \pm 0.27 $ & $\ge 1.94$ & --- & --- & --- & $6.55\pm0.76$ & --- & --- & --- \\         
J0823+0611\tablenotemark{a} & $ 1.68 \pm 0.02$ & $ 2.69 \pm 0.36 $ & $ 1.42 \pm 0.08 $ & $0.92^{+0.10}_{-0.07}$ & $1.75^{+0.49}_{-0.77}$ & $1.89^{+0.08}_{-0.41}$ & $8.75\pm1.18$ & $10.12_{-2.28}^{+2.13}$ & $14.60_{-2.54}^{+2.51}$ & $25.23_{-3.96}^{+3.93}$\\
J0948+0855\tablenotemark{a} & $ 1.02 \pm 0.02$ & $ 6.45 \pm 0.63 $ & $ 1.60 \pm 0.10 $ & $0.51^{+0.07}_{-0.07}$ & $0.78^{+0.10}_{-0.09}$ & $1.13_{-0.17}^{+0.22}$ & $8.45\pm0.86$ & $5.49_{-1.37}^{+1.12}$ & $8.50_{-1.49}^{+1.40}$ & $18.09_{-2.57}^{+2.49}$\\
J1049-0110\tablenotemark{a} & --- &  --- &  --- &  --- & --- & --- & --- & --- & --- & --- \\
J1106-1731\tablenotemark{a} & $ 1.53 \pm 0.01$ & $ 4.18 \pm 0.41 $ &  $\ge 1.65$ & --- & ---& --- & $10.60\pm1.07$ & --- & --- & --- \\
J1111+0237\tablenotemark{a} & $ 1.72 \pm 0.02$ & $ 3.09 \pm 0.30 $ &  $\ge 1.61$ & --- & ---& --- & $4.98\pm0.50$ & --- & --- & --- \\
J1125+0001\tablenotemark{a} & $ 1.54 \pm 0.02$ & $ 3.01 \pm 0.31 $ &  $ 1.41 \pm 0.08 $ & $0.49_{-0.08}^{+0.07}$ & $0.86_{-0.09}^{+0.08}$ & $1.04_{-0.18}^{+0.17}$ & $3.85\pm0.42$ & $3.91_{-0.97}^{+0.87}$ & $5.88_{-1.04}^{+1.02}$ & $10.55_{-1.64}^{+1.66}$ \\
J1144+0327\tablenotemark{a} & $ 1.15 \pm 0.03$ & $ 4.85 \pm 1.92 $ &  $\ge 1.72$ &  --- & --- & --- & $7.57\pm0.76$ & --- & --- & ---\\
\enddata
\end{deluxetable}

%% file: tables/gamma_table.tex
\begin{deluxetable}{lccccc}
	\tablecolumns{6}
	\tablewidth{0pc}
	\tabletypesize{\footnotesize}
	\tablecaption{Distributions of $\Gamma$ in previous studies of RLQ.\label{tab:Gamma_literature}}
	\tablehead{
		\colhead{Study} & \colhead{Redshift} & \colhead{Objects} & \colhead{$\langle\Gamma\rangle$\tablenotemark{a}} & $\sigma_\Gamma$\tablenotemark{b} & \colhead{Energy} \\
		\colhead{} & \colhead{range} & \colhead{in sample} &  &  & \colhead{range [keV]\tablenotemark{c}} \\
		\colhead{(1)} & \colhead{(2)} & \colhead{(3)} & \colhead{(4)} & \colhead{(5)} & \colhead{(6)}
	}
	\startdata
	%\multicolumn{6}{c}{Radio-loud AGN}\\
	This study & $1.5<z<3.6$ & 23 & $1.46\pm0.05$ & 0.23 & 0.3--10, obs. \\
	%This study & $1.5<z<3.6$ & 23 & $1.23\pm0.12$ & 0.59 & 2-10 \\
	\citet{Wu2012}\tablenotemark{d} & $1.5>z>5.5$  & 15 & $1.74\pm0.11$ & 0.42 & 0.3--10, obs. \\
	\citet{Wu2013} & $z\apprge4$ & 16 & $1.60\pm0.06$ & 0.23 & 2--10, obs. \\
	\citet{Scott2011} & $z<6$  & 75 & $1.86\pm0.02$ & $0.20$ & 0.5--12 \\ 
	\citet{Page2005} & $2<z<7$  & 16 & $1.55\pm0.04$ & $0.28$ & 2--10 \\
	%\citet{Porquet2004} & $z\le0.38$ & 2 & $1.74$ & $0.03$ & 2--5 \\
	\citet{Reeves2000} & $z\le4.3$ & 35 & $1.66\pm0.04$ & 0.22 & 2--10 \\
\enddata
	\tablenotetext{a}{The average value of $\Gamma$ for the sample indicated in Column 1. For studies including both RLQs and RQQs, we include $\langle\Gamma\rangle$ for the radio-loud subsample only.}
	\tablenotetext{b}{Sample standard deviation of $\Gamma$.}
	\tablenotetext{c}{The energy range over which $\Gamma$ is measured. Rest frame energies unless noted as `obs.'.}
	\tablenotetext{d}{In this study, $\Gamma$ is only modeled as a free parameter for those objects with source X-ray counts $N_\mathrm{sub}\ge100$, and for objects with $N_\mathrm{sub}\ge50$ and intrinsic absorption consistent with zero. The authors measure $\Gamma$ for 15 of a total of 46 RLQ. This may bias the average value of $\Gamma$ for the measured subsample.}
	%\tablenotetext{e}{This sample includes both Seyfert 1 and Sy2 objects.}
\end{deluxetable}

%% file: tables/uvcorr_analysis_table.tex
% beta_table,F2500_dnu,aox_table,L_int_XRT,L_int_UVOT,L_estimated_UV

\begin{deluxetable}{lccccccccc}
	\rotate
	\tablecolumns{10}
	\tablewidth{0pc}
	\tabletypesize{\footnotesize}
	\tablecaption{Integrated luminosities including preliminary correction for UV emission lines.\label{tab:analysis_uvcorr}}
	\tablehead{
		\colhead{Object} & \colhead{Corr.} & \colhead{$F_\nu$(2500 \AA)} & \colhead{\aox} & \colhead{$L_\mathrm{UV}$} & \colhead{$L_\mathrm{EUV}$} & \colhead{$L_\mathrm{EUV}$} & \colhead{$L_\mathrm{tot}$} & \colhead{$\Delta L_\mathrm{tot}$} & \colhead{$L_{\mathrm{bol}}$} \\
		\colhead{name} & \colhead{\%} & \colhead{} & \colhead{} & \colhead{(1000--3000\AA)} & \colhead{(1000\AA--0.3 keV)} & \colhead{(1000\AA--1 keV)} & \colhead{(3000\AA--25 keV)} & \colhead{\%} & \colhead{(1$\mu$m--10 keV)}\\
		\colhead{(1)} & \colhead{(2)} & \colhead{(3)} & \colhead{(4)} & \colhead{(5)} & \colhead{(6)} & \colhead{(7)} & \colhead{(8)} & \colhead{(9)} & \colhead{(10)}
		}
\startdata  % name      F_nu2500            alpha_ox    LXobs LX0310 L1-25  L_NUV L_FUV03 L_FUV1 L_tot
J014725.50-101439.11 & 10 & $ 3.51 \pm 0.44 $ & $\ge 1.94$ & $5.90\pm0.76$ & --- & --- & --- & ---  & $24.69^{+13.52}_{-12.89}$ \\         
J082328.62+061146.07 & 5  & $ 2.47 \pm 0.35 $ & $ 1.42 \pm 0.08 $ & $7.97\pm1.16$ & $9.71_{-2.27}^{+2.14}$ & $14.00_{-2.56}^{+2.53}$ & $23.87_{-3.96}^{+3.93}$ & 3.8  & $36.70^{+19.63}_{-18.00}$ \\
J094853.60+085514.40 & 10 &$ 6.45 \pm 0.63 $ & $ 1.59 \pm 0.11 $ & $7.61\pm0.86$ & $5.05_{-1.37}^{+1.09}$ & $7.85_{-1.45}^{+1.37}$ & $16.58_{-2.53}^{+2.47}$ & 8.3  & $36.70^{+17.07}_{-17.07}$ \\
J104915.44-011038.18 & ND & --- & --- & --- & --- & --- & --- & ---  & --- \\
J110607.48-173113.60 & 0 & $ 3.97 \pm 0.41 $ &  $\ge 1.63$ & $10.08\pm1.07$ & --- & --- & ---  & --- & $44.37^{+23.25}_{-21.59}$ \\
J111159.70+023719.76 & NB & --- &  --- & --- & --- & --- & ---  & --- & $18.18^{+9.68}_{-8.90}$ \\
J112542.30+000101.33 & 20 & $ 2.35 \pm 0.36 $ &  $ 1.37 \pm 0.08 $ & $2.93\pm0.46$ & $3.41_{-0.95}^{+0.87}$ & $5.14_{-1.05}^{+1.04}$ & $9.11_{-1.64}^{+1.66}$ & 15.2  & $13.06^{+7.16}_{-6.47}$ \\
J114449.32+032751.96 & 10 & $ 4.85 \pm 0.53 $ &  $\ge 1.72$  & $6.80\pm0.76$ & --- & --- & --- & --- & $25.84^{+14.40}_{-12.89}$ \\
\enddata
	\tablecomments{(1) Object name. (2) Estimate of the necessary emission line correction (\S \ref{sec:analysis_integrated_luminosity_uverror}), as a percentage of the original model flux. We are unable to make this estimate for quasars without SDSS photometry (denoted NS), for BOSS spectra without a flux calibration correction (see Appendix \ref{appendix:sed}) (NB) and for objects not detected with UVOT (ND). (3) The flux density at $2500$ \AA, with preliminary emission line correction, in units of $10^{-27}$ erg s$^{-1}$ cm$^{-2}$ Hz$^{-1}$. (4) The corrected X-ray to UV-optical spectral index \aox. (5) The corrected integrated luminosity over the rest-frame interval 1000 \AA\,-- 3000 \AA. Units of $10^{46}$ erg s$^{-1}$. (6,7) Guideline estimates of the extreme-UV luminosity (\S \ref{sec:analysis_integrated_luminosity}), with preliminary emission line correction. We interpolate over the rest-frame intervals 1000 \AA\, to 1 keV and 1000 \AA\,to 0.3 keV. Units of $10^{46}$ erg s$^{-1}$.  (8) Guideline estimate of the integrated luminosity over the rest-frame interval 3000 \AA -- 25 keV, with preliminary emission line correction. Units of $10^{46}$ erg s$^{-1}$. (9) Reduction in $L_\mathrm{tot}$ (3000 \AA -- 25 keV) due to the UV emission line correction, listed as a percentage change in the uncorrected value. (10) The estimated bolometric luminosity (1$\mu$m -- 10 keV), using the bolometric correction from rest-frame 1450 \AA\,as presented by \citet{Runnoe2012}. Units of $10^{46}$ erg s$^{-1}$. }
\end{deluxetable}

%% file: appendixA_sed.tex
\section{Spectral energy distribution figures for sample quasars}\label{appendix:sed}

We present the optical to X-ray spectral energy distributions (SEDs) of the quasars in our sample, along with the UV photometry and continuum modeling, in Figures \ref{fig:sed1} through \ref{fig:sed15}. Where available, we show the SDSS spectra used to guide our selection of UV data (\S \ref{sec:analysis_uv_optical_modeling}) as gray curves in the right-hand panels. For quasars without SDSS spectroscopy, we show the high-redshift quasar template spectrum produced by \citet{Selsing2016}, normalized to the continuum model flux level at 2500 \AA. This template is constructed from spectroscopic observations of seven bright ($M_i\approx-29$ mag) quasars at $1<z<2$, i.e., objects that overlap our sample in terms of luminosity and redshift distribution, but that are somewhat brighter and reside at lower redshift than the average properties of our quasars. For such bright quasars the host galaxy contribution is expected to be small. Note that we do not use this template spectrum directly to model the UV continuum (\S \ref{sec:analysis_uv_optical_modeling}), we merely use it as a rough guide to the amount of emission line flux in a given UV bandpass for a typical bright quasar.

Spectra of $z\apprge2$ quasars observed as part of the SDSS BOSS campaign suffer flux calibration uncertainties due to alterations in the instrumental setup \citep{Dawson2013}, the intent of which was to maximize throughput at short wavelengths. For the BOSS spectra we recalibrate the fluxes using the prescription presented by \citet{Margala2015}, in order to better inform our continuum modeling. For quasars Q1402-012 and Q2350-007, \citet{Margala2015} do not provide a recalibration. In these cases we find significant flux offsets between the SDSS photometry and spectroscopy (see Figures \ref{fig:sed2}, \ref{fig:sed11} and \ref{fig:sed15}). As we do not use the SDSS spectra directly in our SED modeling, we do not attempt additional corrective steps for these spectra. 

%For quasars J111159.70+023719.76, Q1402-012 and Q2350-007, \citet{Margala2015} do not provide a recalibration. In these cases we find significant flux offsets between the SDSS photometry and spectroscopy (see Figures \ref{fig:sed2}, \ref{fig:sed11} and \ref{fig:sed15}).

\subsection*{Comments on individual objects}

\def\sedplotsize{0.495\linewidth}

\begin{figure*}
	\centering
	\includegraphics[width=\sedplotsize]{figures/sedplots/82097_sedplot_nuL_dnu}
	\includegraphics[width=\sedplotsize]{figures/uvotfit_plots/82097_uvotfit}
	\includegraphics[width=\sedplotsize]{figures/sedplots/39807_sedplot_nuL_dnu}
	\includegraphics[width=\sedplotsize]{figures/uvotfit_plots/39807_uvotfit}
	\includegraphics[width=\sedplotsize]{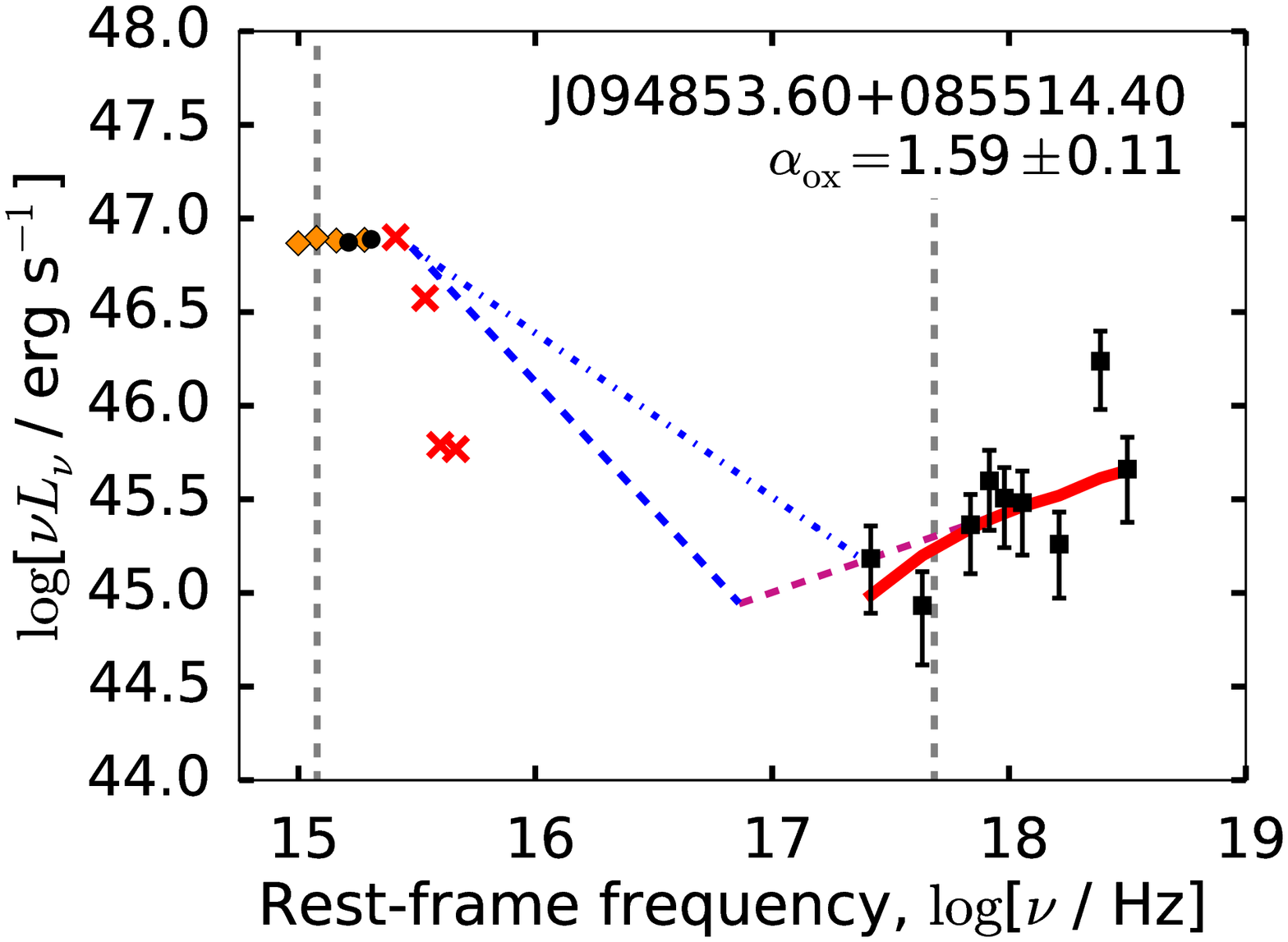}
	\includegraphics[width=\sedplotsize]{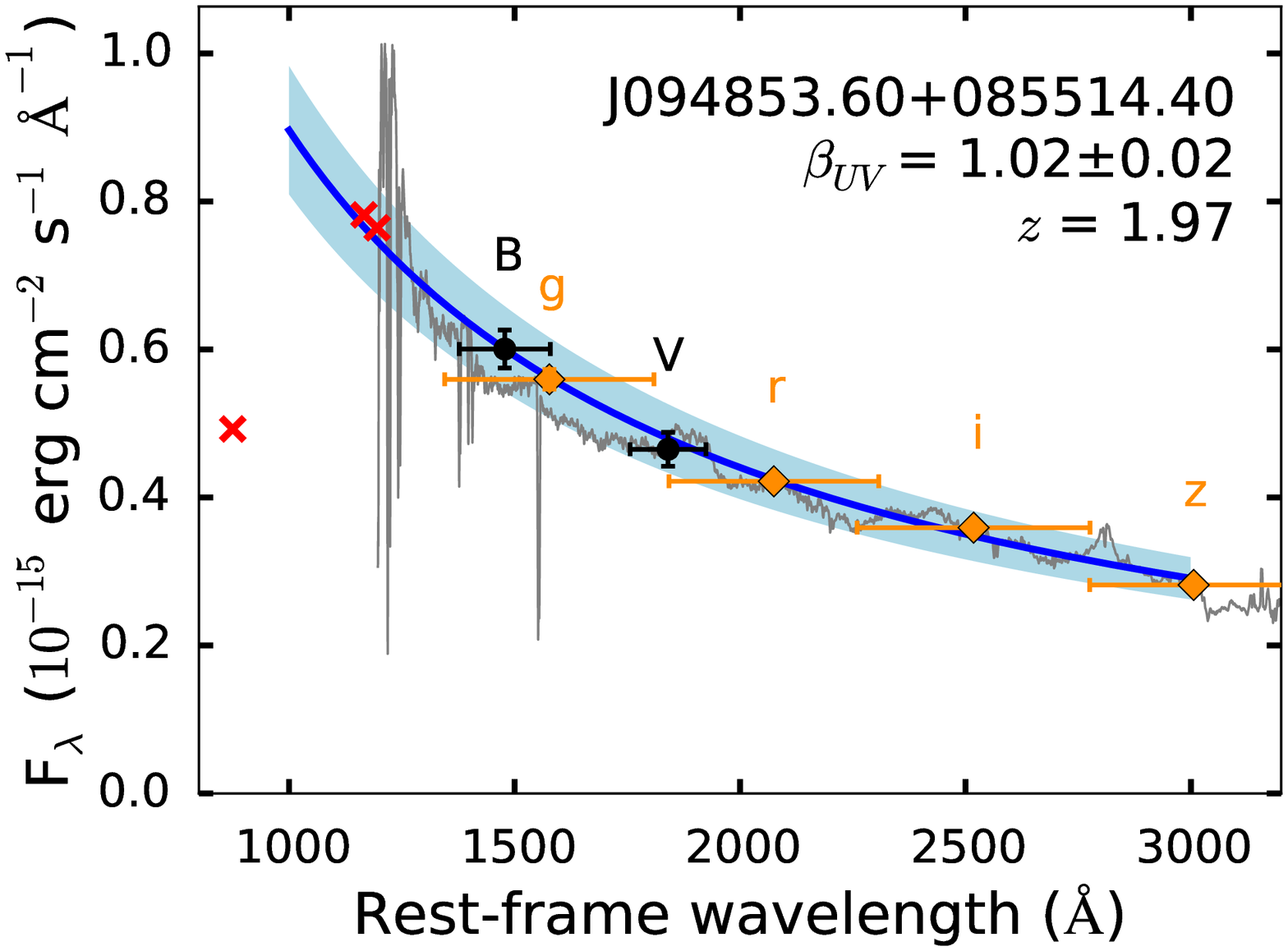}

	\caption{\emph{Left:} Rest-frame UV to X-ray spectral energy distributions (SEDs) of quasars in our sample. \emph{Right}: UV photometry and continuum modeling. See Figure \ref{fig:sed_mainpaper} for symbol and color coding.}
	\label{fig:sed1}
\end{figure*}

\begin{figure*}
	\centering
	\includegraphics[width=\sedplotsize]{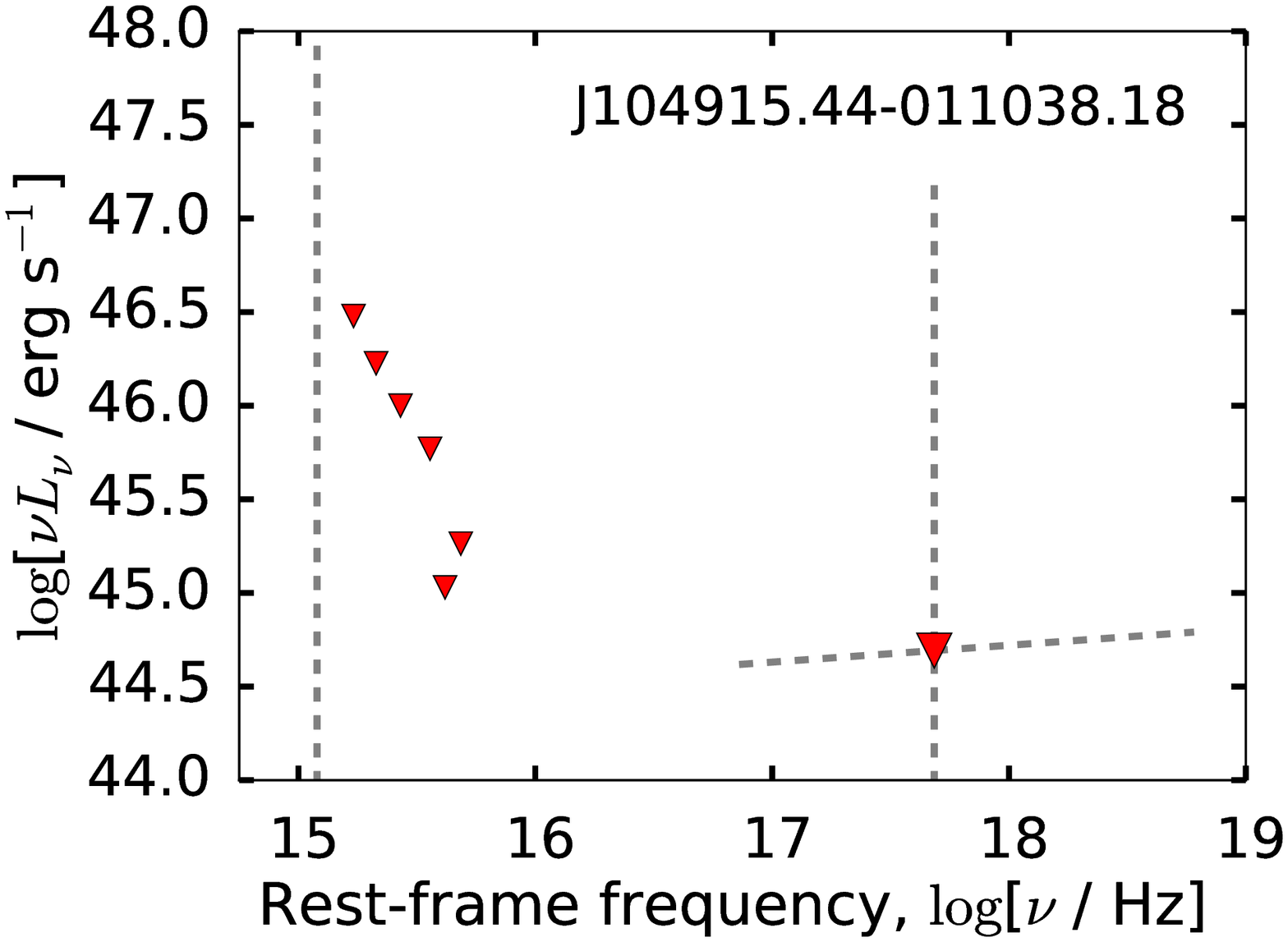}
	\includegraphics[width=\sedplotsize]{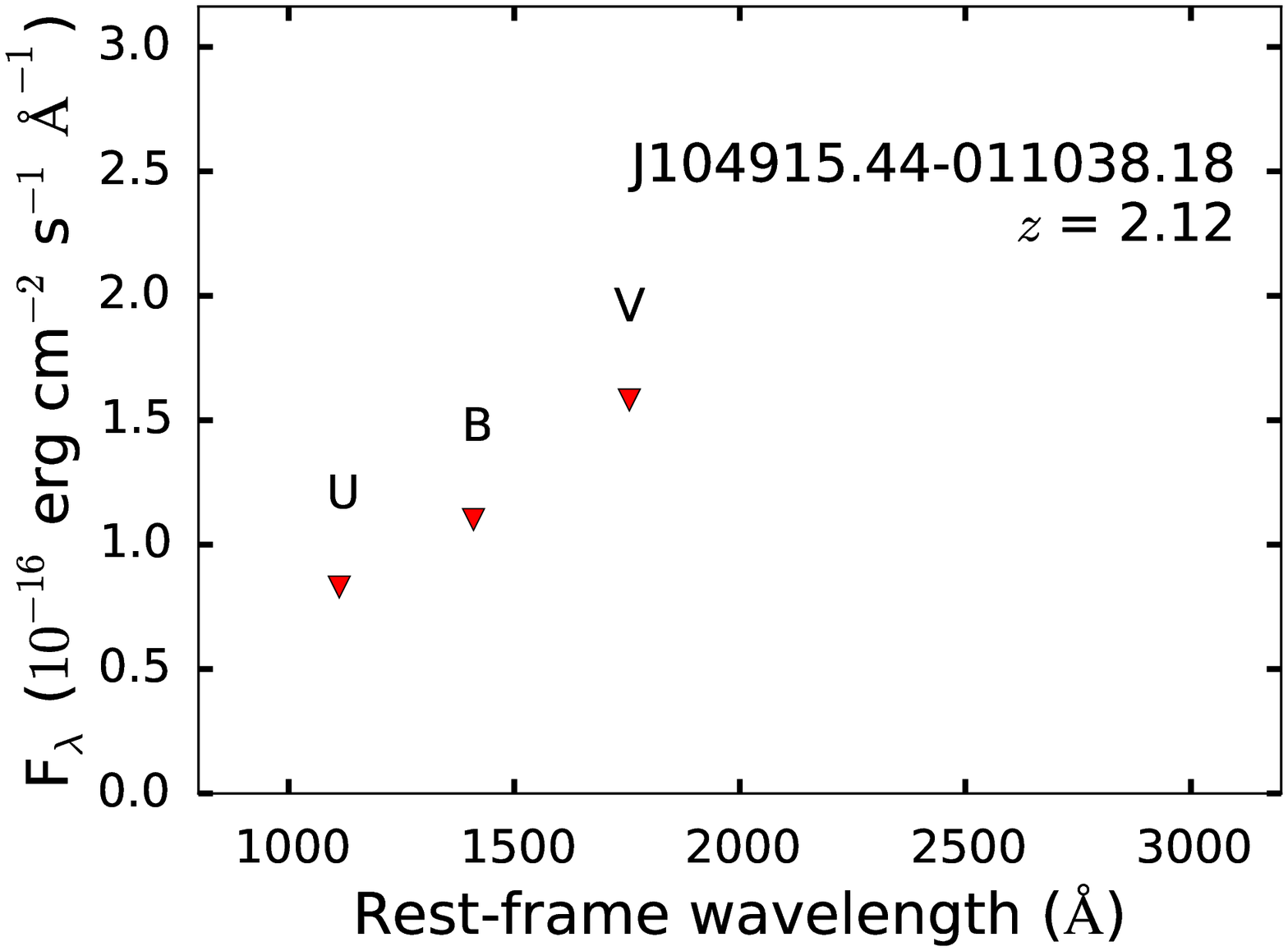}
	\includegraphics[width=\sedplotsize]{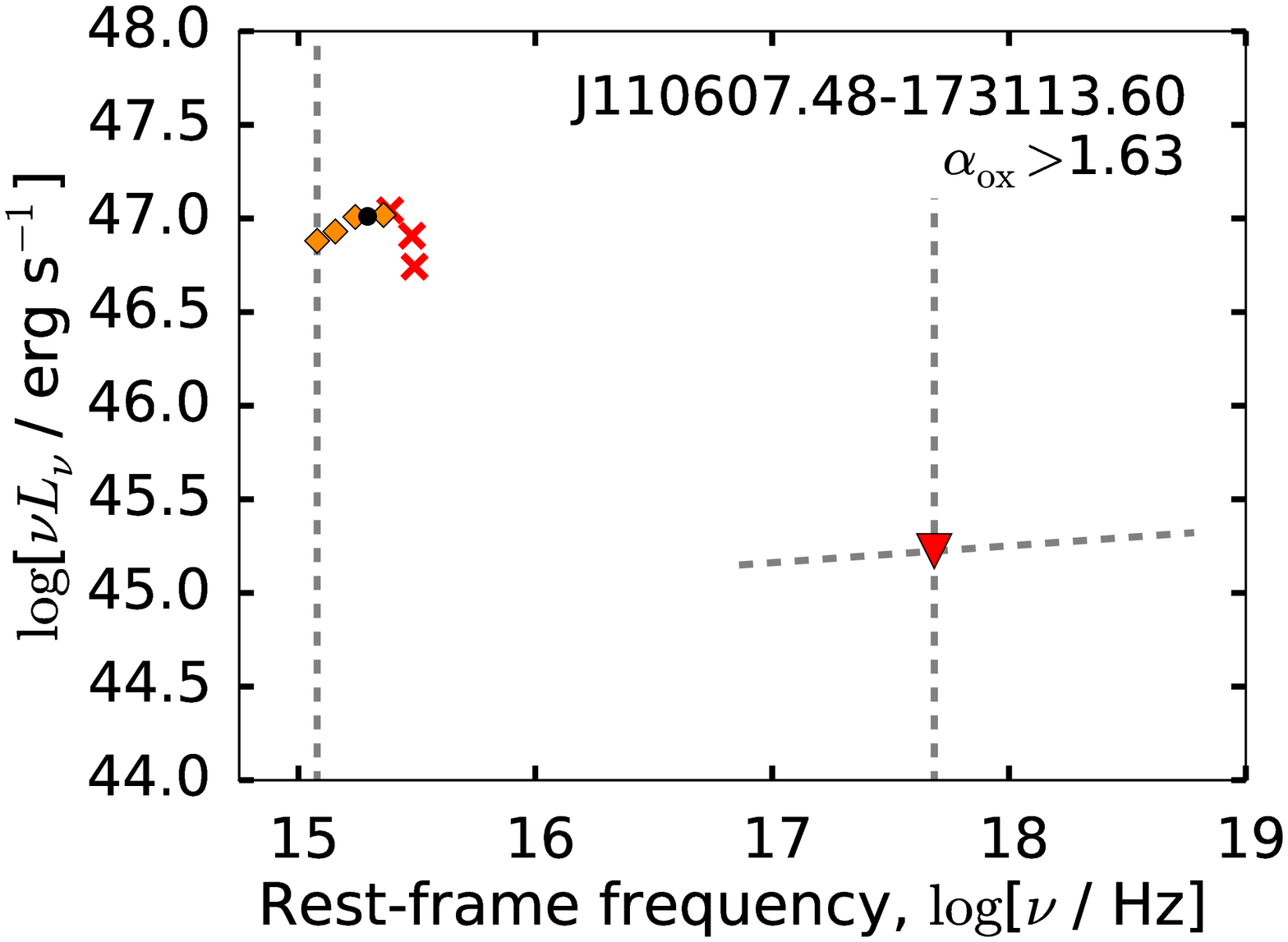}
	\includegraphics[width=\sedplotsize]{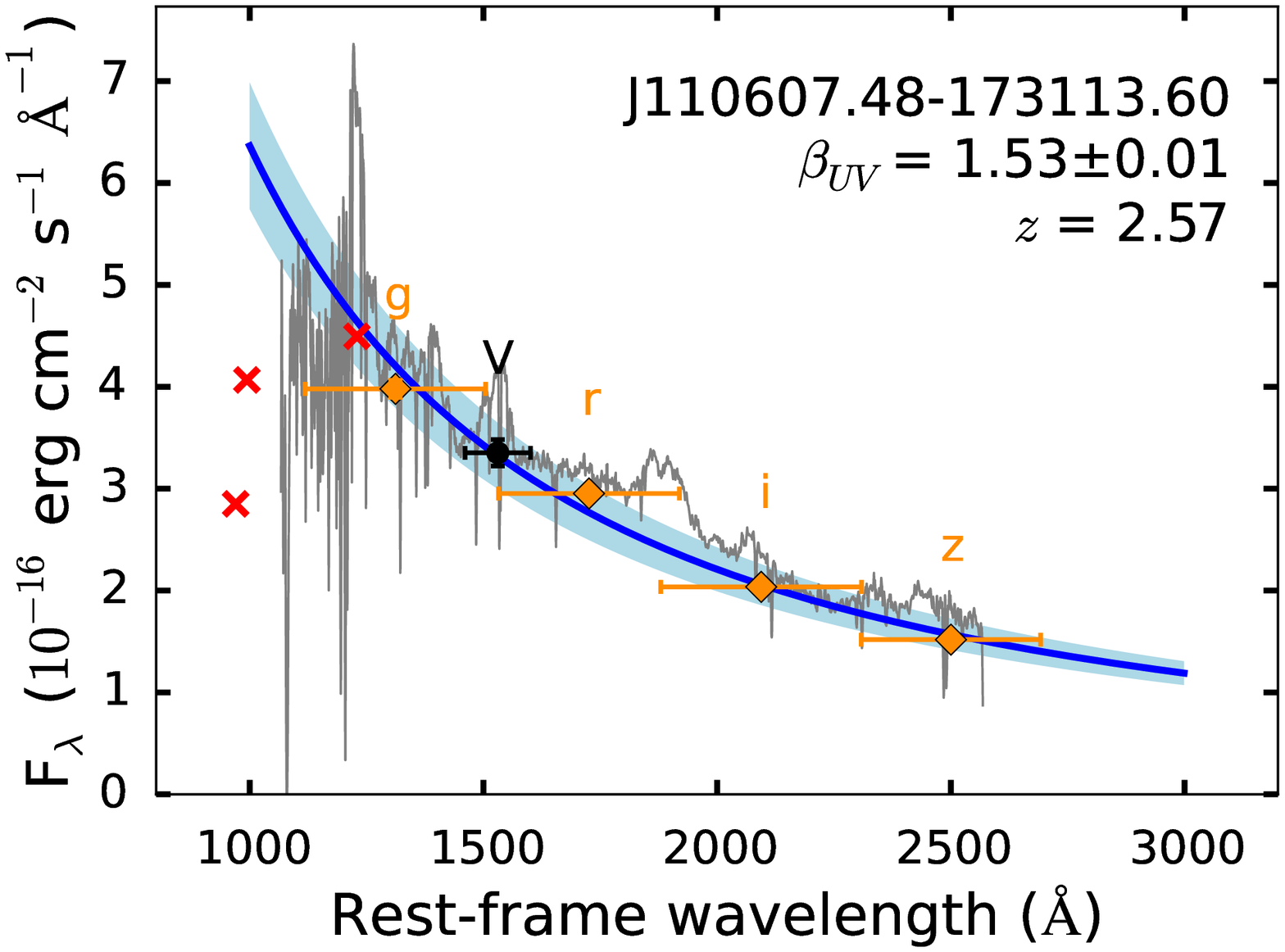}
	\includegraphics[width=\sedplotsize]{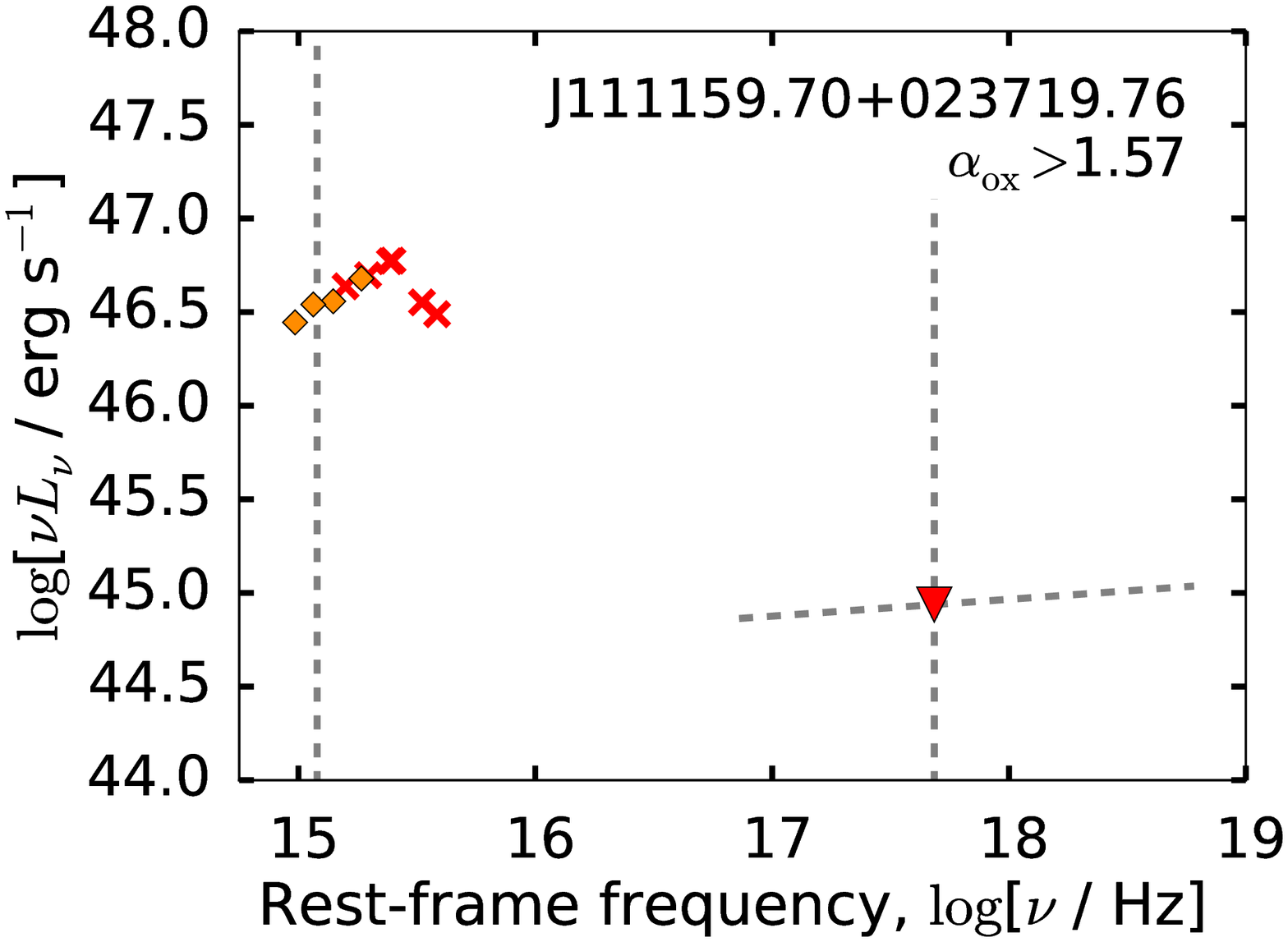}
	\includegraphics[width=\sedplotsize]{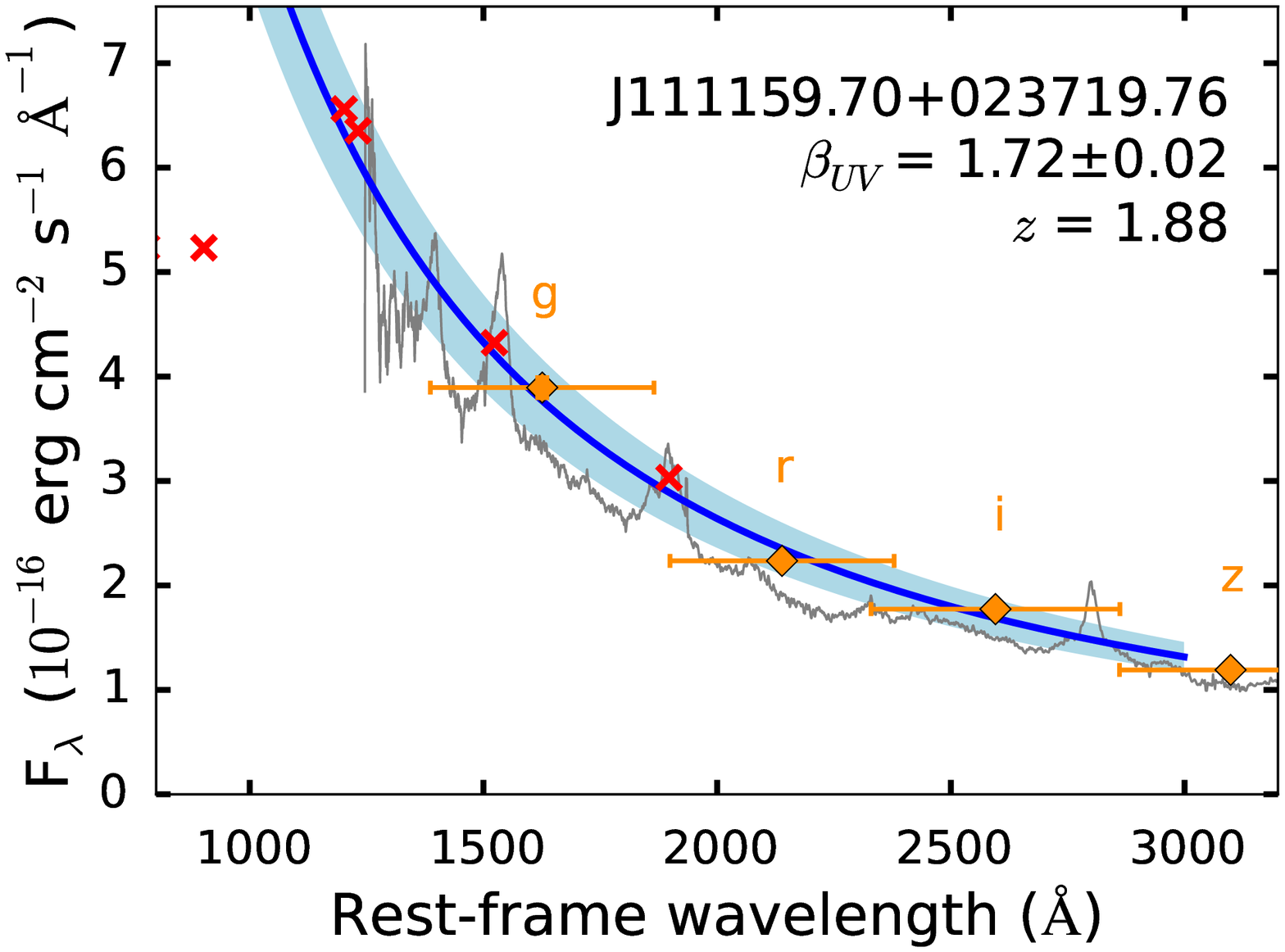}
	
	\caption{\emph{Left:} Rest-frame UV to X-ray spectral energy distributions (SEDs) of quasars in our sample. \emph{Right}: UV photometry and continuum modeling. See Figure \ref{fig:sed_mainpaper} for symbol and color coding.}
	\label{fig:sed2}
\end{figure*}

\paragraph*{J014725.50-101439.11}

We find a $3\sigma$ lower limit on \aox of \aox$>1.94$. Combined with the non-detections in the \emph{Swift} far-UV filters, this may suggest strong absorption.

\paragraph*{J104915.44-011038.18:}

This quasar is a non-detection in our \emph{Swift} observations. We find a $3\sigma$ limiting apparent $V$ magnitude of 18.5 mag. The catalog of \citet{Hewitt1993} lists $m_V=17.9$ for this object; thus it has dimmed significantly over a period of approximately 30 observed-frame years.

\paragraph*{J111159.70+023719.76:}

The UVOT \emph{B} and \emph{B} bandpasses suffer BEL contamination. We therefore rescale the SDSS fluxes based on the UVOT \emph{U} flux, and discard all UVOT data from the final fit. Irrespective of flux rescaling, we find a flux offset between SDSS photometry and spectroscopy. \citet{Margala2015} do not provide a recalibration for this BOSS spectrum.

\begin{figure*}
	\centering
	\includegraphics[width=\sedplotsize]{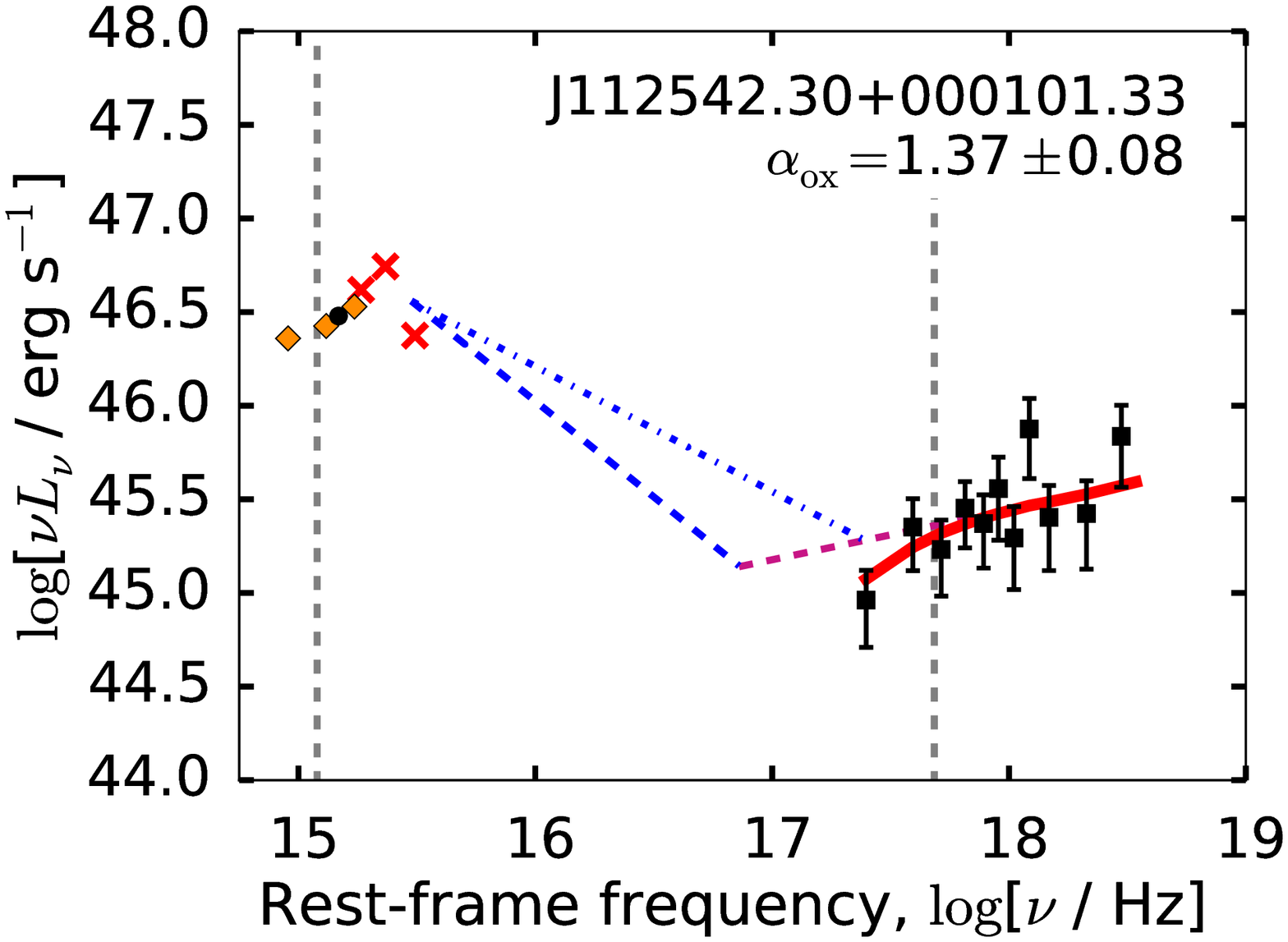}
	\includegraphics[width=\sedplotsize]{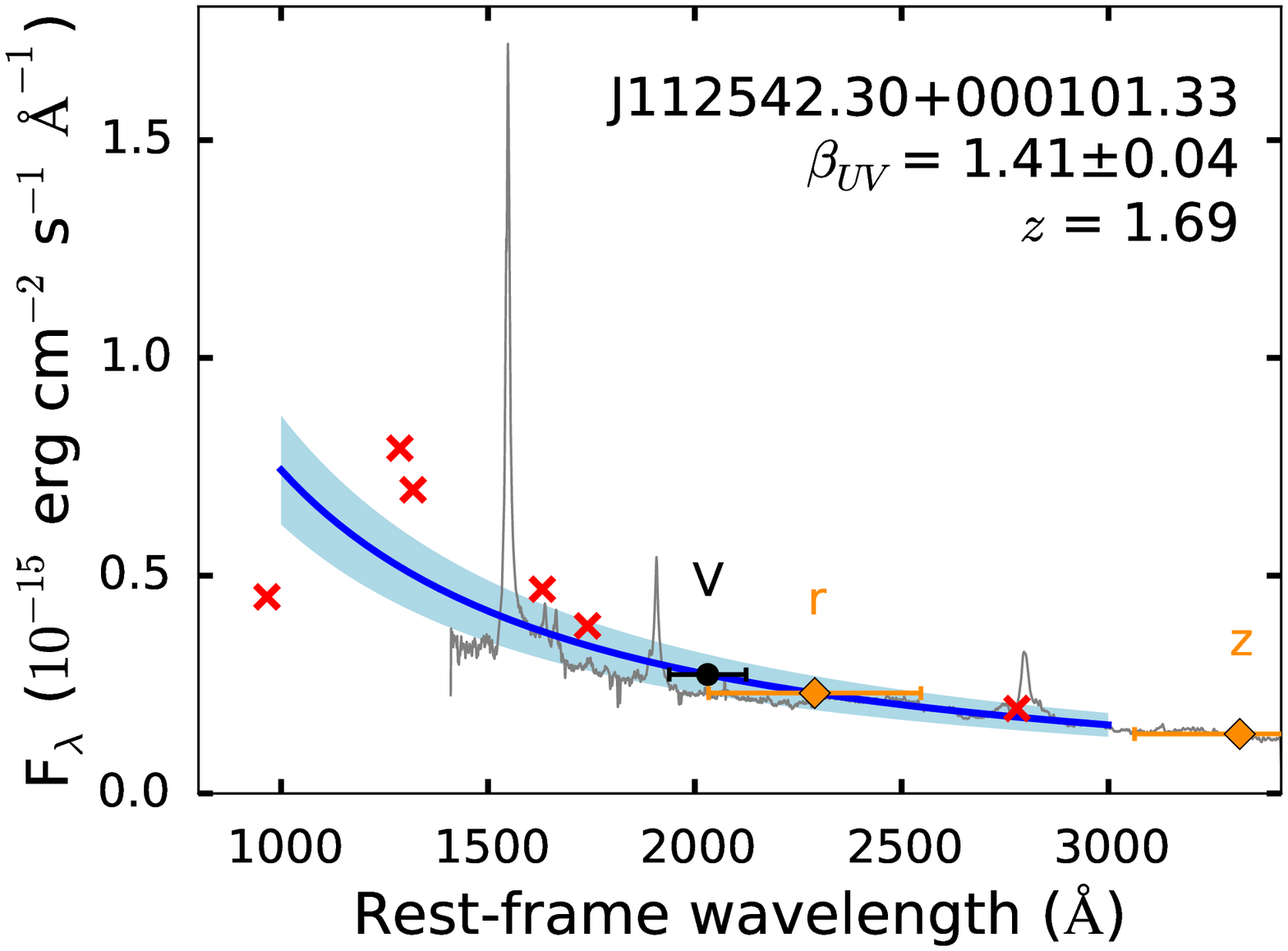}
	\includegraphics[width=\sedplotsize]{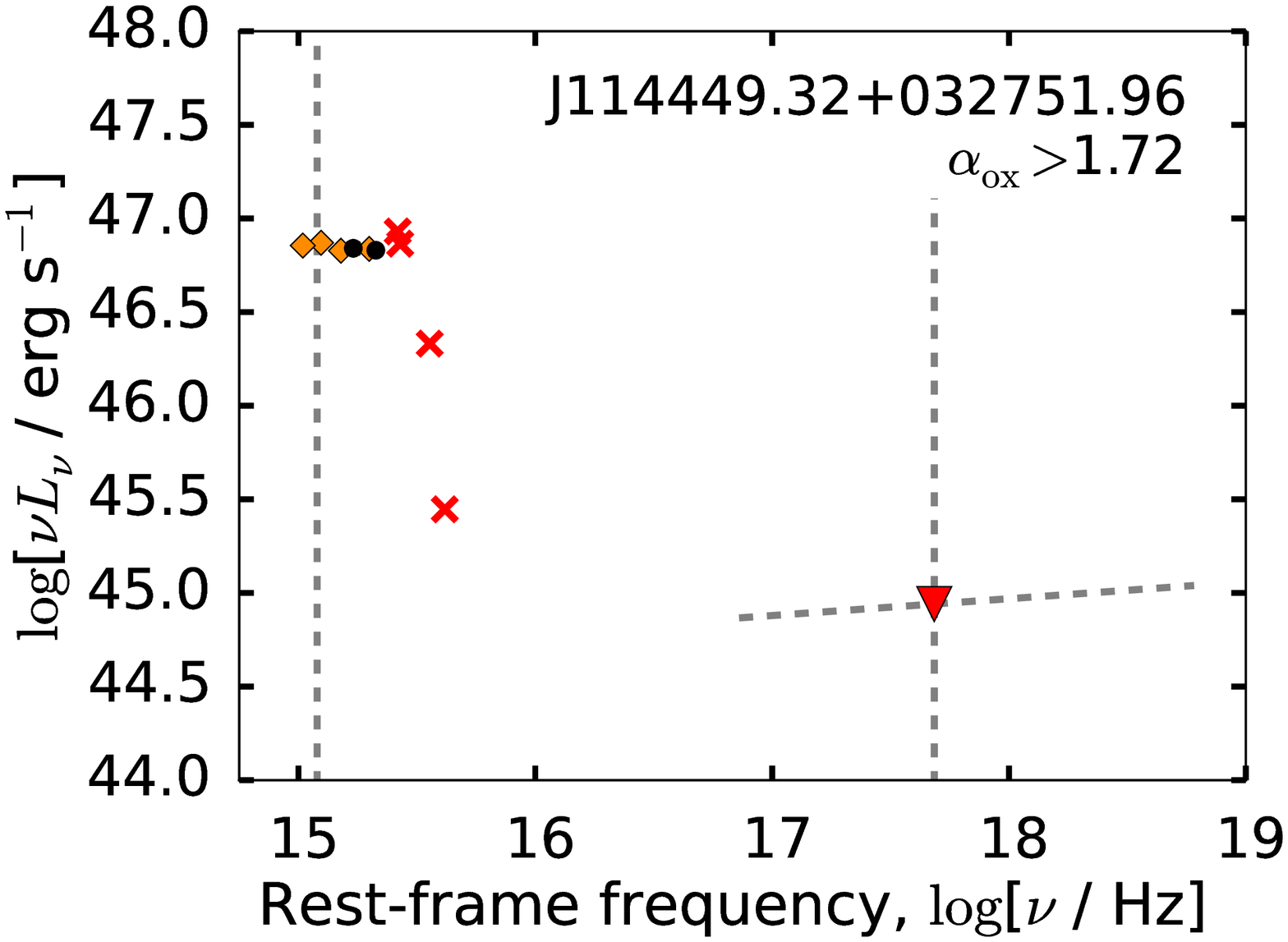}
	\includegraphics[width=\sedplotsize]{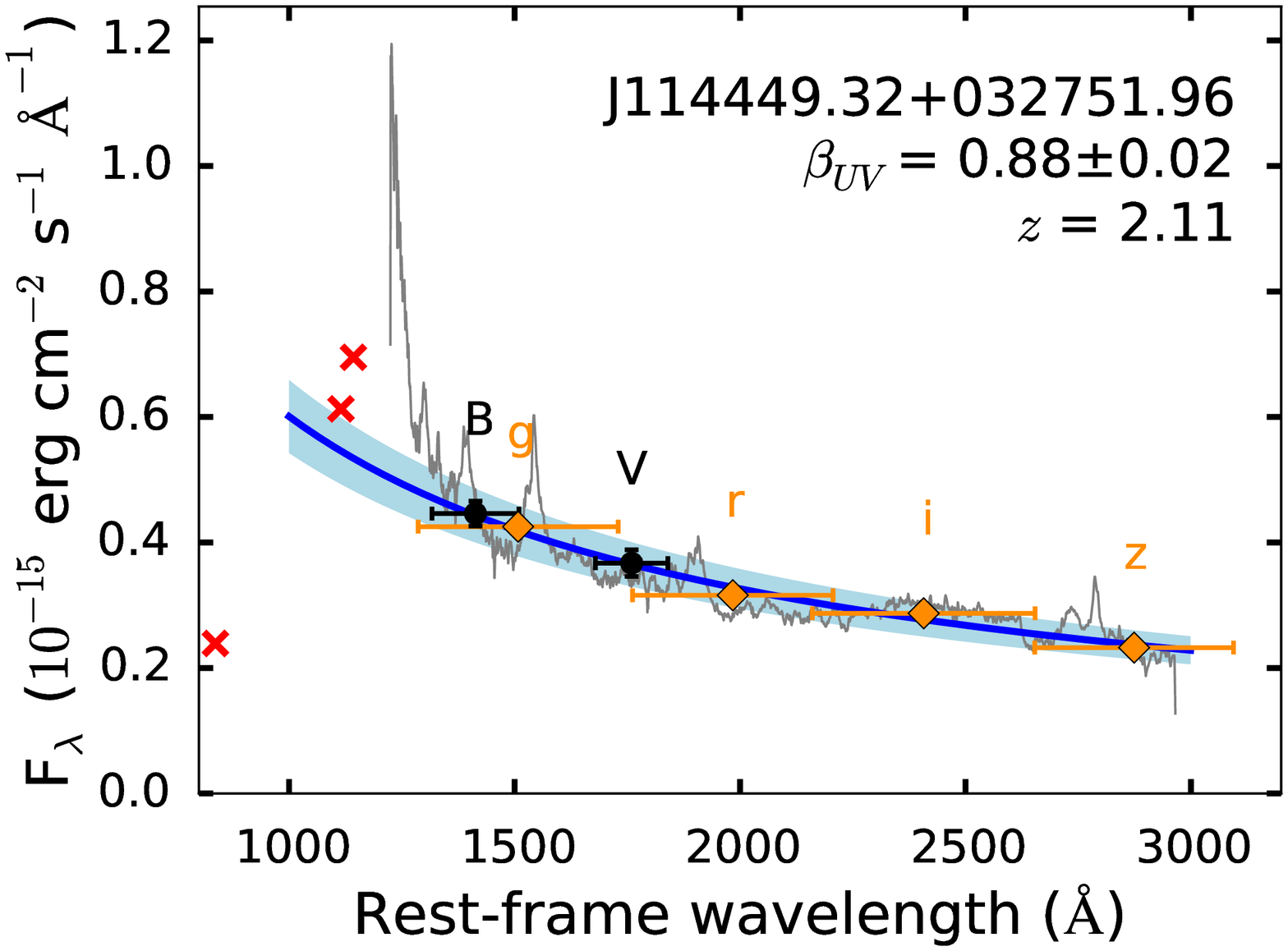}
	\includegraphics[width=\sedplotsize]{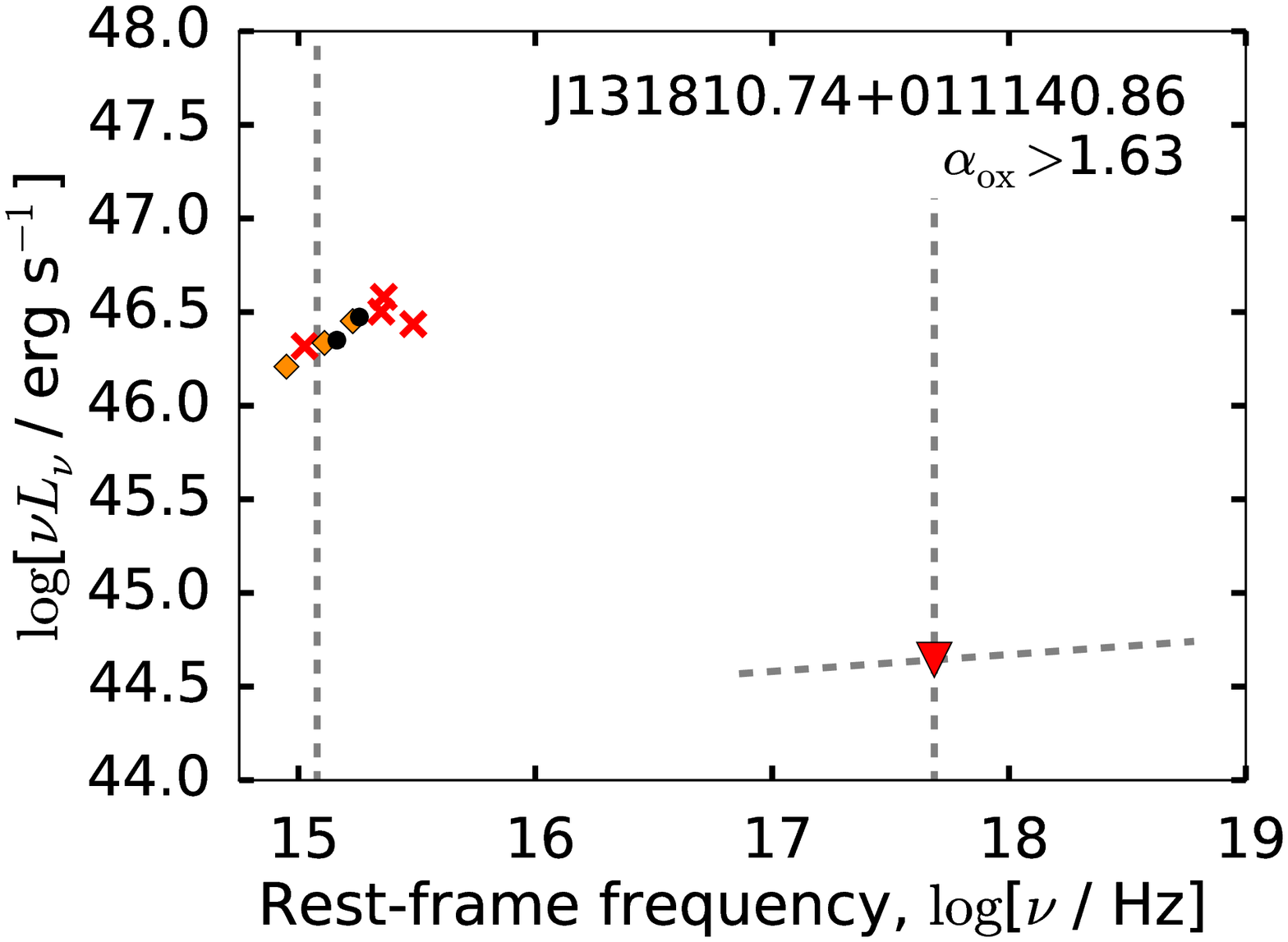}
	\includegraphics[width=\sedplotsize]{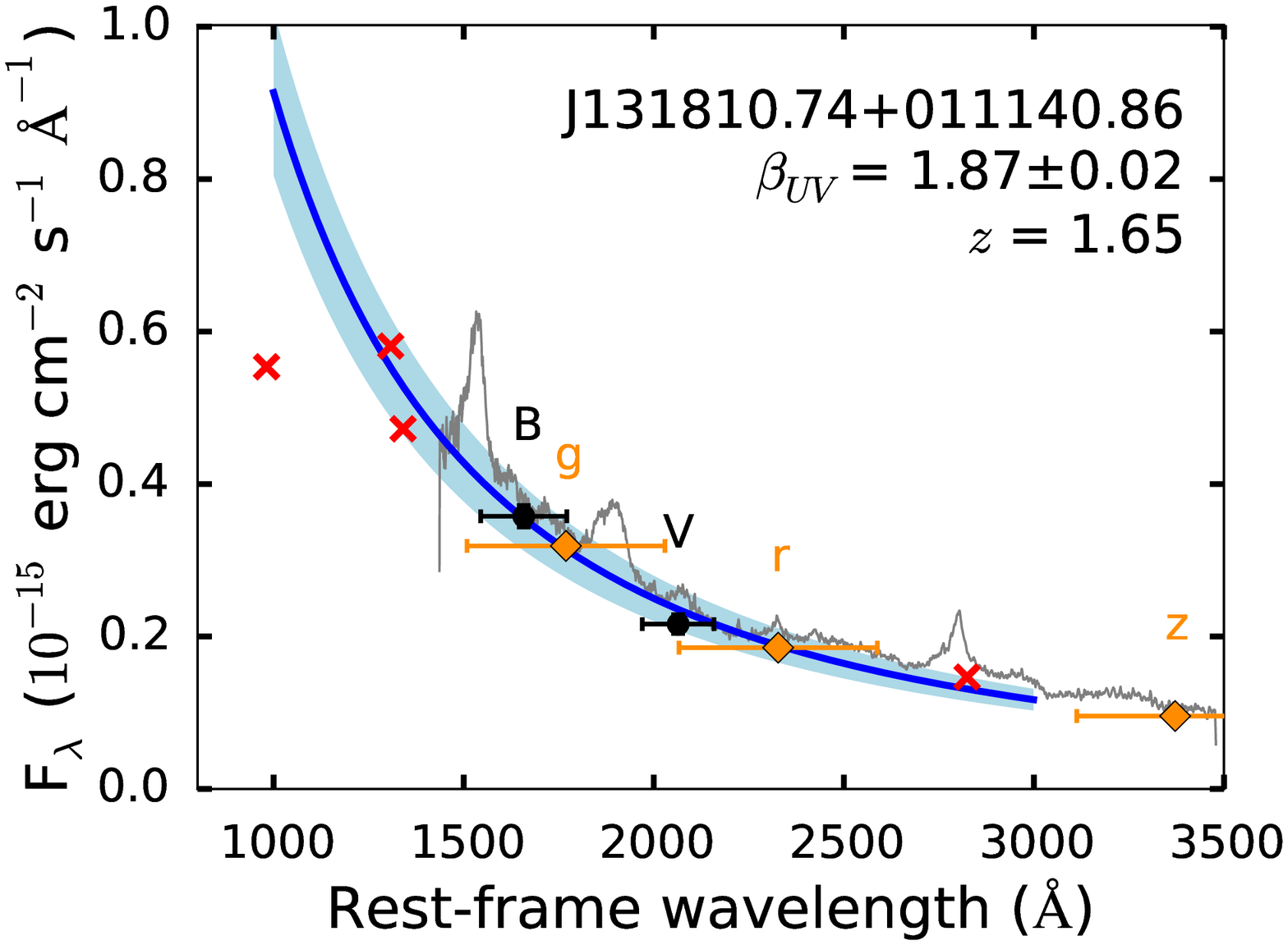}
	
	\caption{\emph{Left:} Rest-frame UV to X-ray spectral energy distributions (SEDs) of quasars in our sample. \emph{Right}: UV photometry and continuum modeling. See Figure \ref{fig:sed_mainpaper} for symbol and color coding.}
	\label{fig:sed3}
\end{figure*}

\begin{figure*}
	\centering
	\includegraphics[width=\sedplotsize]{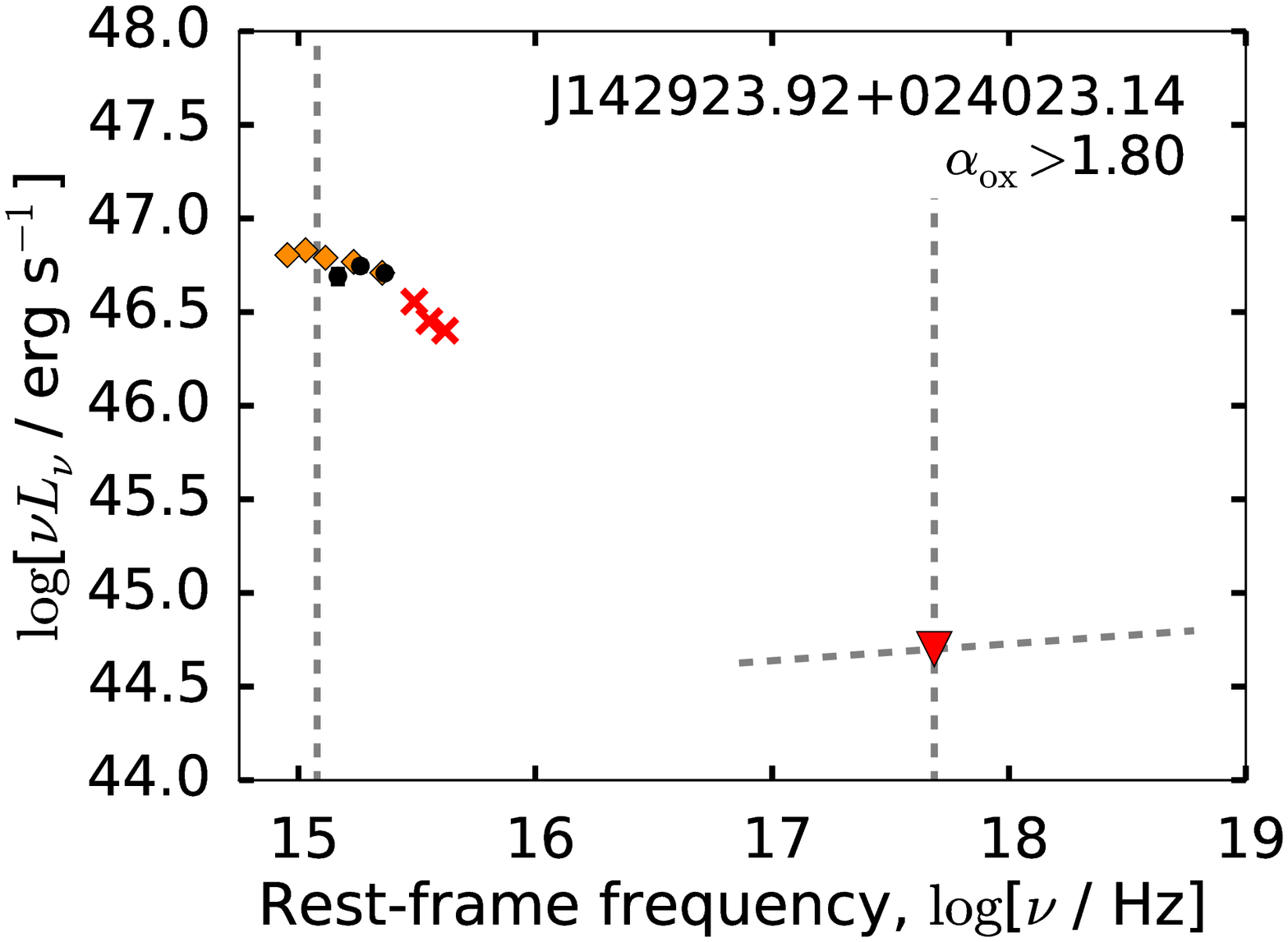}
	\includegraphics[width=\sedplotsize]{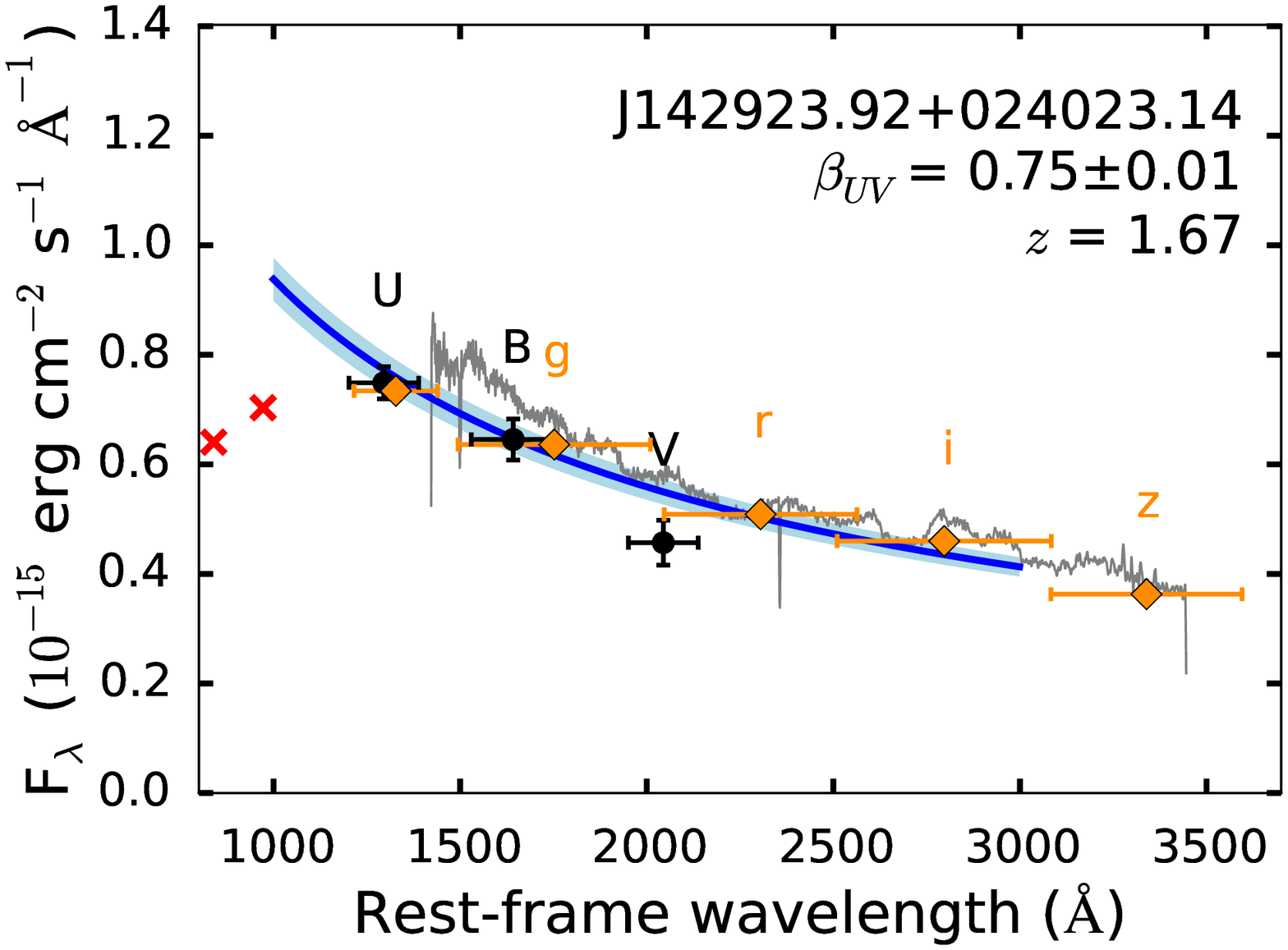}
	\includegraphics[width=\sedplotsize]{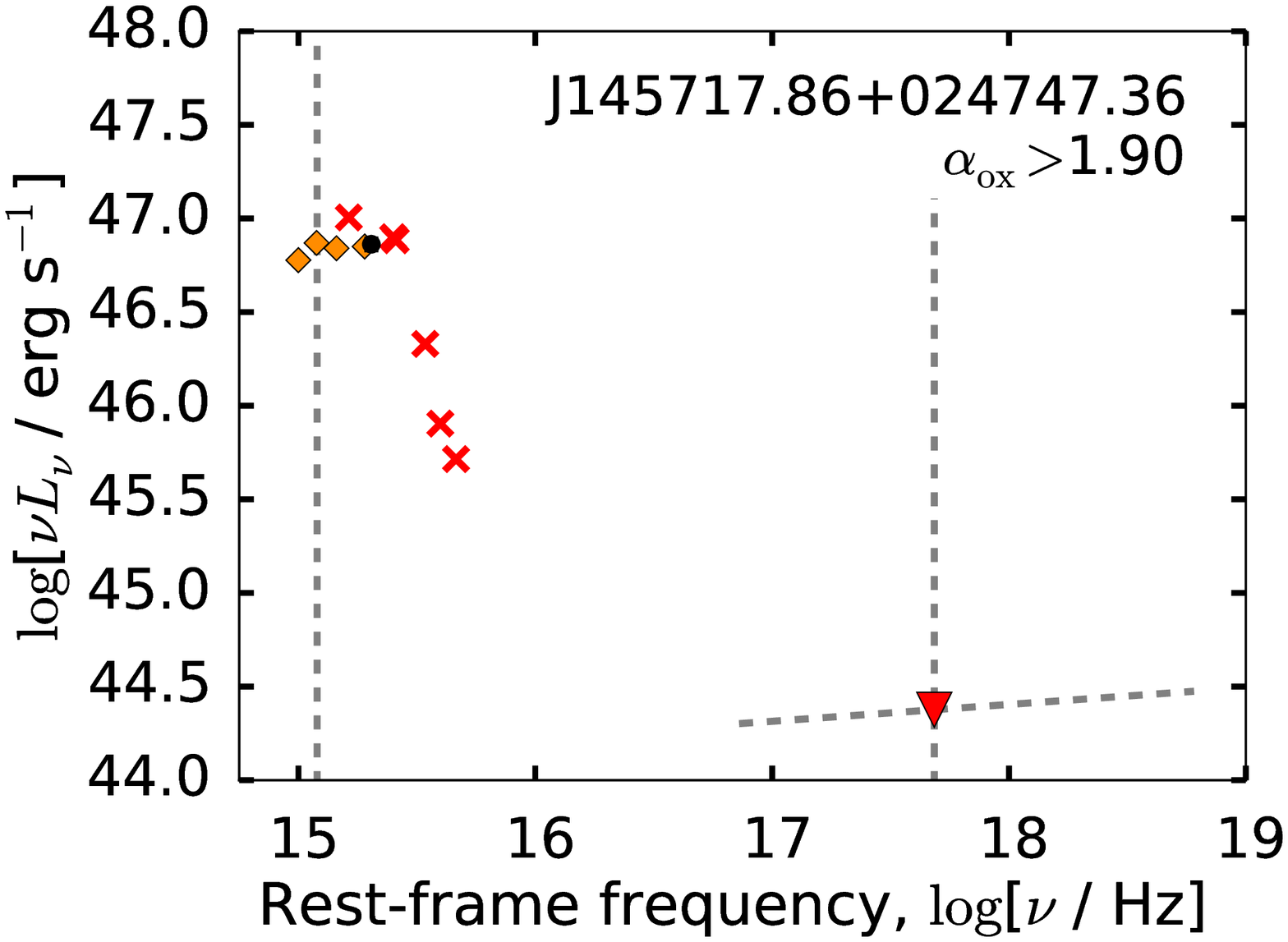}
	\includegraphics[width=\sedplotsize]{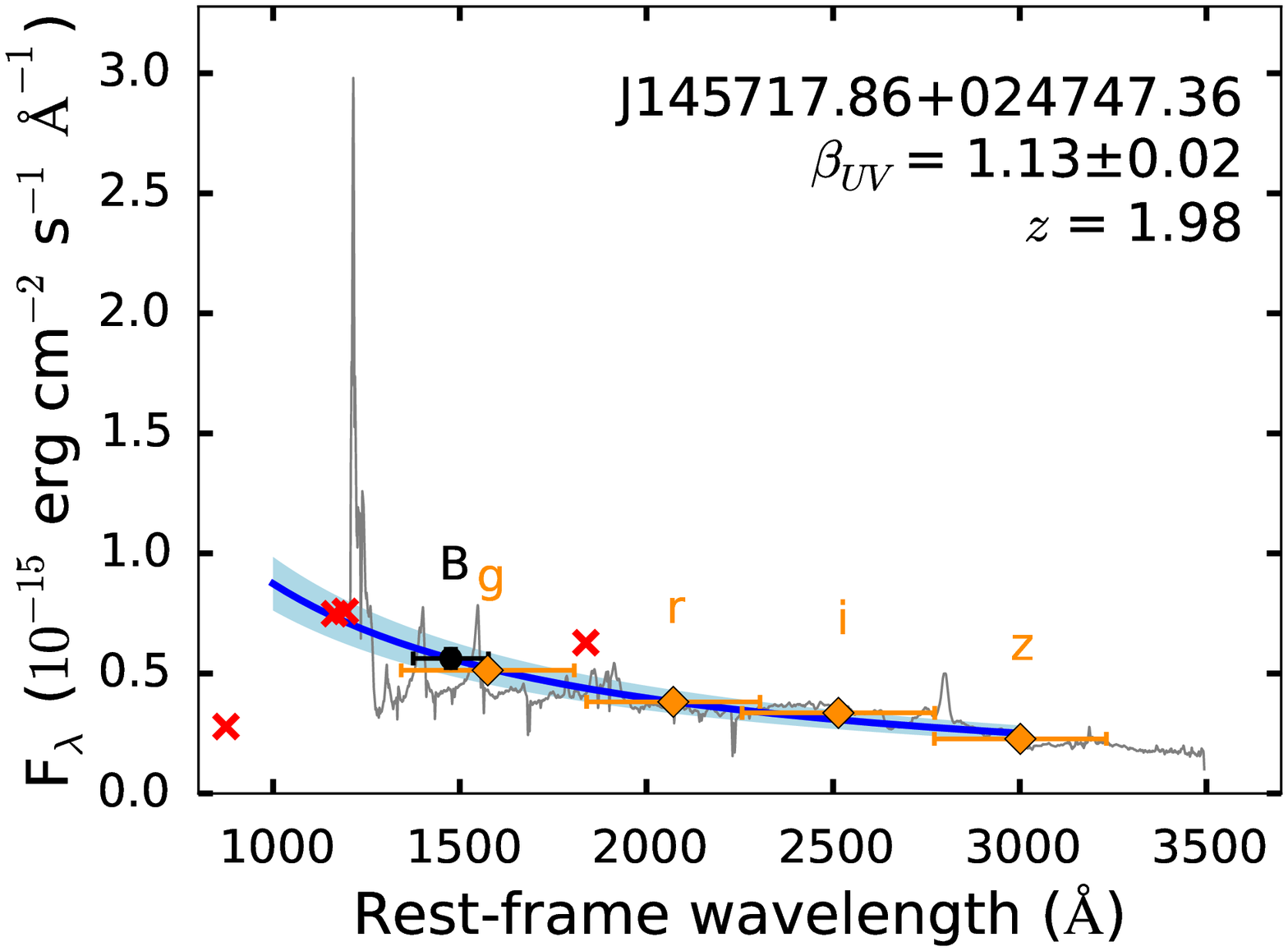}
	\includegraphics[width=\sedplotsize]{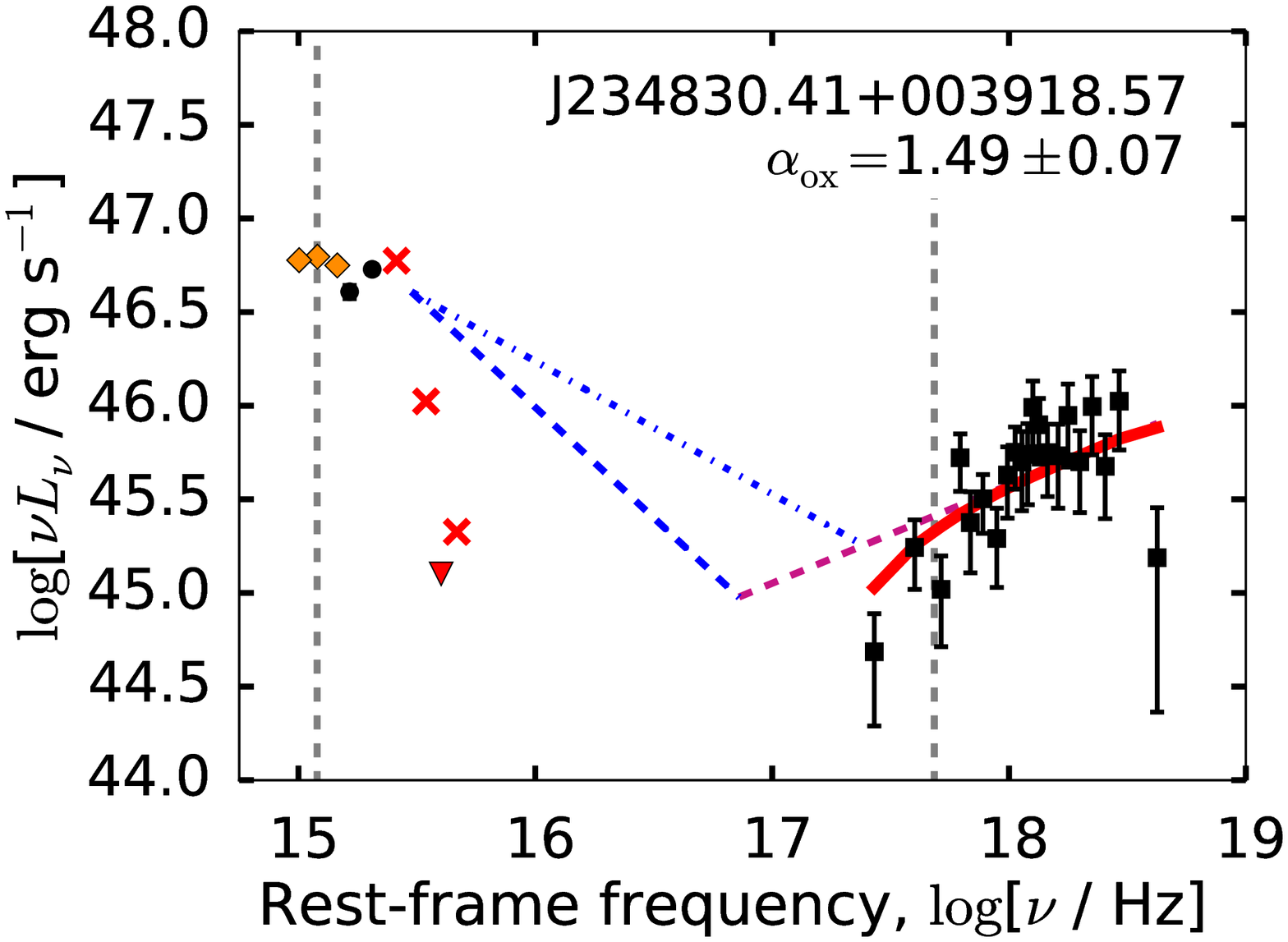}
	\includegraphics[width=\sedplotsize]{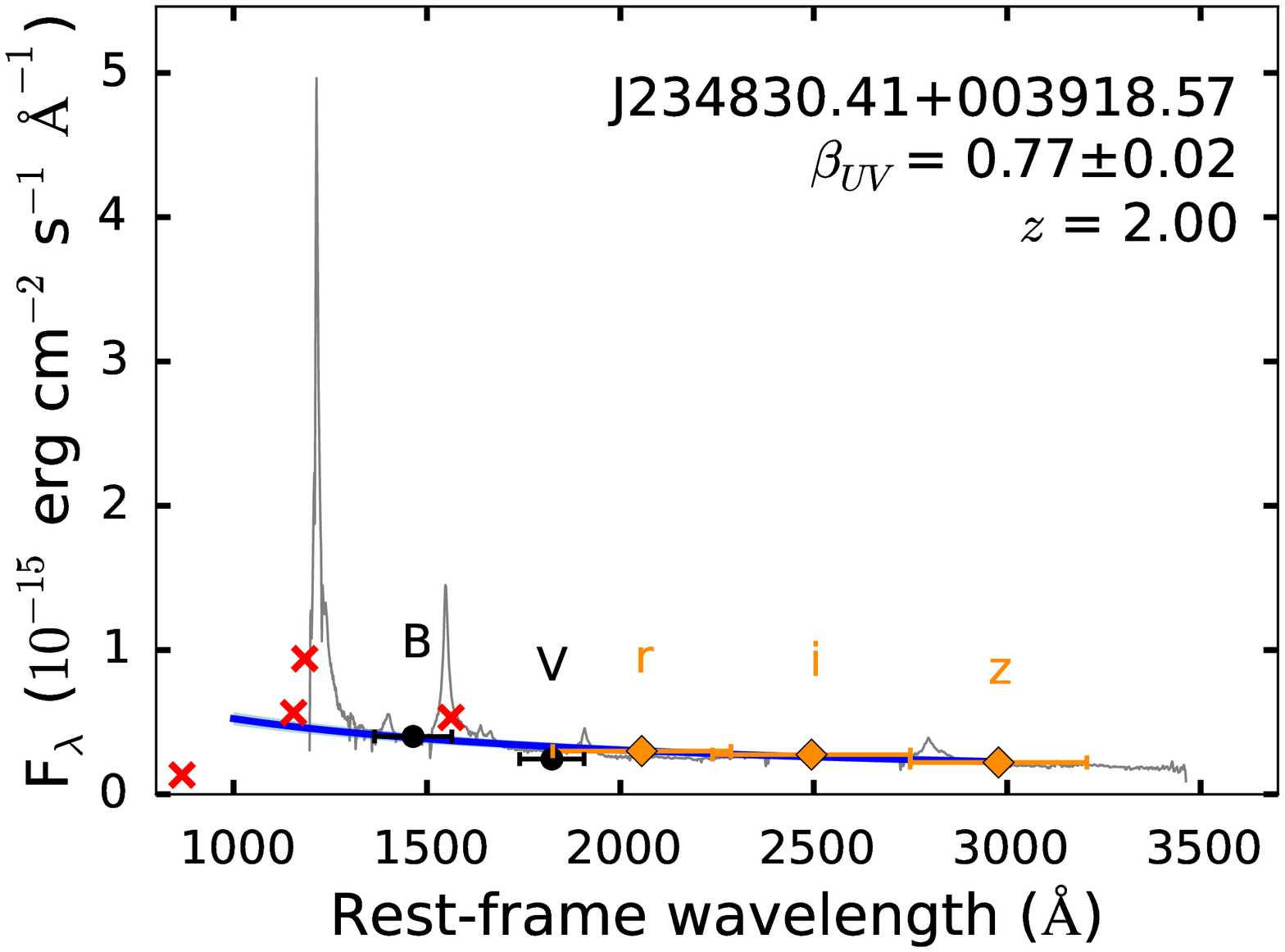}
	
	\caption{Optical to X-ray spectral energy distributions (SEDs) of quasars in our sample. See Figure \ref{fig:sed_mainpaper} for symbol and color coding.}
	\label{fig:sed4}
\end{figure*}

\paragraph*{J145717.86+024747.36:} The SDSS spectrum for this quasar shows a flat continuum. Our continuum model based on SDSS and \emph{Swift} photometric data is somewhat steeper - this may be due to the lack of continuum-dominated photometric data-points towards the blue end of the spectrum. The BOSS flux calibration is known to be erroneous for high-redshift quasars\citep{Dawson2013}. While we apply the flux recalibration of \citet{Margala2015}, they note that the calibration can be inaccurate for individual observations, especially at the blue end of the spectrum.

\begin{figure*}
	\centering
	\includegraphics[width=\sedplotsize]{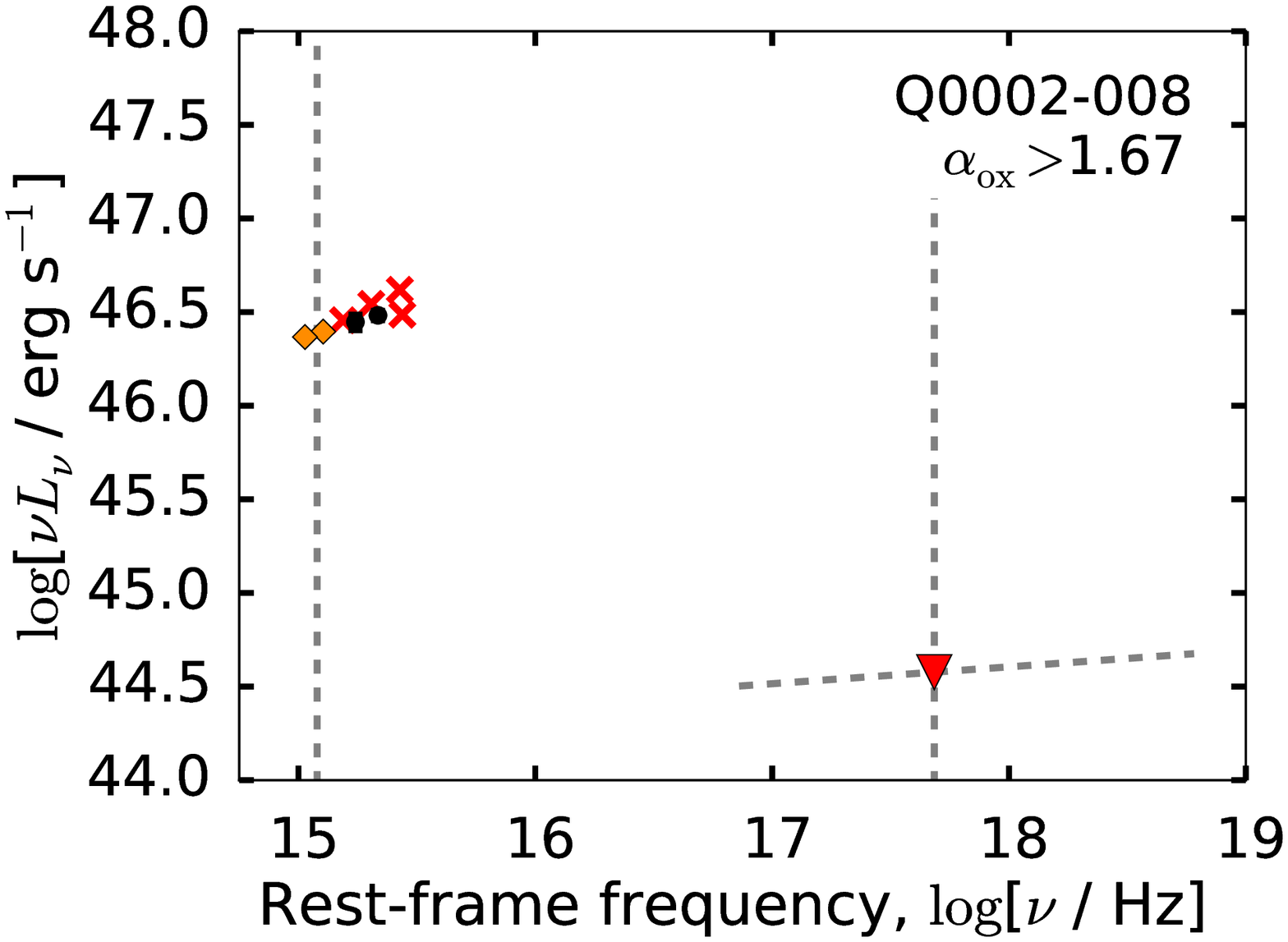}
	\includegraphics[width=\sedplotsize]{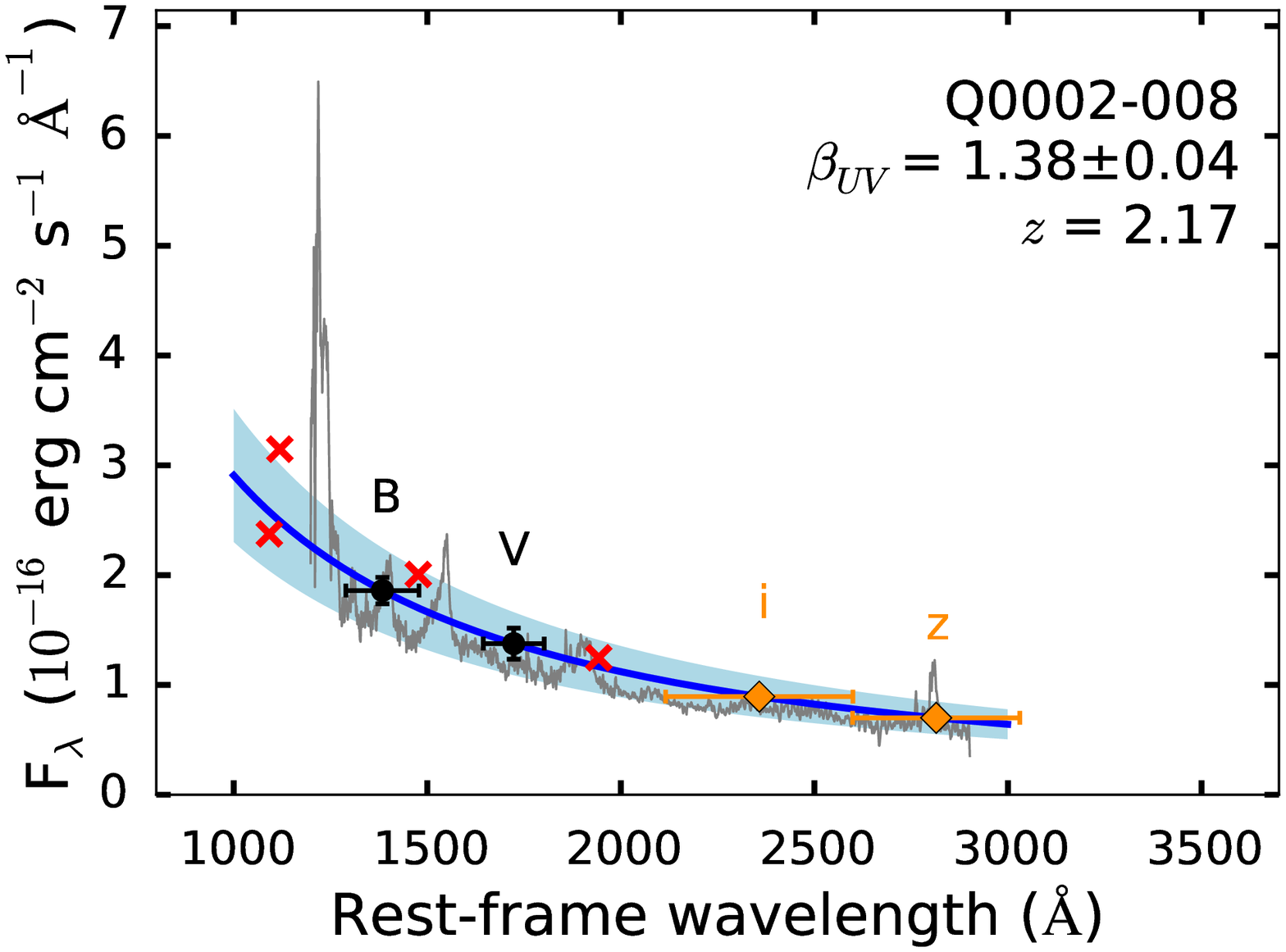}
	\includegraphics[width=\sedplotsize]{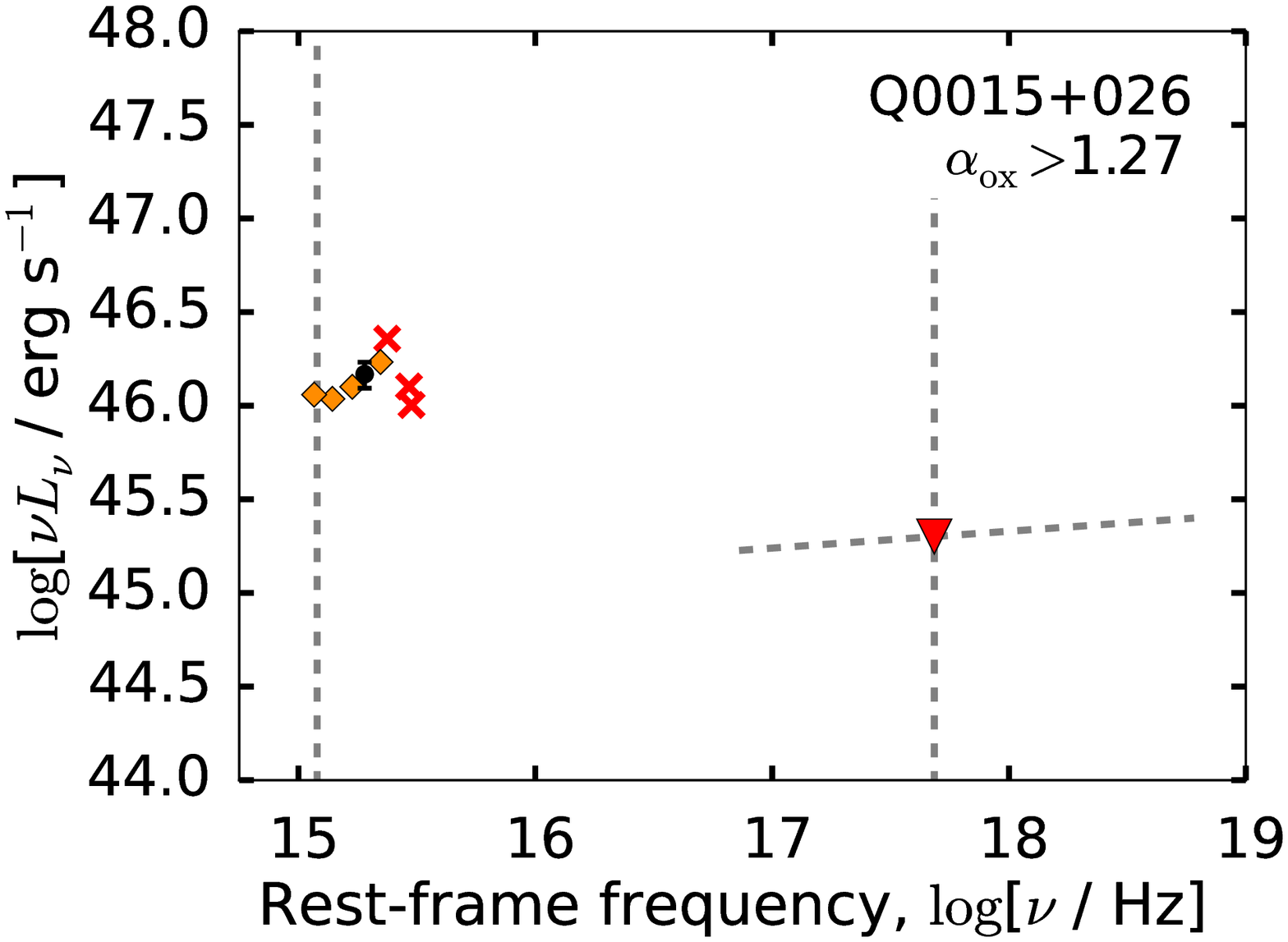}
	\includegraphics[width=\sedplotsize]{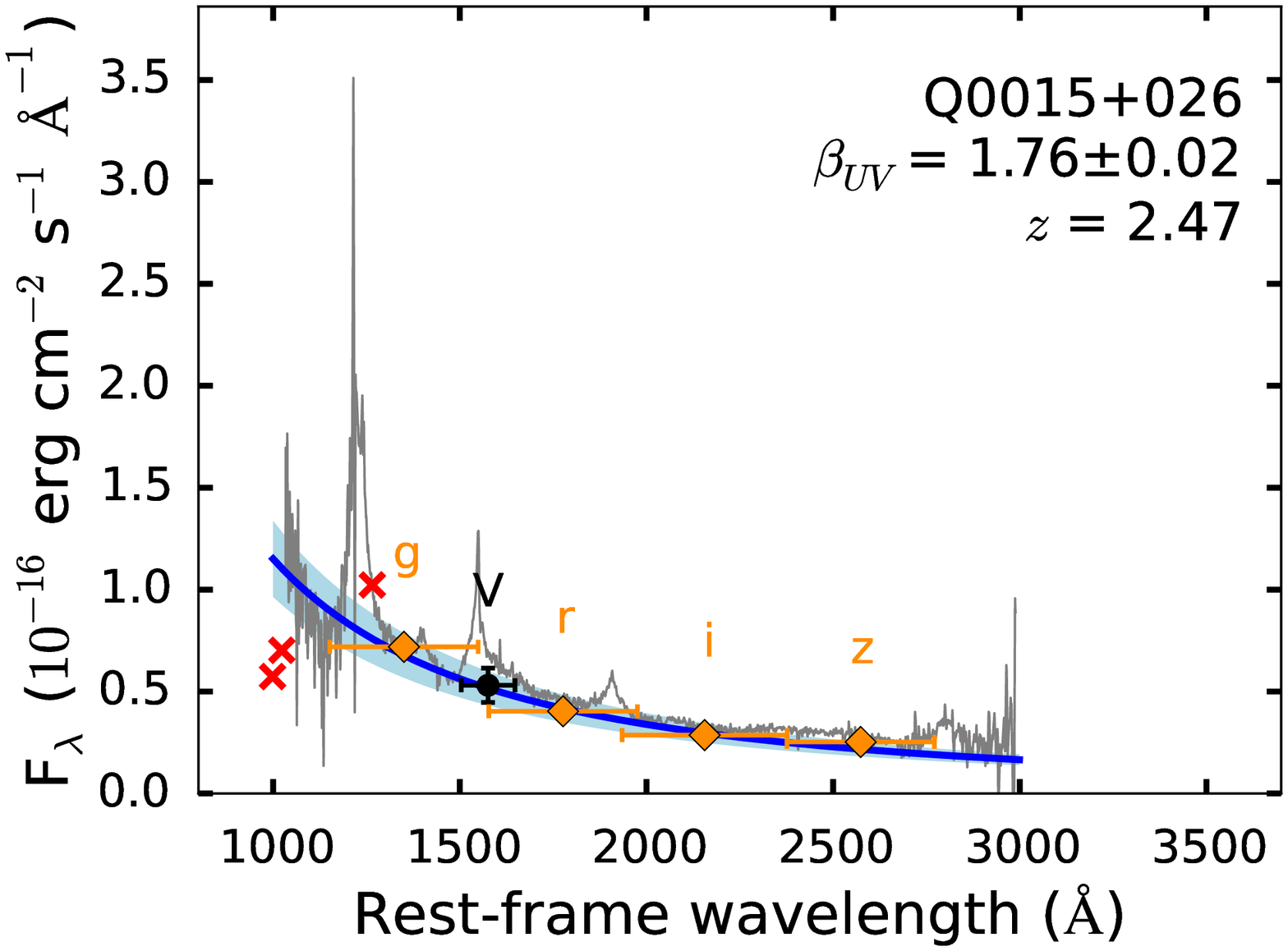}
	\includegraphics[width=\sedplotsize]{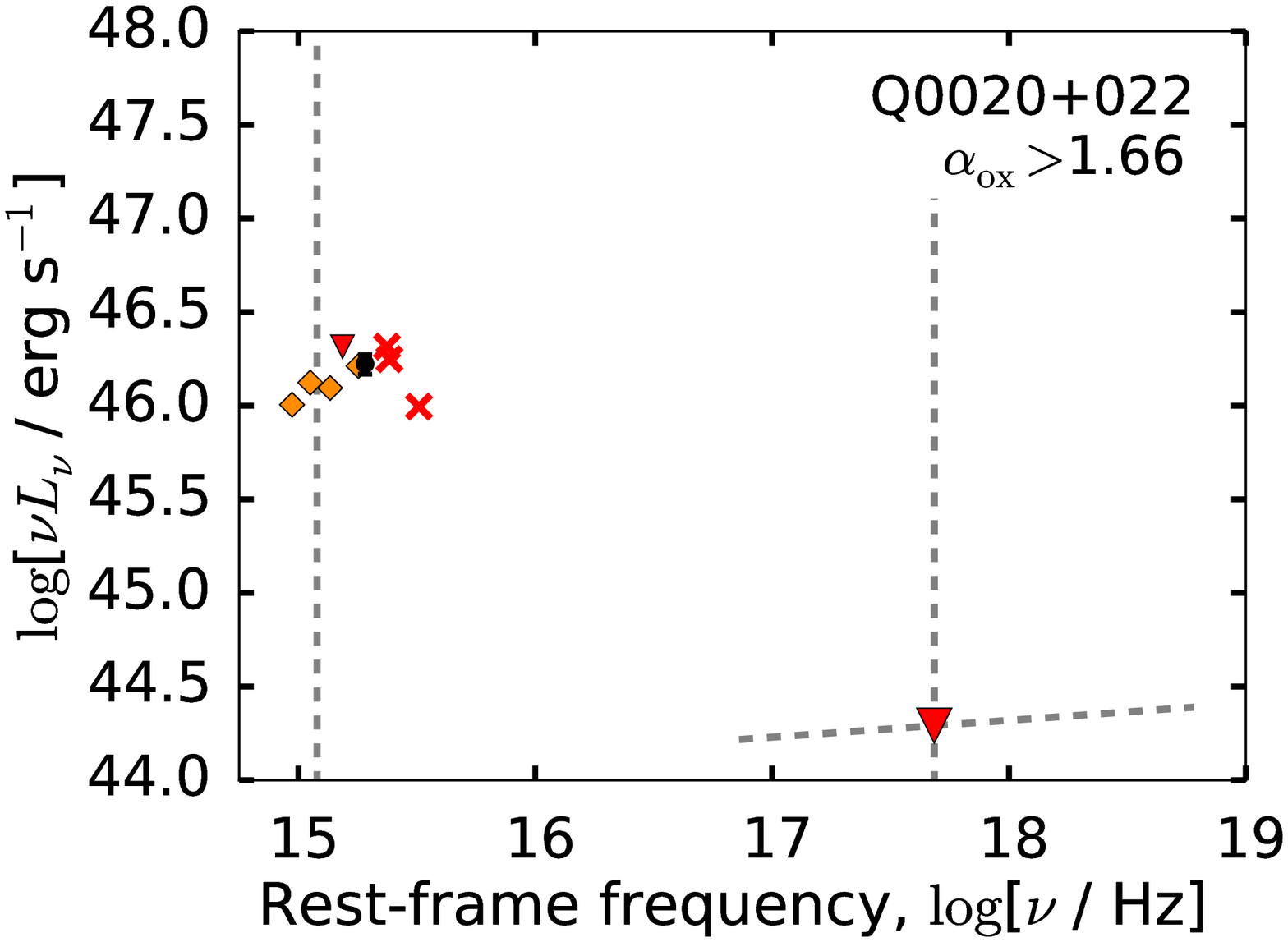}
	\includegraphics[width=\sedplotsize]{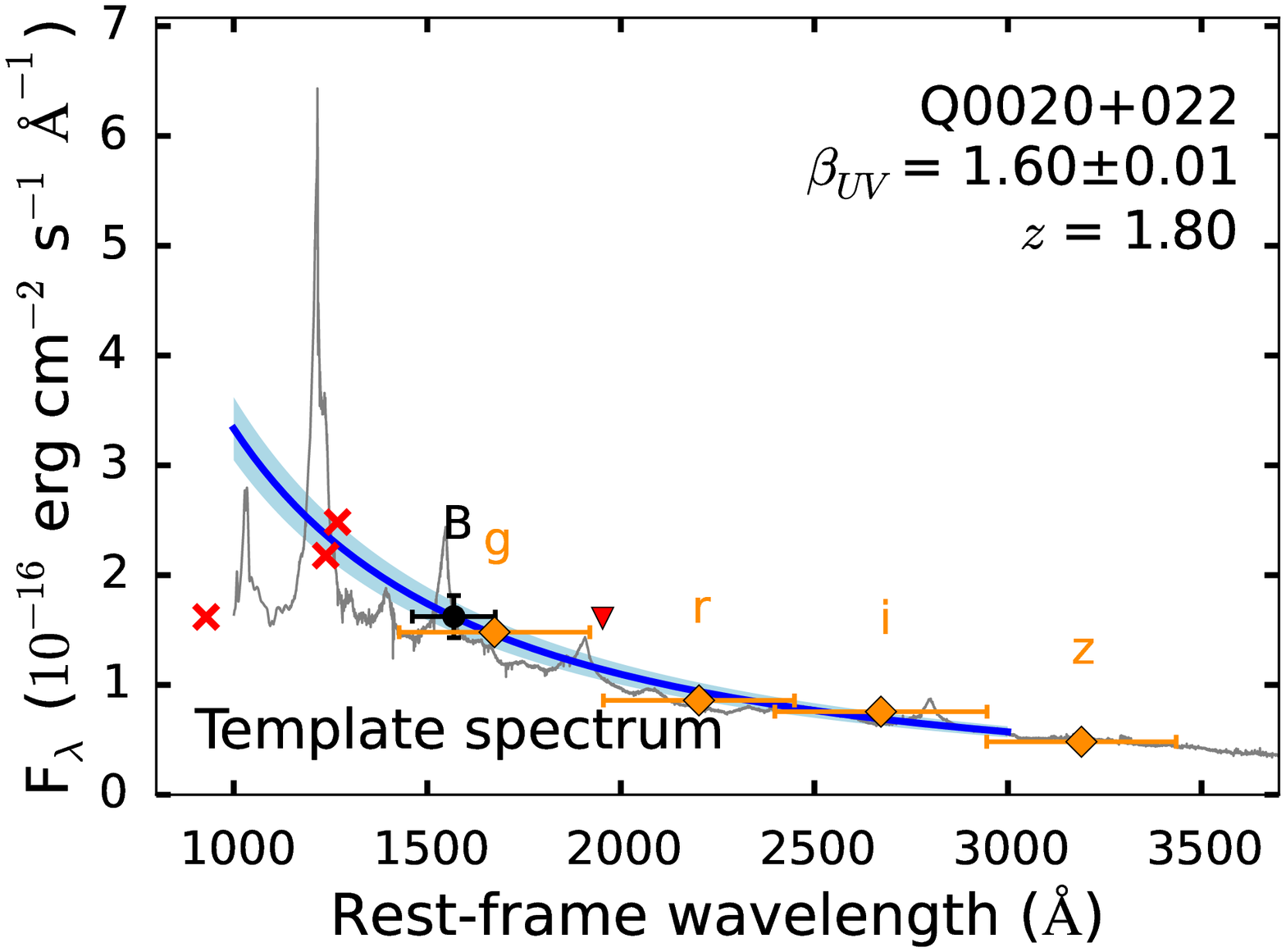}
	
	\caption{\emph{Left:} Rest-frame UV to X-ray spectral energy distributions (SEDs) of quasars in our sample. \emph{Right}: UV photometry and continuum modeling. See Figure \ref{fig:sed_mainpaper} for symbol and color coding.}
	\label{fig:sed5}
\end{figure*}

\paragraph*{Q0015+026, Q0020+022, Q0252+016 and Q0504+030:} The rescaling of the SDSS fluxes may be erroneously large due to the lack of continuum-dominated UVOT bandpasses.

\begin{figure*}
	\centering
	\includegraphics[width=\sedplotsize]{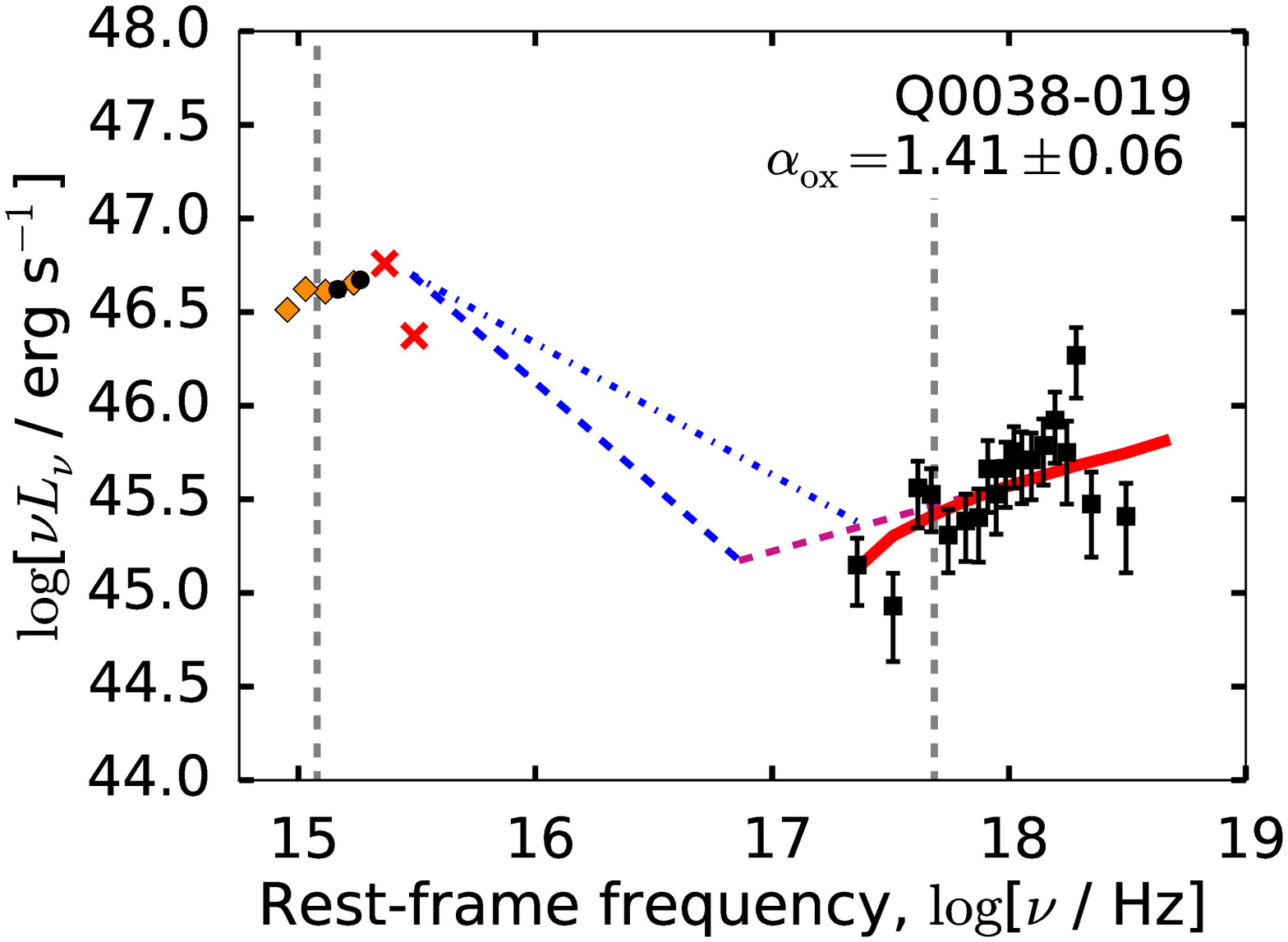}
	\includegraphics[width=\sedplotsize]{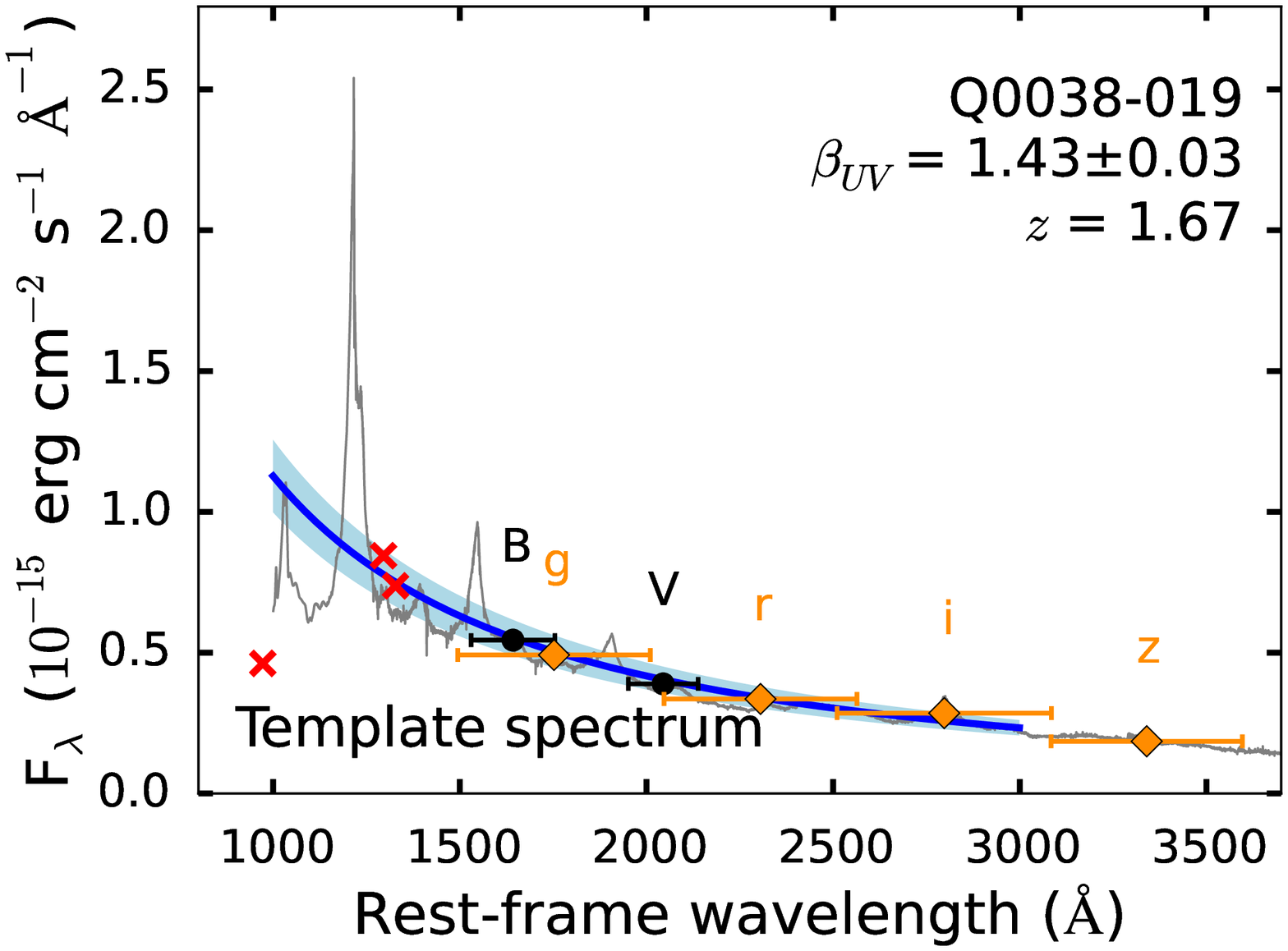}
	\includegraphics[width=\sedplotsize]{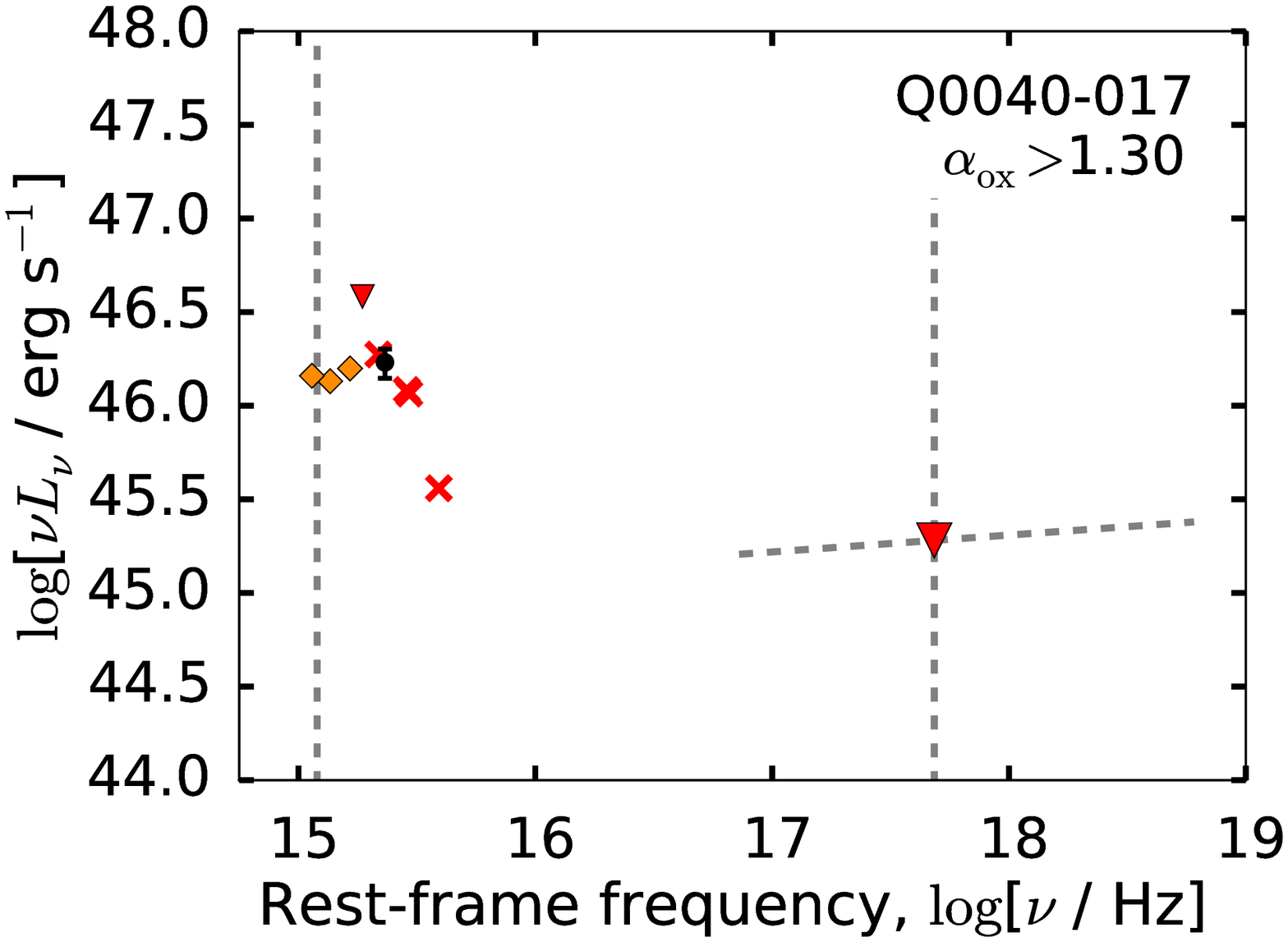}
	\includegraphics[width=\sedplotsize]{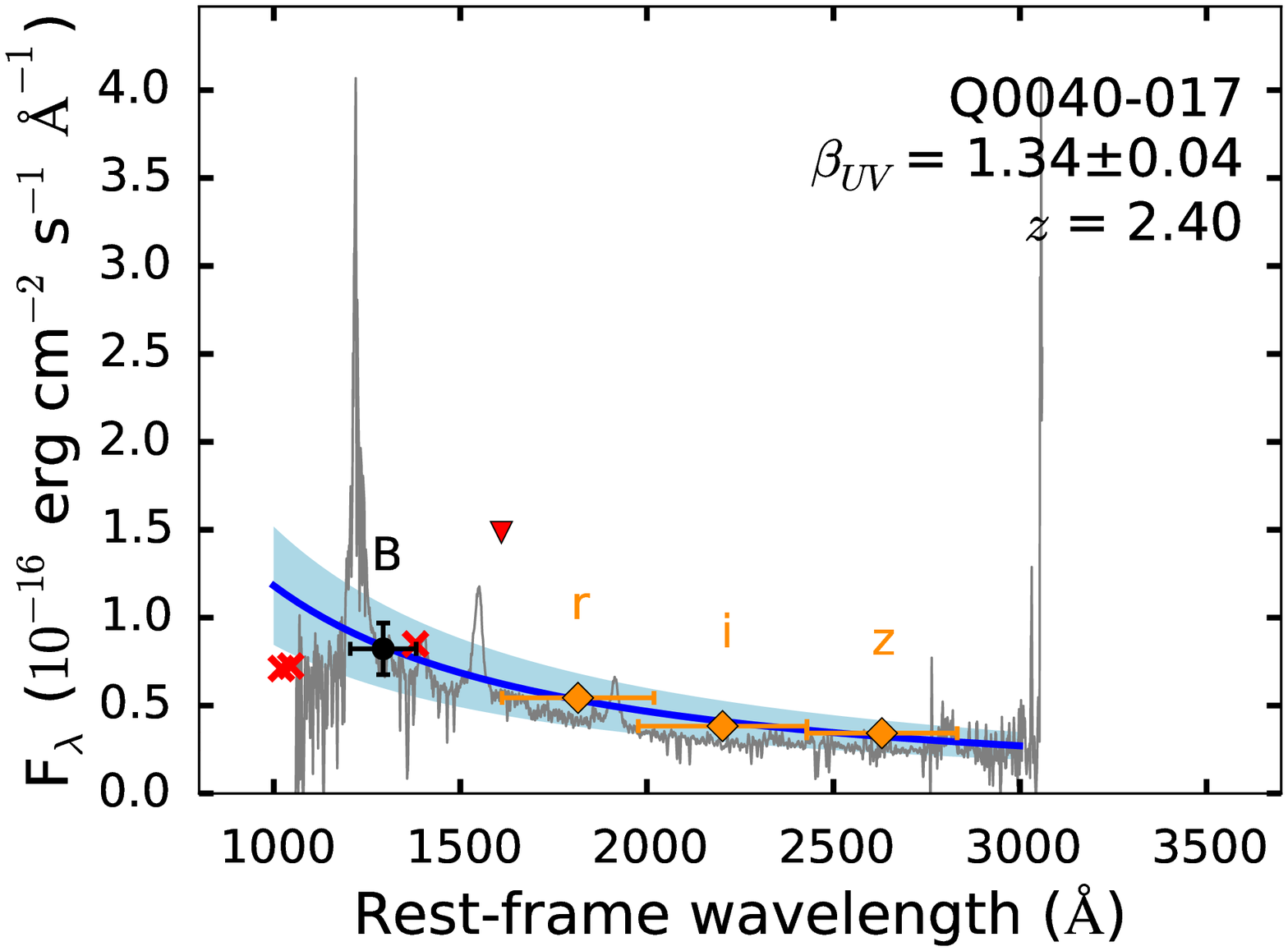}
	\includegraphics[width=\sedplotsize]{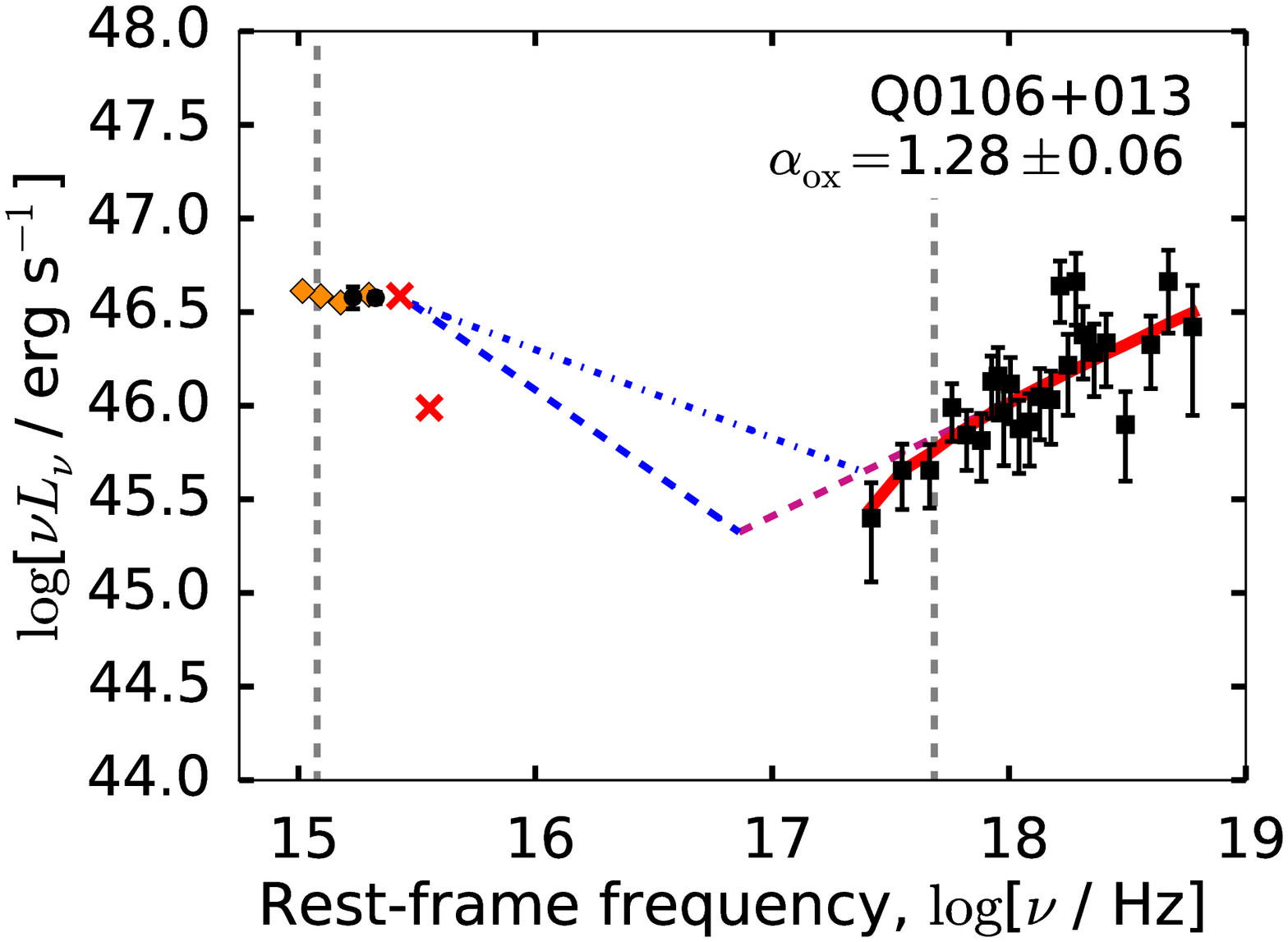}
	\includegraphics[width=\sedplotsize]{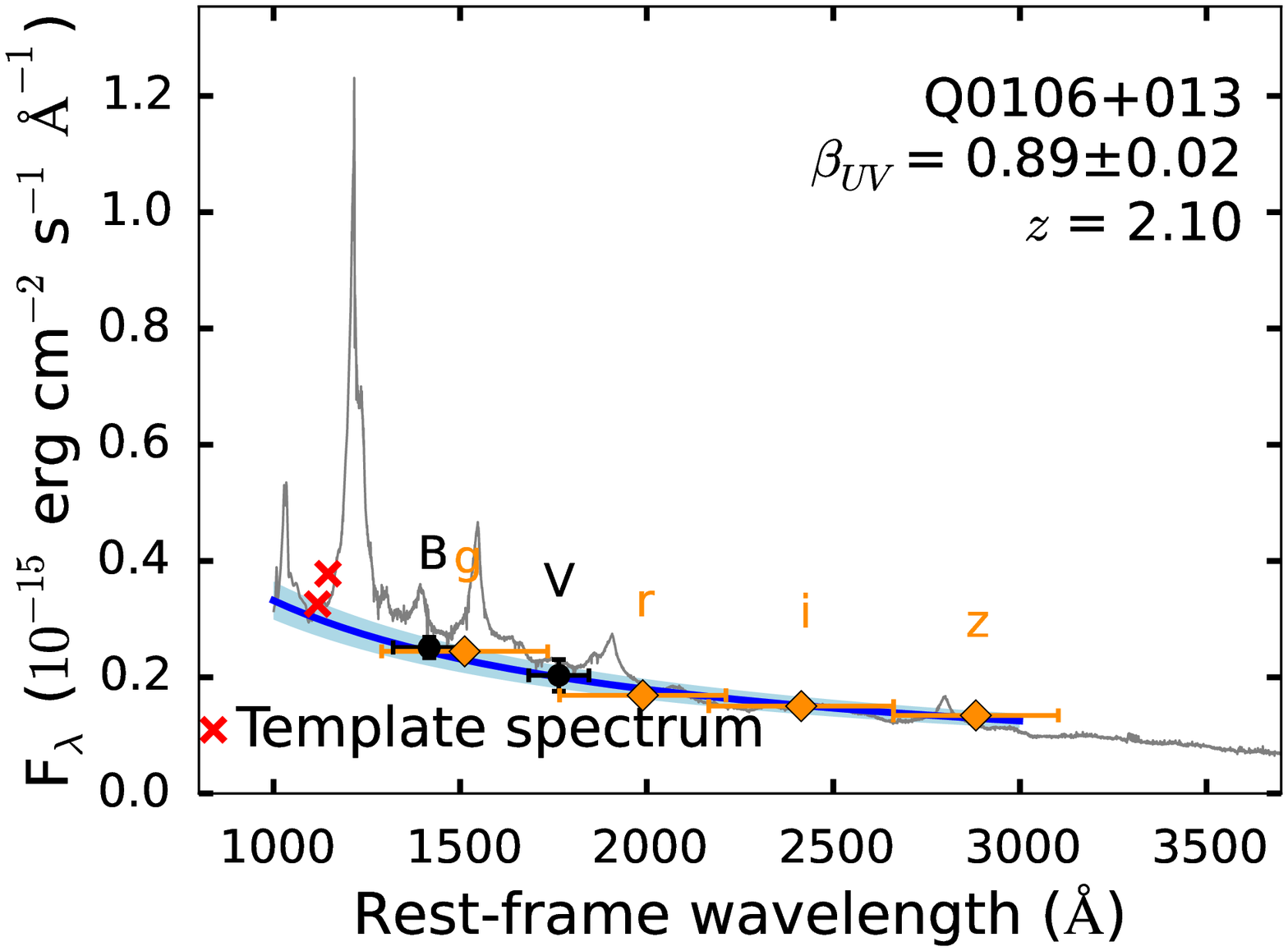}
	
	\caption{\emph{Left:} Rest-frame UV to X-ray spectral energy distributions (SEDs) of quasars in our sample. \emph{Right}: UV photometry and continuum modeling. See Figure \ref{fig:sed_mainpaper} for symbol and color coding.}
	\label{fig:sed6}
\end{figure*}

\begin{figure*}
	\centering
	\includegraphics[width=\sedplotsize]{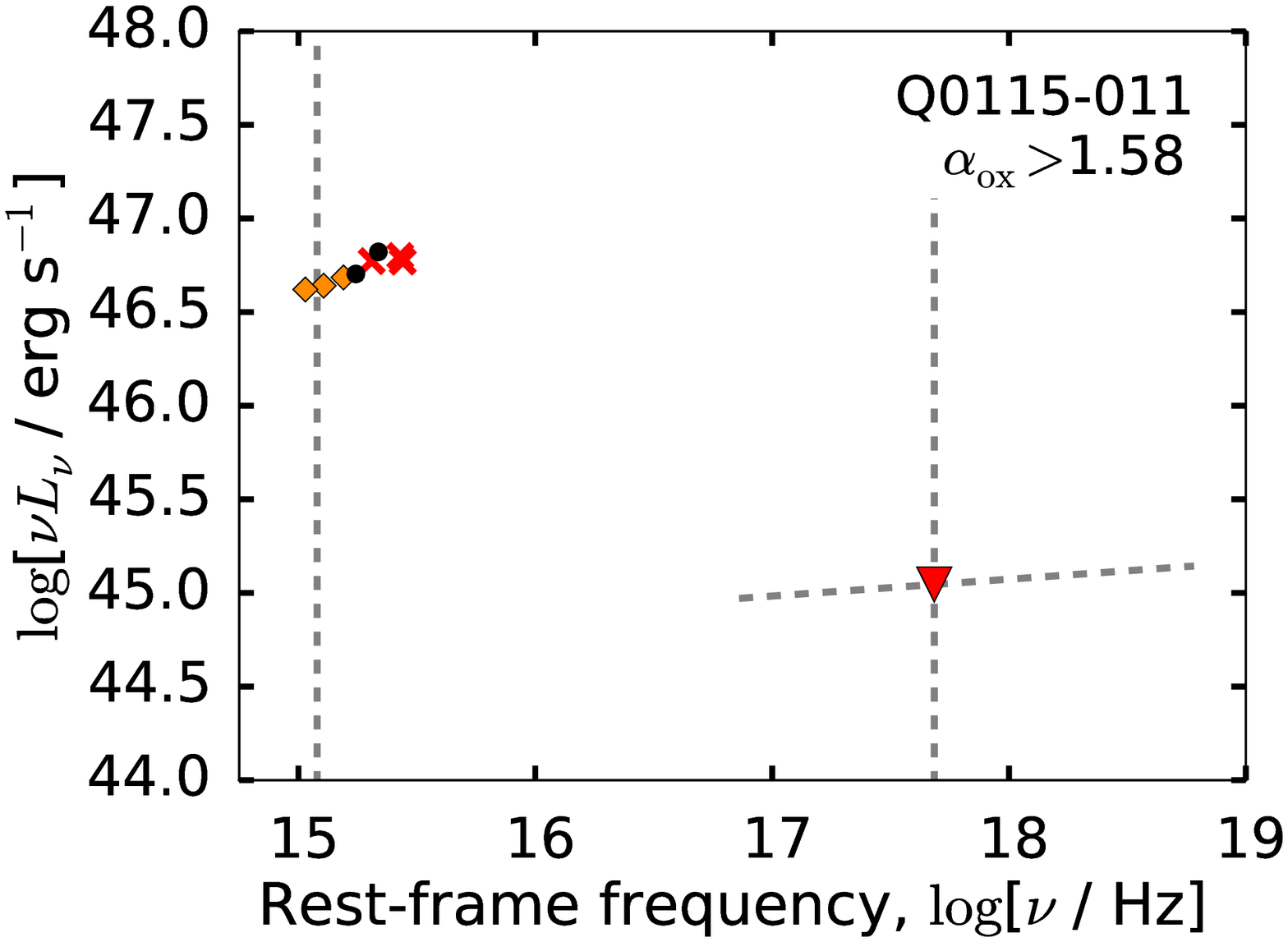}
	\includegraphics[width=\sedplotsize]{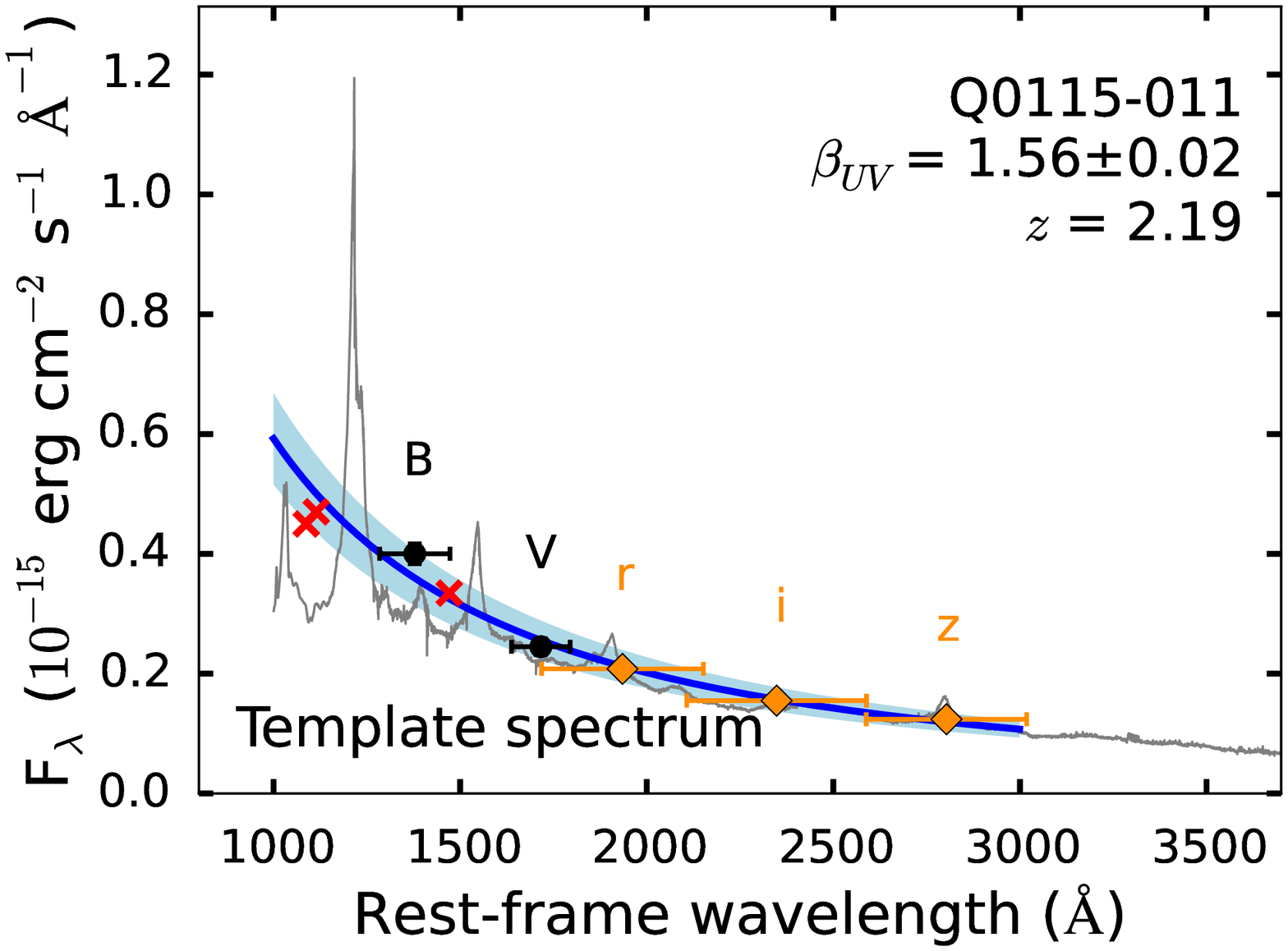}
	\includegraphics[width=\sedplotsize]{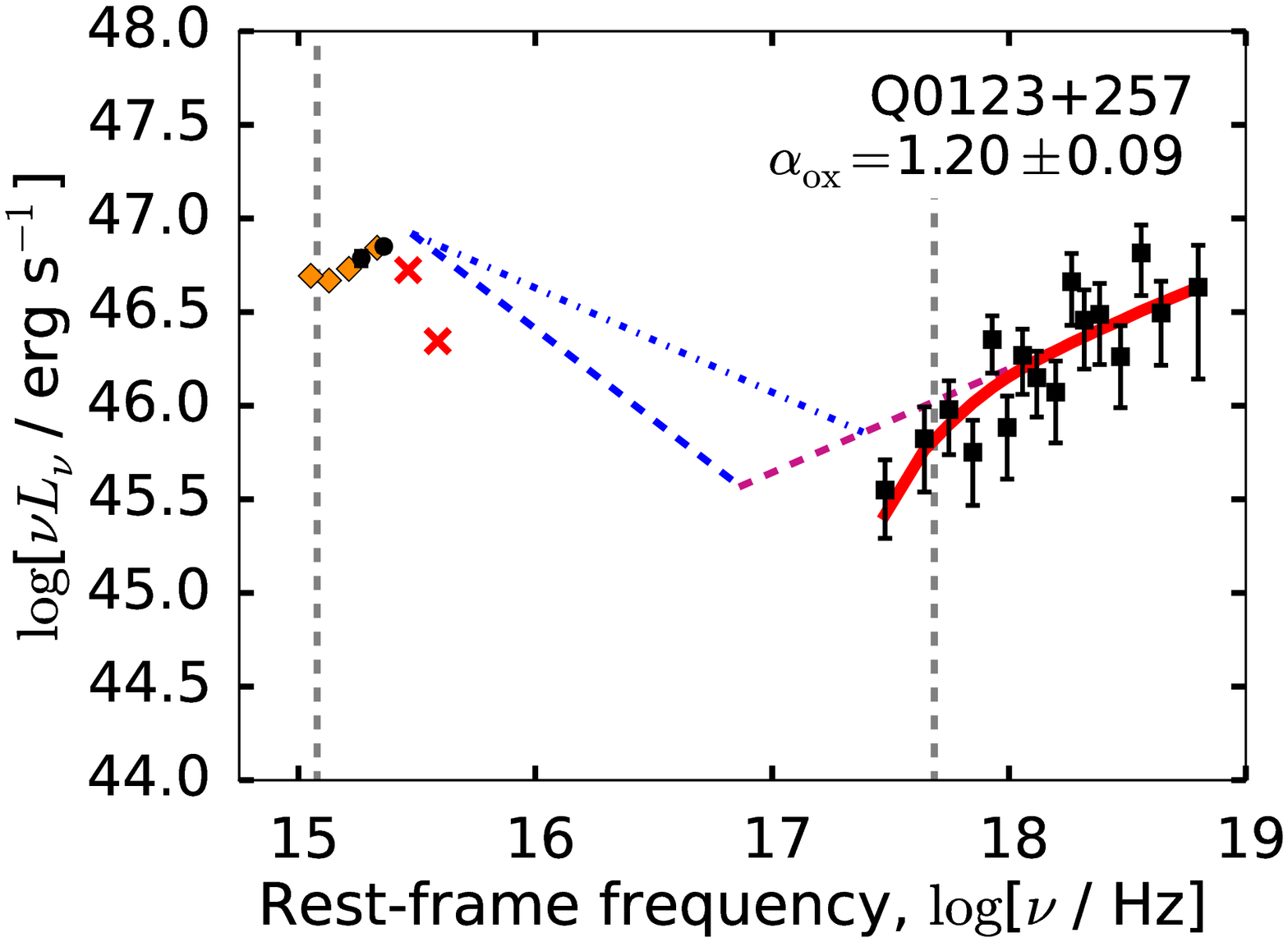}
	\includegraphics[width=\sedplotsize]{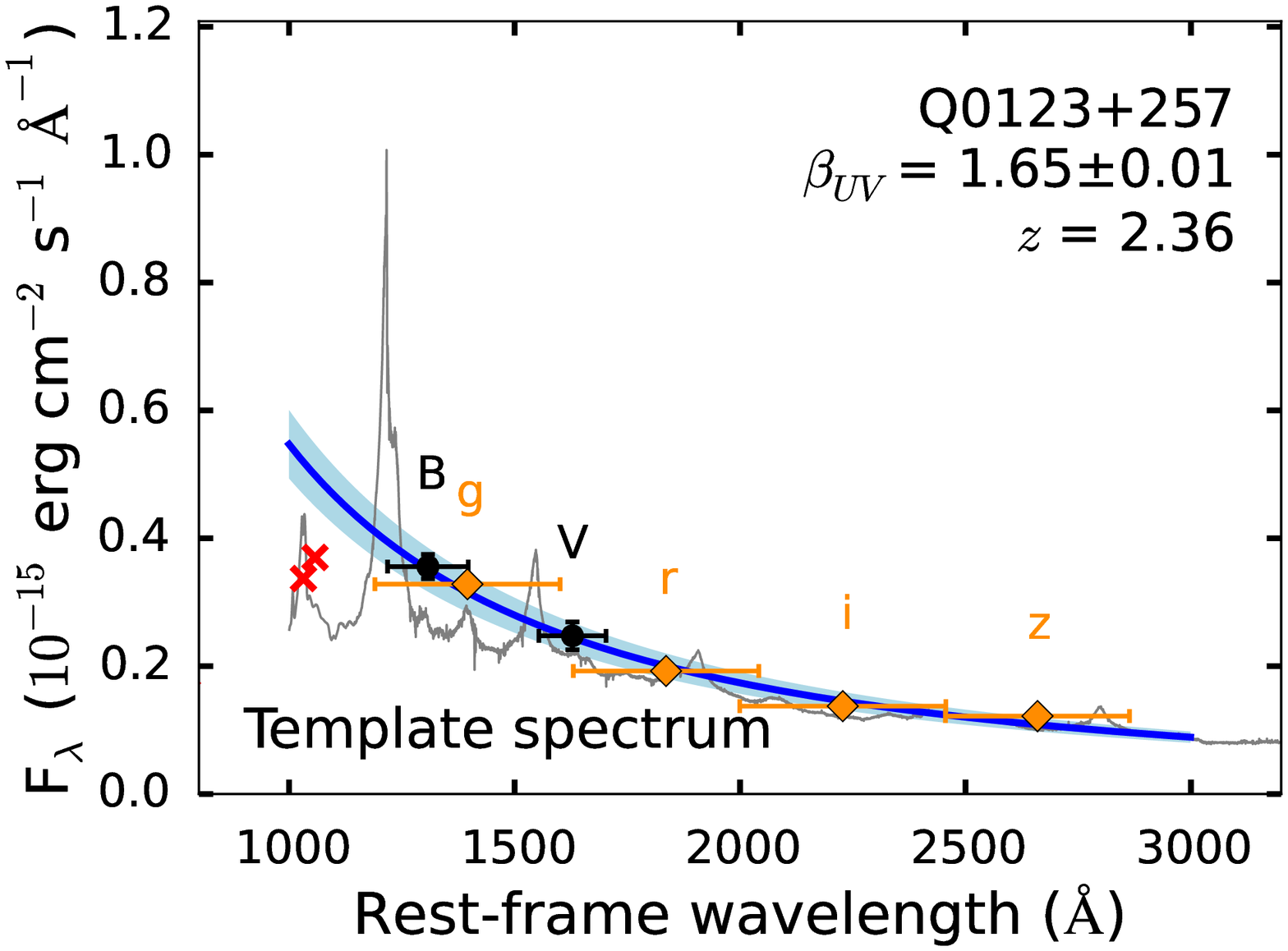}
	\includegraphics[width=\sedplotsize]{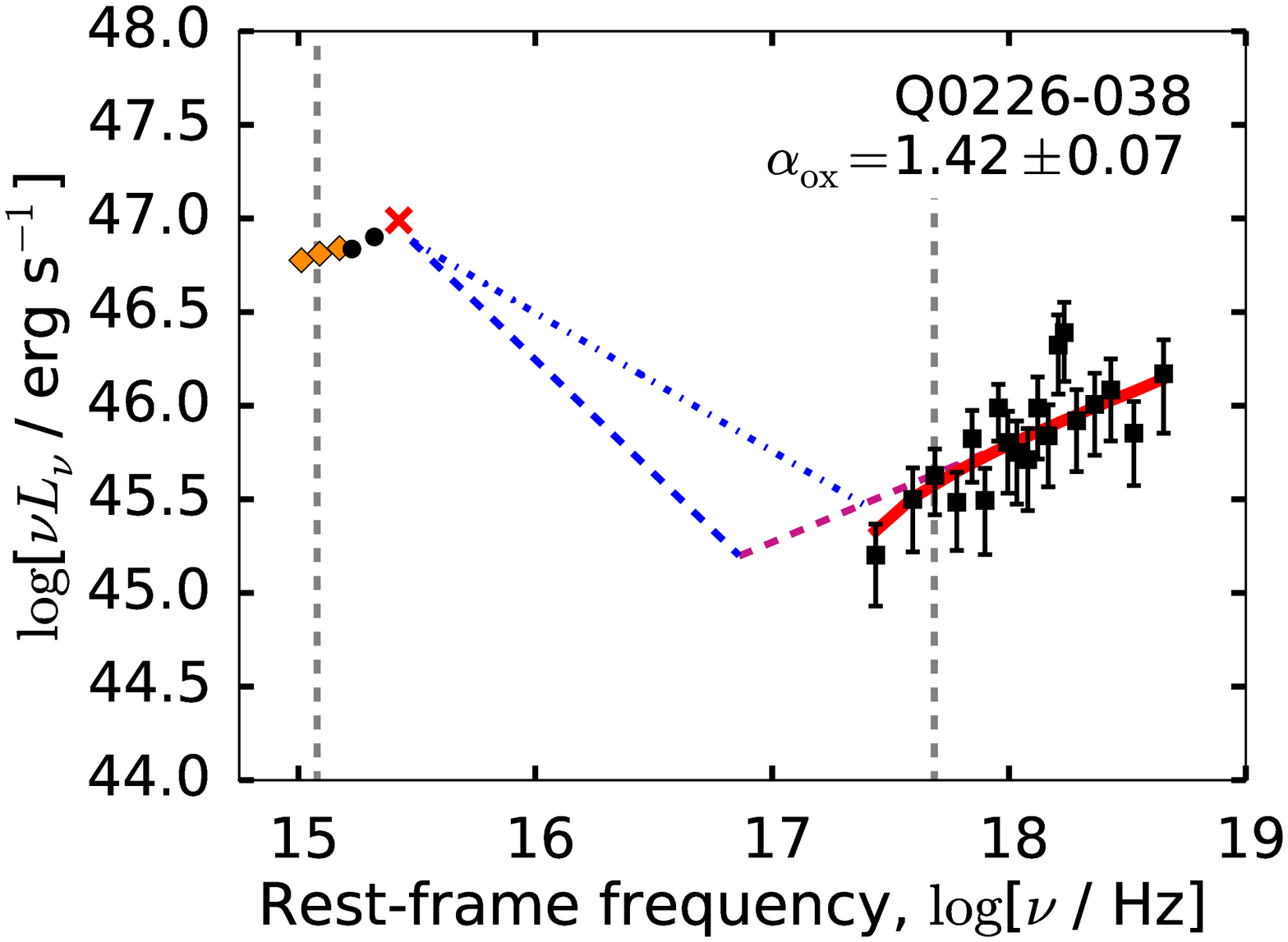}
	\includegraphics[width=\sedplotsize]{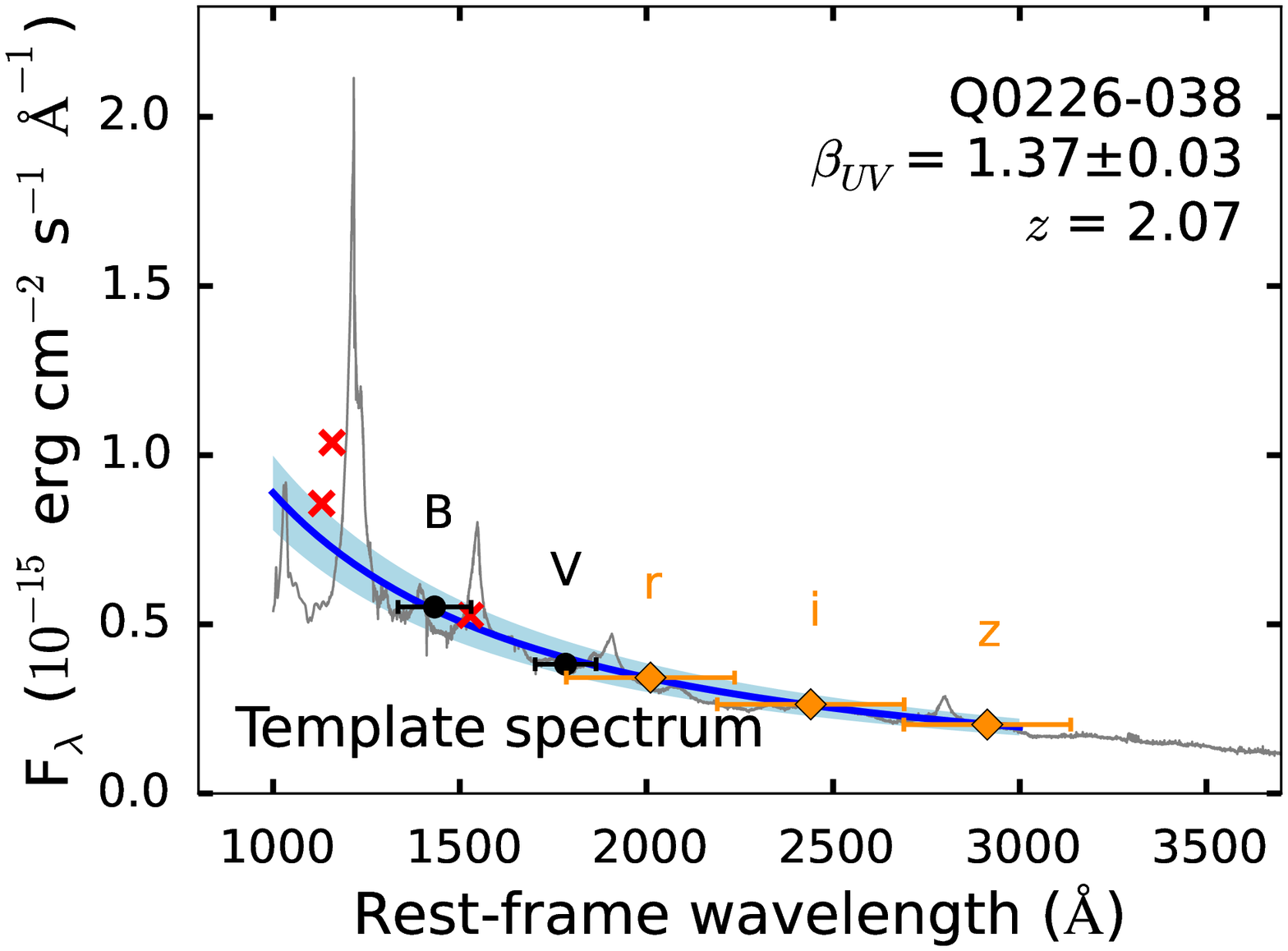}
	
	\caption{\emph{Left:} Rest-frame UV to X-ray spectral energy distributions (SEDs) of quasars in our sample. \emph{Right}: UV photometry and continuum modeling. See Figure \ref{fig:sed_mainpaper} for symbol and color coding.}
	\label{fig:sed7}
\end{figure*}

\begin{figure*}
	\centering
	\includegraphics[width=\sedplotsize]{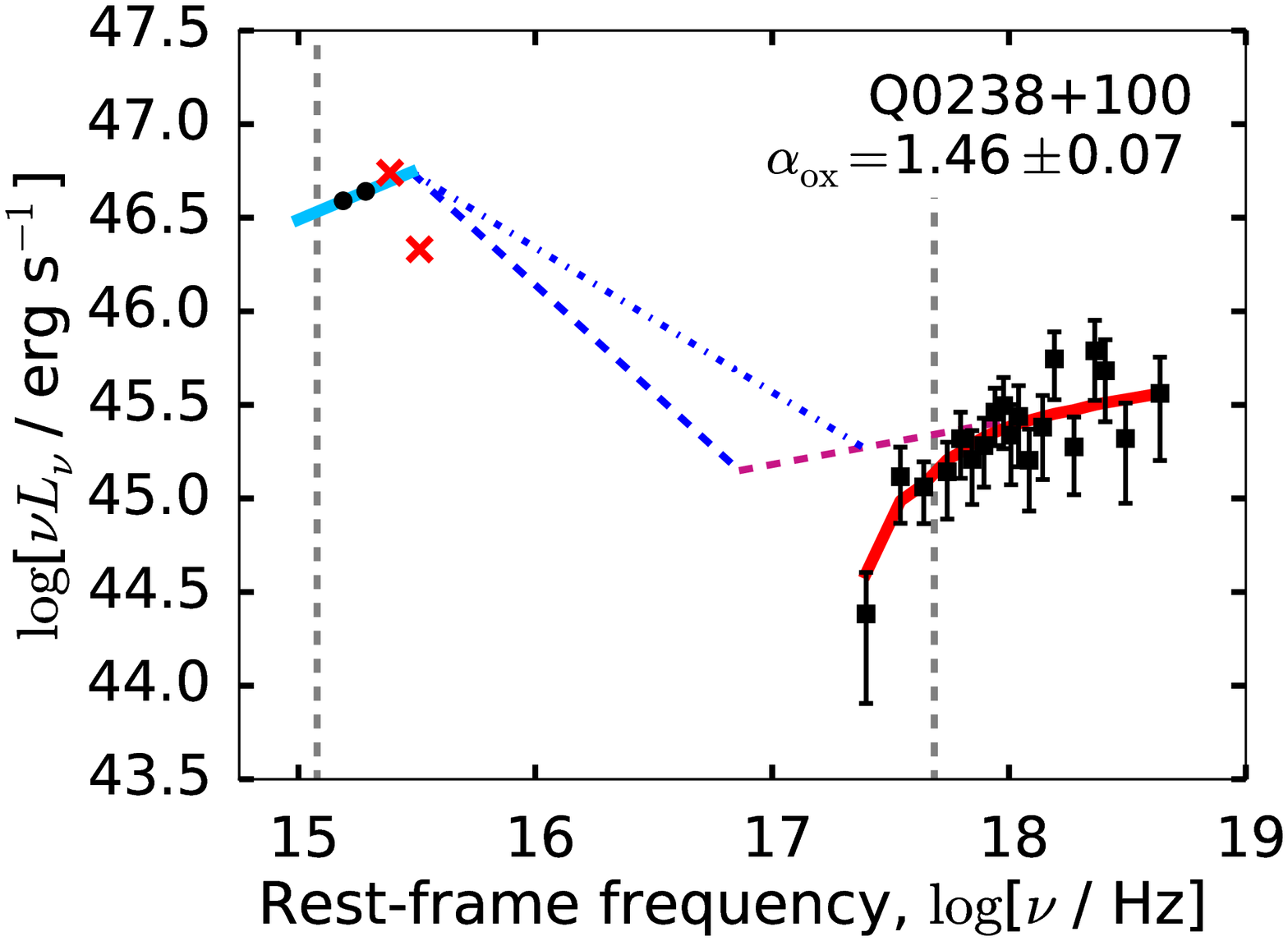}
	\includegraphics[width=\sedplotsize]{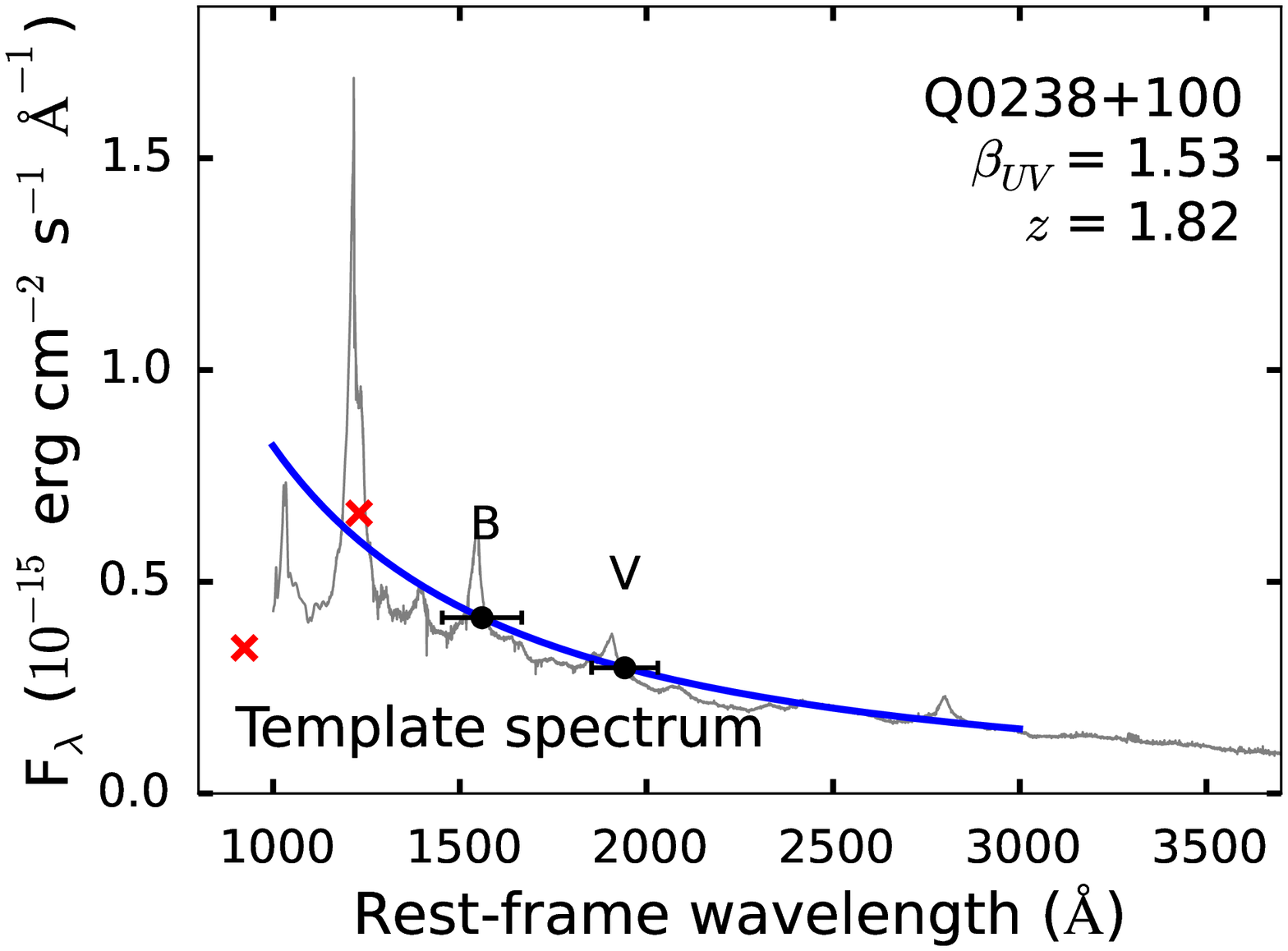}
	\includegraphics[width=\sedplotsize]{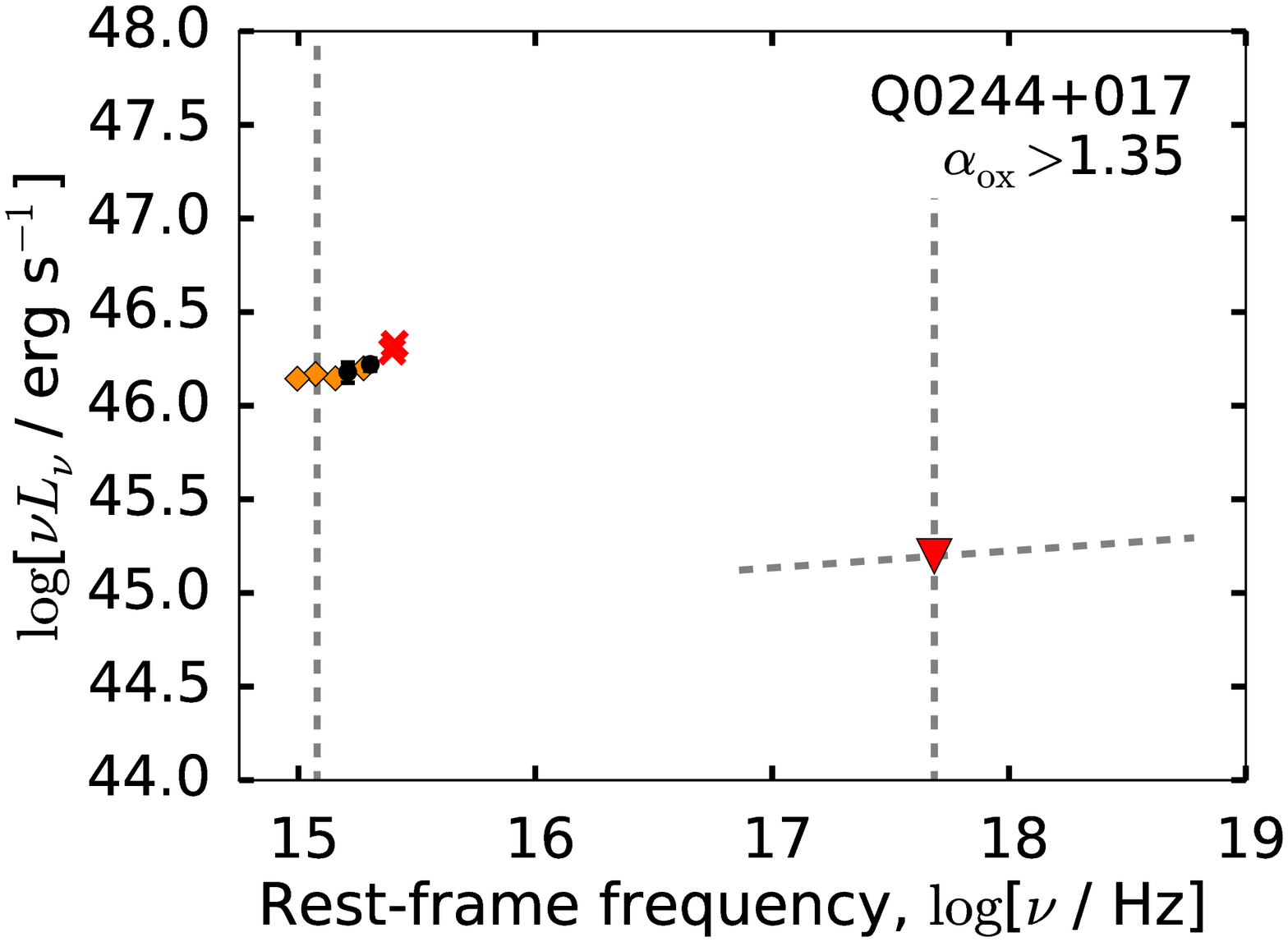}
	\includegraphics[width=\sedplotsize]{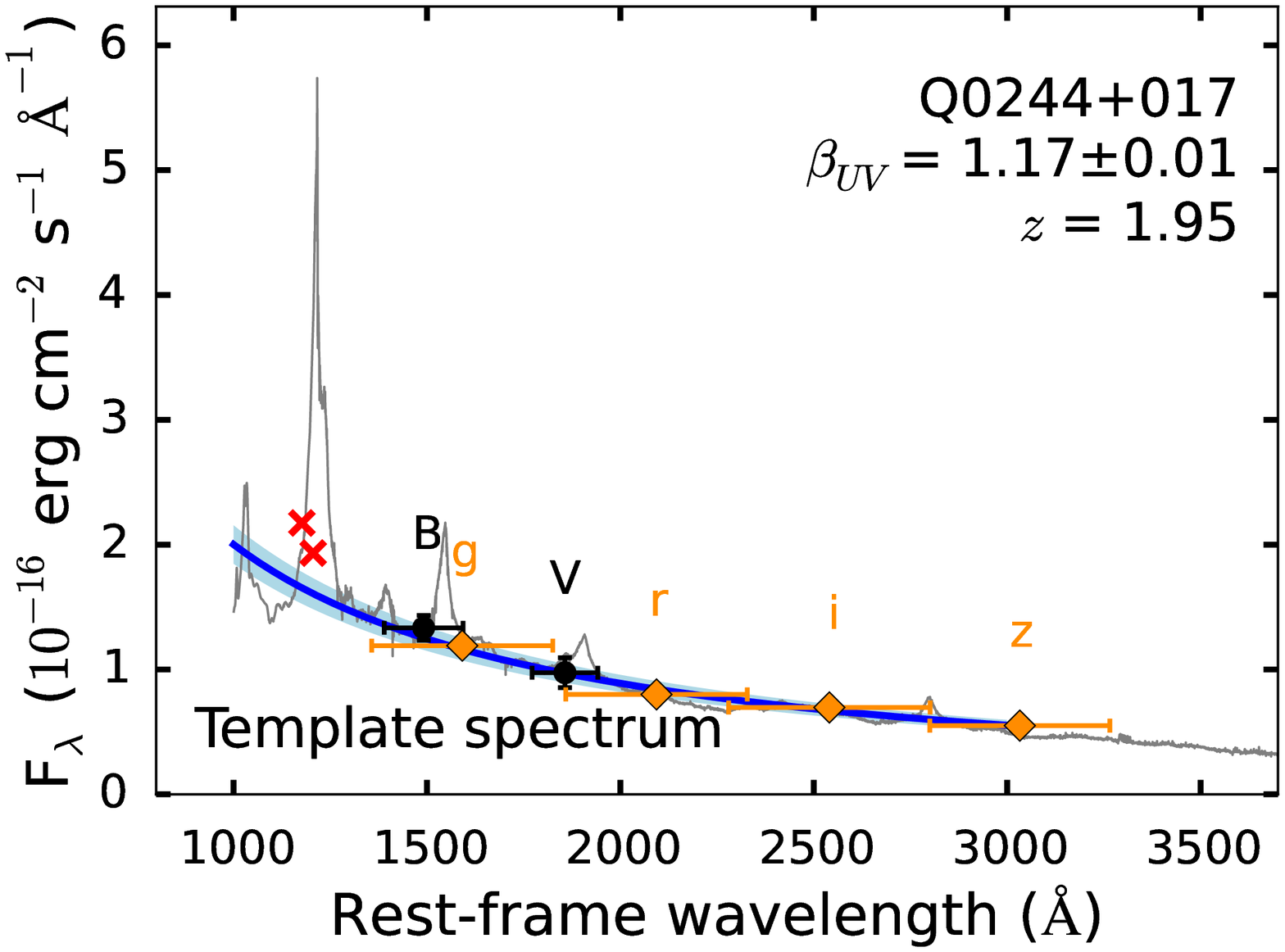}
	\includegraphics[width=\sedplotsize]{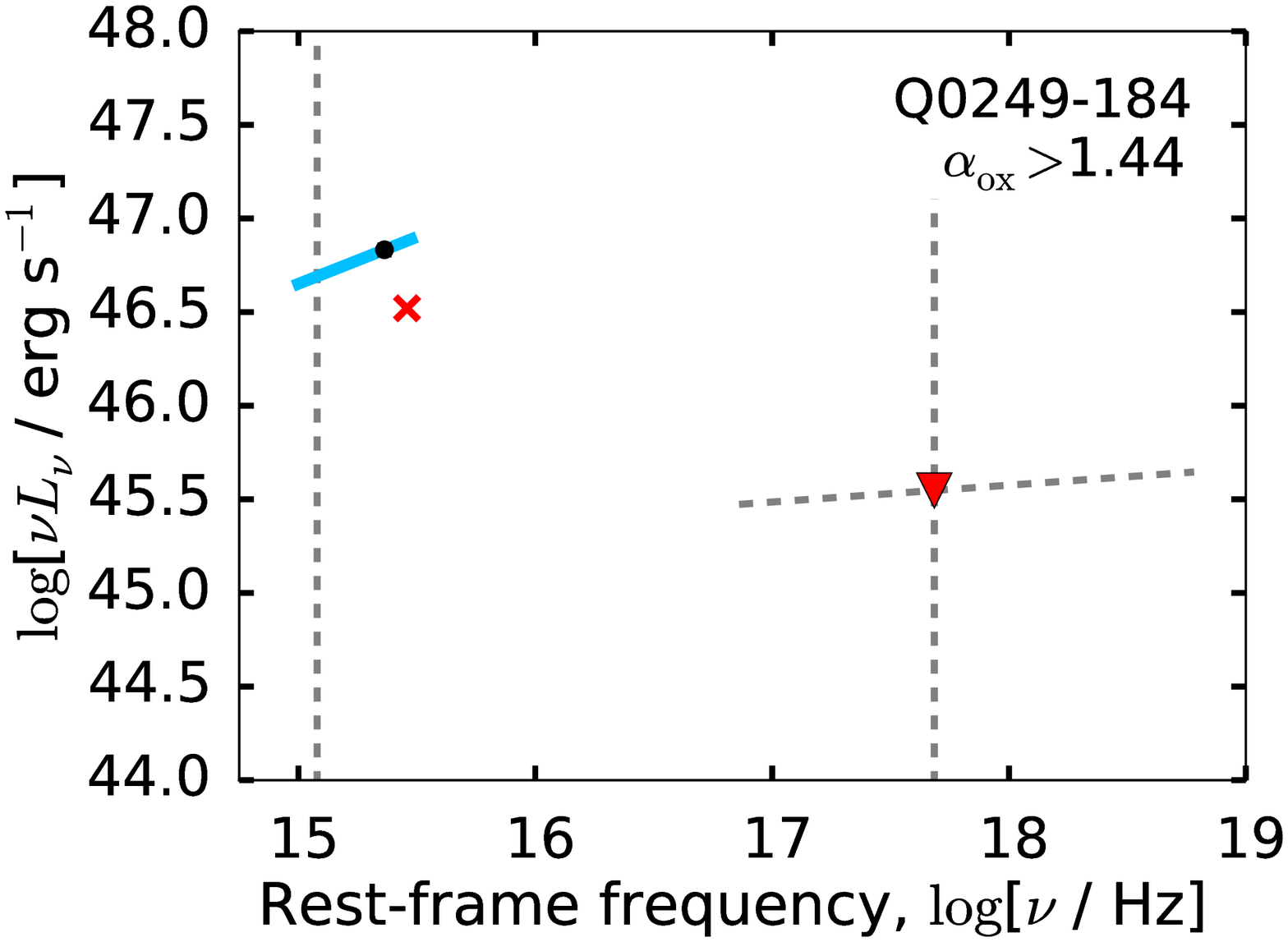}
	\includegraphics[width=\sedplotsize]{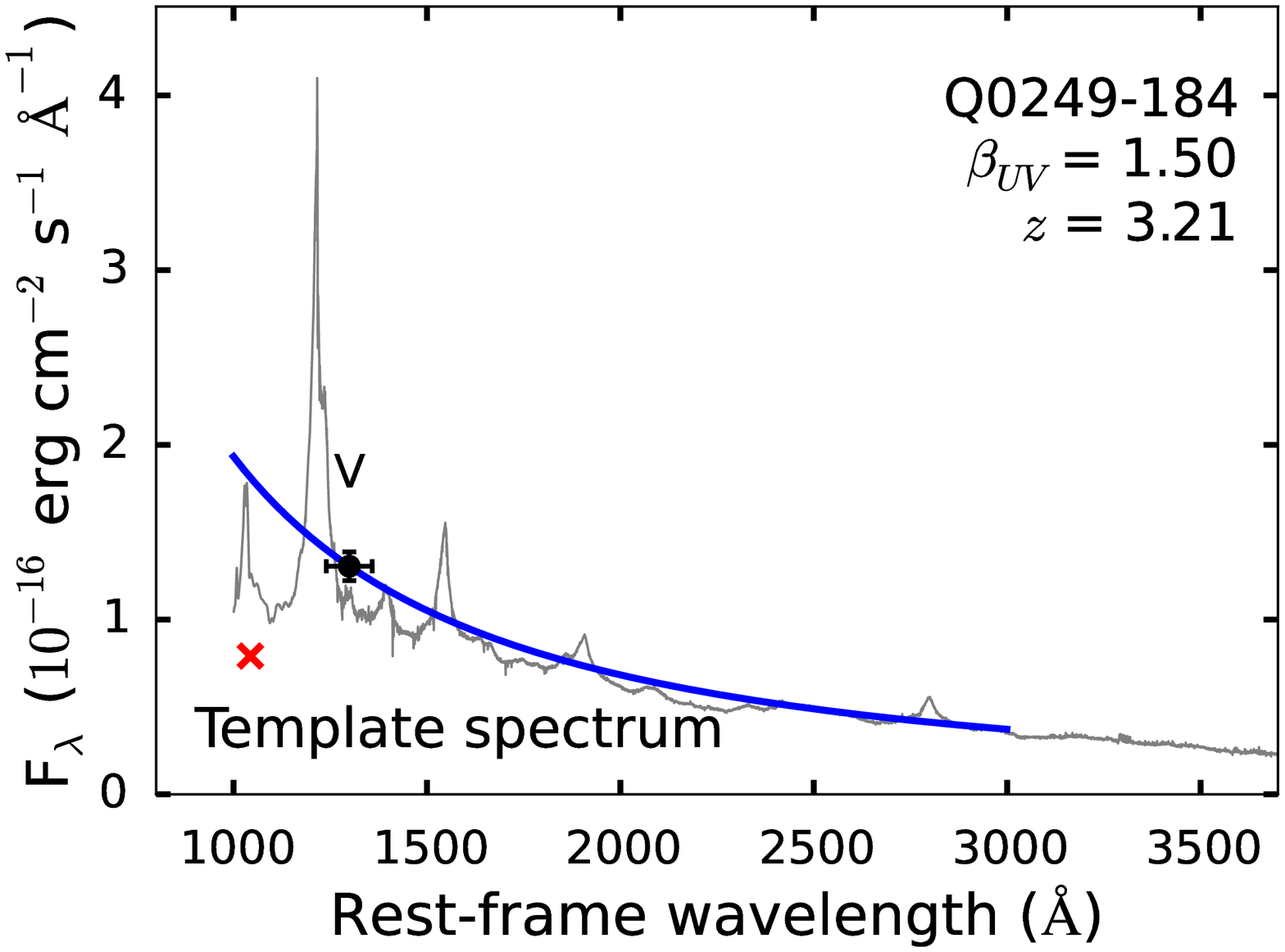}
	
	\caption{\emph{Left:} Rest-frame UV to X-ray spectral energy distributions (SEDs) of quasars in our sample. \emph{Right}: UV photometry and continuum modeling. See Figure \ref{fig:sed_mainpaper} for symbol and color coding.}
	\label{fig:sed8}
\end{figure*}

\begin{figure*}
	\centering
	\includegraphics[width=\sedplotsize]{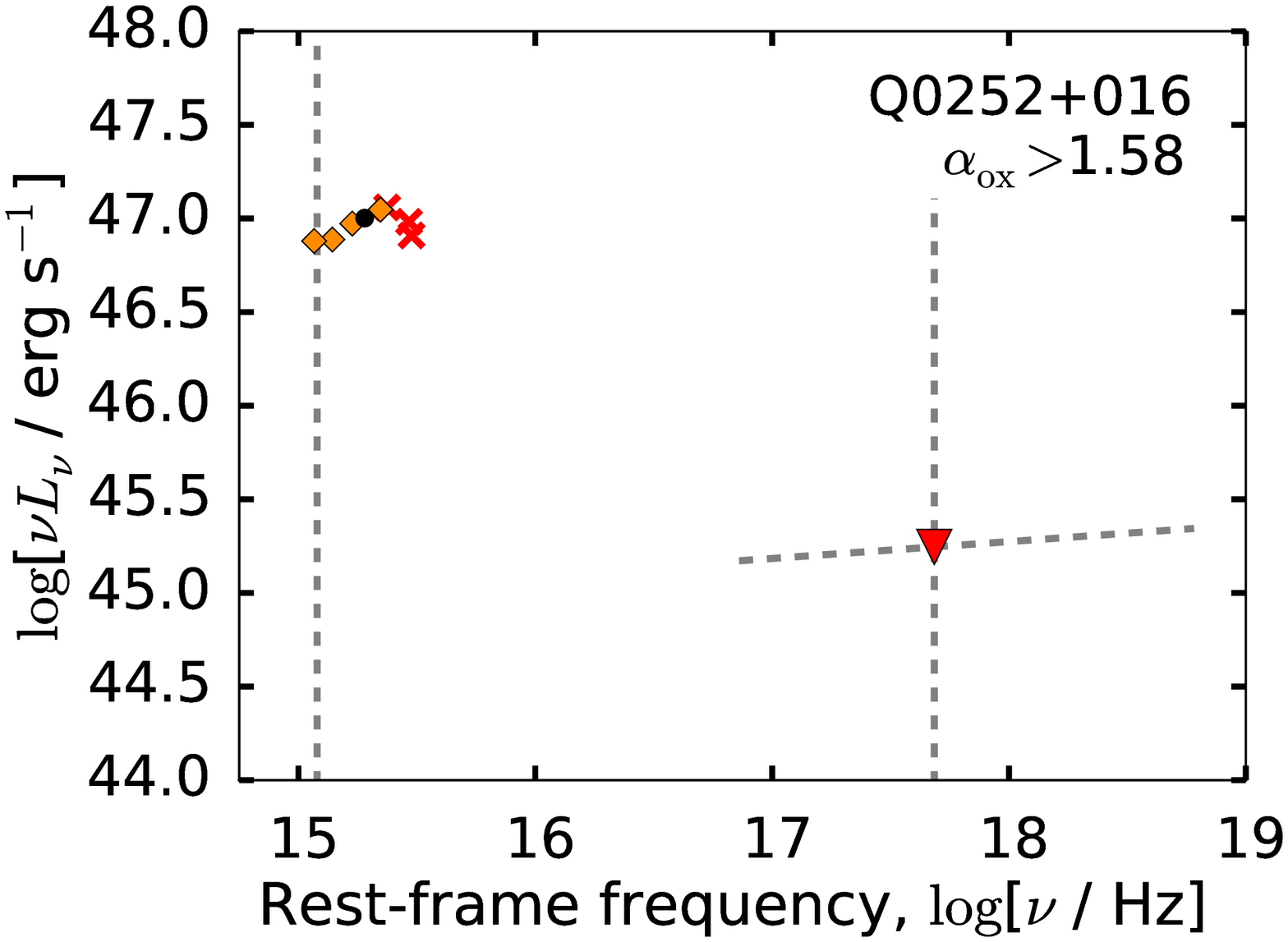}
	\includegraphics[width=\sedplotsize]{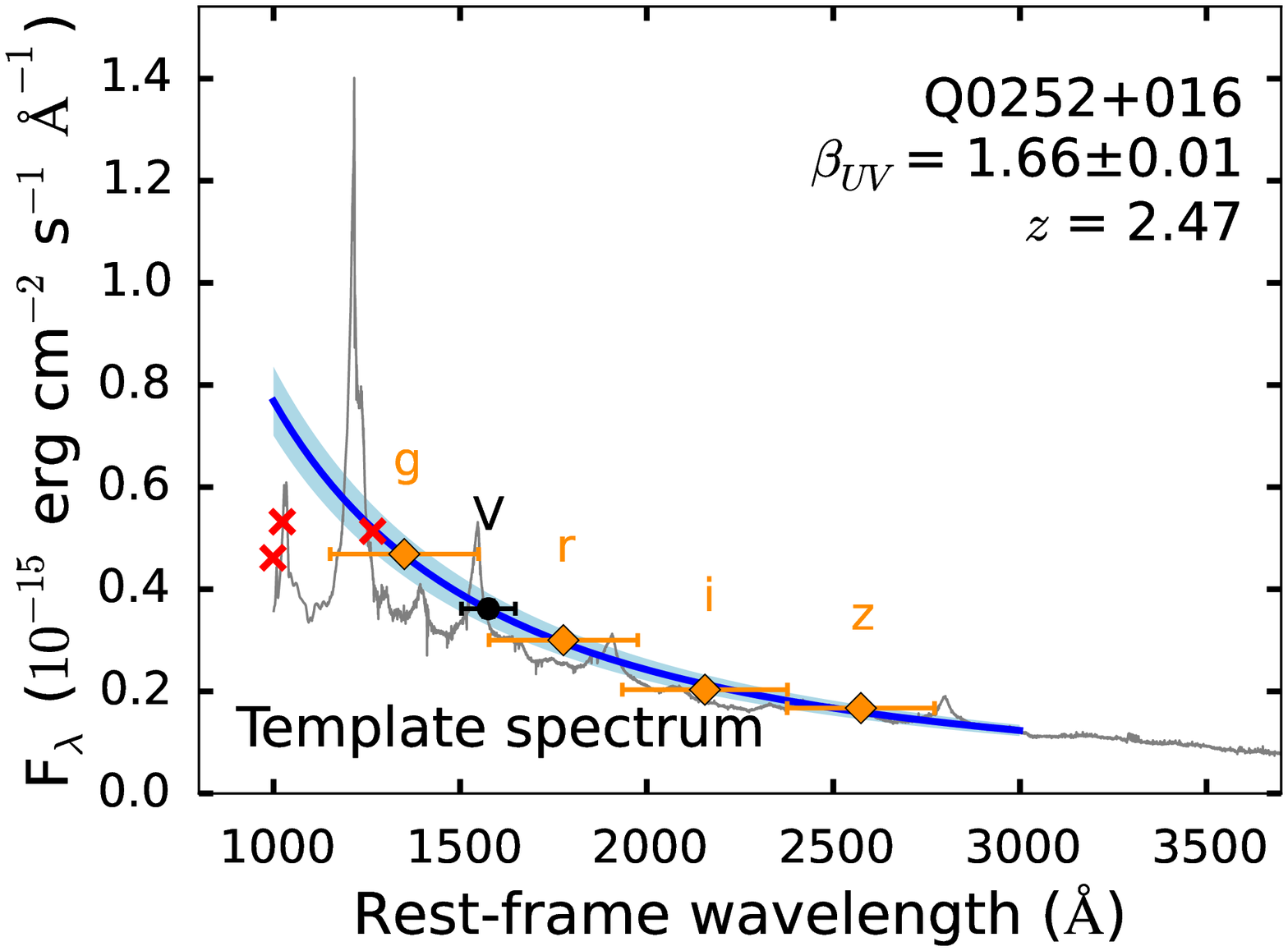}
	\includegraphics[width=\sedplotsize]{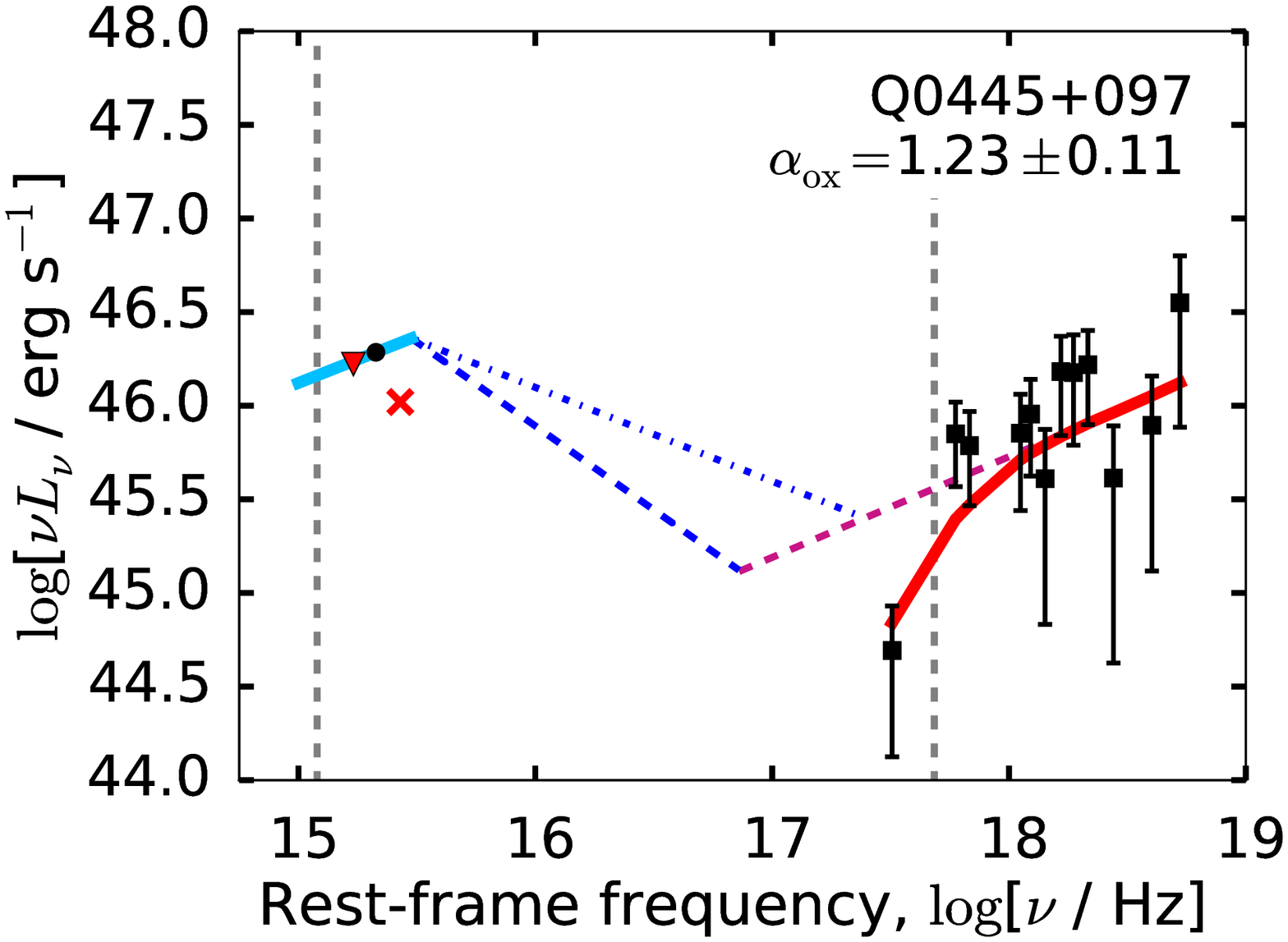}
	\includegraphics[width=\sedplotsize]{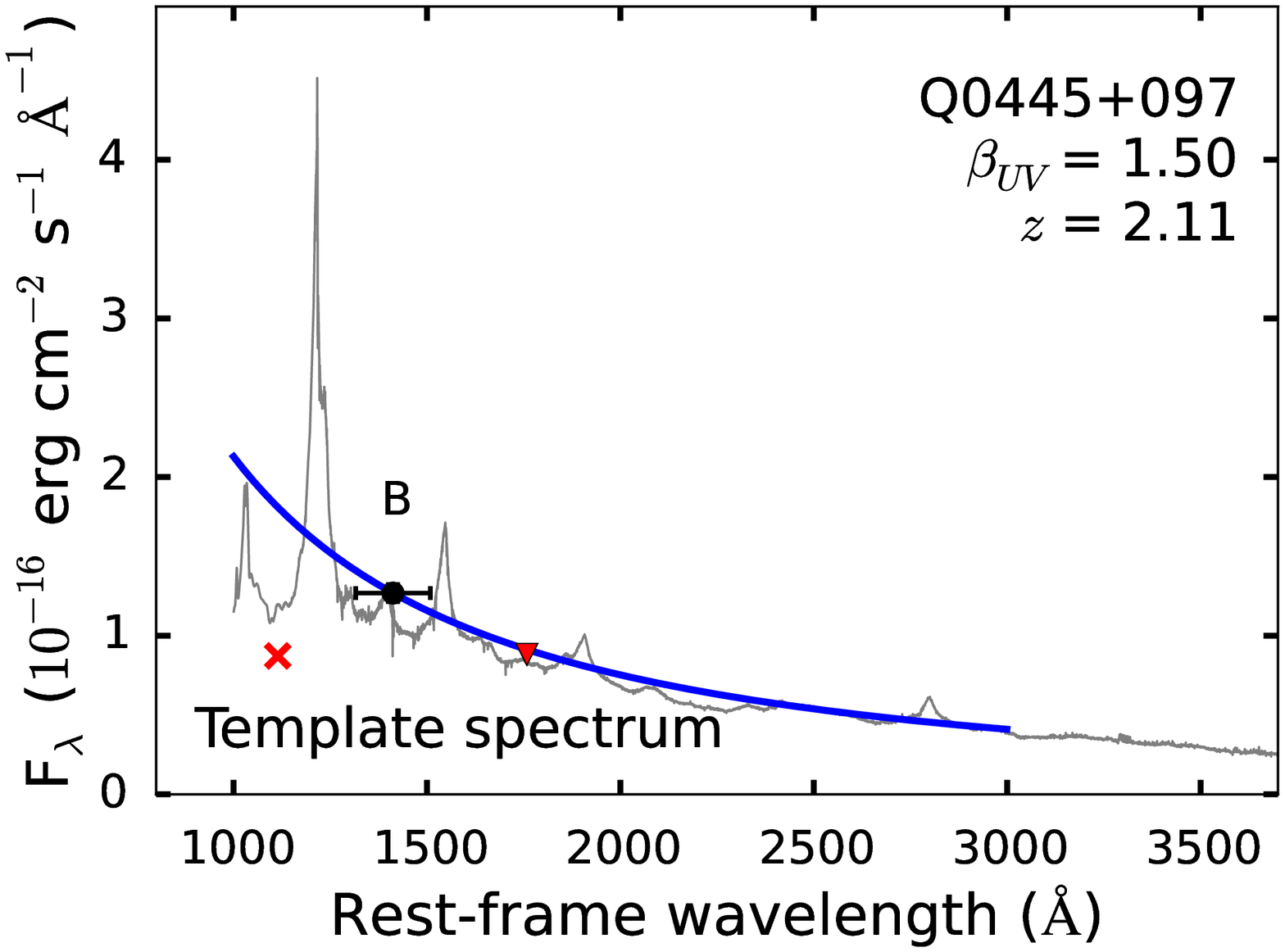}
	\includegraphics[width=\sedplotsize]{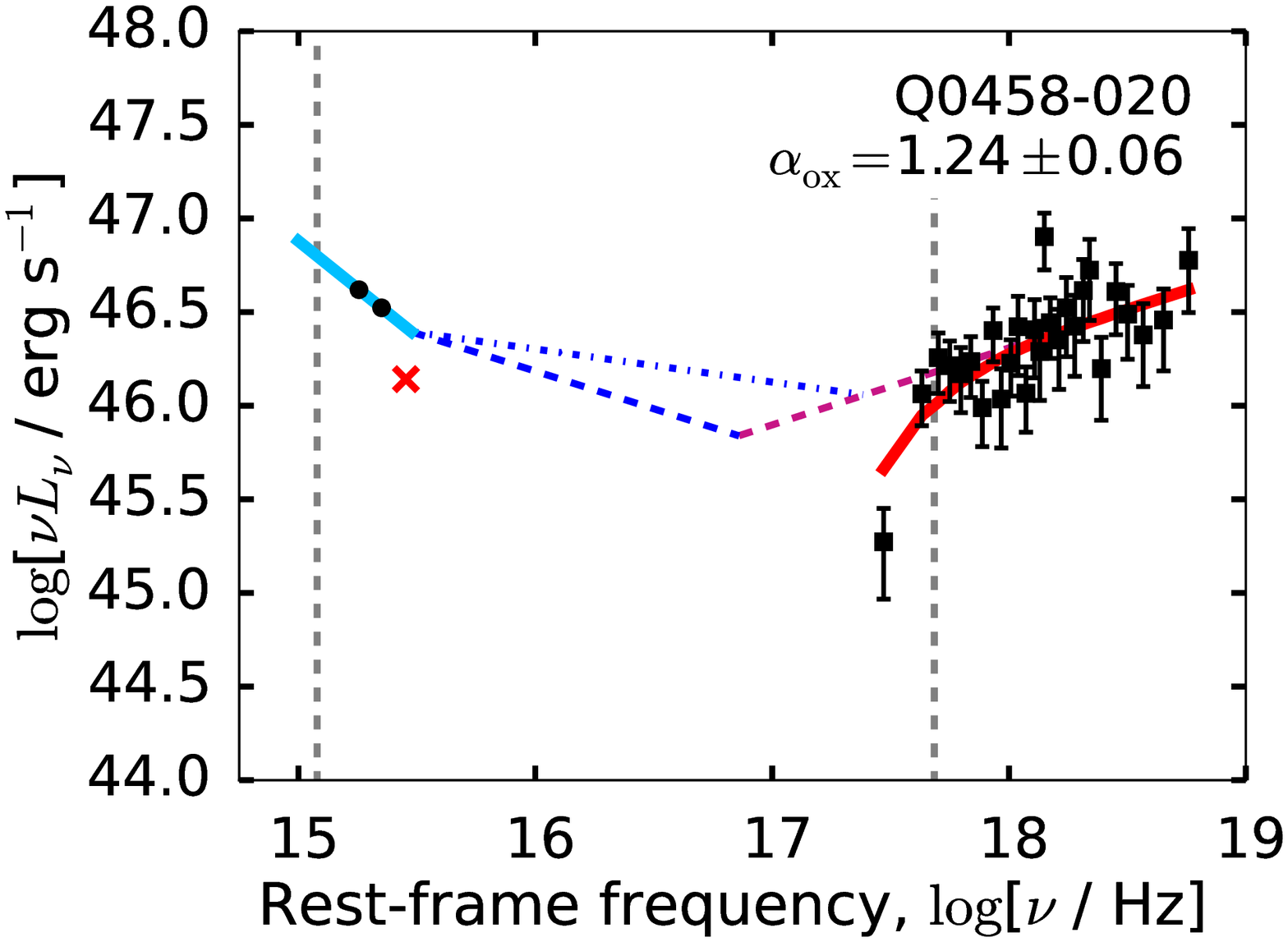}
	\includegraphics[width=\sedplotsize]{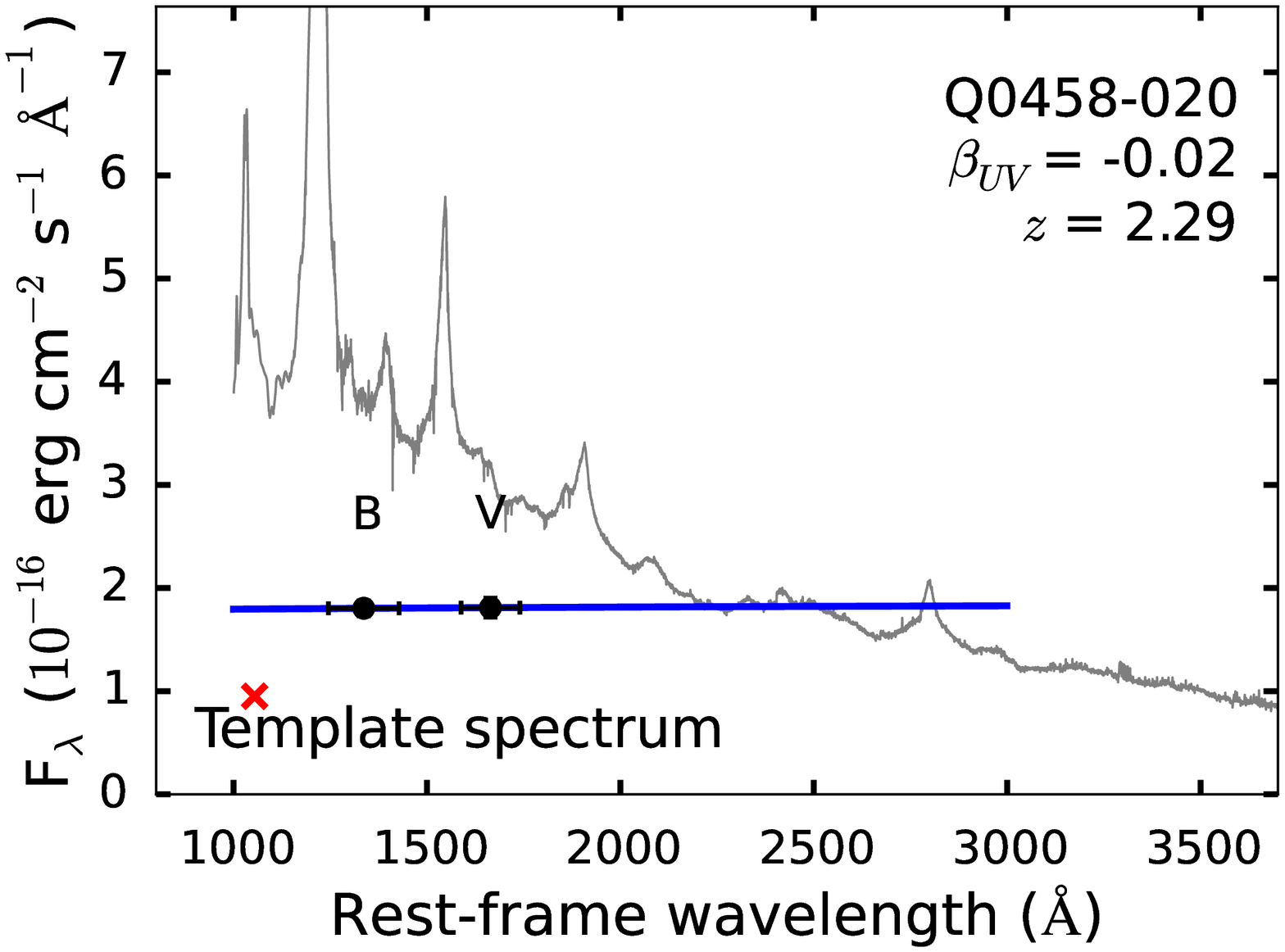}
	
	\caption{\emph{Left:} Rest-frame UV to X-ray spectral energy distributions (SEDs) of quasars in our sample. \emph{Right}: UV photometry and continuum modeling. See Figure \ref{fig:sed_mainpaper} for symbol and color coding.}
	\label{fig:sed9}
\end{figure*}

\paragraph*{Q0458-020:} This quasar appears very red in our UVOT photometry. \citet{Wolfe1993} find a damped Lyman-$\alpha$ absorber at $z=2.04$ towards this quasar, along with at least three other intervening absorption systems. We therefore believe that the intrinsic UV continuum is likely to be significantly brighter and bluer than that inferred from photometry, and regard the measured $L_\mathrm{UV}$ as a lower limit.

\begin{figure*}
	\centering
	\includegraphics[width=\sedplotsize]{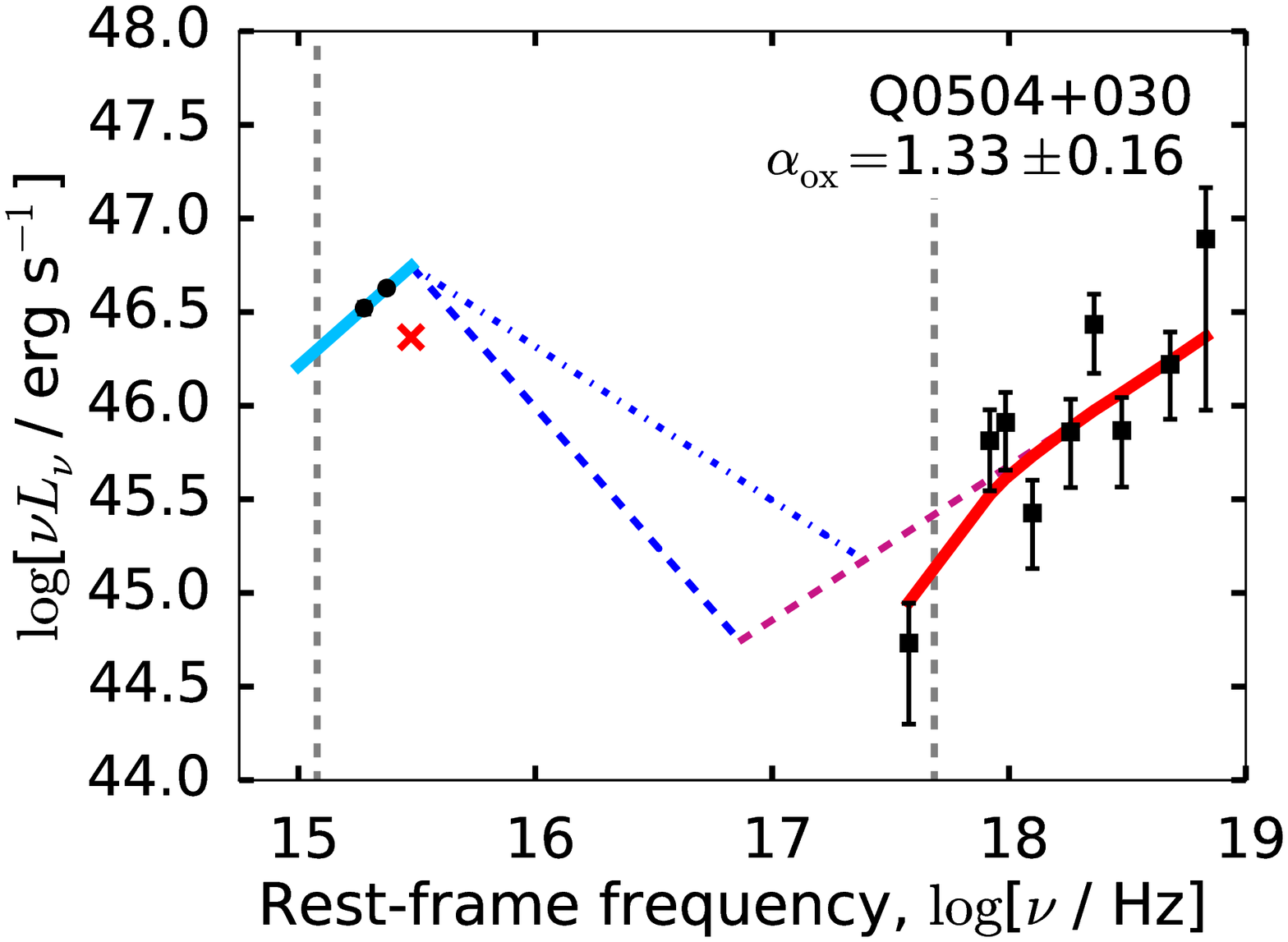}
	\includegraphics[width=\sedplotsize]{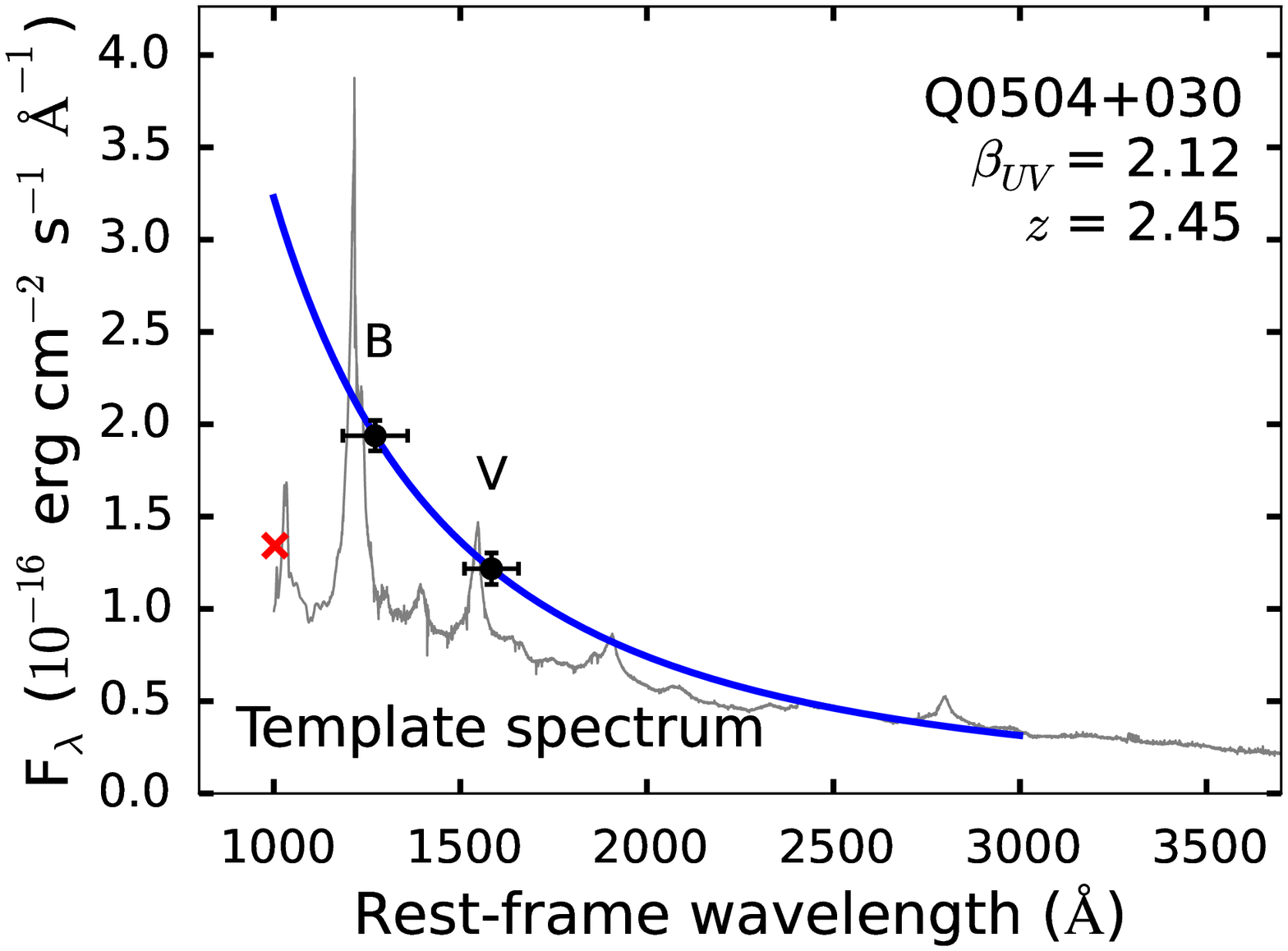}
	\includegraphics[width=\sedplotsize]{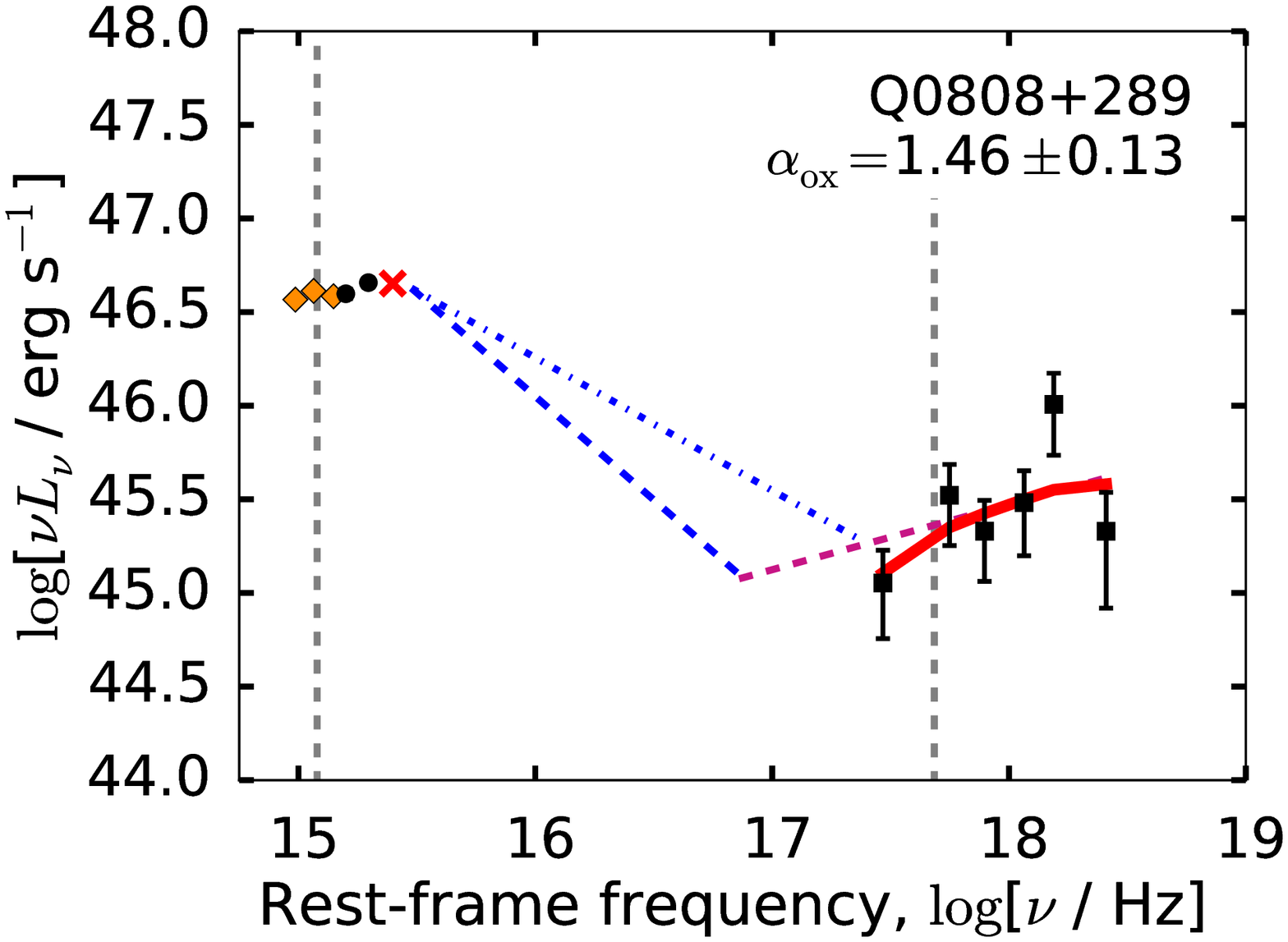}
	\includegraphics[width=\sedplotsize]{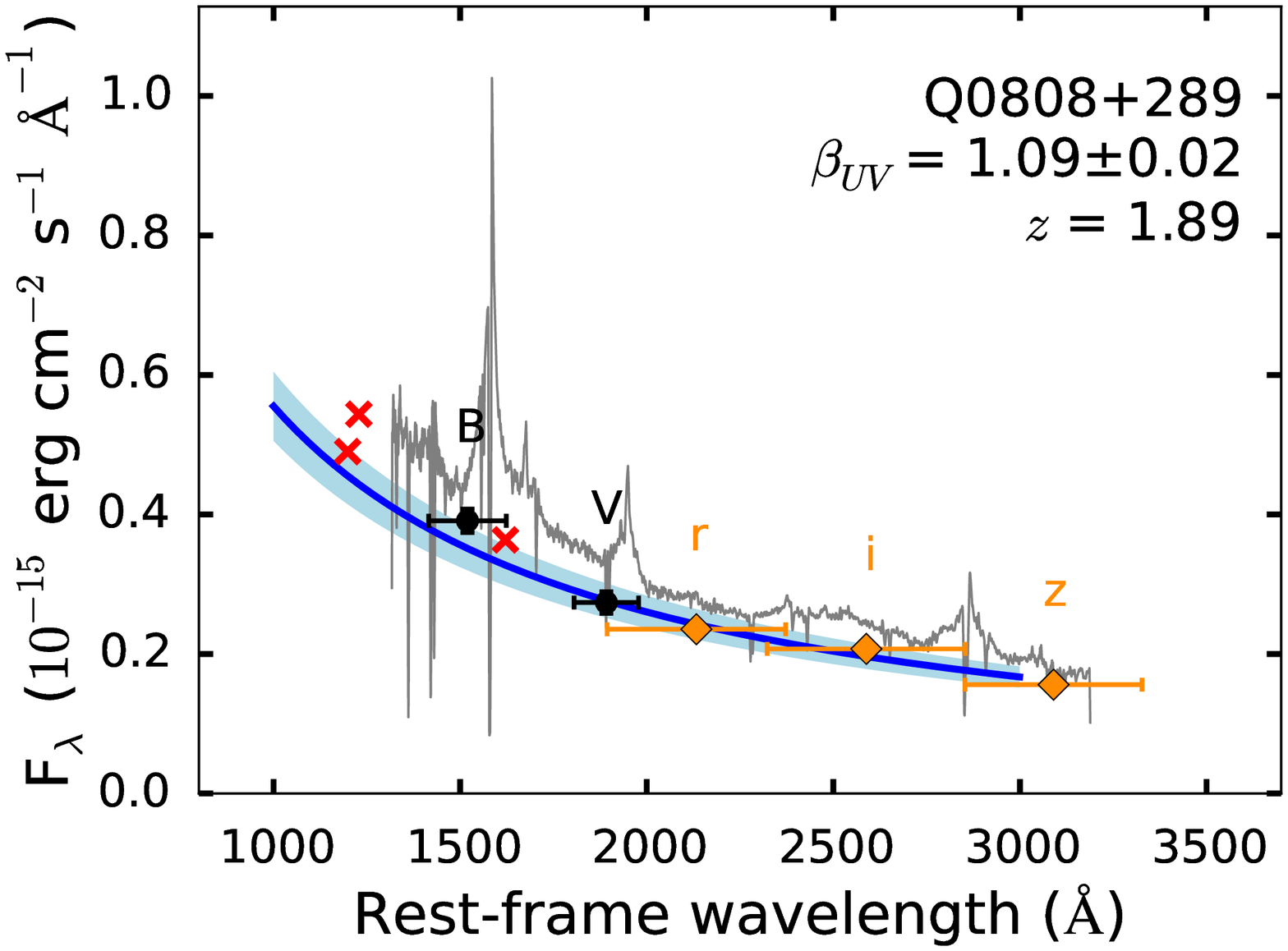}
	\includegraphics[width=\sedplotsize]{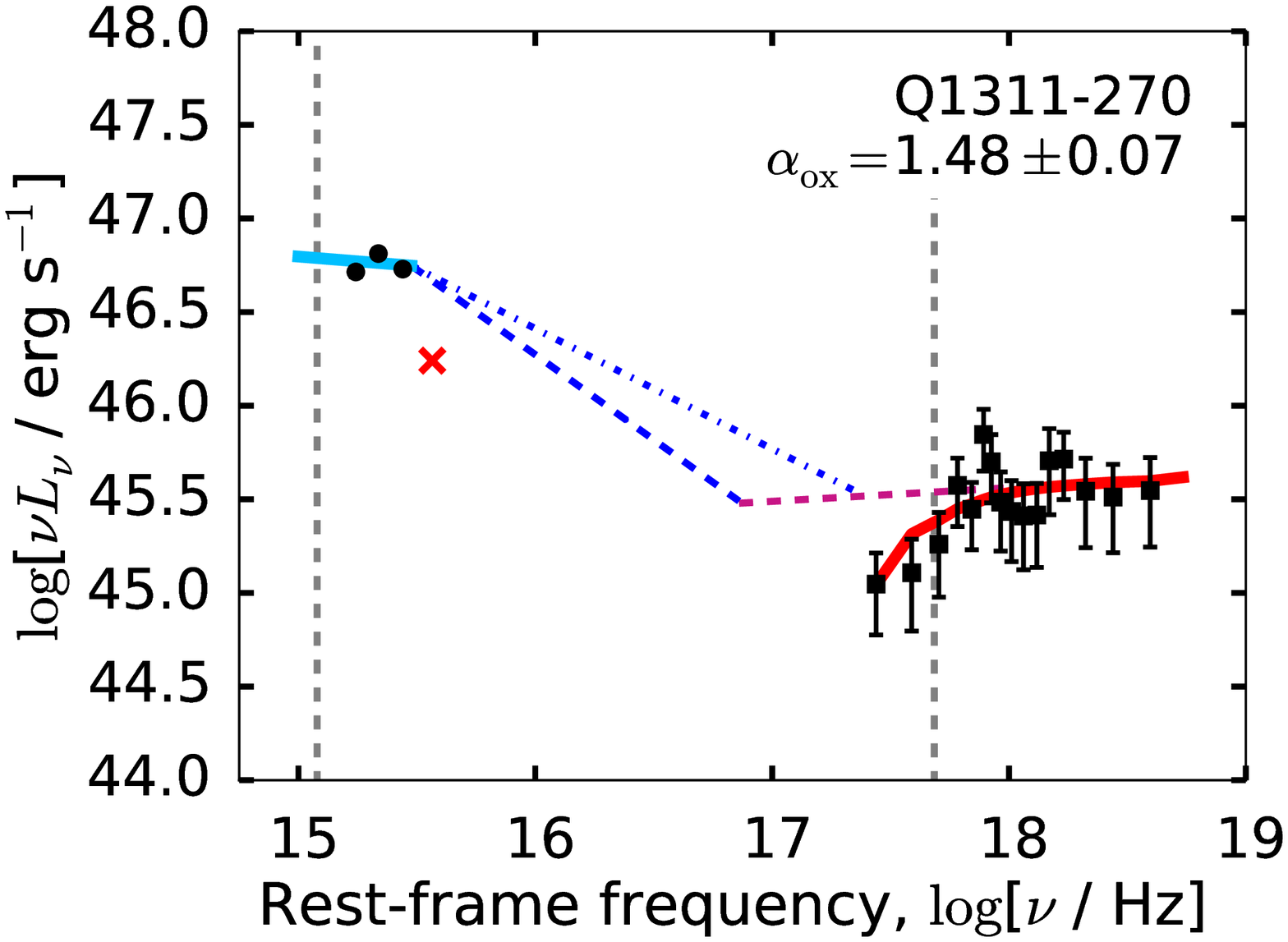}
	\includegraphics[width=\sedplotsize]{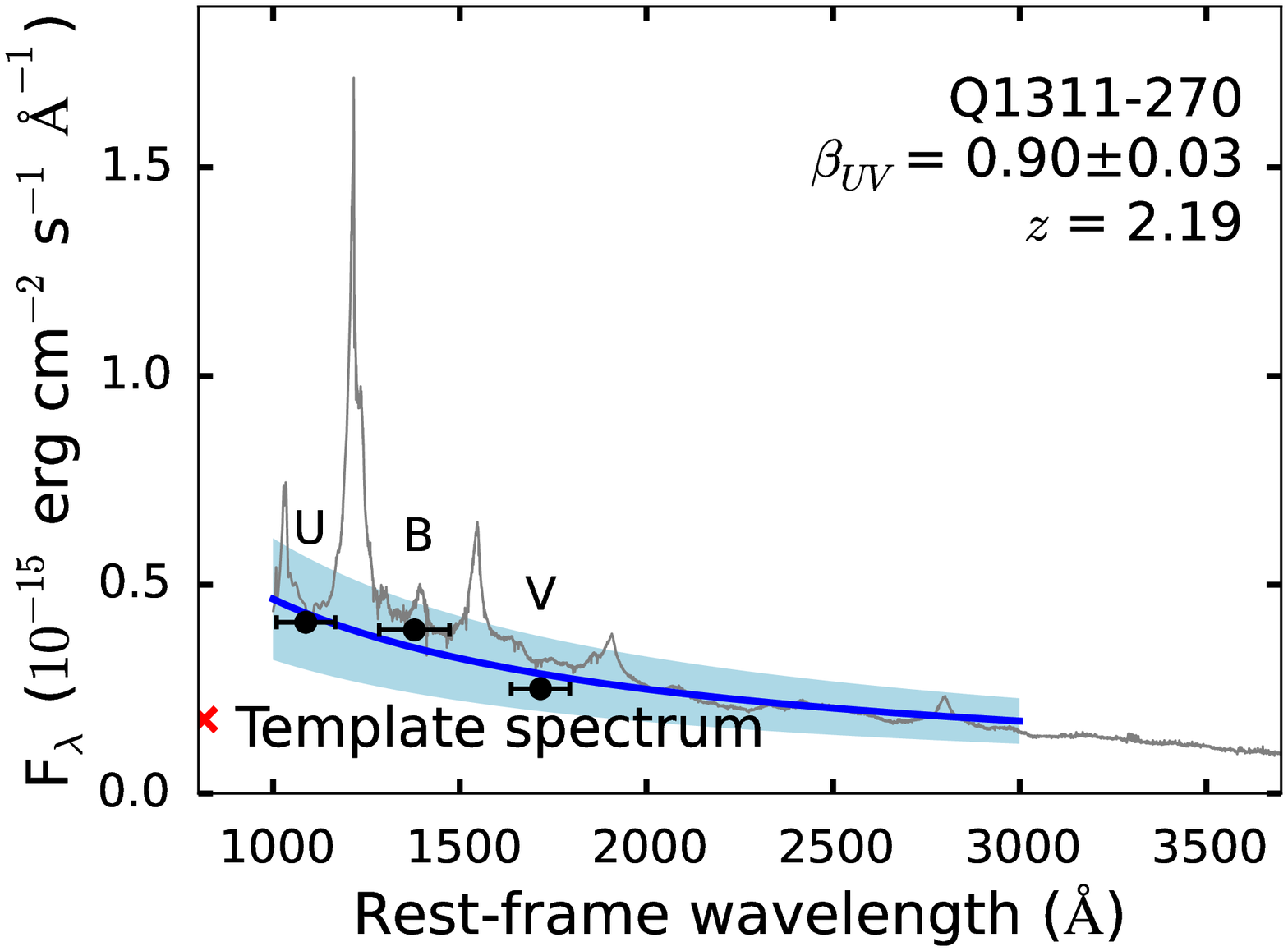}
	
	\caption{\emph{Left:} Rest-frame UV to X-ray spectral energy distributions (SEDs) of quasars in our sample. \emph{Right}: UV photometry and continuum modeling. See Figure \ref{fig:sed_mainpaper} for symbol and color coding.}
	\label{fig:sed10}
\end{figure*}

\paragraph*{Q0504+030:} All UVOT bandpasses likely suffer BEL contamination. We therefore regard all quantities derived from the NUV modeling as highly uncertain. 

\begin{figure*}
	\centering
	\includegraphics[width=\sedplotsize]{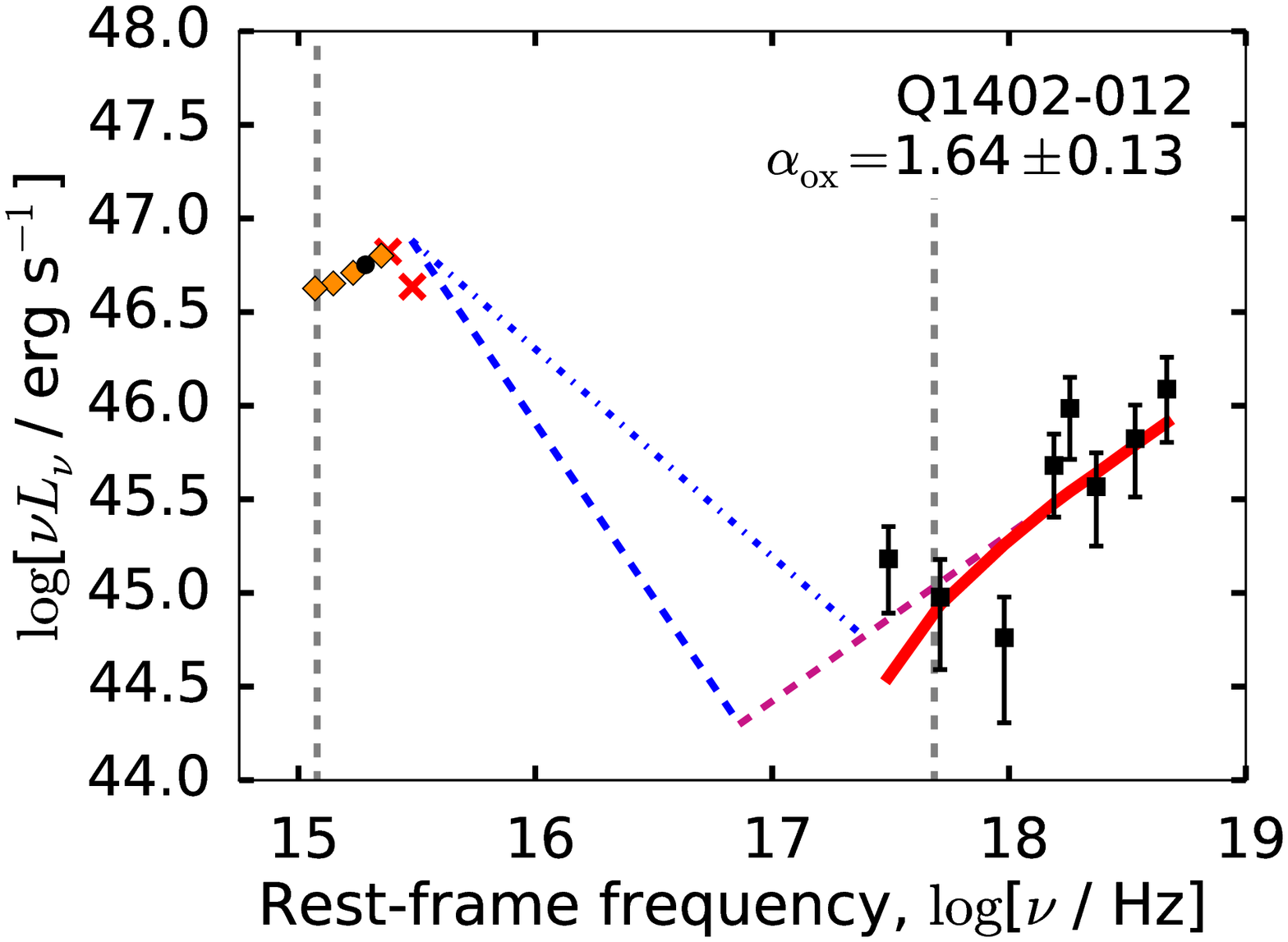}
	\includegraphics[width=\sedplotsize]{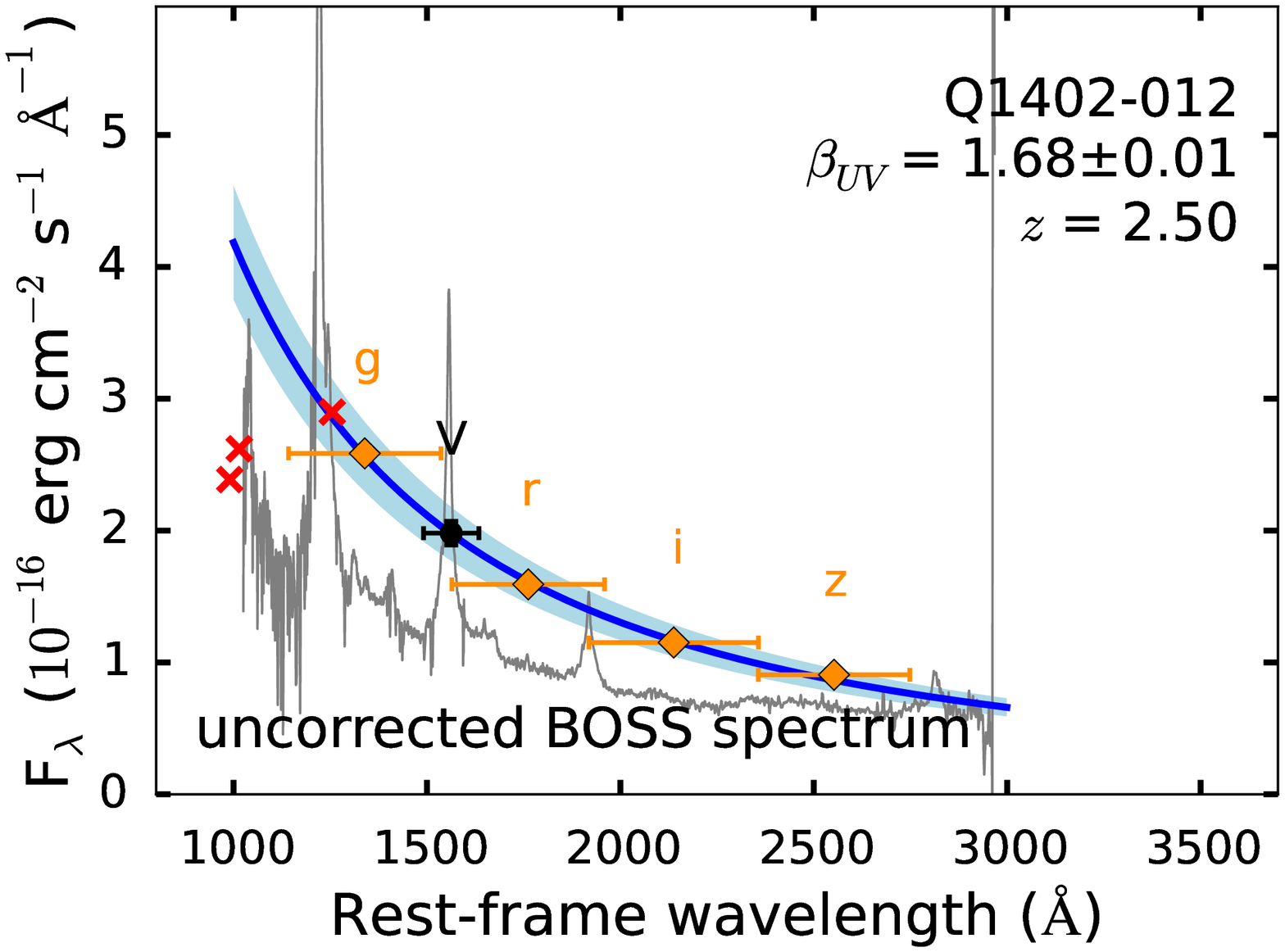}
	\includegraphics[width=\sedplotsize]{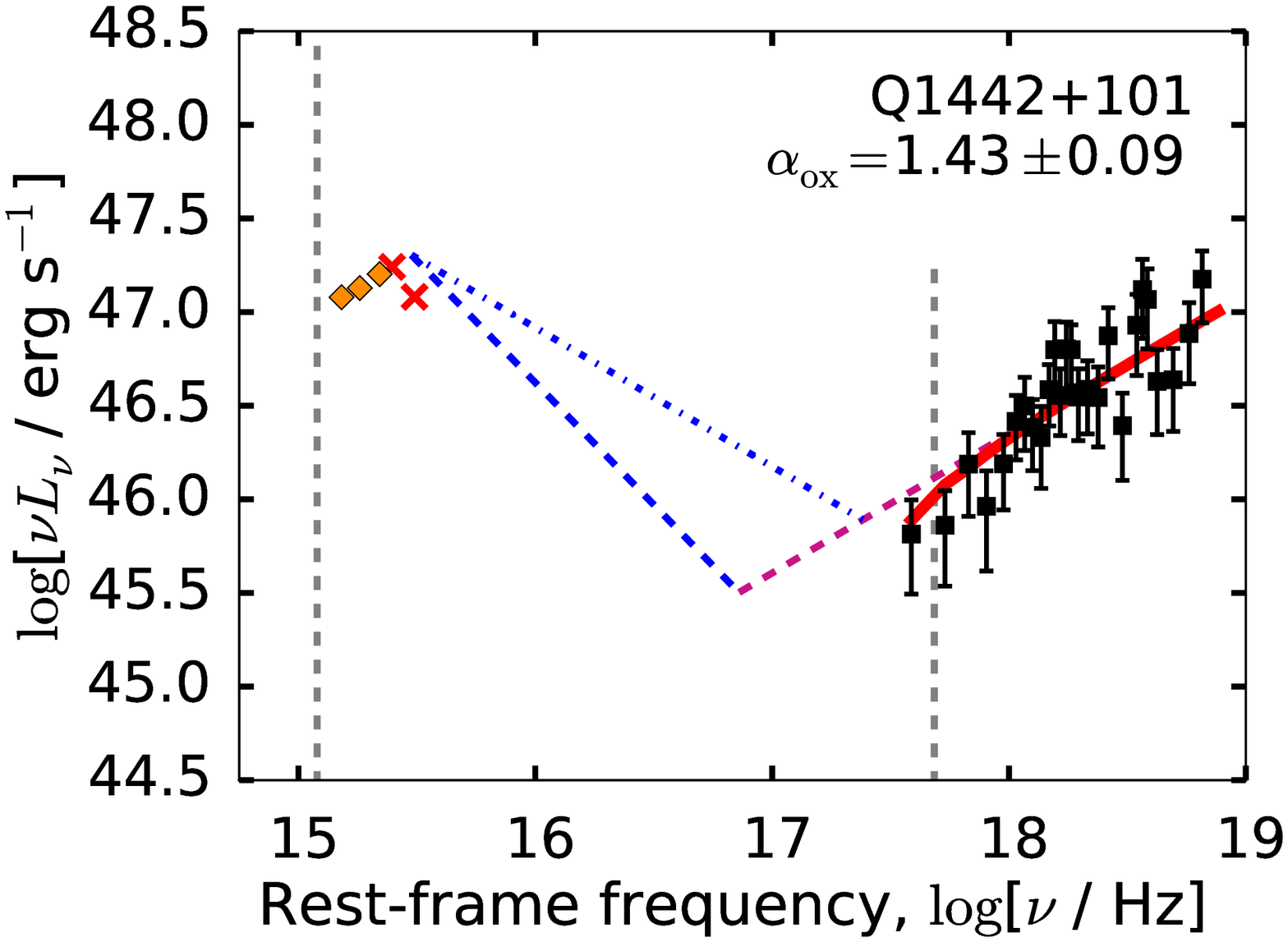}
	\includegraphics[width=\sedplotsize]{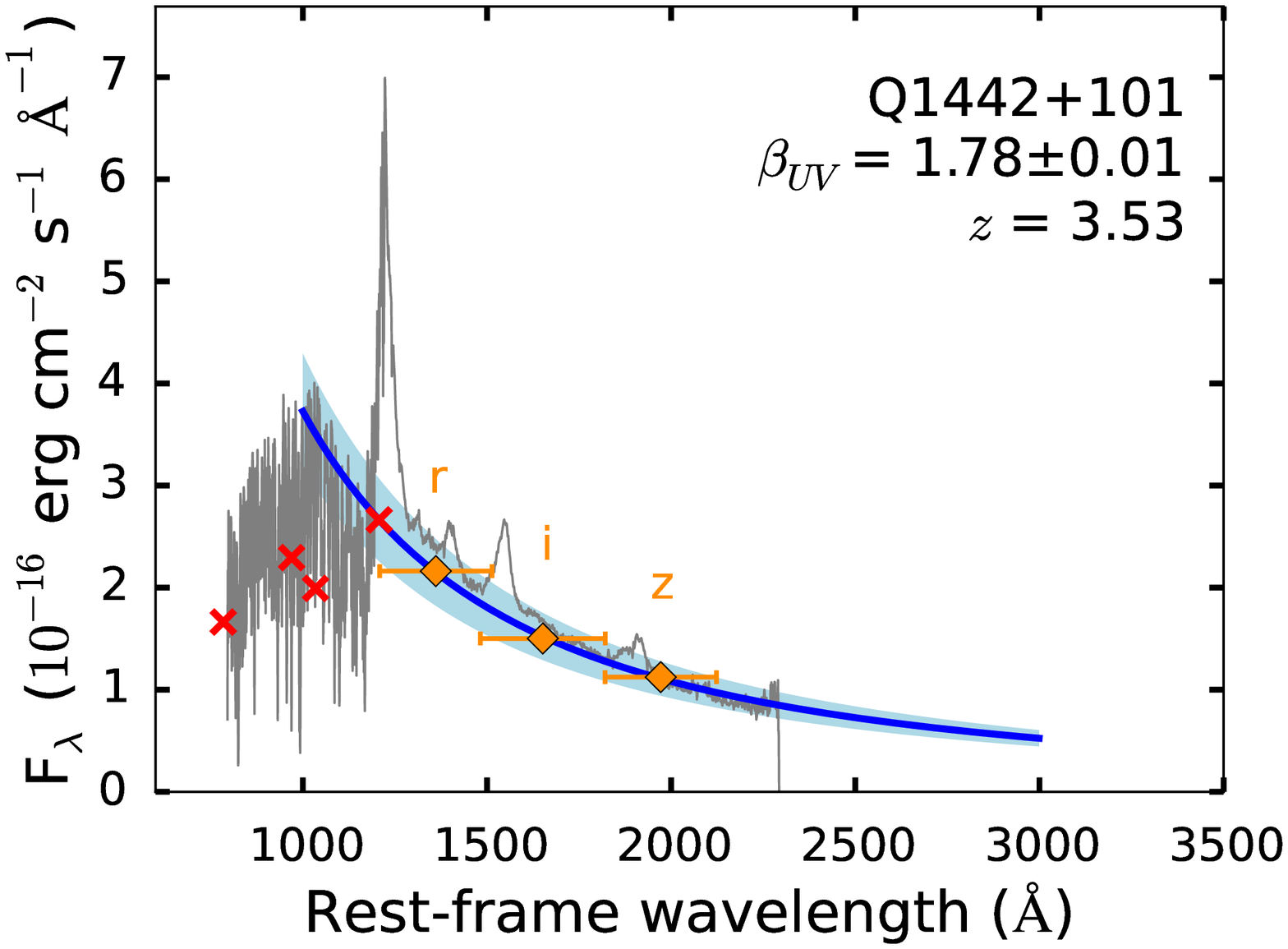}
	\includegraphics[width=\sedplotsize]{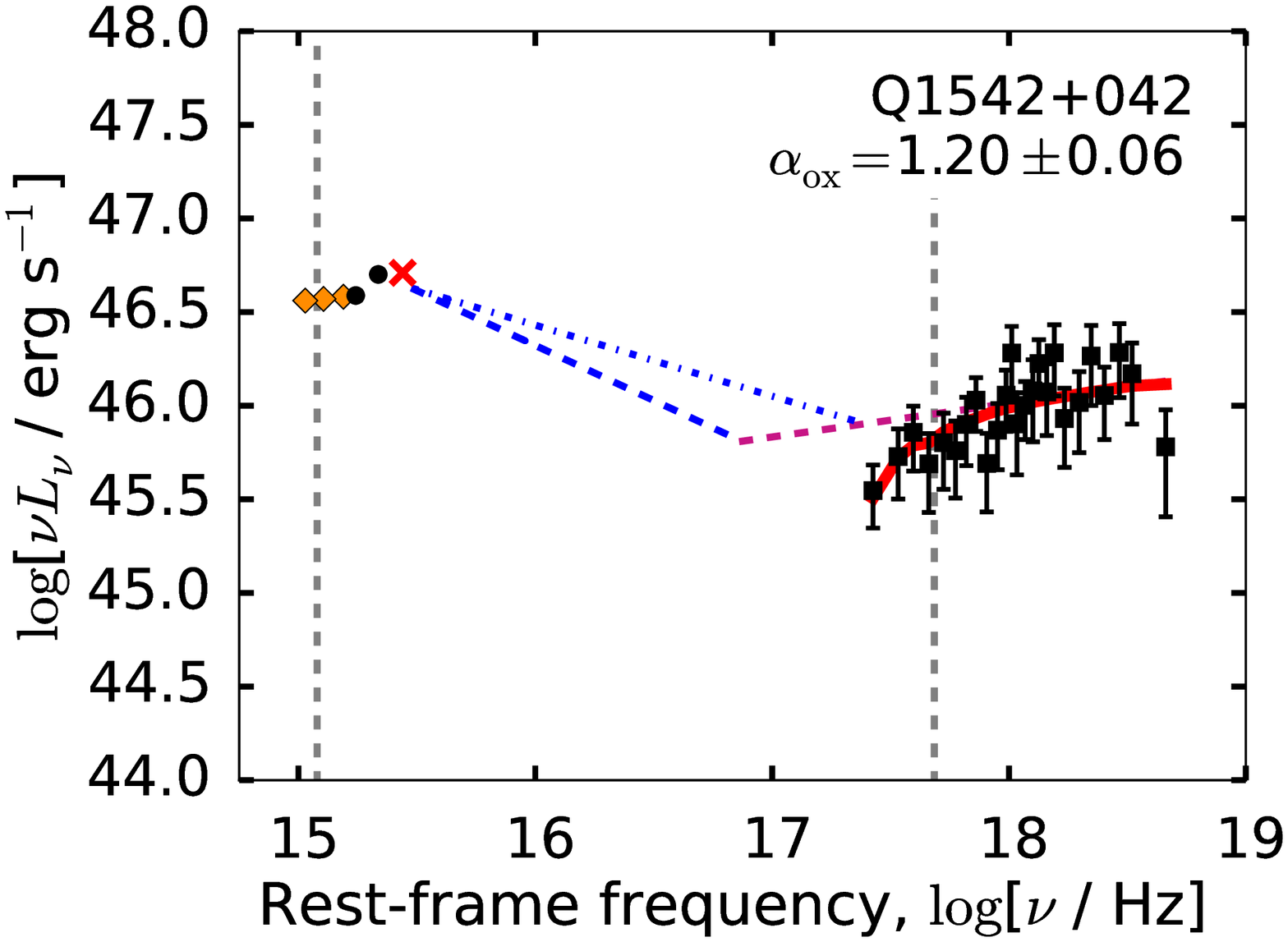}
	\includegraphics[width=\sedplotsize]{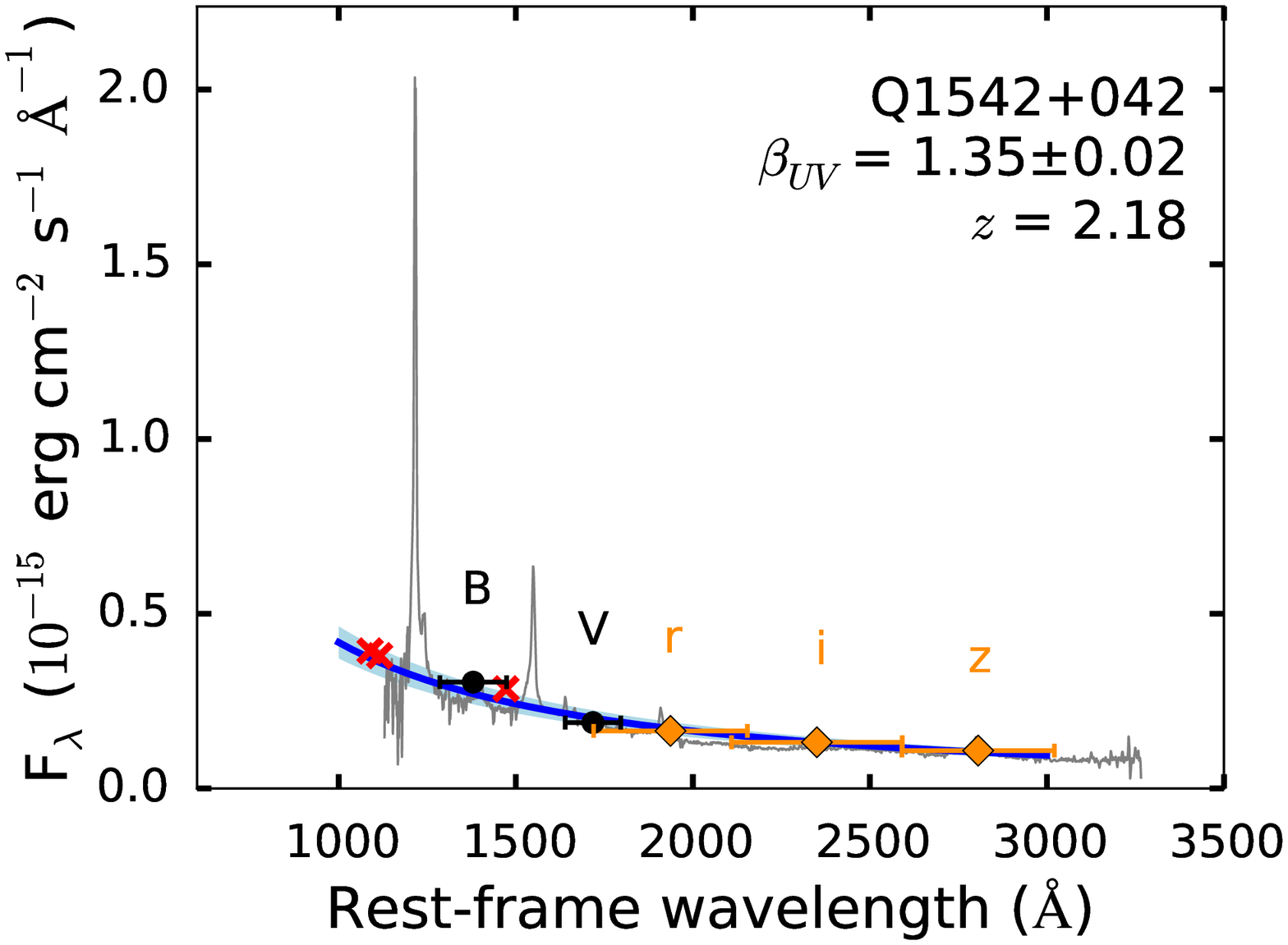}
	
	\caption{\emph{Left:} Rest-frame UV to X-ray spectral energy distributions (SEDs) of quasars in our sample. \emph{Right}: UV photometry and continuum modeling. See Figure \ref{fig:sed_mainpaper} for symbol and color coding.}
	\label{fig:sed11}
\end{figure*}

\paragraph*{Q1402-012:} For this object, all UVOT bandpasses suffer strong BEL contamination. The SDSS-III BOSS spectrum displays a significant offset from the SDSS sphotometry even after applying the flux recalibration of \citet{Margala2015}; this may be due to flux variation between the photometric and spectroscopic SDSS observations, or perhaps the flux recalibration is not accurate for this object.

\paragraph*{Q1442+101:} For this $z=3.53$ quasar, all UVOT photometric bands suffer Ly-$\alpha$ forest absorption. We use the UVOT V band flux to determine a scaling factor between the UVOT and SDSS observations, and model the UV-optical power-law continuum based on the rescaled SDSS \emph{r}, \emph{i} and \emph{z} bands. We regard the UV-optical continuum modeling for this object as highly uncertain.

\begin{figure*}
	\centering
	\includegraphics[width=\sedplotsize]{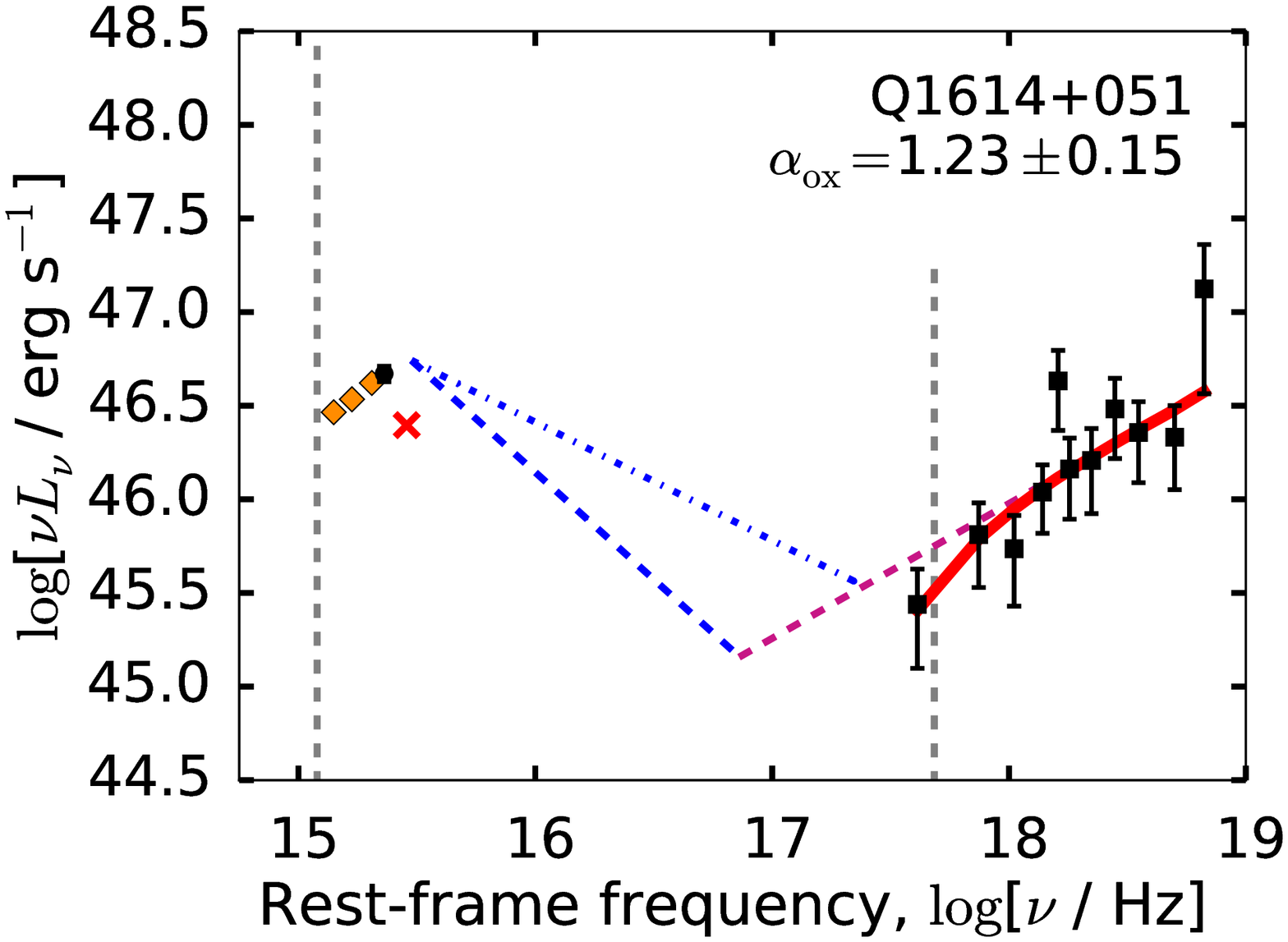}
	\includegraphics[width=\sedplotsize]{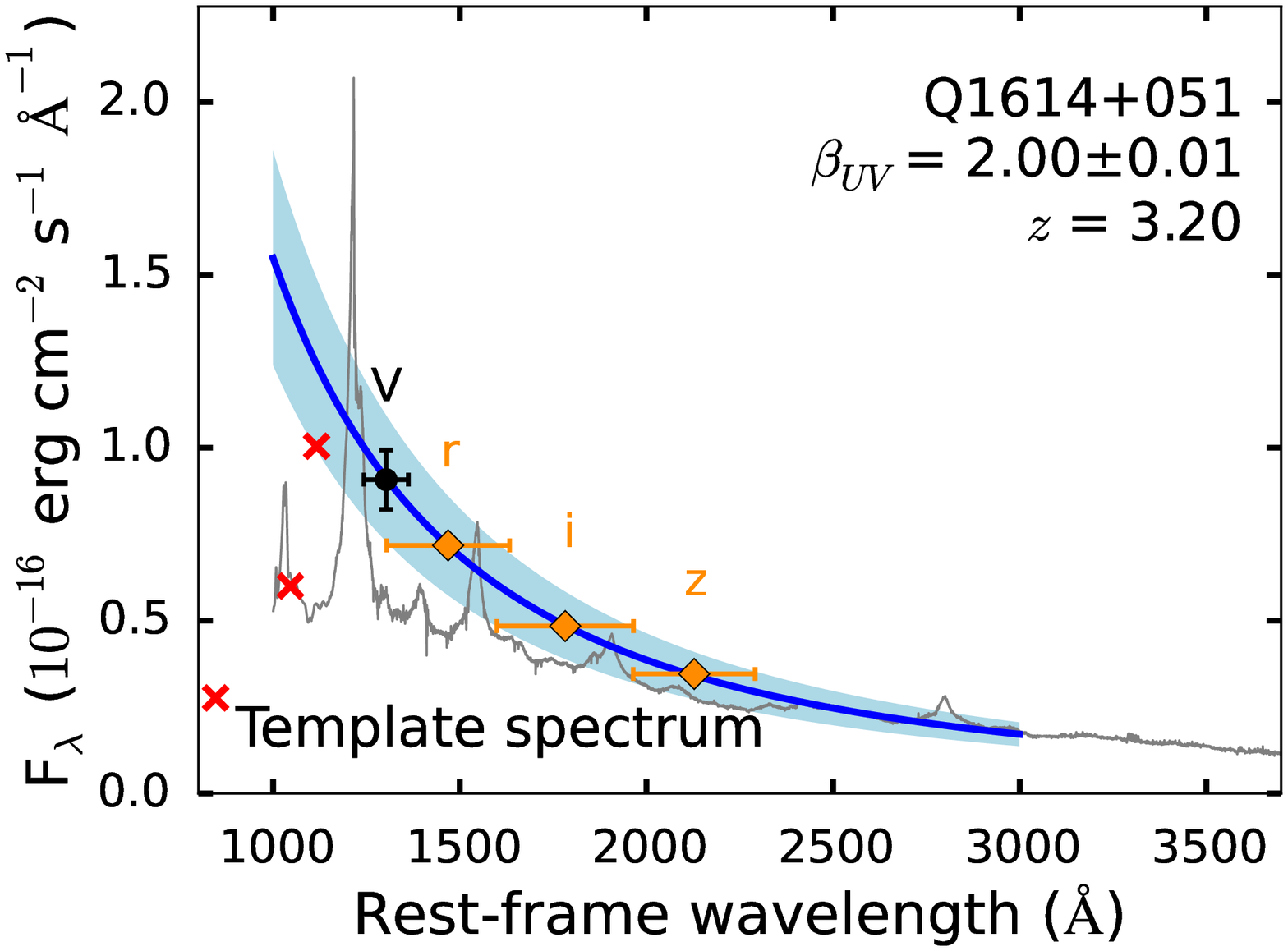}
	\includegraphics[width=\sedplotsize]{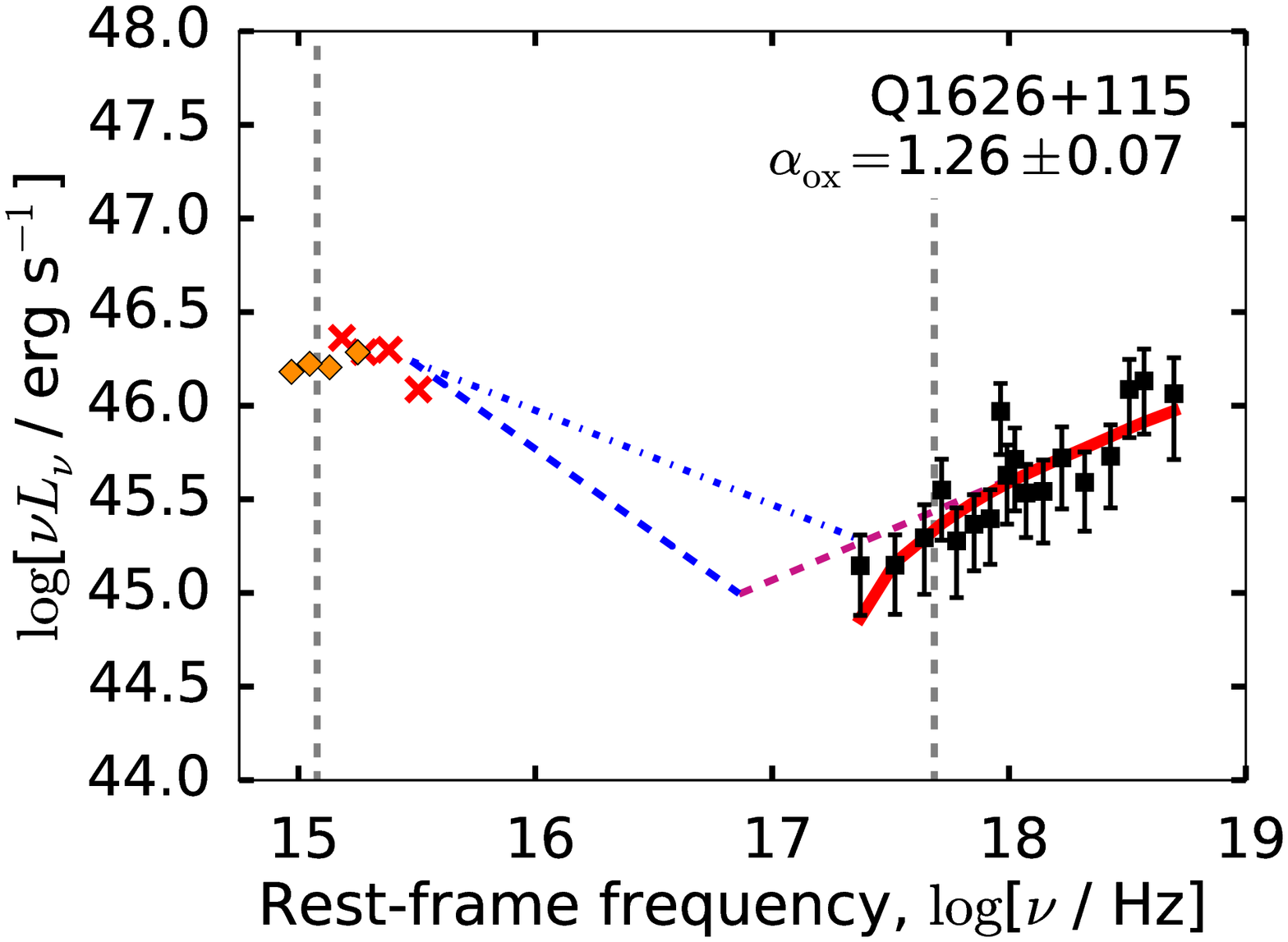}
	\includegraphics[width=\sedplotsize]{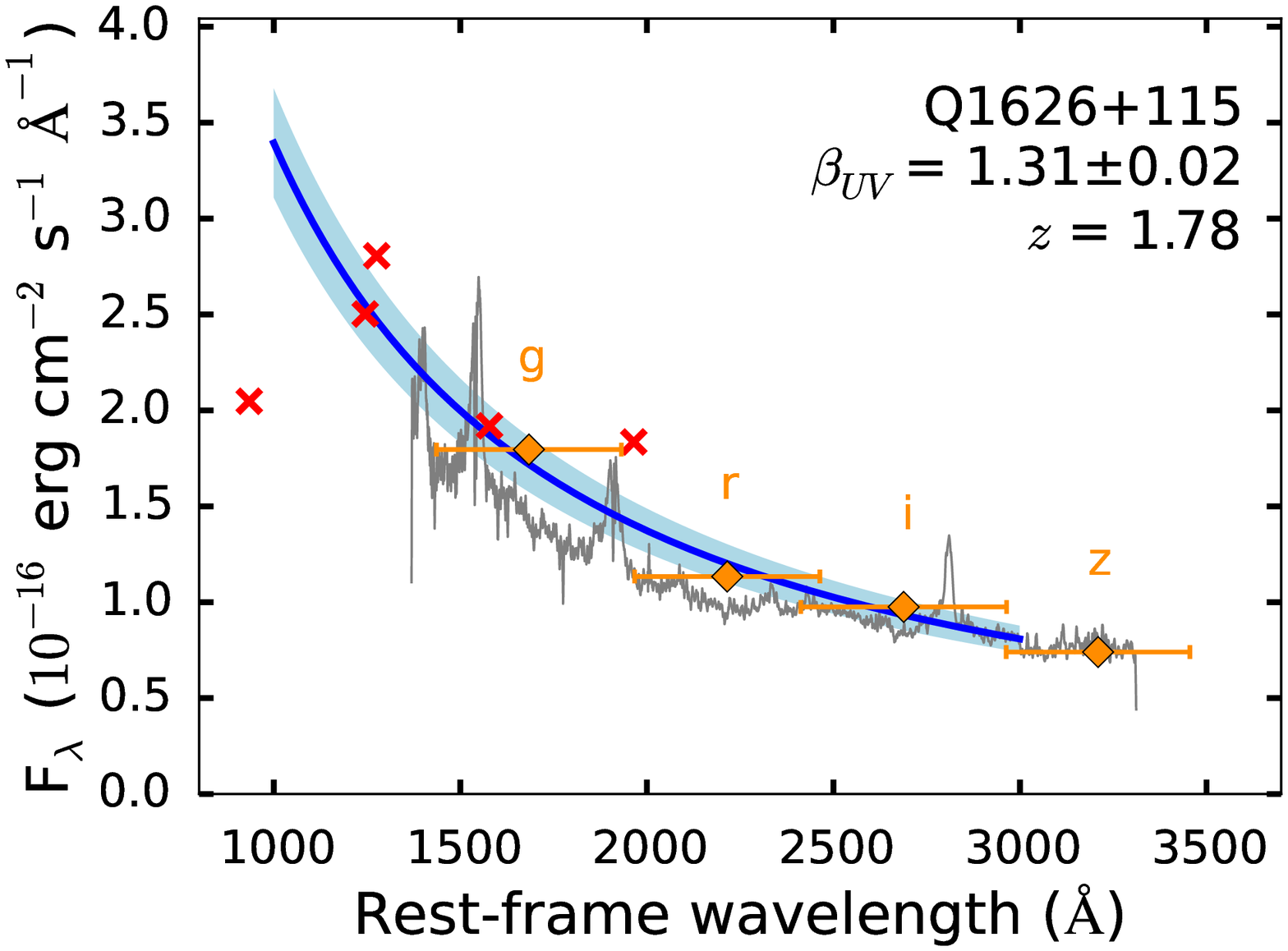}
	\includegraphics[width=\sedplotsize]{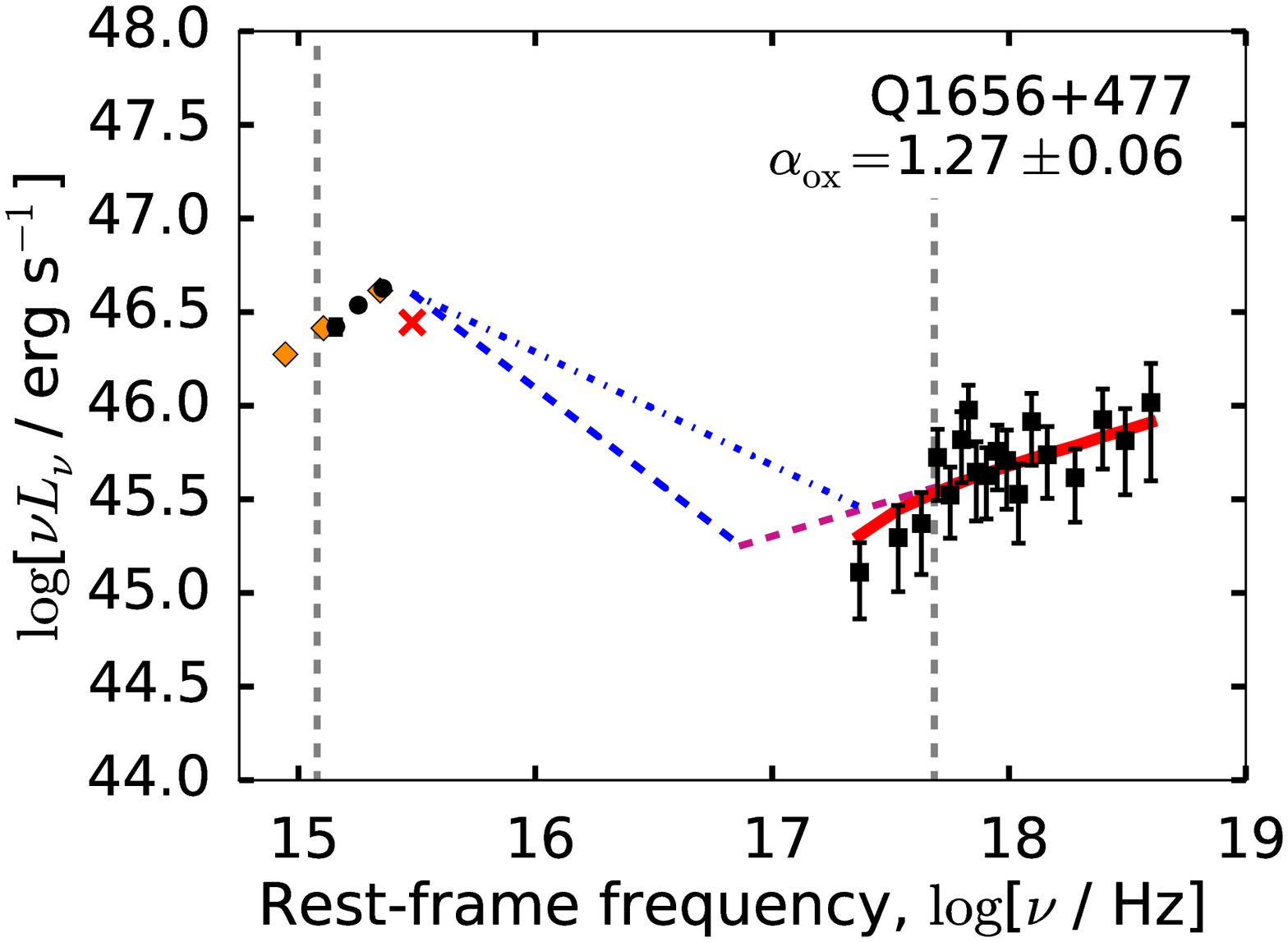}
	\includegraphics[width=\sedplotsize]{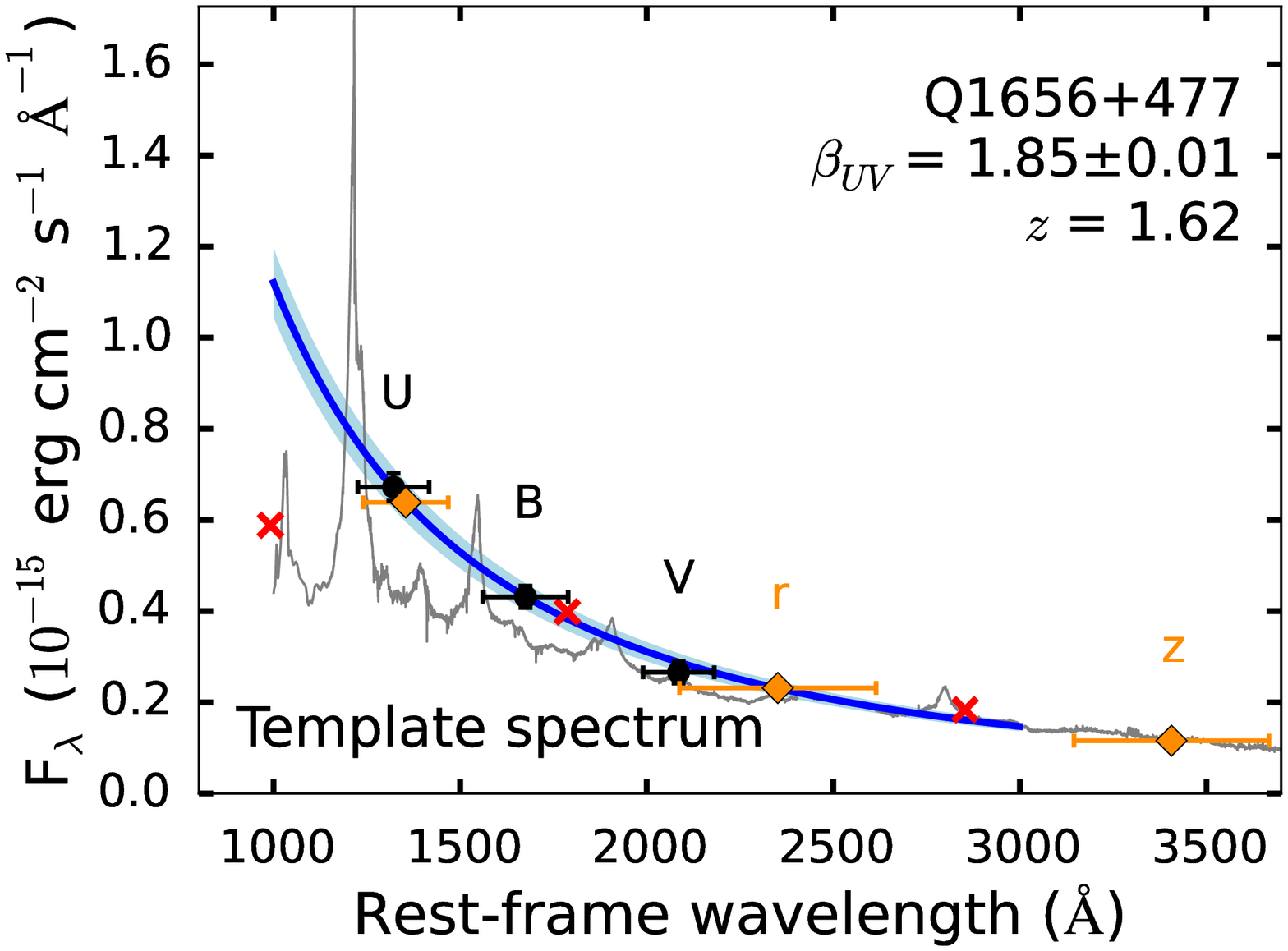}
	
	\caption{\emph{Left:} Rest-frame UV to X-ray spectral energy distributions (SEDs) of quasars in our sample. \emph{Right}: UV photometry and continuum modeling. See Figure \ref{fig:sed_mainpaper} for symbol and color coding.}
	\label{fig:sed12}
\end{figure*}

\paragraph*{Q1626+115:} The SDSS-III BOSS spectrum for this object displays a significant offset from the SDSS photometry even after applying the flux recalibration of \citet{Margala2015}; this may be due to flux variation between the photometric and spectroscopic SDSS observations, or perhaps the flux recalibration is not accurate for this object.

\begin{figure*}
	\centering
	\includegraphics[width=\sedplotsize]{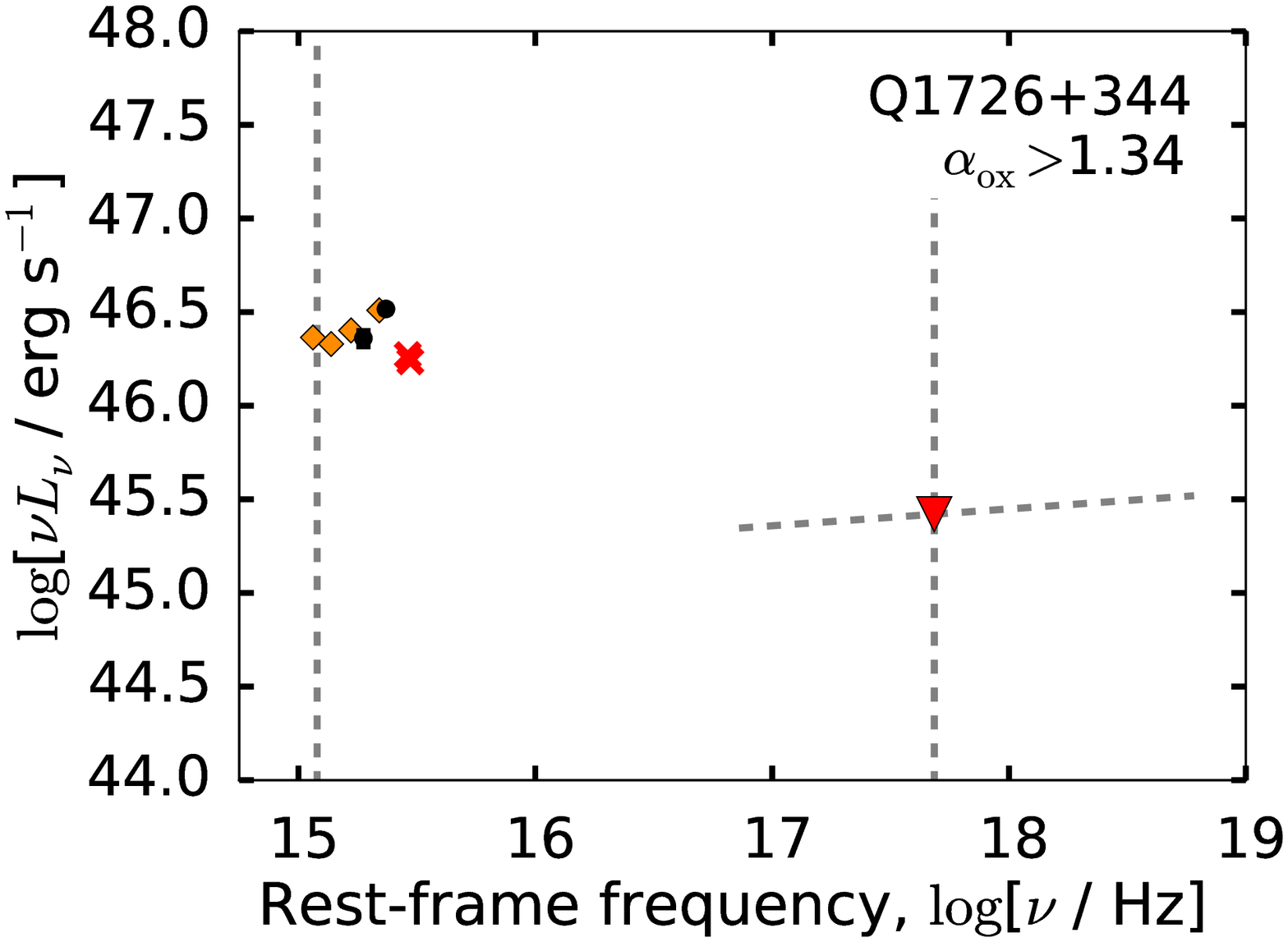}
	\includegraphics[width=\sedplotsize]{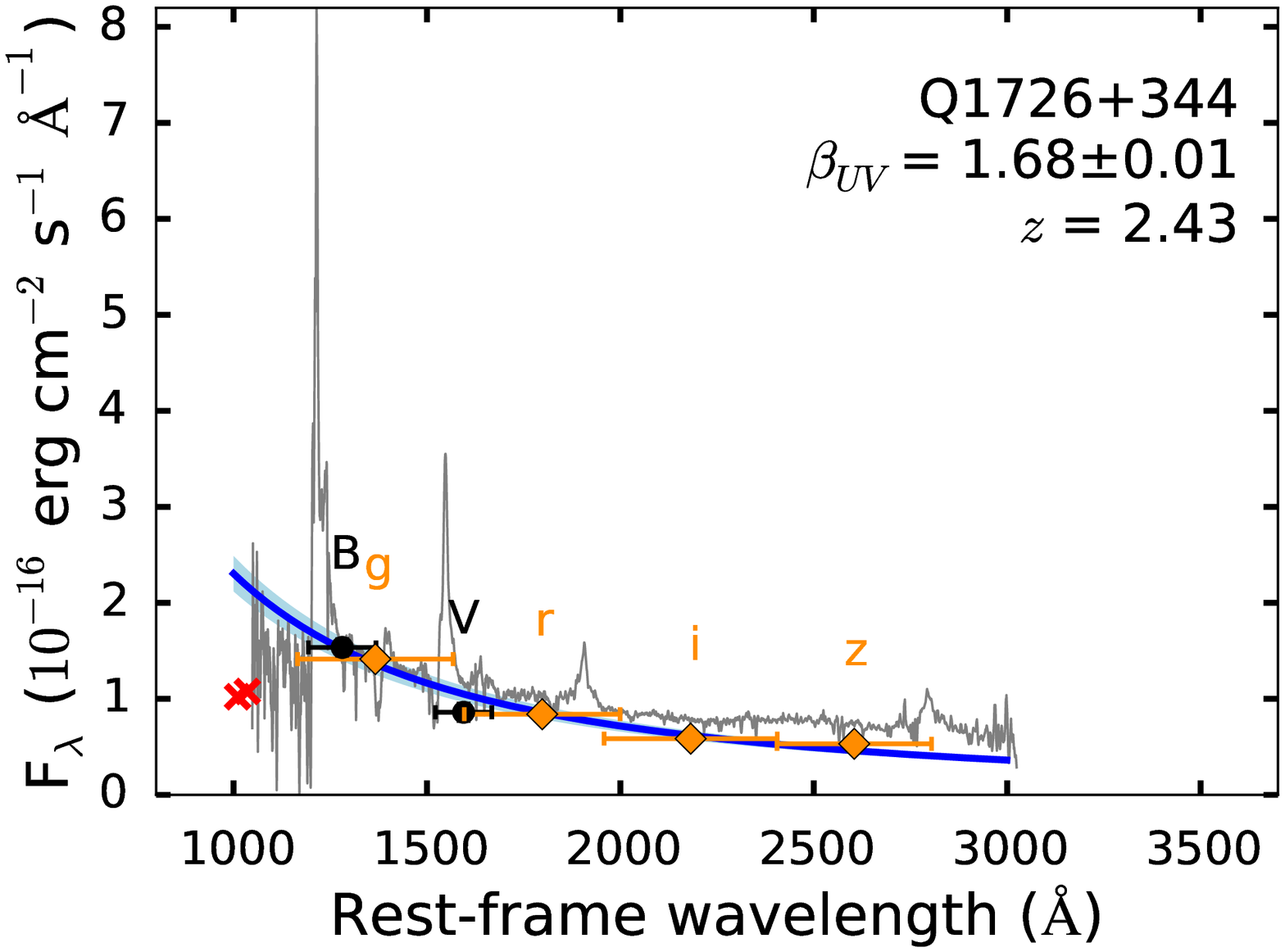}
	\includegraphics[width=\sedplotsize]{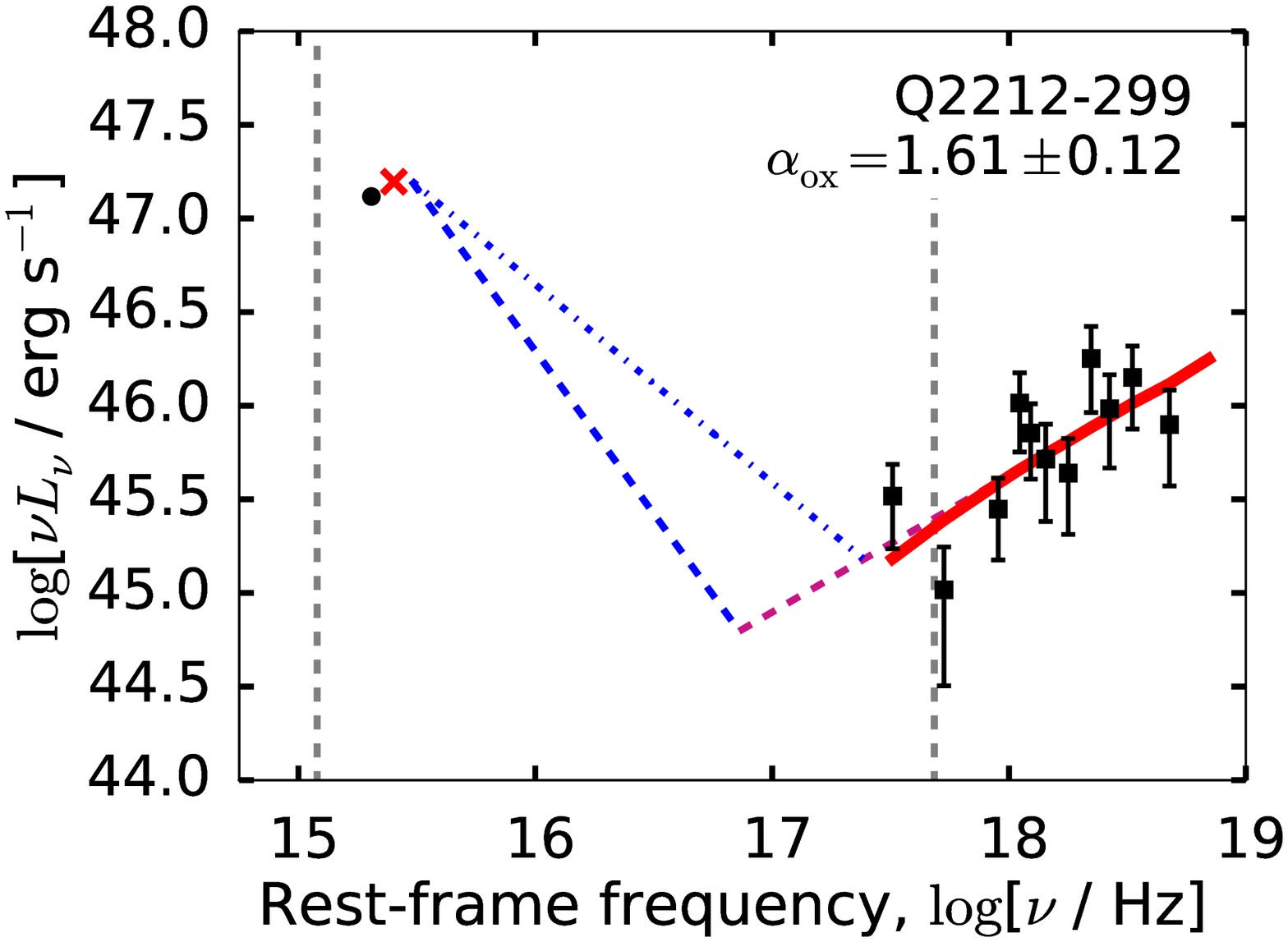}
	\includegraphics[width=\sedplotsize]{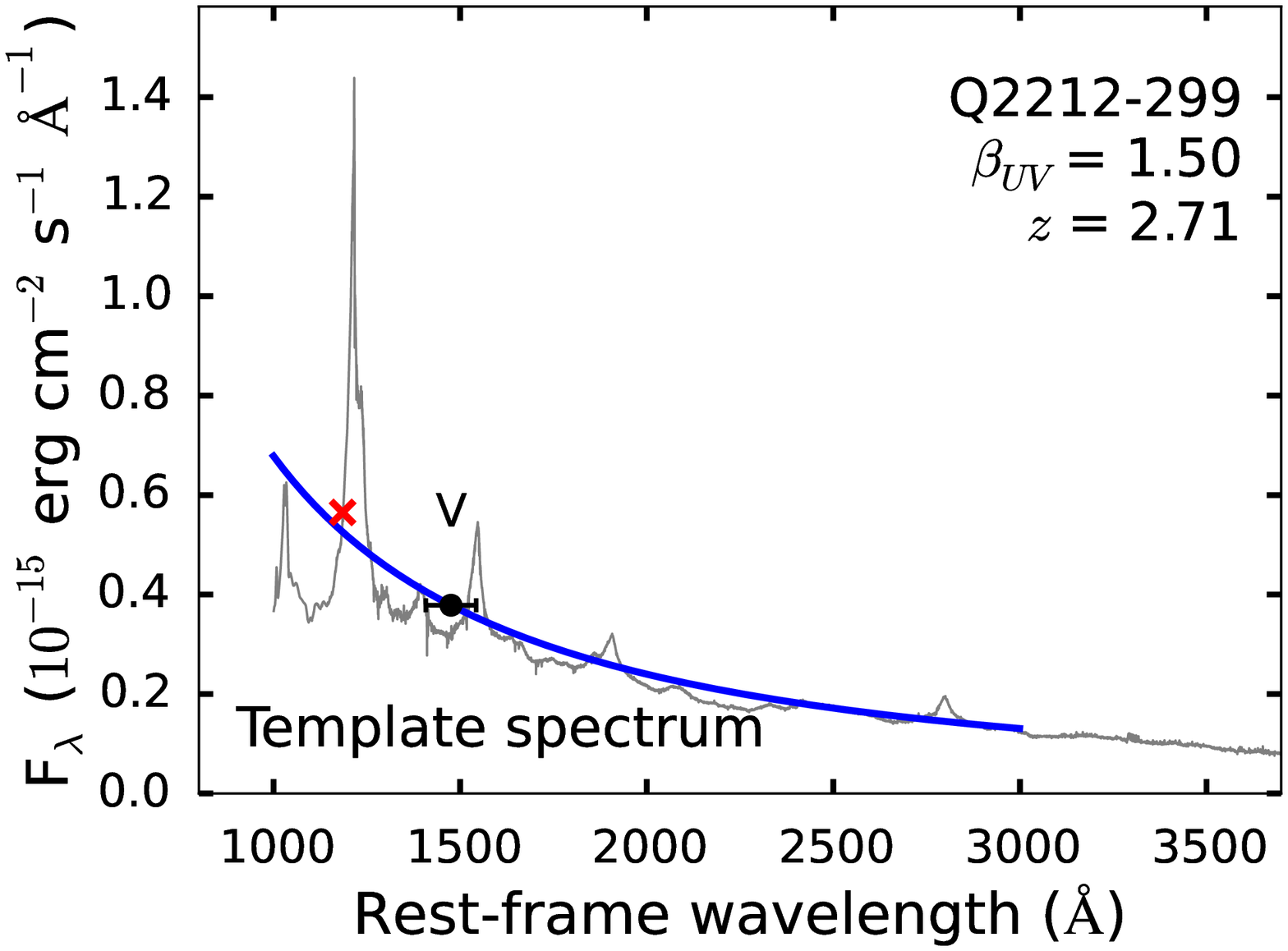}
	\includegraphics[width=\sedplotsize]{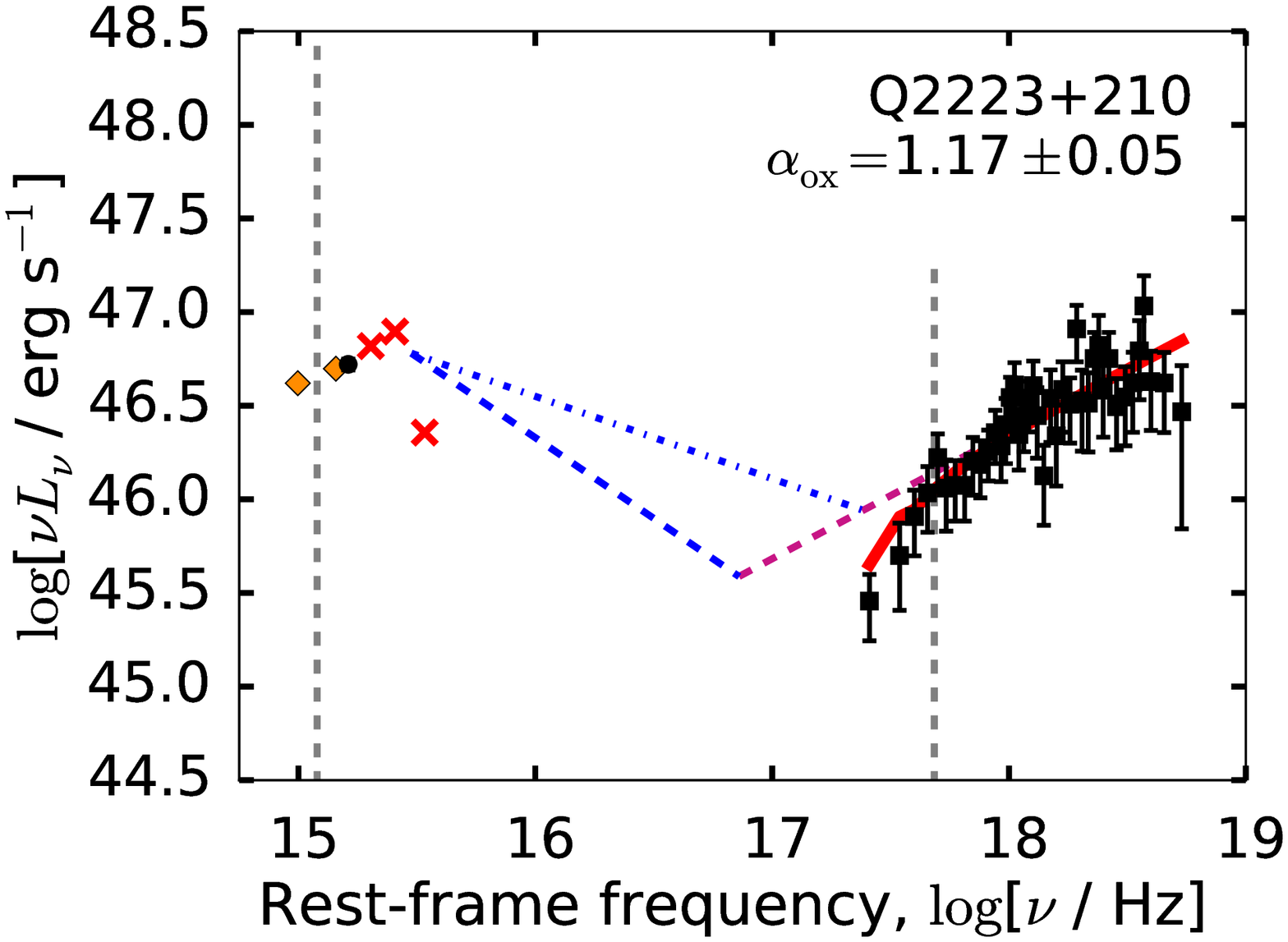}
	\includegraphics[width=\sedplotsize]{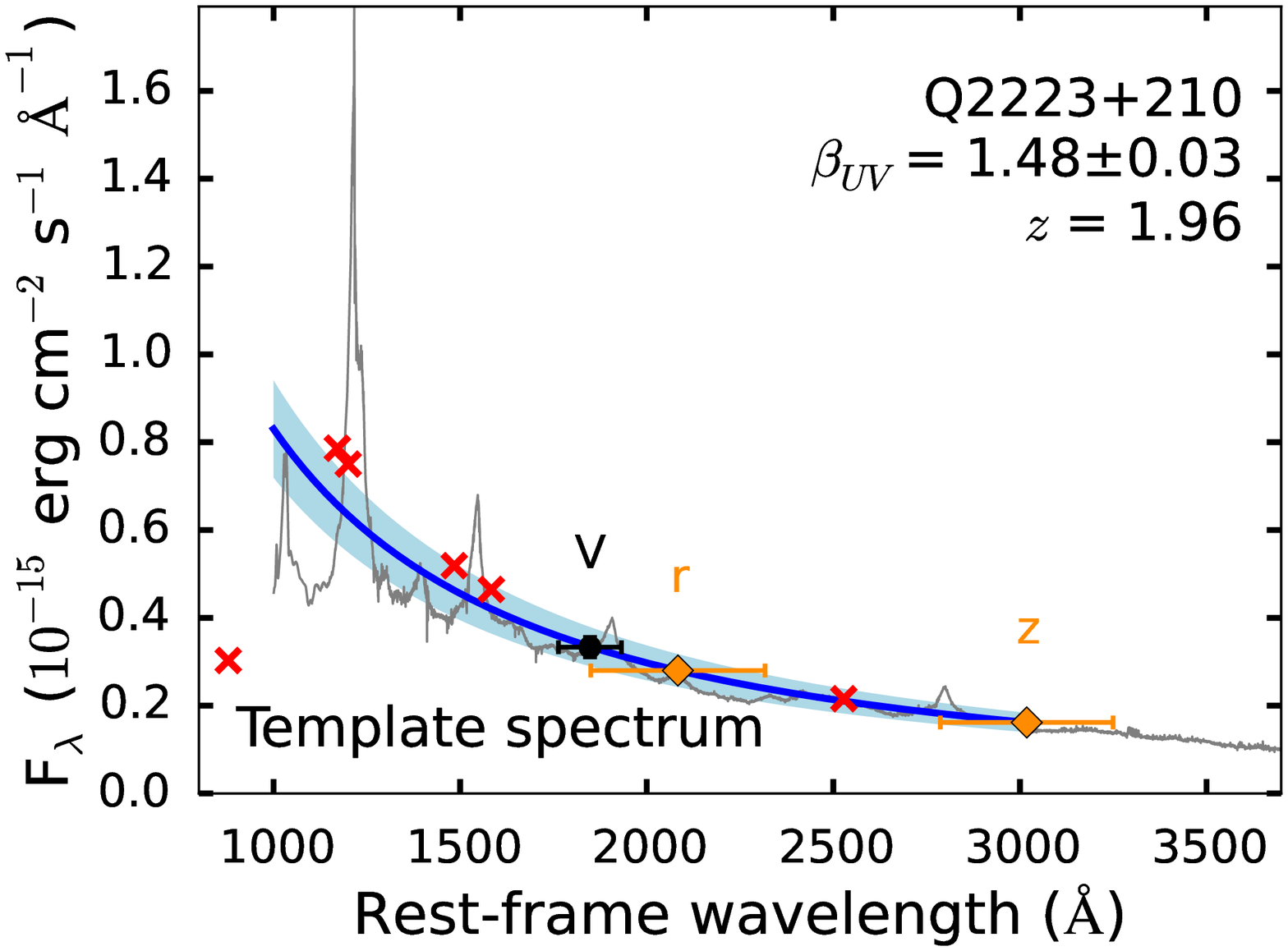}
	
	\caption{\emph{Left:} Rest-frame UV to X-ray spectral energy distributions (SEDs) of quasars in our sample. \emph{Right}: UV photometry and continuum modeling. See Figure \ref{fig:sed_mainpaper} for symbol and color coding.}
	\label{fig:sed13}
\end{figure*}

\begin{figure*}
	\centering
	\includegraphics[width=\sedplotsize]{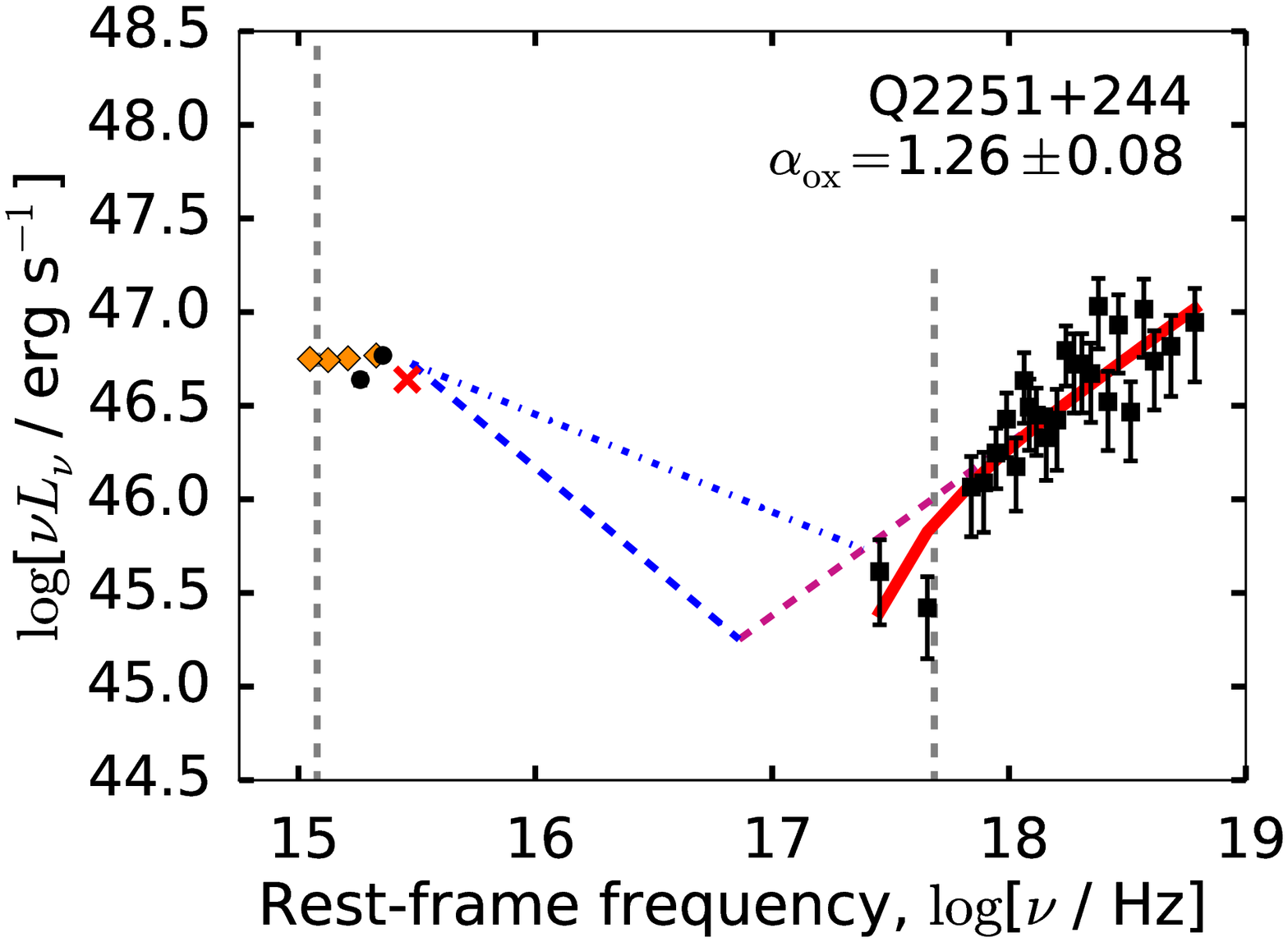}
	\includegraphics[width=\sedplotsize]{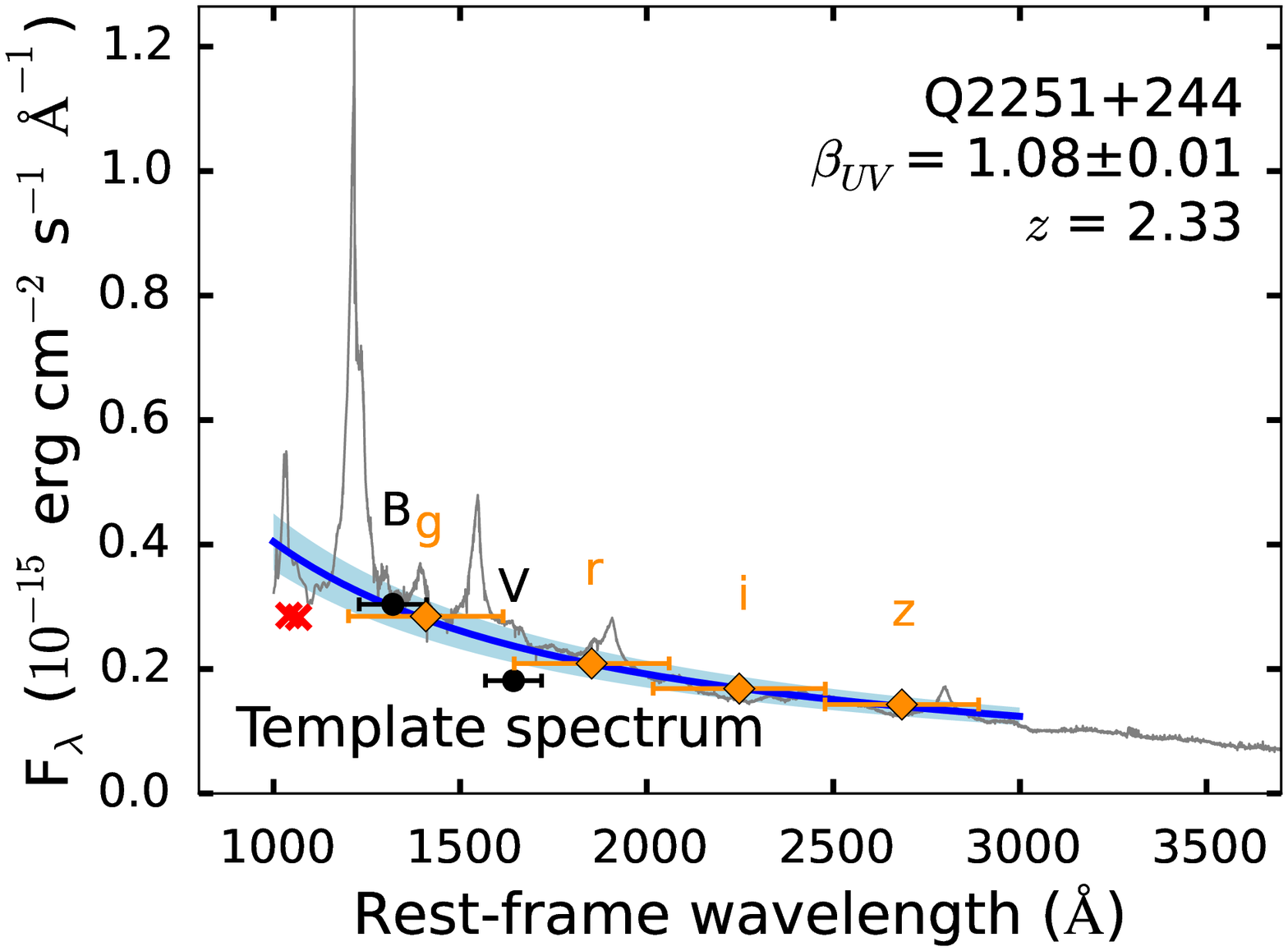}
	\includegraphics[width=\sedplotsize]{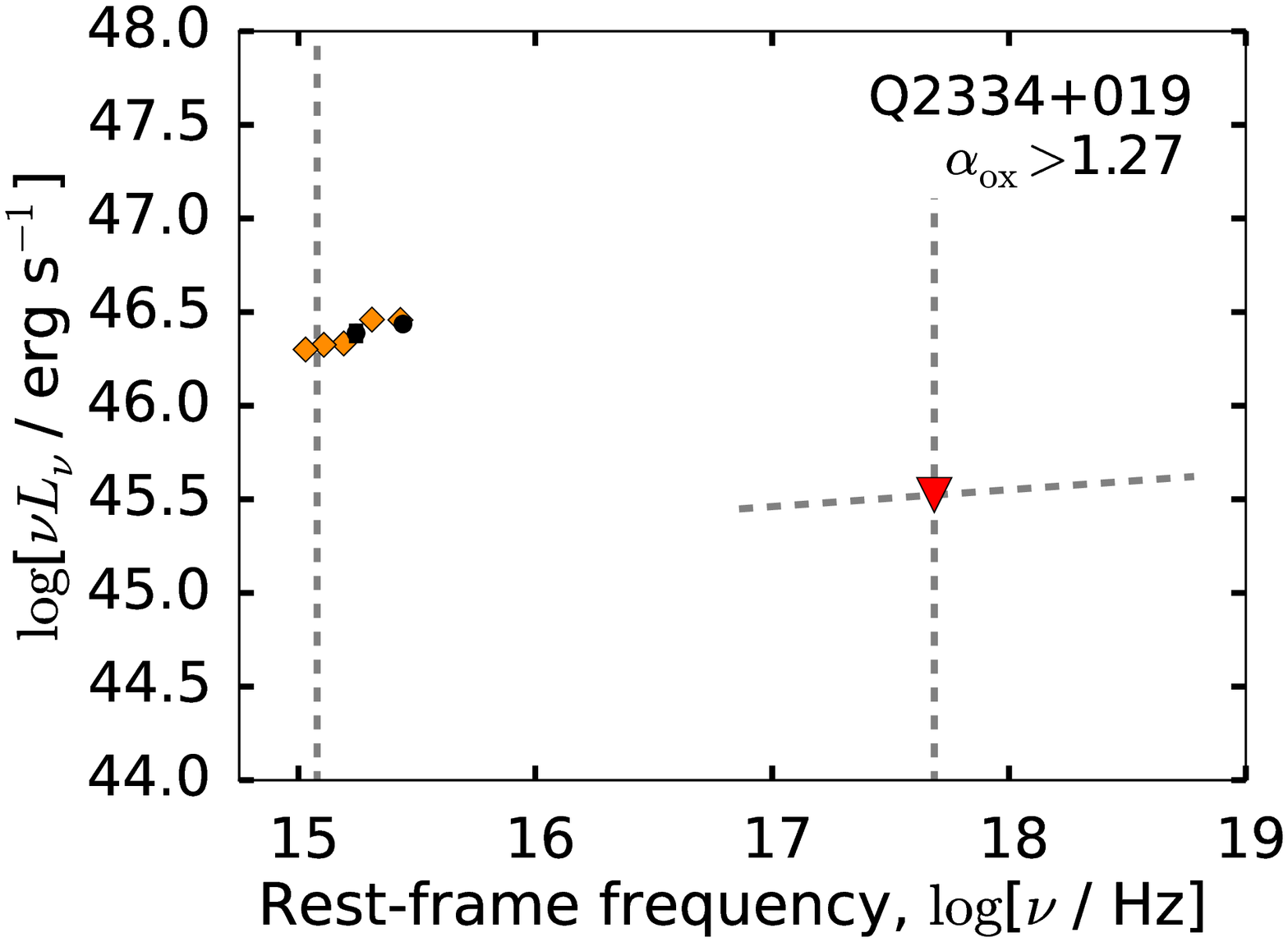}
	\includegraphics[width=\sedplotsize]{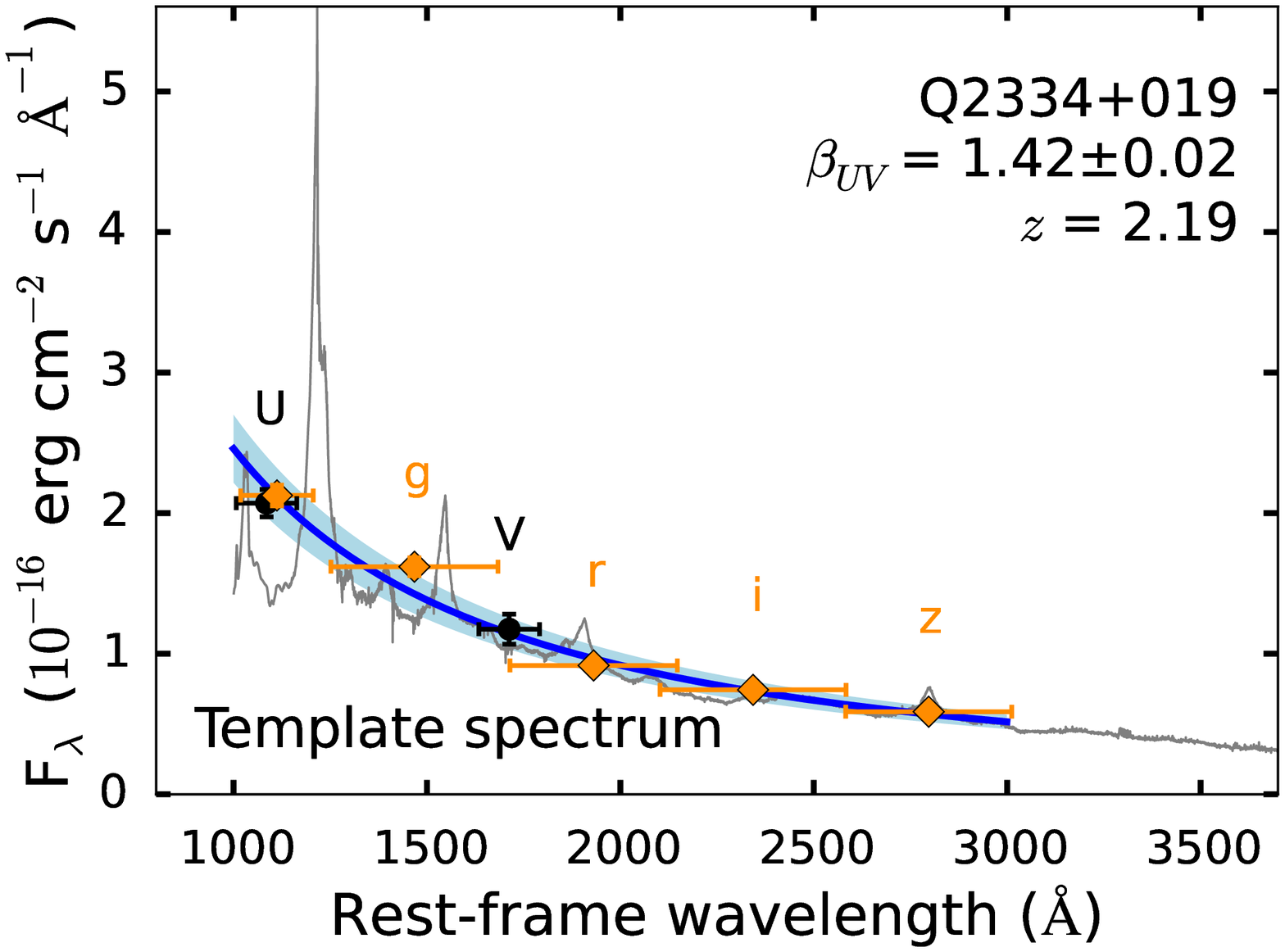}
	\includegraphics[width=\sedplotsize]{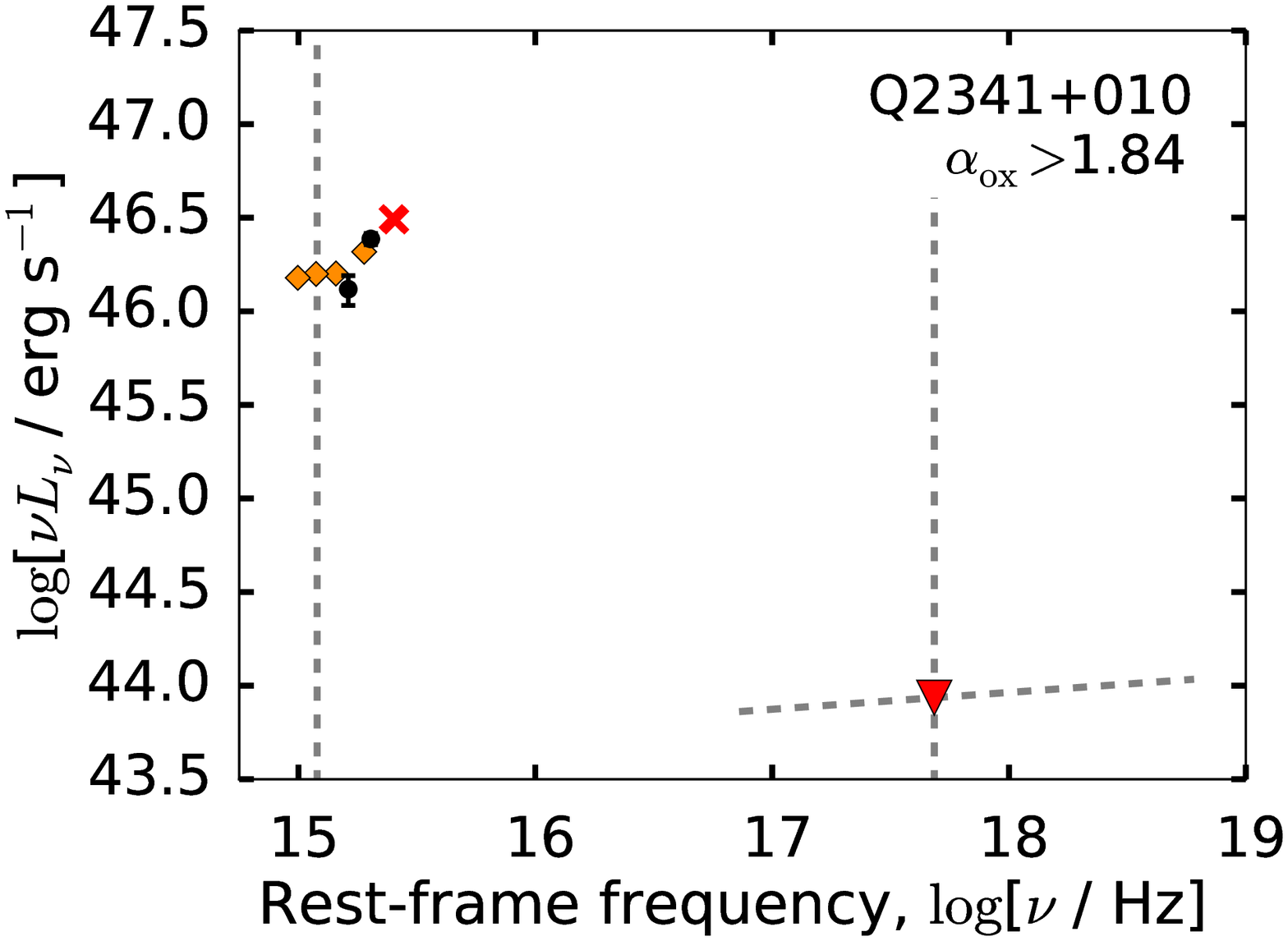}
	\includegraphics[width=\sedplotsize]{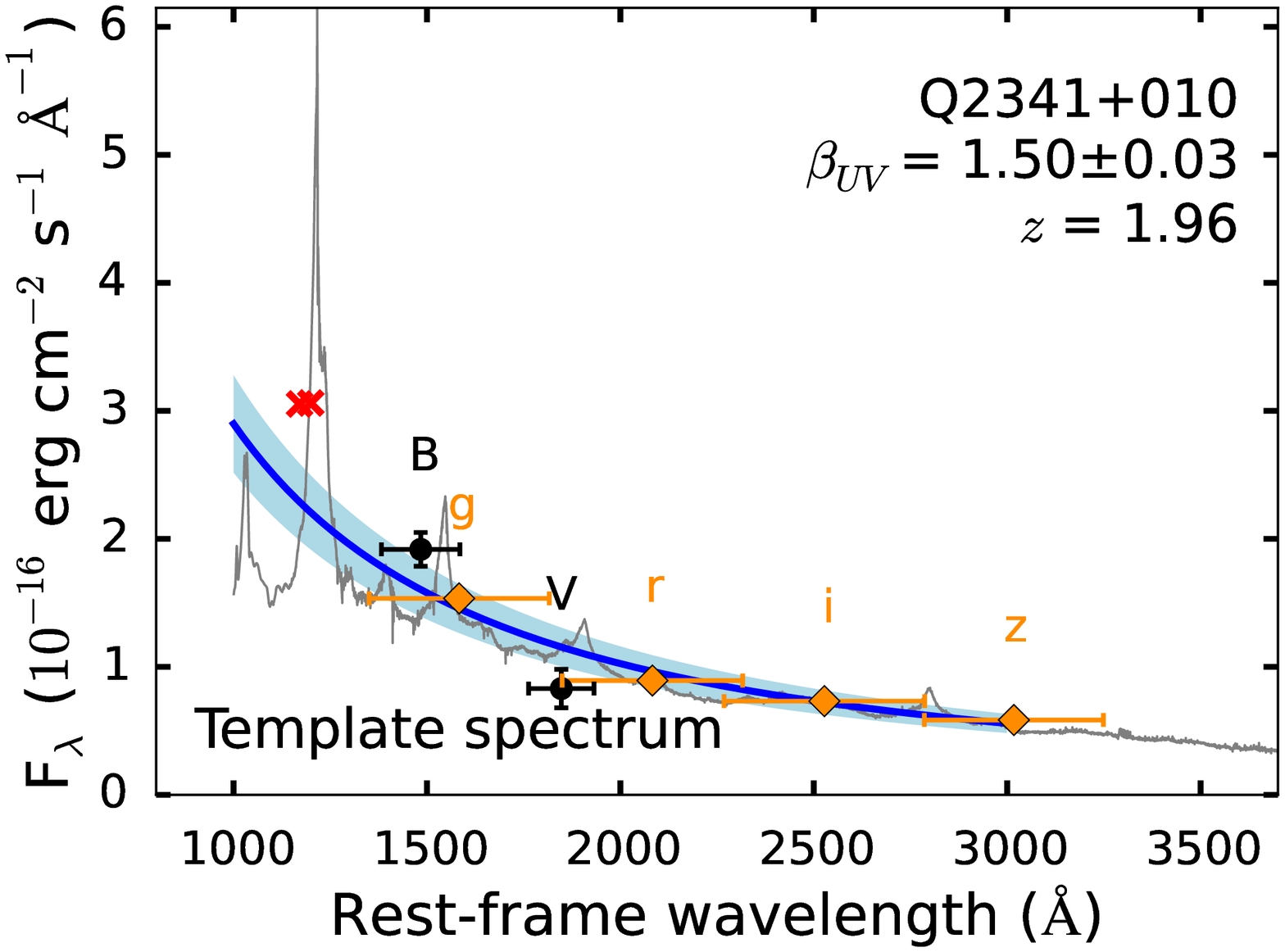}
	
	\caption{\emph{Left:} Rest-frame UV to X-ray spectral energy distributions (SEDs) of quasars in our sample. \emph{Right}: UV photometry and continuum modeling. See Figure \ref{fig:sed_mainpaper} for symbol and color coding.}
	\label{fig:sed14}
\end{figure*}

\begin{figure*}
	\centering
	\includegraphics[width=\sedplotsize]{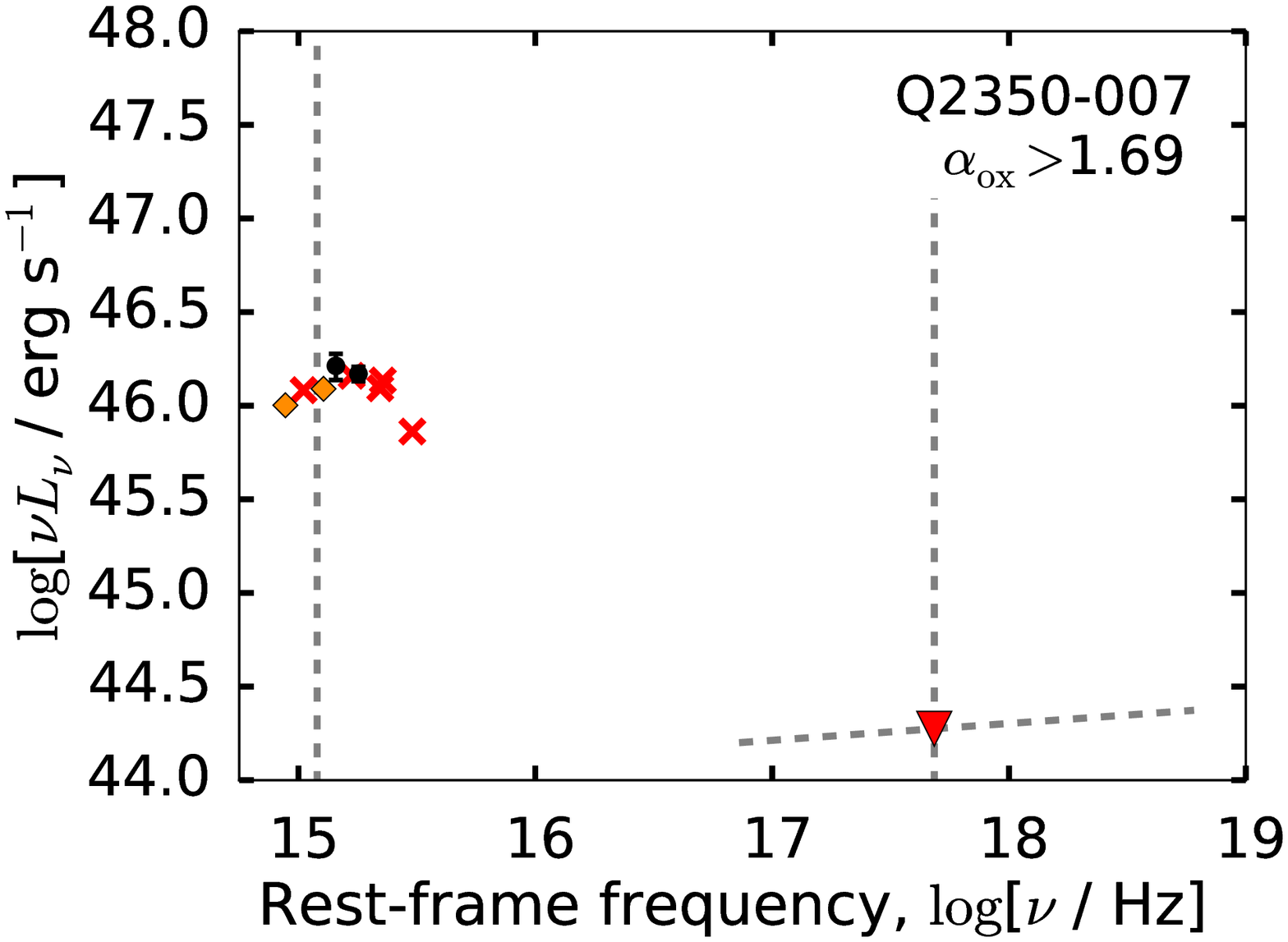}
	\includegraphics[width=\sedplotsize]{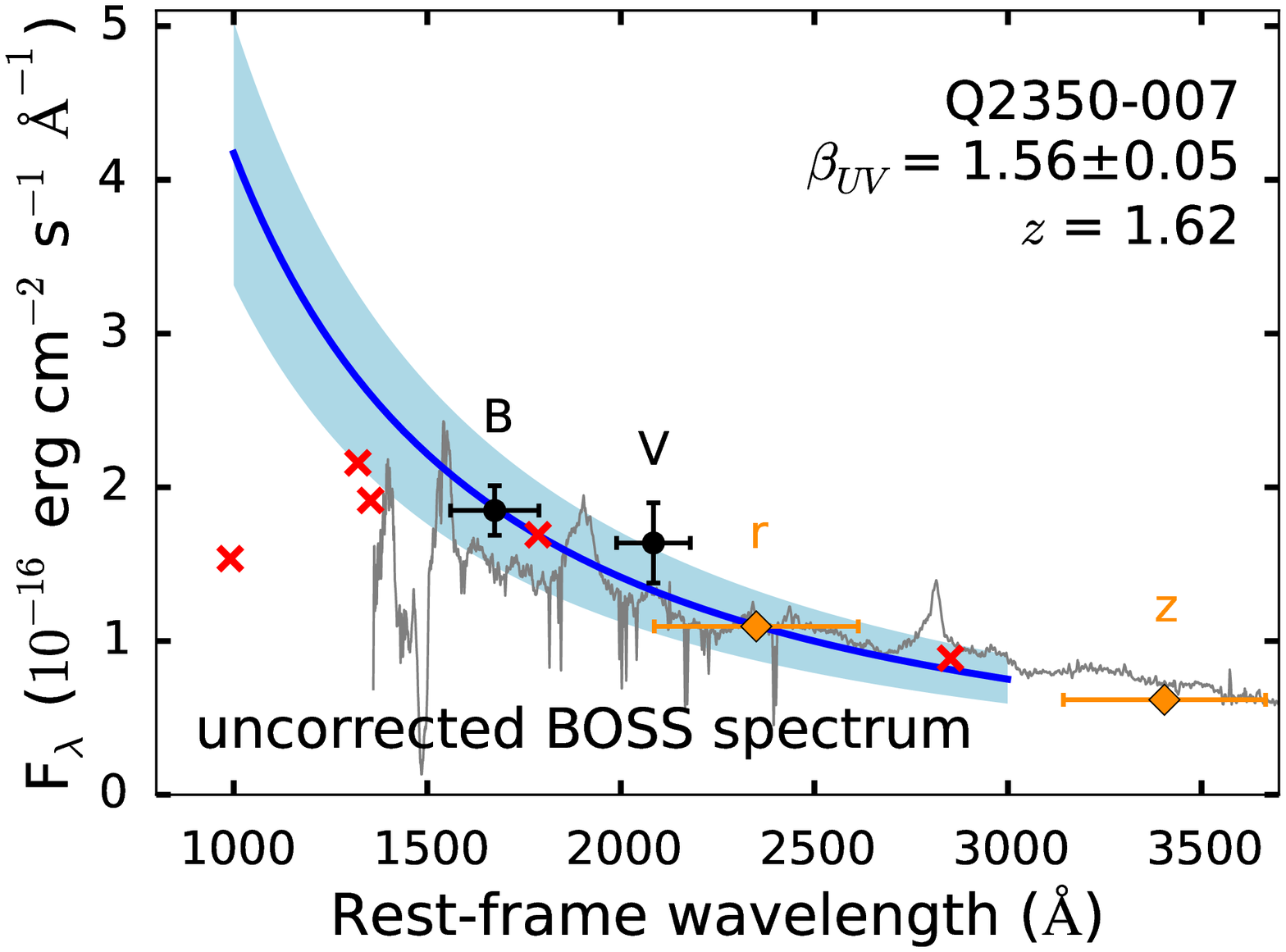}
	\includegraphics[width=\sedplotsize]{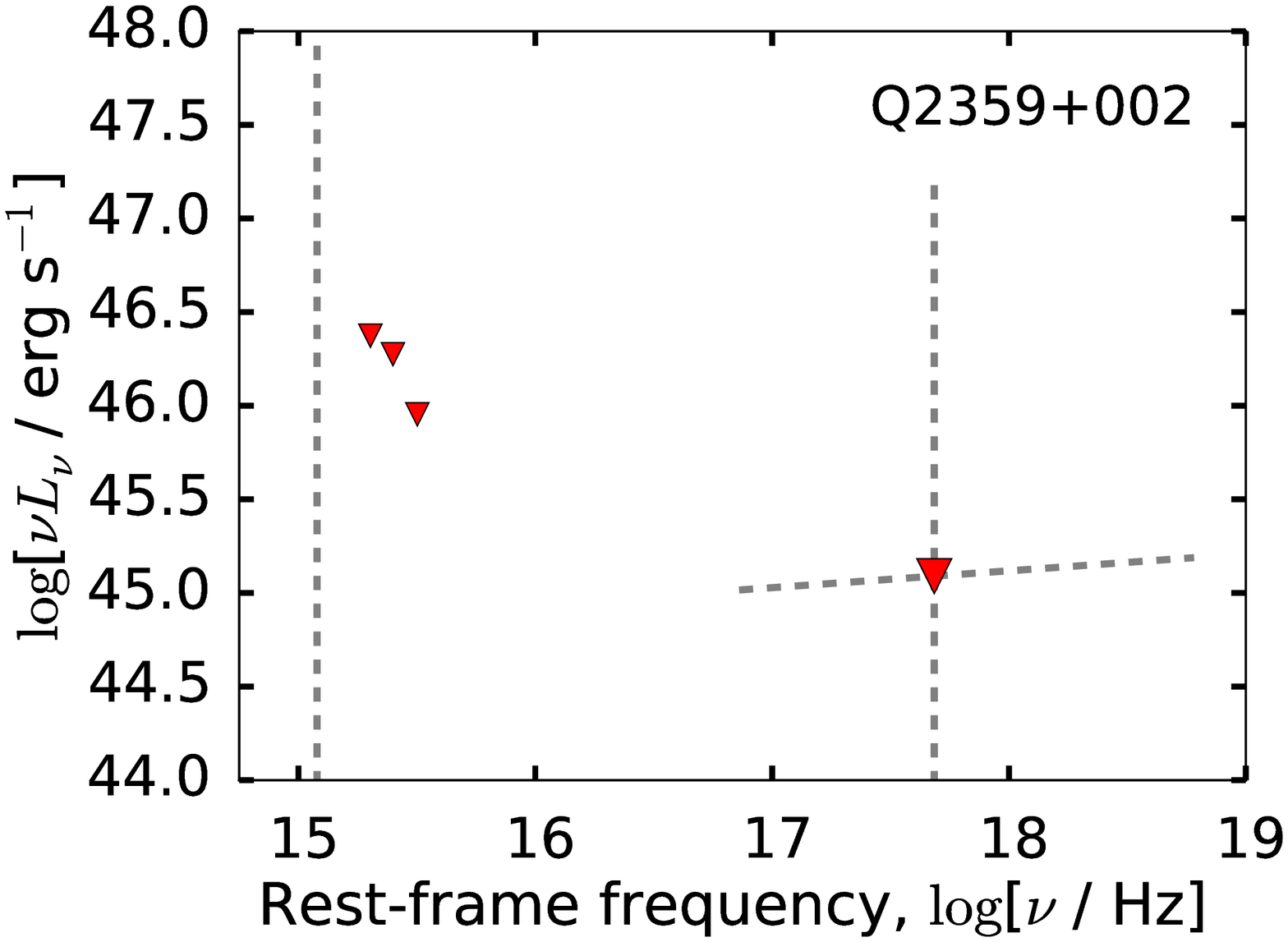}
	\includegraphics[width=\sedplotsize]{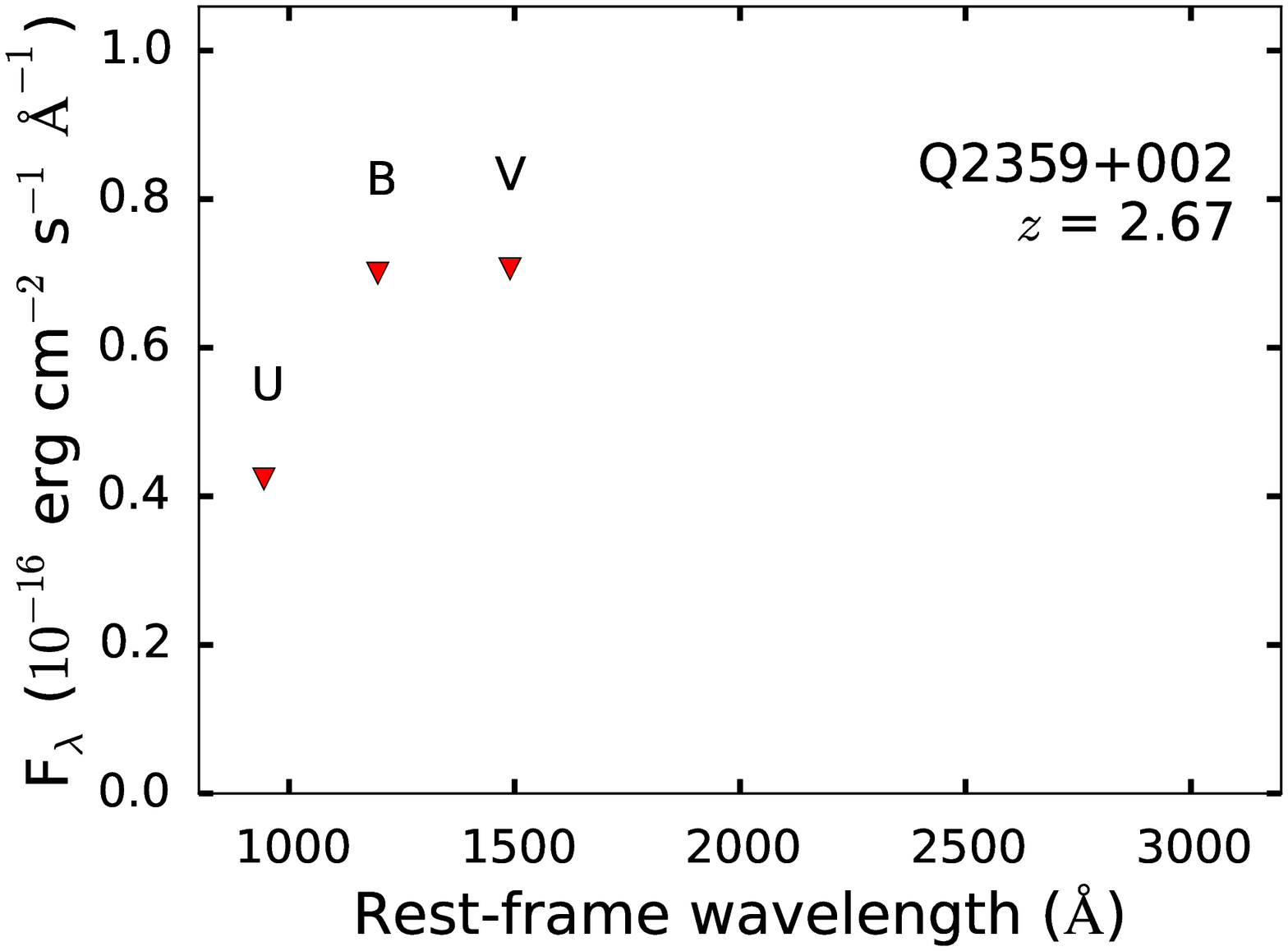}
	
	\caption{\emph{Left:} Rest-frame UV to X-ray spectral energy distributions (SEDs) of quasars in our sample. \emph{Right}: UV photometry and continuum modeling. See Figure \ref{fig:sed_mainpaper} for symbol and color coding.}
	\label{fig:sed15}
\end{figure*}